\documentclass[longauth,traditabstract]{aa}

\usepackage[nonamebreak]{natbib}
\usepackage[stable]{footmisc}
\usepackage{amsmath}
\usepackage{txfonts}
\usepackage{placeins}
\usepackage{natbib}
\bibpunct{(}{)}{;}{a}{}{,}
\usepackage{graphicx}
\usepackage{epstopdf}
\usepackage{ifthen}
\usepackage{mathtools}
\usepackage[breaklinks, colorlinks, citecolor=blue]{hyperref}
\usepackage{fixltx2e}
\usepackage{url}
\usepackage[table,usenames,dvipsnames]{xcolor}
\usepackage{comment} 
\usepackage{datetime} 
\usepackage{soul} 

\def\setsymbol#1#2{\expandafter\def\csname #1\endcsname{#2}}
\def\getsymbol#1{\csname #1\endcsname}

\def\Planck{\textit{Planck}}

\def\HeJT{$^4$He-JT}

\def\allearlypapers{\nocite{planck2011-1.1, planck2011-1.3, planck2011-1.4, planck2011-1.5, planck2011-1.6, planck2011-1.7, planck2011-1.10, planck2011-1.10sup, planck2011-5.1a, planck2011-5.1b, planck2011-5.2a, planck2011-5.2b, planck2011-5.2c, planck2011-6.1, planck2011-6.2, planck2011-6.3a, planck2011-6.4a, planck2011-6.4b, planck2011-6.6, planck2011-7.0, planck2011-7.2, planck2011-7.3, planck2011-7.7a, planck2011-7.7b, planck2011-7.12, planck2011-7.13}}

\def\alltwentythirteenresultspapers{\nocite{planck2013-p01, planck2013-p02, planck2013-p02a, planck2013-p02d, planck2013-p02b, planck2013-p03, planck2013-p03c, planck2013-p03f, planck2013-p03d, planck2013-p03e, planck2013-p01a, planck2013-p06, planck2013-p03a, planck2013-pip88, planck2013-p08, planck2013-p11, planck2013-p12, planck2013-p13, planck2013-p14, planck2013-p15, planck2013-p05b, planck2013-p17, planck2013-p09, planck2013-p09a, planck2013-p20, planck2013-p19, planck2013-pipaberration, planck2013-p05, planck2013-p05a, planck2013-pip56, planck2013-p06b, planck2013-p01a}}

\def\alltwentyfifteenresultspapers{\nocite{planck2014-a01, planck2014-a03, planck2014-a04, planck2014-a05, planck2014-a06, planck2014-a07, planck2014-a08, planck2014-a09, planck2014-a11, planck2014-a12, planck2014-a13, planck2014-a14, planck2014-a15, planck2014-a16, planck2014-a17, planck2014-a18, planck2014-a19, planck2014-a20, planck2014-a22, planck2014-a24, planck2014-a26, planck2014-a28, planck2014-a29, planck2014-a30, planck2014-a31, planck2014-a35, planck2014-a36, planck2014-a37, planck2014-ES}}

\newbox\tablebox    \newdimen\tablewidth
\def\leaderfil{\leaders\hbox to 5pt{\hss.\hss}\hfil}
\def\endPlancktable{\tablewidth=\columnwidth 
    $$\hss\copy\tablebox\hss$$
    \vskip-\lastskip\vskip -2pt}
\def\endPlancktablewide{\tablewidth=\textwidth 
    $$\hss\copy\tablebox\hss$$
    \vskip-\lastskip\vskip -2pt}
\def\tablenote#1 #2\par{\begingroup \parindent=0.8em
    \abovedisplayshortskip=0pt\belowdisplayshortskip=0pt
    \noindent
    $$\hss\vbox{\hsize\tablewidth \hangindent=\parindent \hangafter=1 \noindent
    \hbox to \parindent{$^#1$\hss}\strut#2\strut\par}\hss$$
    \endgroup}
\def\doubleline{\vskip 3pt\hrule \vskip 1.5pt \hrule \vskip 5pt}

\def\L2{\ifmmode L_2\else $L_2$\fi}

\def\DeltaT{\ifmmode \Delta T\else $\Delta T$\fi}
\def\deltat{\ifmmode \Delta t\else $\Delta t$\fi}
\def\fknee{\ifmmode f_{\rm knee}\else $f_{\rm knee}$\fi}
\def\Fmax{\ifmmode F_{\rm max}\else $F_{\rm max}$\fi}
\def\solar{\ifmmode{\rm M}_{\mathord\odot}\else${\rm M}_{\mathord\odot}$\fi}
\def\Msolar{\ifmmode{\rm M}_{\mathord\odot}\else${\rm M}_{\mathord\odot}$\fi}
\def\Lsolar{\ifmmode{\rm L}_{\mathord\odot}\else${\rm L}_{\mathord\odot}$\fi}
\def\inv{\ifmmode^{-1}\else$^{-1}$\fi}
\def\mo{\ifmmode^{-1}\else$^{-1}$\fi}
\def\sup#1{\ifmmode ^{\rm #1}\else $^{\rm #1}$\fi}
\def\expo#1{\ifmmode \times 10^{#1}\else $\times 10^{#1}$\fi}
\def\,{\thinspace}
\def\lsim{\mathrel{\raise .4ex\hbox{\rlap{$<$}\lower 1.2ex\hbox{$\sim$}}}}
\def\gsim{\mathrel{\raise .4ex\hbox{\rlap{$>$}\lower 1.2ex\hbox{$\sim$}}}}

\def\simprop{\mathrel{\raise .4ex\hbox{\rlap{$\propto$}\lower 1.2ex\hbox{$\sim$}}}}
\def\deg{\ifmmode^\circ\else$^\circ$\fi}
\def\pdeg{\ifmmode $\setbox0=\hbox{$^{\circ}$}\rlap{\hskip.11\wd0 .}$^{\circ}
          \else \setbox0=\hbox{$^{\circ}$}\rlap{\hskip.11\wd0 .}$^{\circ}$\fi}
\def\arcs{\ifmmode {^{\scriptstyle\prime\prime}}
          \else $^{\scriptstyle\prime\prime}$\fi}
\def\arcm{\ifmmode {^{\scriptstyle\prime}}
          \else $^{\scriptstyle\prime}$\fi}
\newdimen\sa  \newdimen\sb
\def\parcs{\sa=.07em \sb=.03em
     \ifmmode \hbox{\rlap{.}}^{\scriptstyle\prime\kern -\sb\prime}\hbox{\kern -\sa}
     \else \rlap{.}$^{\scriptstyle\prime\kern -\sb\prime}$\kern -\sa\fi}
\def\parcm{\sa=.08em \sb=.03em
     \ifmmode \hbox{\rlap{.}\kern\sa}^{\scriptstyle\prime}\hbox{\kern-\sb}
     \else \rlap{.}\kern\sa$^{\scriptstyle\prime}$\kern-\sb\fi}
\def\ra[#1 #2 #3.#4]{#1\sup{h}#2\sup{m}#3\sup{s}\llap.#4}
\def\dec[#1 #2 #3.#4]{#1\deg#2\arcm#3\arcs\llap.#4}
\def\deco[#1 #2 #3]{#1\deg#2\arcm#3\arcs}
\def\rra[#1 #2]{#1\sup{h}#2\sup{m}}

\def\dots{\relax\ifmmode \ldots\else $\ldots$\fi}
\def\WHzsr{\ifmmode $W\,Hz\mo\,sr\mo$\else W\,Hz\mo\,sr\mo\fi}
\def\mHz{\ifmmode $\,mHz$\else \,mHz\fi}
\def\GHz{\ifmmode $\,GHz$\else \,GHz\fi}
\def\mKs{\ifmmode $\,mK\,s$^{1/2}\else \,mK\,s$^{1/2}$\fi}
\def\muKs{\ifmmode \,\mu$K\,s$^{1/2}\else \,$\mu$K\,s$^{1/2}$\fi}
\def\muKRJs{\ifmmode \,\mu$K$_{\rm RJ}$\,s$^{1/2}\else \,$\mu$K$_{\rm RJ}$\,s$^{1/2}$\fi}
\def\muKHz{\ifmmode \,\mu$K\,Hz$^{-1/2}\else \,$\mu$K\,Hz$^{-1/2}$\fi}
\def\MJysr{\ifmmode \,$MJy\,sr\mo$\else \,MJy\,sr\mo\fi}
\def\MJysrmK{\ifmmode \,$MJy\,sr\mo$\,mK$_{\rm CMB}\mo\else \,MJy\,sr\mo\,mK$_{\rm CMB}\mo$\fi}
\def\microns{\ifmmode \,\mu$m$\else \,$\mu$m\fi}

\def\muK{\ifmmode \,\mu$K$\else \,$\mu$\hbox{K}\fi}
\def\microK{\ifmmode \,\mu$K$\else \,$\mu$\hbox{K}\fi}
\def\muW{\ifmmode \,\mu$W$\else \,$\mu$\hbox{W}\fi}
\def\kms{\ifmmode $\,km\,s$^{-1}\else \,km\,s$^{-1}$\fi}
\def\kmsMpc{\ifmmode $\,\kms\,Mpc\mo$\else \,\kms\,Mpc\mo\fi}
\providecommand{\sorthelp}[1]{}

\def\LCDM{$\Lambda$CDM}
\def\NHUNIT{\ifmmode {\rm \,cm^{-2}} \else $\rm \,cm^{-2}$ \fi} 

\def\aap{{A\&A}}

\def\apjs{{ApJSupp.}}

\newcommand{\Nside}{\ensuremath{N_{\mathrm{side}}}}

\newif\iflowres 
\lowresfalse

\newcommand{\mksym}[1]{\ifmmode {\rm #1}\else #1\fi}

\newcommand{\planckTTonly}{PlanckTT}

\newcommand{\lowE}{\mksym{lowE}}

\newcommand{\As}{A_{\rm s}}

\newcommand{\ns}{n_{\rm s}}

\newcommand{\lcdm}{{$\rm{\Lambda CDM}$}}

\newcommand{\Alens}{A_{\rm L}}

\newcommand{\lmax}{l_{\text{max}}}

\begin{document}

\title{\vglue -10mm\Planck\ intermediate results. XLVI. 
Reduction of large-scale systematic effects in HFI
polarization maps and estimation of the reionization optical depth}

\author{\small
Planck Collaboration: N.~Aghanim\inst{53}
\and
M.~Ashdown\inst{63, 6}
\and
J.~Aumont\inst{53}
\and
C.~Baccigalupi\inst{75}
\and
M.~Ballardini\inst{29, 45, 48}
\and
A.~J.~Banday\inst{85, 9}
\and
R.~B.~Barreiro\inst{58}
\and
N.~Bartolo\inst{28, 59}
\and
S.~Basak\inst{75}
\and
R.~Battye\inst{61}
% ...
}
\institute{\small
APC, AstroParticule et Cosmologie, Universit\'{e} Paris Diderot, CNRS/IN2P3, CEA/lrfu, Observatoire de Paris, Sorbonne Paris Cit\'{e}, 10, rue Alice Domon et L\'{e}onie Duquet, 75205 Paris Cedex 13, France\goodbreak
\and
African Institute for Mathematical Sciences, 6-8 Melrose Road, Muizenberg, Cape Town, South Africa\goodbreak
\and
Agenzia Spaziale Italiana Science Data Center, Via del Politecnico snc, 00133, Roma, Italy\goodbreak
\and
Aix Marseille Universit\'{e}, CNRS, LAM (Laboratoire d'Astrophysique de Marseille) UMR 7326, 13388, Marseille, France\goodbreak
\and
Aix Marseille Universit\'{e}, Centre de Physique Th\'{e}orique, 163 Avenue de Luminy, 13288, Marseille, France\goodbreak
\and
Astrophysics Group, Cavendish Laboratory, University of Cambridge, J J Thomson Avenue, Cambridge CB3 0HE, U.K.\goodbreak
\and
Astrophysics \& Cosmology Research Unit, School of Mathematics, Statistics \& Computer Science, University of KwaZulu-Natal, Westville Campus, Private Bag X54001, Durban 4000, South Africa\goodbreak
\and
CITA, University of Toronto, 60 St. George St., Toronto, ON M5S 3H8, Canada\goodbreak
\and
CNRS, IRAP, 9 Av. colonel Roche, BP 44346, F-31028 Toulouse cedex 4, France\goodbreak
\and
California Institute of Technology, Pasadena, California, U.S.A.\goodbreak
\and
Centre for Theoretical Cosmology, DAMTP, University of Cambridge, Wilberforce Road, Cambridge CB3 0WA, U.K.\goodbreak
\and
Computational Cosmology Center, Lawrence Berkeley National Laboratory, Berkeley, California, U.S.A.\goodbreak
\and
DTU Space, National Space Institute, Technical University of Denmark, Elektrovej 327, DK-2800 Kgs. Lyngby, Denmark\goodbreak
\and
D\'{e}partement de Physique Th\'{e}orique, Universit\'{e} de Gen\`{e}ve, 24, Quai E. Ansermet,1211 Gen\`{e}ve 4, Switzerland\goodbreak
\and
Departamento de Astrof\'{i}sica, Universidad de La Laguna (ULL), E-38206 La Laguna, Tenerife, Spain\goodbreak
\and
Departamento de F\'{\i}sica, Universidad de Oviedo, Avda. Calvo Sotelo s/n, Oviedo, Spain\goodbreak
\and
Department of Astrophysics/IMAPP, Radboud University Nijmegen, P.O. Box 9010, 6500 GL Nijmegen, The Netherlands\goodbreak
\and
Department of Physics \& Astronomy, University of British Columbia, 6224 Agricultural Road, Vancouver, British Columbia, Canada\goodbreak
\and
Department of Physics and Astronomy, Dana and David Dornsife College of Letter, Arts and Sciences, University of Southern California, Los Angeles, CA 90089, U.S.A.\goodbreak
\and
Department of Physics and Astronomy, University College London, London WC1E 6BT, U.K.\goodbreak
\and
Department of Physics and Astronomy, University of Sussex, Brighton BN1 9QH, U.K.\goodbreak
\and
Department of Physics, Gustaf H\"{a}llstr\"{o}min katu 2a, University of Helsinki, Helsinki, Finland\goodbreak
\and
Department of Physics, Princeton University, Princeton, New Jersey, U.S.A.\goodbreak
\and
Department of Physics, University of California, Berkeley, California, U.S.A.\goodbreak
\and
Department of Physics, University of California, One Shields Avenue, Davis, California, U.S.A.\goodbreak
\and
Department of Physics, University of California, Santa Barbara, California, U.S.A.\goodbreak
\and
Department of Physics, University of Illinois at Urbana-Champaign, 1110 West Green Street, Urbana, Illinois, U.S.A.\goodbreak
\and
Dipartimento di Fisica e Astronomia G. Galilei, Universit\`{a} degli Studi di Padova, via Marzolo 8, 35131 Padova, Italy\goodbreak
\and
Dipartimento di Fisica e Astronomia, Alma Mater Studiorum, Universit\`{a} degli Studi di Bologna, Viale Berti Pichat 6/2, I-40127, Bologna, Italy\goodbreak
\and
Dipartimento di Fisica e Scienze della Terra, Universit\`{a} di Ferrara, Via Saragat 1, 44122 Ferrara, Italy\goodbreak
\and
Dipartimento di Fisica, Universit\`{a} La Sapienza, P. le A. Moro 2, Roma, Italy\goodbreak
\and
Dipartimento di Fisica, Universit\`{a} degli Studi di Milano, Via Celoria, 16, Milano, Italy\goodbreak
\and
Dipartimento di Fisica, Universit\`{a} di Roma Tor Vergata, Via della Ricerca Scientifica, 1, Roma, Italy\goodbreak
\and
Dipartimento di Matematica, Universit\`{a} di Roma Tor Vergata, Via della Ricerca Scientifica, 1, Roma, Italy\goodbreak
\and
Discovery Center, Niels Bohr Institute, Copenhagen University, Blegdamsvej 17, Copenhagen, Denmark\goodbreak
\and
European Space Agency, ESAC, Planck Science Office, Camino bajo del Castillo, s/n, Urbanizaci\'{o}n Villafranca del Castillo, Villanueva de la Ca\~{n}ada, Madrid, Spain\goodbreak
\and
European Space Agency, ESTEC, Keplerlaan 1, 2201 AZ Noordwijk, The Netherlands\goodbreak
\and
Gran Sasso Science Institute, INFN, viale F. Crispi 7, 67100 L'Aquila, Italy\goodbreak
\and
HGSFP and University of Heidelberg, Theoretical Physics Department, Philosophenweg 16, 69120, Heidelberg, Germany\goodbreak
\and
Haverford College Astronomy Department, 370 Lancaster Avenue, Haverford, Pennsylvania, U.S.A.\goodbreak
\and
Helsinki Institute of Physics, Gustaf H\"{a}llstr\"{o}min katu 2, University of Helsinki, Helsinki, Finland\goodbreak
\and
INAF - Osservatorio Astronomico di Padova, Vicolo dell'Osservatorio 5, Padova, Italy\goodbreak
\and
INAF - Osservatorio Astronomico di Roma, via di Frascati 33, Monte Porzio Catone, Italy\goodbreak
\and
INAF - Osservatorio Astronomico di Trieste, Via G.B. Tiepolo 11, Trieste, Italy\goodbreak
\and
INAF/IASF Bologna, Via Gobetti 101, Bologna, Italy\goodbreak
\and
INAF/IASF Milano, Via E. Bassini 15, Milano, Italy\goodbreak
\and
INFN - CNAF, viale Berti Pichat 6/2, 40127 Bologna, Italy\goodbreak
\and
INFN, Sezione di Bologna, viale Berti Pichat 6/2, 40127 Bologna, Italy\goodbreak
\and
INFN, Sezione di Ferrara, Via Saragat 1, 44122 Ferrara, Italy\goodbreak
\and
INFN, Sezione di Roma 1, Universit\`{a} di Roma Sapienza, Piazzale Aldo Moro 2, 00185, Roma, Italy\goodbreak
\and
INFN, Sezione di Roma 2, Universit\`{a} di Roma Tor Vergata, Via della Ricerca Scientifica, 1, Roma, Italy\goodbreak
\and
Imperial College London, Astrophysics group, Blackett Laboratory, Prince Consort Road, London, SW7 2AZ, U.K.\goodbreak
\and
Institut d'Astrophysique Spatiale, CNRS, Univ. Paris-Sud, Universit\'{e} Paris-Saclay, B\^{a}t. 121, 91405 Orsay cedex, France\goodbreak
\and
Institut d'Astrophysique de Paris, CNRS (UMR7095), 98 bis Boulevard Arago, F-75014, Paris, France\goodbreak
\and
Institute of Astronomy, University of Cambridge, Madingley Road, Cambridge CB3 0HA, U.K.\goodbreak
\and
Institute of Theoretical Astrophysics, University of Oslo, Blindern, Oslo, Norway\goodbreak
\and
Instituto de Astrof\'{\i}sica de Canarias, C/V\'{\i}a L\'{a}ctea s/n, La Laguna, Tenerife, Spain\goodbreak
\and
Instituto de F\'{\i}sica de Cantabria (CSIC-Universidad de Cantabria), Avda. de los Castros s/n, Santander, Spain\goodbreak
\and
Istituto Nazionale di Fisica Nucleare, Sezione di Padova, via Marzolo 8, I-35131 Padova, Italy\goodbreak
\and
Jet Propulsion Laboratory, California Institute of Technology, 4800 Oak Grove Drive, Pasadena, California, U.S.A.\goodbreak
\and
Jodrell Bank Centre for Astrophysics, Alan Turing Building, School of Physics and Astronomy, The University of Manchester, Oxford Road, Manchester, M13 9PL, U.K.\goodbreak
\and
Kavli Institute for Cosmological Physics, University of Chicago, Chicago, IL 60637, USA\goodbreak
\and
Kavli Institute for Cosmology Cambridge, Madingley Road, Cambridge, CB3 0HA, U.K.\goodbreak
\and
LAL, Universit\'{e} Paris-Sud, CNRS/IN2P3, Orsay, France\goodbreak
\and
LERMA, CNRS, Observatoire de Paris, 61 Avenue de l'Observatoire, Paris, France\goodbreak
\and
Laboratoire Traitement et Communication de l'Information, CNRS (UMR 5141) and T\'{e}l\'{e}com ParisTech, 46 rue Barrault F-75634 Paris Cedex 13, France\goodbreak
\and
Laboratoire de Physique Subatomique et Cosmologie, Universit\'{e} Grenoble-Alpes, CNRS/IN2P3, 53, rue des Martyrs, 38026 Grenoble Cedex, France\goodbreak
\and
Laboratoire de Physique Th\'{e}orique, Universit\'{e} Paris-Sud 11 \& CNRS, B\^{a}timent 210, 91405 Orsay, France\goodbreak
\and
Lawrence Berkeley National Laboratory, Berkeley, California, U.S.A.\goodbreak
\and
Max-Planck-Institut f\"{u}r Astrophysik, Karl-Schwarzschild-Str. 1, 85741 Garching, Germany\goodbreak
\and
Mullard Space Science Laboratory, University College London, Surrey RH5 6NT, U.K.\goodbreak
\and
Nicolaus Copernicus Astronomical Center, Bartycka 18, 00-716 Warsaw, Poland\goodbreak
\and
Niels Bohr Institute, Copenhagen University, Blegdamsvej 17, Copenhagen, Denmark\goodbreak
\and
Nordita (Nordic Institute for Theoretical Physics), Roslagstullsbacken 23, SE-106 91 Stockholm, Sweden\goodbreak
\and
SISSA, Astrophysics Sector, via Bonomea 265, 34136, Trieste, Italy\goodbreak
\and
School of Chemistry and Physics, University of KwaZulu-Natal, Westville Campus, Private Bag X54001, Durban, 4000, South Africa\goodbreak
\and
School of Physics and Astronomy, University of Nottingham, Nottingham NG7 2RD, U.K.\goodbreak
\and
Simon Fraser University, Department of Physics, 8888 University Drive, Burnaby BC, Canada\goodbreak
\and
Sorbonne Universit\'{e}-UPMC, UMR7095, Institut d'Astrophysique de Paris, 98 bis Boulevard Arago, F-75014, Paris, France\goodbreak
\and
Space Research Institute (IKI), Russian Academy of Sciences, Profsoyuznaya Str, 84/32, Moscow, 117997, Russia\goodbreak
\and
Space Sciences Laboratory, University of California, Berkeley, California, U.S.A.\goodbreak
\and
Sub-Department of Astrophysics, University of Oxford, Keble Road, Oxford OX1 3RH, U.K.\goodbreak
\and
The Oskar Klein Centre for Cosmoparticle Physics, Department of Physics,Stockholm University, AlbaNova, SE-106 91 Stockholm, Sweden\goodbreak
\and
UPMC Univ Paris 06, UMR7095, 98 bis Boulevard Arago, F-75014, Paris, France\goodbreak
\and
Universit\'{e} de Toulouse, UPS-OMP, IRAP, F-31028 Toulouse cedex 4, France\goodbreak
% ...
}

\date{\vglue -1.5mm \today \vglue -5mm}
\abstract{\vglue -3mm 
This paper describes the identification, modelling, and removal of previously unexplained systematic effects in the polarization data of the \Planck\ High Frequency Instrument (HFI) on large angular scales, including new mapmaking and calibration procedures, new and more complete end-to-end simulations, and a set of robust internal consistency checks on the resulting maps. These maps, at 100, 143, 217, and 353\,GHz, are early versions of those that will be released in final form later in 2016.
The improvements allow us to determine the cosmic reionization optical depth $\tau$ using, for the first time, the low-multipole $EE$ data from HFI, reducing significantly the central value and uncertainty, and hence the upper limit. Two different likelihood procedures are used to constrain $\tau$ from two estimators of the CMB $E$- and $B$-mode angular power spectra at 100 and 143\,GHz, after debiasing the spectra from a small remaining systematic contamination. These all give fully consistent results.
A further consistency test is performed using cross-correlations derived from the Low Frequency Instrument maps of the \Planck\ 2015 data release and the new HFI data. For this purpose, end-to-end analyses of systematic effects from the two instruments are used to demonstrate the near independence of their dominant systematic error residuals. 
The tightest result comes from the HFI-based $\tau$ posterior distribution using the maximum likelihood power spectrum estimator from $EE$ data only, giving a value $0.055\pm 0.009$. In a companion paper these results are discussed in the context of the best-fit \Planck\ $\Lambda$CDM cosmological model and recent models of reionization.
} 
 

\authorrunning{Planck Collaboration}
\titlerunning{Large-scale polarization and reionization}
\maketitle

\allearlypapers

\alltwentythirteenresultspapers

\alltwentyfifteenresultspapers

\section{Introduction}

The $E$-mode polarization signal of the cosmic microwave background (CMB) at multipoles less than 15 is sensitive to the value of the Thomson scattering optical depth $\tau$. In polarization at large angular scales, the extra signal generated by reionization dominates over the signal from recombination. CMB polarization measurements thus provide an important constraint on models of early galaxy evolution and star formation, providing the integrated optical depth of the entire history of reionization, which is complementary information to the lower limit on the redshift of full reionization provided by Lyman-$\alpha$ absorption in the spectra of high redshift objects \citep[see][for recent results and reviews]{Dijkstra2014,2016MNRAS.455.2101M,2015ApJ...802L..19R,bouwens2015,zitrin2015}. The reionization parameter $\tau$ is difficult to constrain with CMB temperature measurements alone, as the $TT$ power spectrum depends on the combination $A_{\rm s}e^{-2\tau}$, and $\tau$ is degenerate with $A_{\rm s}$, the amplitude of the initial cosmological scalar perturbations.

The $B$-mode polarization signal at low multipoles is created by the scattering of primordial tensor anisotropies in the CMB by reionized matter. This signal scales roughly as $\tau^2$. The value of $\tau$ is thus also important for experiments constraining the tensor-to-scalar ratio, $r$, using the reionization peak of the $B$ modes.

Due to the large angular scale of the signals, $\tau$ has been measured only in full-sky measurements made from space. \citet{hinshaw2012} report $\tau=0.089\pm0.014$ from the WMAP9 analysis. \citet{planck2014-a13} used \Planck\footnote{\Planck\ (\url{http://www.esa.int/Planck}) is a project of the European Space Agency (ESA), with instruments provided by two scientific consortia funded by ESA member states and led by Principal Investigators from France and Italy, telescope reflectors provided through a collaboration between ESA and a scientific consortium led and funded by Denmark, and additional contributions from NASA (USA).}%
polarized 353-GHz data to clean the WMAP Ka, Q, and V maps of polarized dust emission, and WMAP K-band as a template to remove polarized synchrotron emission, lowering $\tau$ from WMAP data alone by about $1\,\sigma$ to $\tau=0.075 \pm 0.013$. \citet{planck2014-a15} used 70-GHz polarization data at low multipoles, which were assumed to be nearly noise-limited (see Sect.~\ref{sec:LFI} below for new estimates of systematic effects) and were cleaned using 353\,GHz and 30\,GHz to remove dust and synchrotron emission. This low-multipole likelihood alone gave $\tau=0.067\pm0.022$.  In combination with the \Planck\ high-multipole $TT$ likelihood, it gave $\tau=0.078\pm0.019$.  The combination PlanckTT+lensing gave a comparable value of $\tau=0.070\pm0.024$.

The HFI 100 and 143-GHz low-multipole polarization data alone could provide tighter constraints on $\tau$, reducing uncertainties by nearly a factor of about two with respect to the results quoted above; however, systematic errors remaining in the HFI polarized data at large angular scales in the 2015 \Planck\ release led the \Planck\ collaboration to delay their use.  Since the time when the data for the 2015 release were frozen, we have made substantial improvements in the characterization and removal of HFI systematic errors, which we describe in this paper. The maps, simulations, and sometimes computer codes used here are referred to as ``pre-2016,'' and are nearly identical to those in the 2016 data release to come.

The paper is structured in the following way.
Section~\ref{sec:TOIsqualitytests} analyses systematic errors affecting the HFI time-ordered information processing, describes tests of these errors, and describes new versions of polarized mapmaking and calibration.
Section~\ref{sec:hfimaps} discusses global tests of HFI polarization at the power spectrum level.
Section~\ref{sec:LFI} discusses quality tests of Low Frequency Instrument (LFI) polarization data. 
Section~\ref{sec:Tests} describes component separation and cross-spectra and their suitability for the measurement of $\tau$. 
Section~\ref{sec:analysis} describes the likelihood analysis of the $\tau$ parameter.  Finally
Section~\ref{sec:cosmo} discusses the implications of our new $\tau$ value.

These data described here are also used in an accompanying paper \citep{planck2014-a25} to constrain different models for the reionization history of the Universe.

\section{Improvements of HFI mapmaking and calibration for low multipoles}
\label{sec:TOIsqualitytests}

\subsection{Overview}
\label{sec:mapoverview}

Analysis of HFI polarization systematic effects is complex and technical in nature. This paper, together with \cite{planck2014-a08}, completes the characterization of the HFI time-ordered information (TOI) products, extending analysis of the polarization power spectra to the lowest multipoles.

The HFI TOI used in this paper are described in the 2015 data release \citep{planck2014-a08}, and a calibrated version is publicly available in the Planck Legacy Archive.\footnote{\url{http://archives.esac.esa.int/pla}} The spacecraft pointing solution \citep{planck2014-a01} and all instrumental parameters, including the focal plane geometry, cross-polar leakage parameters, and polarization angle parameters, are identical to those used in the 2015 release.

HFI systematic errors have been discussed extensively in earlier papers \citep{planck2013-p03e, planck2014-a08} and were shown to be under control on small angular scales.  Large angular scales, however, were still affected by not-fully-understood systematic effects. 

The reduction of systematic errors in HFI polarization measurements at large angular scales reported in this paper is the result of applying a new mapmaking code, {\tt SRoll}.   {\tt SRoll} is a polarized destriping algorithm, similar to the 2015 HFI mapmaking pipeline \citep{planck2014-a09}, but it solves simultaneously for absolute calibration, leakage of temperature into polarization, and various other sources of systematic error. Systematic residuals are removed via template fitting, using {\tt HEALPix}-binned rings (HPR) to compress the time-ordered information. Each HPR consists of data from a given stable pointing period (ring) binned into {\tt HEALPix} pixels \citep{gorski2005}.  A version of this code with minor additions will be used for the final \Planck\ data release in 2016. In its present form, {\tt SRoll} reduces the systematic errors in the HFI $EE$ polarization spectra at multipoles $\ell \ge 2$ to levels close to the nominal instrumental noise. This allows a measurement of the reionization optical depth $\tau$ from the HFI alone, which was not possible at the time of the 2015 \Planck\ data release.

The rest of this section discusses each of the specific improvements.
Section~\ref{sec:intro2e2e} introduces the end-to-end simulations used in this paper.
Section~\ref{sec:detnoise} analyses the detector system noise, which sets a fundamental limit to the measurement of the polarization spectra and to $\tau$ with HFI data. 
(Appendix~\ref{TOIproc} summarizes a number of effects removed in the TOI processing that make a negligible contribution to the systematic error budget of polarization measurements at large angular scales.)
Section~\ref{sec:skycomponents} describes removal of the zodiacal light, and far sidelobes. 
Section~\ref{sec:adc} characterizes the systematic errors induced by nonlinearities in the analogue-to-digital converters (ADCs) used in the bolometer readout electronics. These nonlinearities, if uncorrected, would be the dominant source of systematic errors in polarization at large angular scales.
Section~\ref{sec:error_map} discusses all other systematic errors not corrected after the TOI processing, and mitigated by {\tt SRoll}.

\subsection{End-to-end simulations}
\label{sec:intro2e2e}

End-to-end (E2E) HFI simulations, introduced for the first time in 2015 \citep{planck2014-a13},  include an accurate representation of the response of the instrument to sky signals, as well as known instrumental systematic errors at the TOI or ring level, all of which are propagated through the entire analysis pipeline to determine their effect on the final maps. The version used in this paper runs the pre-2016 {\tt Sroll} mapmaking code, described in detail in Sect.~\ref{sec:E2E}.  Simulations built with this pre-2016 pipeline are referred to as HFI focal plane simulations (HFPSs).  Three sets of HFPSs are used in this paper: HFPS1 contains 83 realizations; HFPS2 and HFPS3 contain 100 realizations each, which can be used either separately or together.

\subsection{Detector noise}
\label{sec:detnoise}

Detector noise is determined through a multi-step process, starting with first-order correction of the TOI for ADC nonlinearity, demodulation, deglitching, 4-K line removal, and time-response deconvolution. The sky signal is then estimated (using the redundancy of multiple scans of the same sky during each stable pointing period) and removed, leaving the noise, which can then be characterized \citep[see][for details]{planck2014-a08}.

\begin{figure}[htbp!] 
\includegraphics[width=\columnwidth]{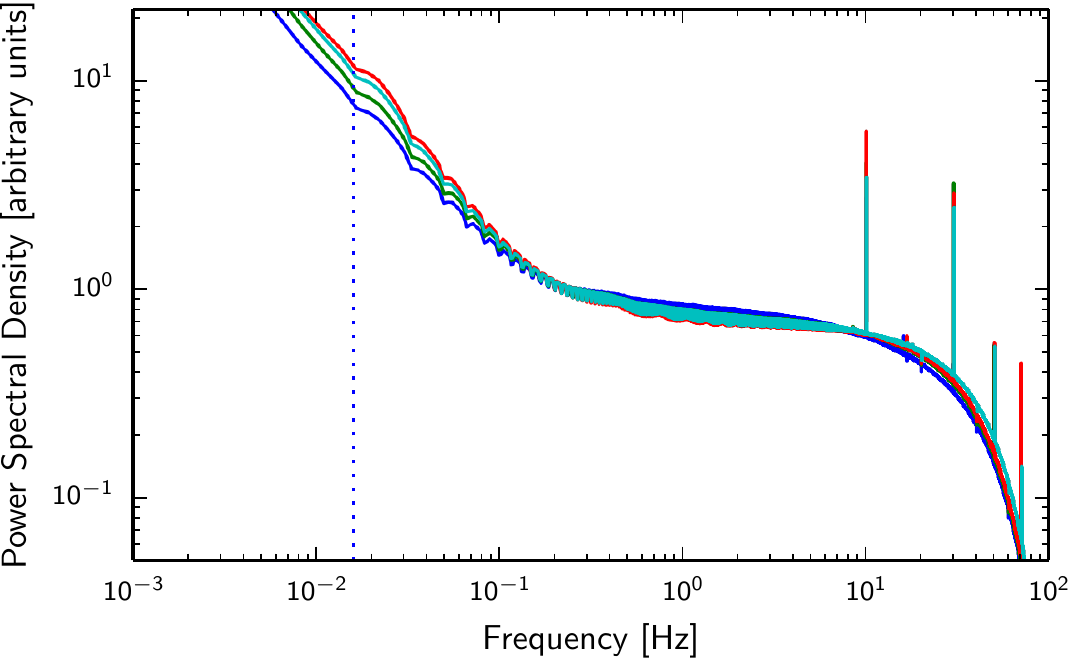} 
\caption{Mean power spectra of the signal-subtracted, time-ordered data from Survey~2 for each polarization-sensitive HFI frequency channel. The spectra are normalized at 0.25\,Hz. Blue, green, red, and cyan represent 100, 143, 217, and 353\,GHz, respectively. The vertical dashed line marks the spacecraft spin frequency. The sharp spikes at high frequencies are the so-called 4-K cooler lines. These noise spectra are built before the time transfer function deconvolution.}
\label{fig:A08meannoisescaled}
\end{figure}

Within a frequency band, the noise power spectrum of each bolometer has a similar shape below 5\,Hz. Figure~\ref{fig:A08meannoisescaled} shows the mean noise spectrum for each of the polarization-sensitive HFI channels, normalized to unity at 0.25\,Hz (we are interested here in the shapes of the noise spectra, not their absolute amplitude). 
\begin{figure}[htbp!]
\includegraphics[width=\columnwidth]{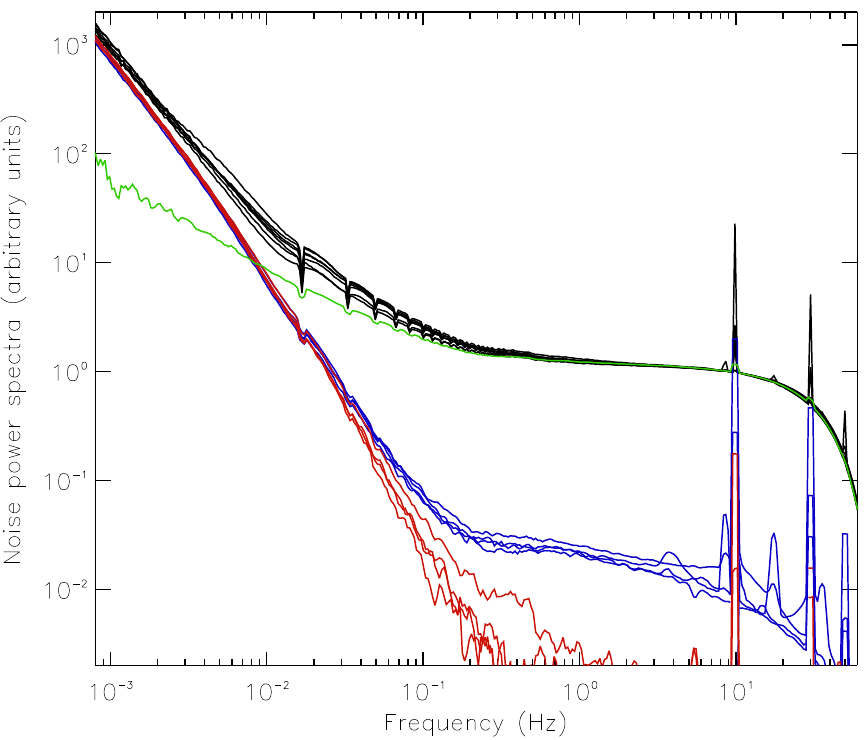} 
\caption{
Noise cross-power spectra of the 143-GHz bolometers, with the unpolarized spider-web bolometers (SWBs) in red and the polarization-sensitive bolometers (PSBs) in blue. The low-level correlated white noise component of the PSB noise is associated with common glitches below the detection threshold. Auto-spectra are shown in black. The uncorrelated noise is in green.}
\label{fig:Darkcorrelationsurvey2}
\end{figure}

Detector noise can be divided into three components (see Fig.~\ref{fig:Darkcorrelationsurvey2}), with spectra varying approximately as $f^0$ (i.e., white), $f^{-1}$, and $f^{-2}$.  The $f^0$ and $f^{-1}$ components are uncorrelated between detectors; however, the $f^{-2}$ component is correlated between detectors. This component dominates below $10^{-2}$\,Hz, and thus below the spin frequency.  Noise levels at frequencies around 0.002\,Hz (much below the spin frequency) changed during the mission, reflecting variations of the Galactic cosmic ray flux with Solar modulation and their effect on the temperature of the bolometer plate.   These effects are corrected using a template built from the signal of the dark bolometers smoothed on minute timescales \citep{planck2013-p03}.  The $f^{-2}$ component is different for spider-web bolometers (SWBs, red curves) and polarization-sensitive bolometers (PSBs, blue curves). Common glitches seen in the two silicon wafers of a PSB produce correlated noise with a knee frequency of about 0.1\,Hz \citep{planck2013-p03e}. Above 0.2\,Hz, the correlated noise is at a level of 1\,\% of the total detector noise; at twice the spin frequency (multipole $\ell=2$ in the maps), it contributes only 10\,\% of the total noise.

\begin{figure}[htbp!]
\includegraphics[width=\columnwidth]{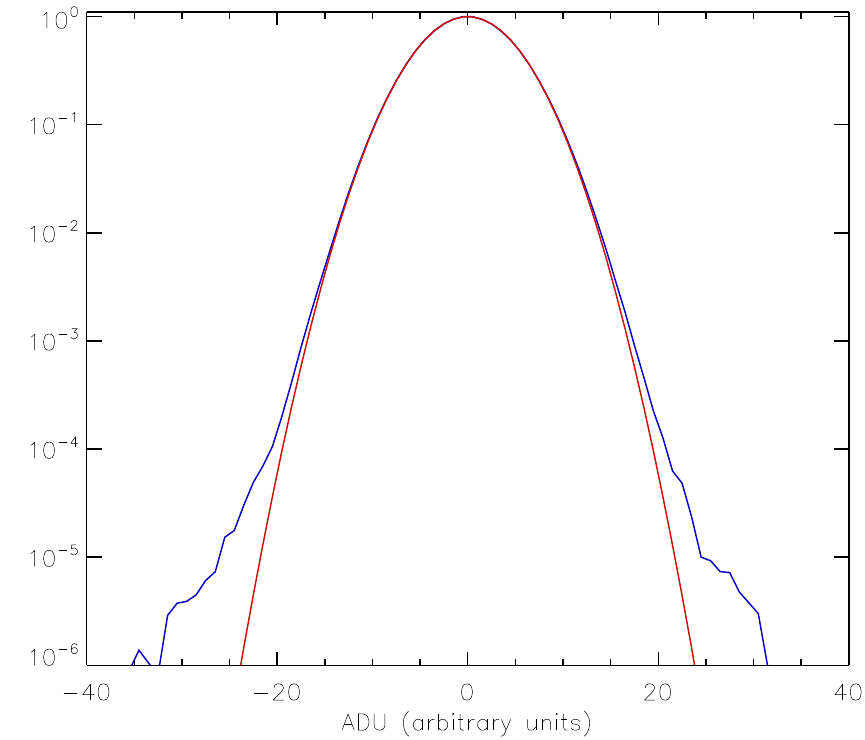} 
\caption{Histogram of the noise between 0.018 and 0.062\,Hz (frequencies at which it is dominated by the uncorrelated $1/f$ noise) for detector 143-1a in blue,
together with the best-fit Gaussian distribution in red. }
\label{fig:HistoBruit1sf1431a}
\end{figure}

After removal of the correlated part, an uncorrelated $f^{-1}$ component remains (green curve in Fig.~\ref{fig:Darkcorrelationsurvey2}) that has $f_{\rm knee} \approx 0.2$\,Hz for all HFI detectors (see, e.g., figure~13 of \citealt{planck2013-p03e}). Above this frequency the noise is predominantly white, with amplitude in good agreement with ground-based measurements. The 10-M$\Omega$ resistor and the capacitor in the focal plane, read out through the same electronics as the bolometers, show only $1/f$ noise below 3\,mHz, showing that this additional $1/f$ component is not generated in the readout but is intrinsic to the bolometers and connectors.\footnote{Telegraphic noise is seen on some detectors and has been shown to be affected by connections and disconnections in ground tests of the focal plane.} The $1/f$ component of the noise is discussed in Sect.~\ref{sec:glitches}, and is well approximated by a Gaussian statistical distribution, as seen in Fig.~\ref{fig:HistoBruit1sf1431a}. The low level ($< 10^{-3}$) non-Gaussian wings, not seen on 100-ring averages, suggest that they are caused by rare events.

$1/f$ noise around 0.06\,Hz is largely uncorrelated, and is constant during the mission within uncertainties. This confirms that undetected glitches do not contribute significantly, apart from a small contribution near the spin frequency.
 No intrinsic $1/f$ noise was detected in individual bolometer ground tests, but a $1/f$ component was seen in HFI focal plane unit ground calibration measurements below 0.1\,Hz \citep[see, e.g., figure~10 of][]{lamarre2010}, comparable to those seen in flight. 

In summary, the correlated component ($f^{-2}$) between all bolometers is mostly removed by the dark bolometer baseline removal. The weak correlated white noise ($1\,\%$) between pairs of bolometers within a PSB, caused by undetected glitches, is not removed by this baseline removal procedure.

In conclusion, the component of TOI noise that is uncorrelated between detectors (except near the spin frequency) can be modelled with two Gaussian components, white and $1/f$.  These are taken from version~8 of the FFP8 noise model \cite{planck2014-a14}, adjusted to the power spectrum observed in the half-ring null tests. The TOI noise, and its propagation to maps and power spectra, is taken in this paper as the fundamental limitation of HFI, and the maps built with FFP8 noise give the reference for maps and power spectra in simulations.

\subsection{Sky components removed in the ring making}
\label{sec:skycomponents}

\subsubsection{Removal of zodiacal dust emission}

Thermal emission from interplanetary dust---the zodiacal emission---varies not only with frequency and direction, but also with time as \Planck\ moves through the Solar system. The size of the effect on temperature is small, as estimated in \cite{planck2014-a09}, and it was not removed from the TOI released in 2015. Rather, we reconstruct the emission using the COBE Zodiacal Model \citep{kelsall1998} and the zodiacal emissivity parameters found in \cite{planck2014-a09}, and subtract it from the TOI of each detector prior to mapmaking.

The zodiacal emission is not expected to be intrinsically polarized; however, HFI measures polarization by differencing the signals from quadruplets of PSBs. Since each detector has a slightly different spectral response, any component that has a different spectrum from the blackbody spectrum of the primordial CMB anisotropies, such as zodiacal emission, can introduce leakage of temperature into polarization. By subtracting a model of zodiacal emission from the TOI of each detector, this source of leakage is strongly suppressed.

Differences between $Q$ and $U$ maps made with and without removal of zodiacal emission show the size of the effect. At 100, 143, and 217\,GHz, the zodiacal correction is less than about $100\,\mathrm{nK}_\mathrm{CMB}$. At 353\,GHz, it is an order of magnitude larger. If our correction was so poor that it left 25\,\% of the emission, temperature leakage would contribute errors of order $25\,\mathrm{nK}_\mathrm{CMB}$ to the polarization maps at 100, 143, and 217\,GHz, and perhaps $100\,\mathrm{nK}_\mathrm{CMB}$ at 353\,GHz. {\it The effect on $\tau$ is negligible.}

\subsubsection{Far sidelobes}
\label{sec:farsidelobe}
HFI far sidelobes (FSLs) are discussed in detail in \citet{planck2014-a08}. They are dominated by radiation from the feedhorns spilling over the edge of the secondary mirror, and by radiation reflected by the secondary mirror spilling over the edge of primary mirrorr or the main baffle.\footnote{Here we adopt the convention of following the light from the detectors outwards, as used in the simulations.} The effects of diffraction by edges other than the mouths of the feedhorns themselves is important only at 100\,GHz, where diffraction can affect very low multipoles (see below). FSL variations inside a frequency band are negligible at HFI frequencies. The \Planck\ optical system is modelled using the {\tt GRASP} software.\footnote{\url{http://www.ticra.com/products/software/grasp}}  Higher-order effects with more than one reflection and diffraction by sharp edges can be computed by {\tt GRASP}, but the complexity and computational time is a strong function of the order number. We have computed the first-order FSLs for each bolometer, and then checked for a few representative bolometers that the addition of the next seven orders gives small corrections to the first-order computations.

Convolution of the sky maps with the FSL model predicts small contributions to the maps. In addition, the contributions from FSLs close to the spin axis are reduced by destriping during the mapmaking process (a contribution of FSLs on the spin-axis is completely removed by the destriper). The Galactic contributions though the FSLs are similar at 100, 143, and 217\,GHz, since the decrease in the FSL amplitude and the increase in Galactic emission at higher frequencies roughly compensate each other. The Solar dipole FSL contributions decrease with frequency, as summarized in Table~\ref{tab:fsl}.  The FSLs produce a direct dipole calibration shift by increasing the  effective beam etendue, though the shift is reduced by the destriping. The dispersion between detectors is caused by the variation in the main spillover (the rays that miss the secondary mirror), which is dominated by the position of the horns in the focal plane: the bolometers further away from the symmetry axis experience a larger spillover. This effect can be used as an additional test of the fidelity of the {\tt GRASP} calculations used to model the FSLs.

The impact of FSLs on the HFI maps depends on the scanning strategy, and so must be corrected in either the TOI processing or in building the HPRs. A FSL model has been computed from the first-order {\tt GRASP} calculations. Higher-order effects are absorbed, together with other residuals, into the empirical complex transfer functions discussed in Sect.~\ref{sec:timeresponse}. The parameters of the transfer functions are then determined from the data.

\begin{figure}[htbp!]
\includegraphics[width=\columnwidth]{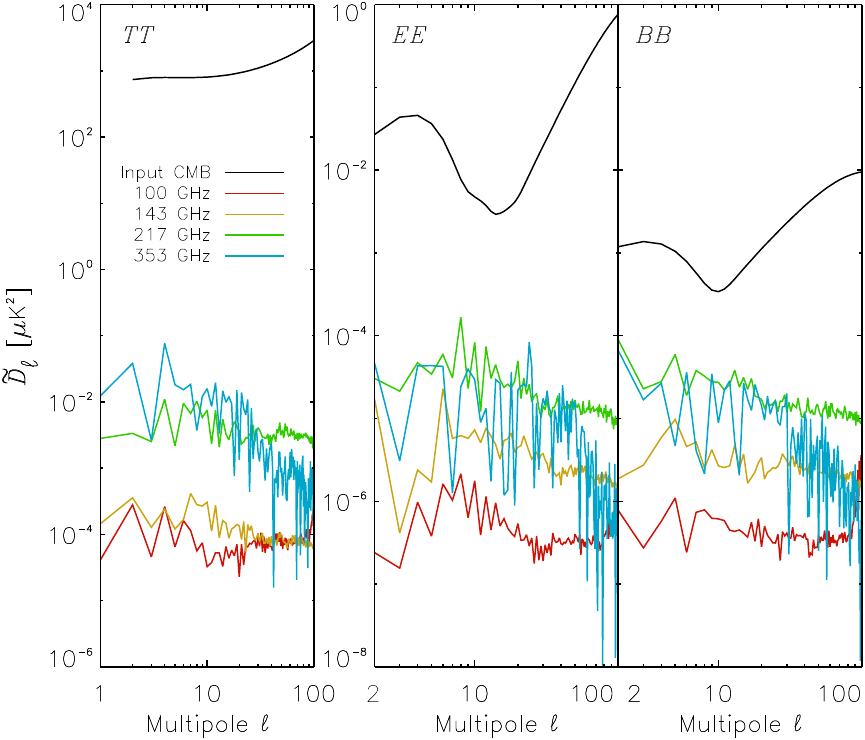} 
\caption{Auto-power spectra, showing the level of the simulated FSL projected on the maps predicted using the {\tt GRASP} model. At $\ell<10$, the FSL signal at all frequencies is at least two orders of magnitude smaller than the expected cosmological $EE$ signal.}
\label{fig:DIFF_FSL}
\end{figure}

Angular power spectra\footnote{Throughout this paper, we denote by $\tilde C_{\ell}$ (or $\tilde D_{\ell}$) the undeconvolved power spectra, and $\hat C_{\ell}$ (or $\hat D_{\ell}$) the deconvolved power spectra.} from HFI pre-2016 E2E simulations of the FSLs are displayed in Fig.~\ref{fig:DIFF_FSL}. At all HFI frequencies, the FSL effects are smaller than the fiducial $EE$ power spectrum by two orders of magnitude, where we define the fiducial $EE$ spectrum to be the base $\Lambda$CDM power spectrum from best-fit \Planck\ 2015 cosmological parameters \citep{planck2014-a15}.

The FSL signals are removed using {\tt GRASP} first-order predictions, leaving residuals due to higher orders and uncertainties. These residuals are smaller than the effect displayed in Fig.~\ref{fig:DIFF_FSL}.

Table~\ref{tab:fsl} provides estimates of the effects of FSLs on the relative inter-calibration with respect to the average of all detectors within the same frequency band. As discussed in Sect.~\ref{sec:farsidelobe}, the effect of the FSLs depends on the scanning strategy, and must be removed at the HPR level. The effect on the dipole requires propagation of the FSLs through the HFI pre-2016 E2E simulations to take into account the filtering by the destriper. These numbers can be compared directly with the main beam dipole amplitudes measured from each bolometer. At 100 and 143\,GHz, the rms dispersions of the relative dipole calibration measured from individual bolometers are $5\times 10^{-6}$ and $9\times 10^{-6}$, respectively (see Fig.~\ref{fig:interbol}). Comparing with the rms variation of the FSL contribution to the dipole calibration listed in Table~\ref{tab:fsl} ($1\times10^{-4}$ and $2\times10^{-5}$) shows that corrections using the {\tt GRASP} model are accurate to better than 5 and 2\,\%, respectively, for 100 and 143\,GHz. This is not surprising, since the first order (spillover) dominates, as discussed earlier in this section.
 
The asymmetries of the FSLs are mainly caused by the secondary mirror spillover, which depends on the position of the detector in its line parallel to the scan direction. Smaller asymmetries are generated by the small higher-order effects in the FSL {\tt GRASP} calculations discussed above. Uncertainties in the very long time constants can also leave small transfer function residuals. These can be tested on the data redundancies, and when detected, corrected by an empirical complex (phase and amplitude) transfer function (see Sect.~\ref{sec:timeresponse}) at low frequencies.
 
\begin{table}[tb]
\newdimen\tblskip \tblskip=5pt
\caption{Estimates of the relative impact of the FSLs per bolometer within a frequency band. The table lists the minimum, maximum, average, and rms values between the bolometers of one frequency band. These numbers are computed by convolving the FSLs with the dipole and propagating the resulting signal to the maps and power spectra to include the filtering of the destriper.}
\label{tab:fsl}
\vskip -6mm
\footnotesize
\setbox\tablebox=\vbox{
 \newdimen\digitwidth 
 \setbox0=\hbox{\rm 0}
 \digitwidth=\wd0
 \catcode`*=\active
 \def*{\kern\digitwidth}
\newdimen\signwidth
\setbox0=\hbox{.}
\signwidth=\wd0
\catcode`!=\active
\def!{\kern\signwidth}
\halign{\hbox to 2.0cm{#\leaderfil}\tabskip 2em&
\hfil#\hfil&
\hfil#\hfil&
\hfil#\hfil&
\hfil#\hfil\tabskip 0em\cr
\noalign{\doubleline}
\omit\hfil Frequency\hfil&Min&Max&Mean&rms\cr
\noalign{\vskip 3pt} 
\omit\hfil [GHz]\hfil&[\%]&[\%]&[\%]&[\%]\cr
\noalign{\vskip 3pt\hrule\vskip 5pt}
\noalign{\vskip 2pt}
100&0.074&0.101&0.086&*0.010\cr
143&0.051&0.059&0.054&*0.002\cr
217&0.048&0.053&0.051&*0.002\cr
353&0.015&0.016&0.016&$<$0.001\cr
\noalign{\vskip 3pt\hrule\vskip 5pt}}}
\endPlancktable
\end{table}


\subsection{ADC nonlinearity systematic effects}
\label{sec:adc}

Nonlinearity in the ADCs in the bolometer readouts (\citealt{planck2013-p03}; \citealt{planck2014-a08}) introduces systematic errors in the data.
\begin{figure}[htbp!]
\includegraphics[width=88mm]{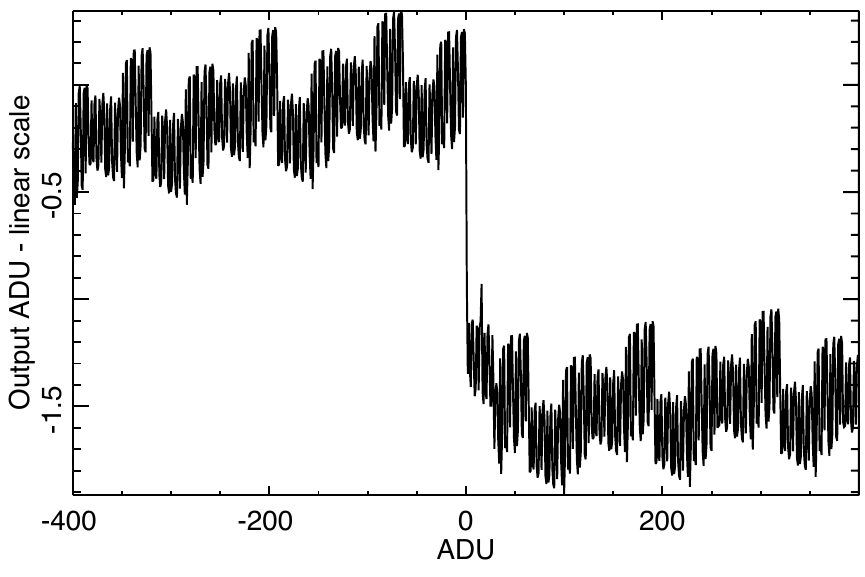} 
\caption{Relationship between input and output, for one spare ADC. The plot shows the difference between the measured digitized output signal level and the one with a perfectly linear ADC, as a function of the output level, over a signal range appropriate for the sky signals. Thus, on the one hand, a perfectly linear device would be a horizontal line at zero; in a real device such as shown here, on the other hand, the relationship between input and output is complicated and nonlinear everywhere, especially near the middle of the range around 0\,ADU.}
\label{fig:ADCerrors}
\end{figure}
Figure~\ref{fig:ADCerrors}, for example, shows the deviations from linearity measured in a flight spare ADC.   A perfectly linear device, in contrast, would lead to a horizontal line at zero.  The strongest nonlinearity in Fig.~\ref{fig:ADCerrors} at zero ADU, lies in the middle of the ADC range, and drives the main ADC nonlinearity effect.  ADC deviations from linearity are the dominant source of systematic errors in HFI low-$\ell$ polarization data. They create a first-order variable gain for the readout electronics of each detector, as well as higher-order effects as described below. 

\begin{figure*}[htbp!]
\includegraphics[width=180mm]{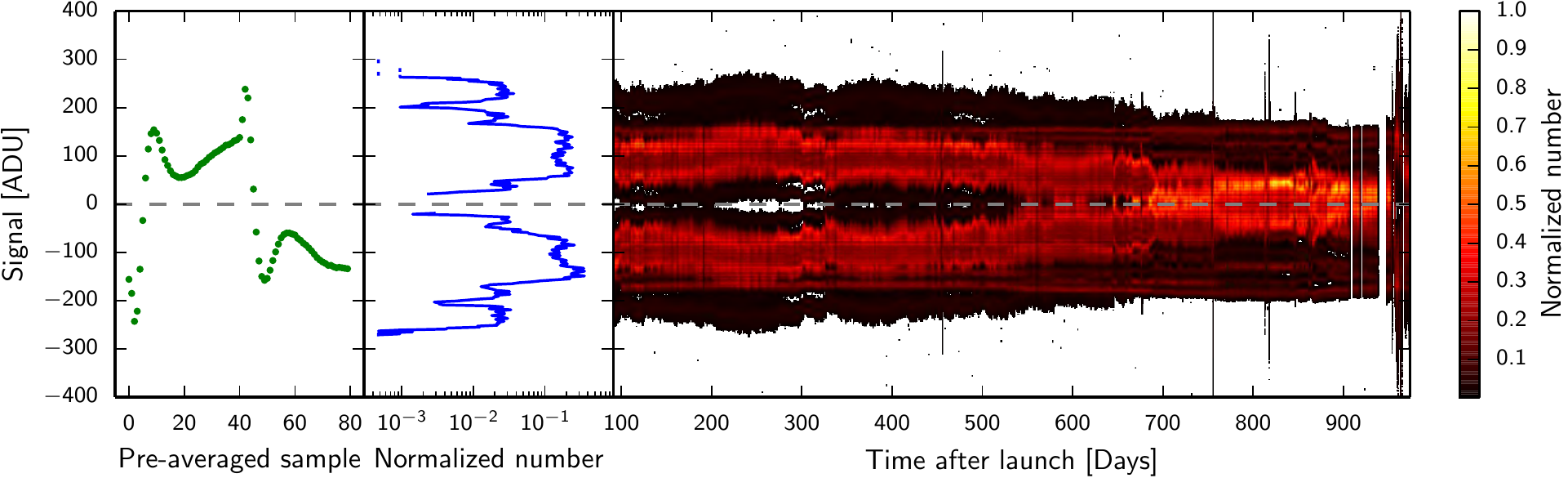} 
\caption{Fast-sampled signal from bolometer 143-1a.  {\it Left\/}: Eighty samples from early in the mission, corresponding to one cycle of the 90-Hz square-wave modulation of the bias current across the bolometer. Because of the square-wave modulation, the first and last sets of 40\,samples are nearly mirror images of one another across the $x$-axis. The units are raw analogue-to-digital units (ADU).  {\it Middle\/}: Normalized histogram of the fast-sample signal values for day 91 of the mission, in ADU. The signal spread is dominated by the combination of the noise and the square-wave modulation of the bias current across the bolometer (left panel), with additional contributions from the CMB Solar dipole.  {\it Right\/}: Normalized histograms for each day, starting with day~91, stacked left to right for the entire mission. Histogram values are given by colour, as indicated. The obvious symmetric trends during the mission are caused by drifts in temperature of the bolometer plate. Isolated days with large deviations from zero, seen as narrow black vertical lines, are due to Solar flares.}
\label{fig:HISTOADU}
\end{figure*}

In the 2013 data release, the effects of ADC nonlinearities were partially corrected (during mapmaking) by application of a time-variable gain for each detector (i.e., calibration to astrophysics units; \citealt{planck2013-p03}), with residual effects at levels of a few percent or less in the TOI. This simplified approach to ADC nonlinearity was adequate for the analysis of temperature anisotropies; however, it leaves residual effects in the polarization maps at the lowest multipoles at levels about 30 times higher than the noise.  

In 2013, a proper correction that would take into account the detailed characteristics of the ADCs themselves at the TOI processing level was not possible, because the pre-launch measurements of the ADCs had inadequate precision \citep{planck2014-a08}.  Accordingly, during the extended \Planck\ observations with LFI, after the $^3$He for the 0.1-K cooler ran out (February 2012 to August 2013), we conducted an in-flight measurement campaign to characterize the HFI ADC nonlinearities more accurately. The resulting model of ADC nonlinearity was used to correct the 2015 HFI data (in TOI processing). Residual effects were much reduced, to a level of 0.2--0.3\,\% in the TOI, and had negligible effect on the 2015 polarization maps at multipoles $\ell \ga 30$.  No additional correction of the type made in 2013 was performed, because it did not bring clear improvement.  (This was subsequently understood to be a consequence of degeneracies with other corrections.)  However, a better treatment of ADC nonlinearities is required to achieve noise-limited polarization maps at multipoles less than~30.

The results in this paper are based on an improved treatment of ADC nonlinearities, made possible by the new mapmaking code ({\tt Sroll}), which  simultaneously solves for temperature-to-polarization leakage and residual gain variations from ADC nonlinearities.  Specifically, the 2015 correction in the TOI (using the model of ADC nonlinearity derived from the extended-mission measurements) is followed by a step in {\tt Sroll} that calculates a multiplicative gain correction of the residuals left by the first step. 

In this section, we explain why application of a time-variable gain worked reasonably well for temperature in 2013, why a model of ADC nonlinearity worked well enough for high-$\ell$ polarization in 2015, and why a global determination of leakage levels, together with the residual gain variations induced by the ADC nonlinearity, works better yet, and can be used in this paper for low-$\ell$ polarization.  In addition, we show that a higher-order (but non-negligible)x of ADC non-linearity acting on the CMB dipole also has to be taken into account.

To begin, consider how the signal levels at the inputs of the ADCs change with time. Figure~\ref{fig:HISTOADU} shows the fast-sampled signal from a single bolometer (143-1a) for three different time periods.
The left panel shows the 80 fast samples in one cycle of the 90-Hz square-wave modulation of the bias current across the bolometer. \citet{planck2014-a08} and Fig.~13 of \citet{lamarre2010} present details and an explanation of the shape, which varies from bolometer to bolometer and also changes slowly throughout the mission. A square-wave compensation voltage is subtracted from the signal at the input of the readout electronics to bring both modulations close to zero, in order to limit nonlinearity effects in the analogue amplification stages.  The middle panel of Fig.~\ref{fig:HISTOADU} shows a normalized histogram of all the fast-sample signal values for the first day of observations, day 91.\footnote{As described in \citet{planck2014-a08}, the downlink bandwidth allowed one and only one fast-sampled detector signal to be sent to the ground at a time.  One set of 80~samples was transferred to the ground for any given bolometer every 101.4\,s. For all bolometers, the 40 fast samples from each half of the square-wave-modulated signal were summed on-board before being sent to the ground.}
The signal spread is dominated by the noise, combined with the square-wave modulation of the bias current across the bolometer (left panel), with additional contributions from the CMB Solar dipole. The right panel shows daily normalized histograms of the fast samples of detector 143-1a for each day in the mission, starting with day 91, stacked left to right, with the histogram values colour-coded, as indicated. The two modulation states of the signal and their evolution on the ADC are clearly seen as positive and negative bands. The large, long-term, symmetric trends during the mission are caused mainly by the slow temperature drift in the bolometer plate. Additional asymmetric drifts are due to long-term variations in the readout electronics. 

\begin{figure}[htbp!]
\includegraphics[width=\columnwidth]{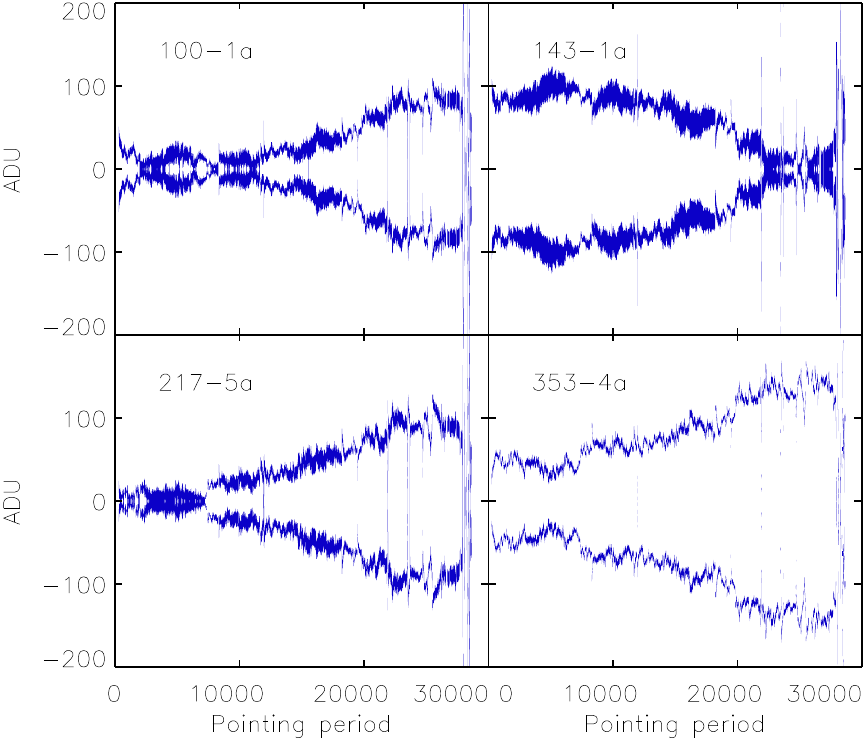} 
\caption{Simulation of the modulated noiseless sky signal per ring shows the variable amplitude of the dipole and thermal drift, which dominates the signal, as a function of time during the entire mission for four representative bolometers. The different signs of the drifts derivative depends on the level of the compensation of the modulation, which can bring one state of the modulation on either side of the middle of the ADC.}
\label{fig:dip}
\end{figure}

Because the spread of the signal is broader than the sky signal by an order of magnitude (comparing Fig.~\ref{fig:HISTOADU} and Fig.~\ref{fig:dip}), it combines the various discontinuities shown in Fig.~\ref{fig:ADCerrors} into a small but complex relationship between the signal and the power on the detectors. Nevertheless, approximations are possible. In 2013, as mentioned at the beginning of this section, this effective gain was calculated for every pointing period. The TOI were then corrected with these gains smoothed by a boxcar average over 50~pointing periods. This corrected the main gain effects of the ADC nonlinearity as the signal level drifts slowly throughout the mission. In addition to the gain variation the higher-order nonlineariries distort significantly the shape of the dipole.

 The shape of the input signal due to the 90-Hz modulation discussed above, and an estimate of how it changed throughout the HFI lifetime, were established for each bolometer \citep[see figures~2, 3, and 4 of][]{planck2014-a08}. This permitted an approximate empirical reconstruction of the input signal at the level of the TOI. 

The effects of ADC nonlinearity on the dipole signal amplitude give an excellent measure of the gain, which can be applied linearly to all signals. Note that the distortion of the shape of the large-scale CMB anisotropies is negligible, but not the dipole distortion, which leaves non-negligible, additive, large-angular-scale residuals. Figure~\ref{fig:dip} shows a simulation that contains only the dipole signal and the thermal baseline drift throughout the mission, for four representative bolometers.
The amplitude of the dipole signal in a given ring changes with the offset between the spin axis of \Planck\ and the axis of the Solar dipole. When the two are nearly aligned, the amplitude on a ring is small. When the two are far apart, the amplitude can reach 5--30\,ADU units or so (noticeably less at 353\,GHz).

In 2015, the model of ADC nonlinearities developed during the warm mission was applied to the TOI data \citep{planck2014-a08}. The correction takes into account the shape of the bolometer modulation (left panel of Fig.~\ref{fig:HISTOADU}), and corrects for the ADC nonlinearity induced by long-term drifts, resulting in residual effects in the maps an order of magnitude smaller than the 2013 correction, and good enough to be usable for high-$\ell$ polarization. However, the detailed shape of the dipole signal after passage through the bolometer readout circuits, and at the input of the ADC, is still not taken into account, leaving residuals in the data that are too large for low-$\ell$ polarization .

Figure 19 of \cite{planck2014-a08} shows the angular power spectra of difference maps, made with and without the ADC nonlinearity correction, for individual bolometers at 100, 143, 217, and 353\,GHz. Before the ADC nonlinearity correction, errors caused by nonlinearities scale approximately as $\ell ^{-2}$ below multipoles $\ell \la 100$. After correction, the power spectra are reduced in amplitude by factors of between 10 and 100 at low multipoles. 

Clearly the ADC nonlinearity correction performed in the 2015 analysis is a big improvement over the variable-gain adjustment used in the 2013 results.  Nevertheless, the residual errors from the nonlinearity correction alone are still too large for accurate polarization measurements at low multipoles \citep{planck2014-a09}.  However,  correction of these residual errors with a residual gain adjustment is now possible in the {\tt Sroll} mapmaking, which solves simultaneously for temperature-to-polarization leakage levels and gain variations, something that was not possible in the previous mapmaking algorithm. The implementation of this approach is described in the following sub-section.


\subsection{Correction of temperature-to-polarization leakage}
\label{sec:error_map}

Once the HPRs are built with far sidelobes and zodiacal emission removed, 6\,\% of the HPRs are discarded using the criteria described in \cite{planck2014-a08}. Using a generalized destriper, which takes advantage of the redundancies in the scanning strategy, we solve self-consistently for all temperature-to-polarization leakage terms:
\begin{itemize}
\item ADC nonlinearity-induced gain variation;
\item additional empirical complex transfer functions;
\item calibration factors;
\item bandpass mismatch coefficients associated with foregrounds.
\end{itemize}

The destriper baseline solution avoids regions of the sky with strong gradients that could bias the baselines. We construct one sky mask per frequency using a threshold in temperature (Sect.~\ref{sec:srollmask}). The foreground templates for the bandpass mismatch leakage are the {\tt Commander} dust, CO, and free-free maps \citep{planck2014-a11}. The detailed description of the method, performance from simulations, and tests are to be found in Appendix~\ref{sroll}.

\subsubsection{ADC-induced gain correction}

\begin{figure*}[htbp!]
\includegraphics[width=\textwidth]{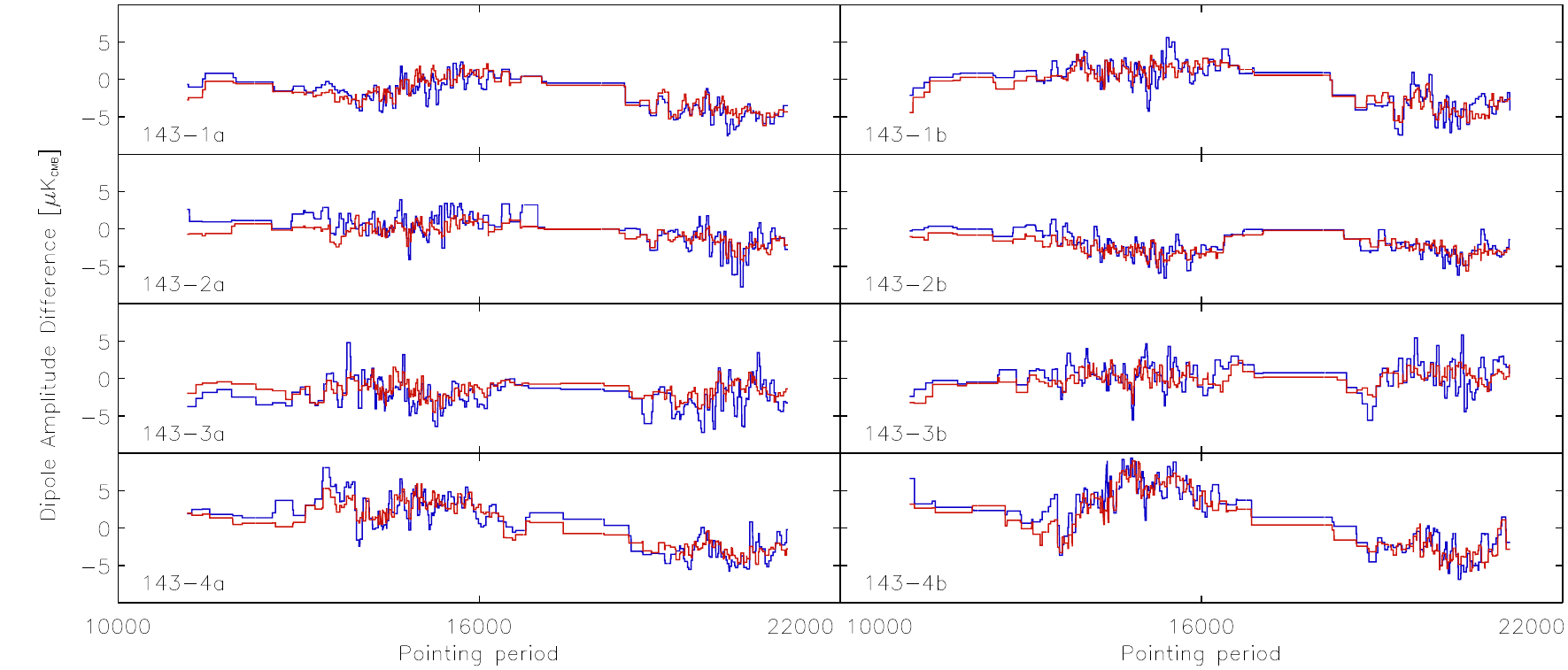} 
\caption{Dipole amplitude difference on the rings observed twice, one year apart, for each of the 143-GHz polarization-sensitive bolometers. This detects the time-dependent response associated with the excursions of the signal on the ADC, illustrated in Fig.~\ref{fig:dip}. The blue curve shows the dipole differences in units of $\mu$K after ADC correction in the TOI processing, with no other processing. The red curve shows the measured
dipole amplitude solved by {\tt Sroll}, demonstrating the reliability of the model, which can then be applied to small signals.
}
\label{fig:GAINADU}
\end{figure*}

A less biased measure of gain changes is given in Fig.~\ref{fig:GAINADU}, which shows differences in dipole amplitudes measured on the same ring one year apart.
Application of the ADC nonlinearity model to the TOI in the 2015 data release reduces the rms dipole amplitude variations from around 1\,\% (no correction) to $2\times10^{-3}$ (blue curve). The red curve shows how well {\tt SRoll} is able to solve for these residual gain adjustments, which are then applied. Algorithmic details and accuracies are described in Appendix~\ref{sroll}.

\begin{figure}[htbp!]
\includegraphics[width=\columnwidth]{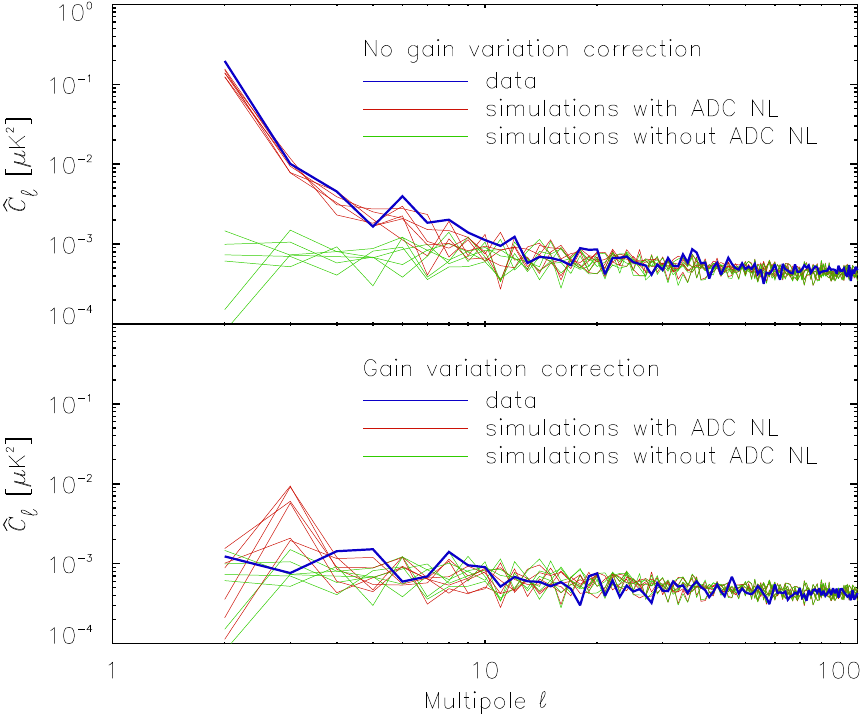} 
\caption{Power spectra of the 100-GHz half-mission null-test maps shown in Fig.~\ref{fig:plmapresidumaps}, which are dominated by ADC nonlinearity effects. We compare six simulations at 100\,GHz drawn from the {\tt Sroll} uncertainties (red lines), and the average for the full frequency 100-GHz data (blue line). In the top panel only the TOI nonlinearity correction is applied. In the bottom panel, both the TOI correction and the time-varying gain correction are applied. This leaves only the dipole distortion. For both panels the green lines are for simulations without the ADC nonlinearity effect (only noise and other systematic residuals).}
\label{fig:plmapresiduspectre}
\end{figure}

Distortions of the dipole shape caused by ADC nonlinearities within each modulation state, however, are still not accounted for. Because the drift of the signal on the ADC is large between the two halves of the mission, half-mission null tests both in the data and in simulations give an excellent test of the quality of the ADC nonlinearity corrections and of this residual dipole distortion. These are shown for 100\,GHz in Fig.~\ref{fig:plmapresidumaps} (see Sect.~\ref{sec:perfadc} for details) and their associated $EE$ power spectra in Fig.~\ref{fig:plmapresiduspectre}.
These spectra are calculated for three cases: (i)~simulated data from six realizations corrected with the ADC TOI correction alone (top panel, red lines); (ii)~simulated data corrected with the additional correction of {\tt Sroll} residual gain variation (bottom panel, red lines); and (iii)~simulated data without any ADC nonlinearity (i.e., the ``ideal'' case; both panels, green lines). In both panels, blue lines correspond to the data themselves processed in the same way. 

The difference between the red lines in the top and bottom panels of Fig.~\ref{fig:plmapresiduspectre} shows the efficiency of the {\tt Sroll} residual gain variation correction.  The difference between the red and green lines in the bottom panel shows the level of the residuals after this correction, which is dominated by dipole distortion. At $\ell \ge4$, the red and green lines are at the same level within the noise plus small contributions from other systematic residuals.  They all stay below $1\times 10^{-3}\muK^2$.  The only significant deviations are for the quadrupole and octopole terms.  Although this is the largest systematic effect left uncorrected in the HFI polarization maps, it is rather weak ($\le 20\,\%$) compared to the expected $EE$ signal. This effect has been well-simulated, and the simulations will be used to remove it from the $EE$ power spectra in the science analysis.

\subsubsection{Empirical complex transfer function}
\label{sec:timeresponse}
 
In the TOI processing, the time response of the bolometers and associated readout electronics is modelled as a Fourier filter, which is determined from observations of Saturn, Jupiter, and stacked glitches. The data are deconvolved from this response function \citep{planck2014-a08}. In addition, an in-scan, phase-shifted dipole caused by very long time constants (VLTC) is subtracted from the timeline, a necessary step for the convergence of the orbital dipole calibration, as described in \cite{planck2014-a09}. There are unavoidable uncertainties in the transfer function used to correct the timelines. We fit an additional empirical complex transfer function in the mapmaking, taking advantage of the redundancies in the data to capture any residuals from the above-mentioned bolometer/electronics time response deconvolution.  The fit also captures residuals from FSLs, which shift dipoles and low frequency signals (in-scan and cross-scan).  

The empirical transfer function is composed of four complex quantities associated with four bands of spin frequency harmonics ([0.017\,Hz], [0.033-0.050\,Hz], [0.067-0.167\,Hz], and [0.133-0.250\,Hz]), which are adjusted to minimize the residuals in the global mapmaking (see Appendix ~\ref{sec:TimeResponse}).  Including higher spin frequency harmonics provides only negligible improvements to the maps.  In the four spin-harmonic frequency bands, the phase shifts of the transfer functions are easily fitted, because they are not degenerate with the sky signal. However, for 100--217\,GHz, the CMB+foreground signals are too weak to fix the amplitudes of the transfer functions accurately and no amplitude correction is applied.

\begin{figure}[htbp!]
\includegraphics[width=\columnwidth]{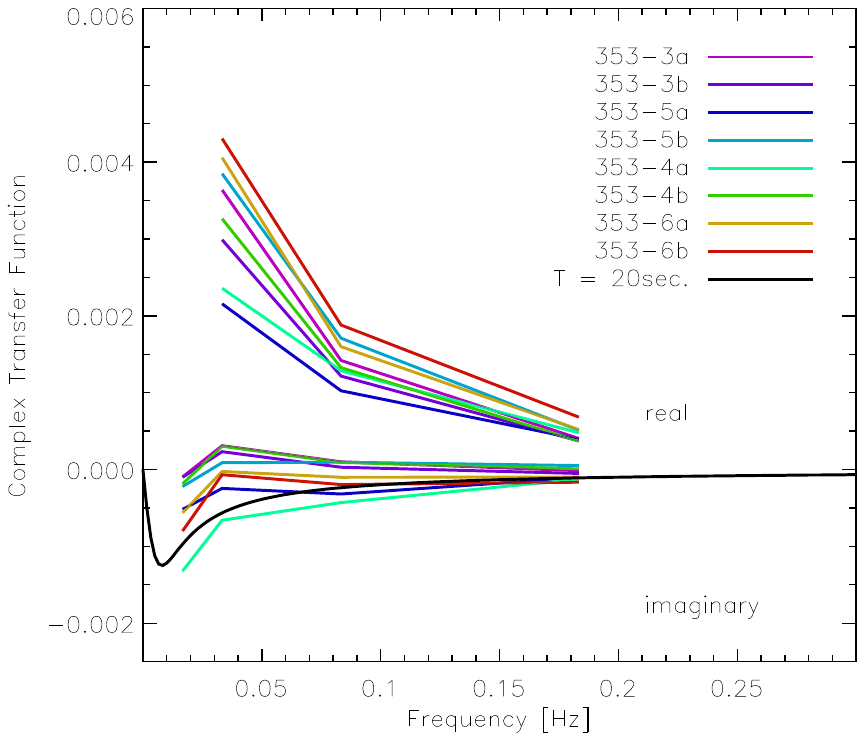} 
\caption{Best-fit solutions for the real and imaginary parts of the empirical additional transfer function as a function of frequency, for the 353-GHz bolometers.}
\label{fig:vltc353}
\end{figure}

Figure~\ref{fig:vltc353} shows this transfer function for the eight PSBs at 353\,GHz, where the strong Galactic dust signal allows an accurate determination.   The real part of the function measures the asymmetry between the cross-scan and the in-scan residuals. The imaginary part measures the shift along the scan.  At 353\,GHz, the imaginary part is almost negligible.

\begin{figure}[htbp!]
\includegraphics[width=\columnwidth]{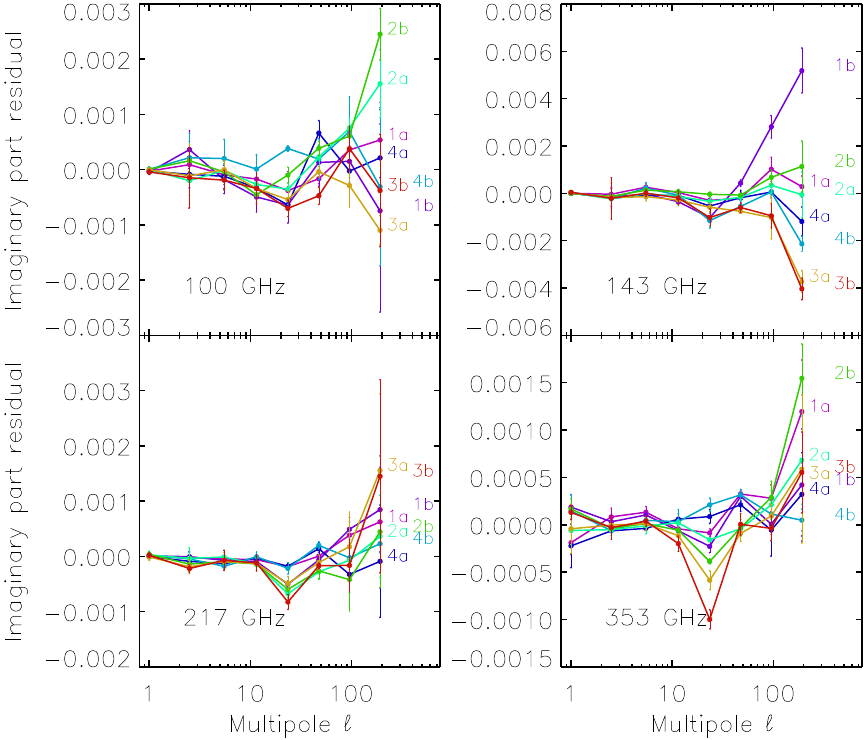} 
\caption{Ratio of the fitted data to simulated patterns detecting the residual imaginary part of the empirical transfer function, measured in odd minus even Survey difference maps averaged for sets of harmonics. The transfer function correction has been applied only over the four first sets of harmonic ranges ($\ell<15$); higher harmonics have not been corrected by the empirical transfer function.}
\label{fig:VLTCharmonicfit}
\end{figure}

To estimate the accuracy of the empirical transfer function, we use odd-minus-even Survey map differences, which are sensitive to phase shifts at low harmonics of the spin frequency.  We compute a pattern map associated with a phase shift of signal. The correlation of the data with this pattern gives the residual error left in the signal after correction with the empirical transfer function.  These relative errors on the signal are shown in Fig.~\ref{fig:VLTCharmonicfit}.   Comparing the residual errors at the four lowest sets of multipoles to those at higher multipoles, we see only upper limits below $10^{-3}$ at 100 and 143\,GHz, and three times lower at 217 and 353\,GHz.  This clearly demonstrates that the additional complex transfer function works well to correct phase shift residuals at the low multipoles that have been fitted. The mapmaking does not include corrections for temporal frequencies higher than 0.250\,Hz, corresponding roughly to $\ell>15$, and the odd$\,-\,$even Survey difference test still detects some shifts in the data.

\begin{figure}[htbp!]
\includegraphics[width=\columnwidth]{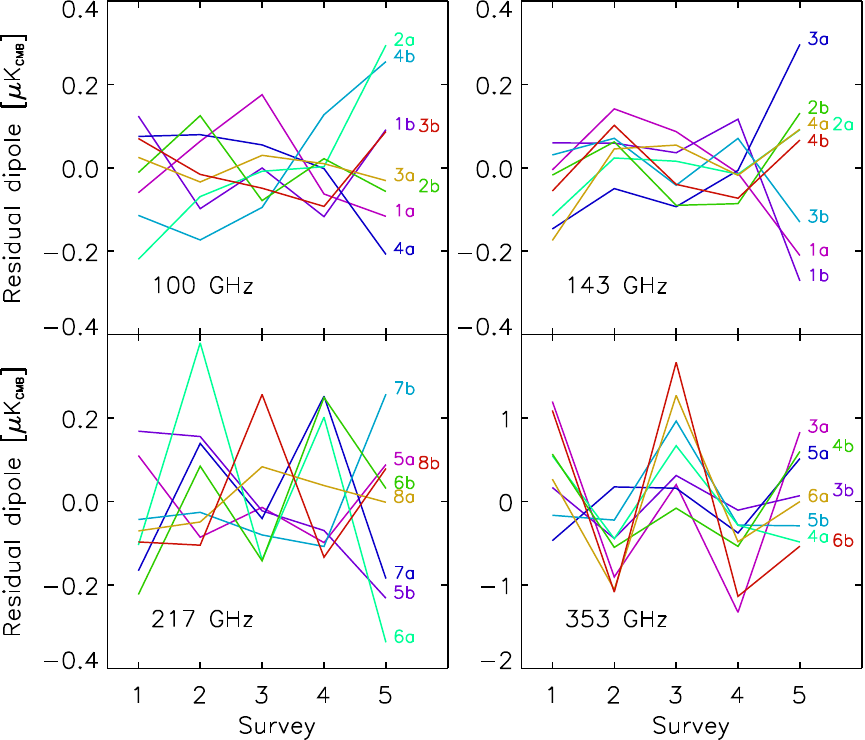} 
\caption{Residual Solar dipole amplitude for each bolometer, by Survey.  The average dipole at each frequency is subtracted.  For 100 and 143\,GHz (top panels), the variations are compatible with the relative calibration uncertainty of $10^{-4}$.  At 353\,GHz, the scale is expanded by a factor of five, and all detectors show an obvious odd/even pattern, which is marginally apparent at 217\,GHz.}
\label{fig:gainvar}
\end{figure}

Transfer function residuals also induce leakage of the Solar dipole into the orbital dipole. This leakage affects calibration differently in odd and even surveys. The Solar dipole residual amplitudes per detector with respect to the average per frequency are displayed in Fig.~\ref{fig:gainvar}.
The residual amplitude provides a strong test of the improvement provided by the transfer function correction in reducing the leakage between dipoles and gain differences between odd and even Surveys. At CMB frequencies (100 and 143\,GHz), this figure does not show any systematic odd/even Survey behaviour at the level of $0.2\,\mu$K. This translates into an upper limit on dipole-leakage-induced miscalibration better than 0.01\,\% for each bolometer. Nevertheless the odd$\,-\,$even differences of dipole amplitudes at 353\,GHz are apparent for all bolometers, with an amplitude up to $\pm 1.5\,\mu$K or approximately 0.1\,\% in odd$\,-\,$even miscalibration. The empirical real part of the transfer function cannot be determined for the dipole. Note that Survey~5 is affected by residuals from Solar flares and end-of-life tests; the last part of Survey~5 will be removed from the 2016 data release.

We propagate the residual uncertainties of the empirical transfer function (as determined from the fitting procedure; Appendix~\ref{sec:TimeResponse}) to the maps and power spectra using the E2E simulations. The auto-power spectra (Fig.~\ref{fig:DIFF_H0}) show that over the reionization peak the residuals are at a very low level: $D(\ell) < 10^{-4}\,\mu$K$^2$ for CMB channels\footnote{Throughout this paper, we call CMB channels the 100, 143, and 217\,GHz channels.}; and $D(\ell) \approx 10^{-3}\,\mu$K$^2$ for 353\,GHz, except at $\ell=2$, where it reaches $10^{-2}\,\mu$K$^2$. This demonstrates that the residuals of these systematic effects have a negligible effect on $\tau$ measurements, except possibly at $\ell=2$.

\subsubsection{Inter-detector calibration of the polarization-sensitive bolometers}
\label{INTER_DETECT_CALIB}

\begin{figure}[htbp!]
\includegraphics[width=\columnwidth]{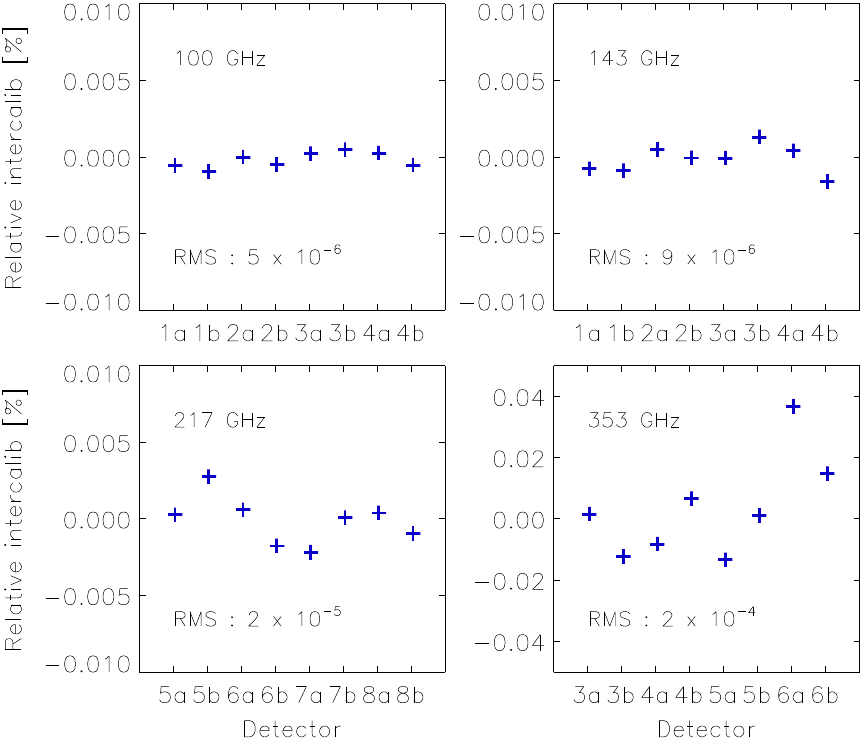} 
\caption{Relative calibration measured by the dipole amplitude for each polarization-sensitive detector with respect to the mean dipole across a frequency band.}
\label{fig:interbol}
\end{figure}

Calibration mismatches between detectors produce leakage of temperature to polarization.  The inter-detector relative calibration of the PSBs within a frequency band can be tested on single-detector, temperature-only maps (the polarized signal of the full-frequency map is subtracted from the detector TOI).   The best-fit Solar dipole (determined from the 100 and 143-GHz full-frequency maps) is removed.  The residual dipoles in the maps measure the relative calibration of each detector with respect to the average over the frequency band. The results are shown in Fig.~\ref{fig:interbol}.

The low level of variations constrain the residual calibration mismatch between detectors that could lead to leakage of temperature to polarization. 
The relative calibration factors averaged over the full mission show rms dispersions of $\pm 5\times10^{-6}$ for 100\,GHz, $9\times10^{-6}$ for 143\,GHz, $2\times10^{-5}$ for 217\,GHz, and $2\times10^{-4}$ for 353\,GHz.  At 100\,GHz, 143\,GHz, and 217\,GHz, these dispersions are very small, and unprecedented for a CMB experiment.  Absolute calibrations of band averages are discussed in Sect.~\ref{sec:interfrequencycal}. At CMB frequencies, results are still affected by gain errors between bands and by residual gain variations over time.  At these frequencies, the gain mismatch is consistent with statistical errors \citep{tristram2011}, therefore, it is not possible to improve the gain mismatch any further at these frequencies.  At 353\,GHz, the gain mismatch is larger than the statistical errors.  The worst outliers (bolometers 353-6a and 353-6b) also show large odd$\,-\,$even discrepancies in Fig.~\ref{fig:VLTCharmonicfit}.  Therefore there is hope for improving the relative calibration.

\subsubsection{Tests of bandpass mismatch leakage coefficients on gain:
 comparisons with ground tests}

\begin{figure}[htbp!]
\includegraphics[width=\columnwidth]{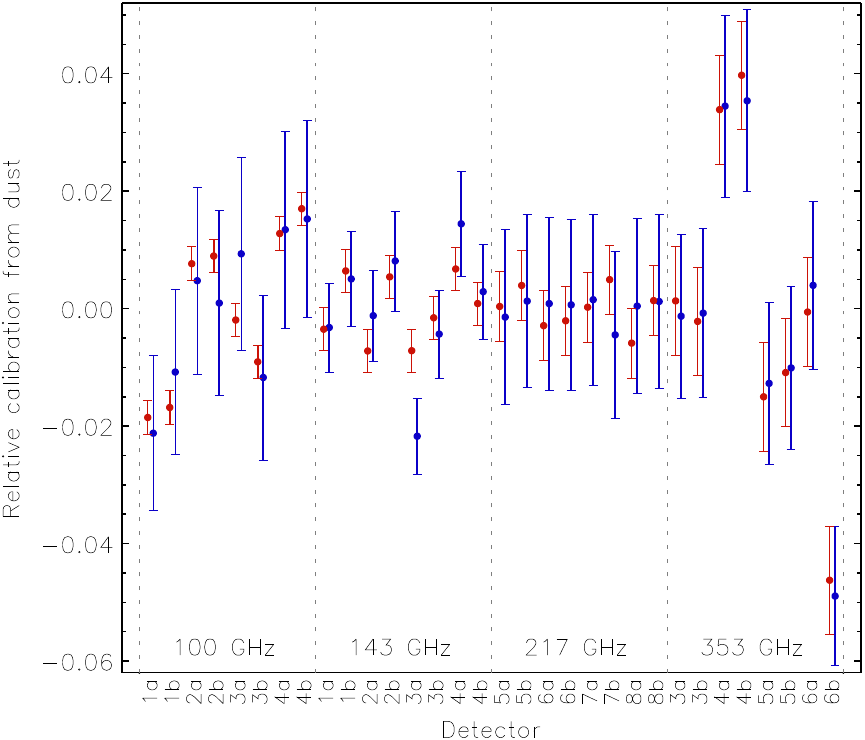} 
\caption{Comparison of the response of each detector to dust as measured on the ground (blue) and as solved by {\tt SRoll} (red).}
\label{fig:leak_rd12ll}
\end{figure}

The {\tt SRoll} mapmaking procedure (Appendix~\ref{sroll}) solves for temperature-to-polarization leakage resulting from the different response that each bolometer has to a foreground with an SED different from that of the CMB anisotropies. The solved bandpass mismatch coefficient associated with thermal dust emission is compared in Fig.~\ref{fig:leak_rd12ll} to that expected from pre-launch measurements of the detector bandpass \citep{planck2013-p03d}. The statistical uncertainties on the ground measurements are dominated by systematic effects in the measurements. For all bolometers except 143-3b, the two are consistent to within the error bars estimated using simulations. The smaller error bars are the sky determinations by {\tt SRoll} and are in closer agreement with ground measurements than the conservative estimates of the systematic errors on the ground measurements would predict. The only exception is bolometer 3b, for which the ground and sky measurements differ by nearly $3 \times$ the more accurate sky uncertainty. 

\subsubsection{Summary of improvements}

The generalized destriper solution, solving simultaneously for bandpass-mismatch leakage, intercalibration errors, and ADC-induced gain variations and dipole distortions, has been shown to be necessary to achieve a nearly complete correction of the ADC nonlinearities. This leads to much improved maps at low multipoles compared to previous releases.  In Sect.~\ref{sec:debias} and Appendix~\ref{sroll} we will demonstrate that {\tt SRoll} mapmaking does not filter or affect the CMB signal itself.

At 100, 143, and 217\,GHz, we are now close to being noise-limited on all angular scales, with small remaining systematic errors due to the empirical ADC corrections at the mapmaking level.  These separate corrections should be integrated in the TOI/HPR processing for better correction.   At 353\,GHz, and to a lesser extent 217\,GHz, we observe  residual systematic calibration effects, as seen in Fig.~\ref{fig:gainvar} and \ref{fig:interbol}, but we will show in the following sections that the residuals are small enough to have negligible effect on the determination of $\tau$.  The origin of this effect is not fully understood; correction algorithms are in development.

\section{Consistency tests of the HFI polarization maps}
\label{sec:hfimaps}

As described above and in Appendix~\ref{sroll}, the {\tt SRoll} mapmaking algorithm corrects simultaneously for several sources of temperature-to-polarization leakage that were not previously corrected.

Section~\ref{sec:nulltests_powerspectra} gives the results of null tests that show how systematic effects at large angular scales are very significantly reduced compared to those in the HFI polarization maps from the 2015 \Planck\ data release. As a test of the accuracy of this process, the results are also compared with the HFI pre-2016 E2E simulations. We begin in the next sub-section (Sect.~\ref{sec:intensity-to-polarization_leakage_residuals}) by showing that detection of the a posteriori cross-correlations of the final maps with leakage templates
cannot work because of the degeneracy with the dipole distortion.

\subsection{Temperature-to-polarization leakage}
\label{sec:intensity-to-polarization_leakage_residuals}

As discussed in Sect.~\ref{sec:TOIsqualitytests}, any bandpass and calibration mismatch between bolometers induces temperature-to-polarization leakage, and hence spurious polarization signals. Each leakage pattern on the sky for each bolometer is fully determined by the scanning strategy, along with a set of leakage coefficients, and the temperature maps involved (dipole and foregrounds). The {\tt SRoll} approach improves on the 2013 and 2015 HFI mapmaking pipeline by correcting all temperature-to-polarization mismatches to levels where they are negligible. Detector inter-calibration has been much improved, as shown in Sect.~\ref{INTER_DETECT_CALIB}. Similarly the residual bandpass leakage (mainly due to dust and CO) in the $Q$ and $U$ HFI pre-2016 maps is also greatly reduced (Appendix~\ref{sec:srollbandpasses}).

In the 2015 data release, we used leakage template fitting \citep{planck2014-a09} to check a posteriori the level of temperature-to-polarization leakage residuals, although this leakage was not removed from the maps. From Sect.~\ref{sec:error_map} and the appendices, we expect temperature-to-polarization leakage to be very much reduced for the HFI pre-2016 data set. Contrary to expectations, however, the template-fitting test (expanded to account for synchrotron polarized emission) on the HFI pre-2016 release used in this paper still reveals significant leakage. Suspecting that the problem is residual dipole distortion induced by ADC nonlinearity, which is not yet removed, we performed leakage tests on simulations both with and without the effect. This is done by forcing the gain to be constant within a pointing period, which removes only the dipole distortion within the ADC nonlinearity effects.

\begin{figure}[htbp!]
\includegraphics{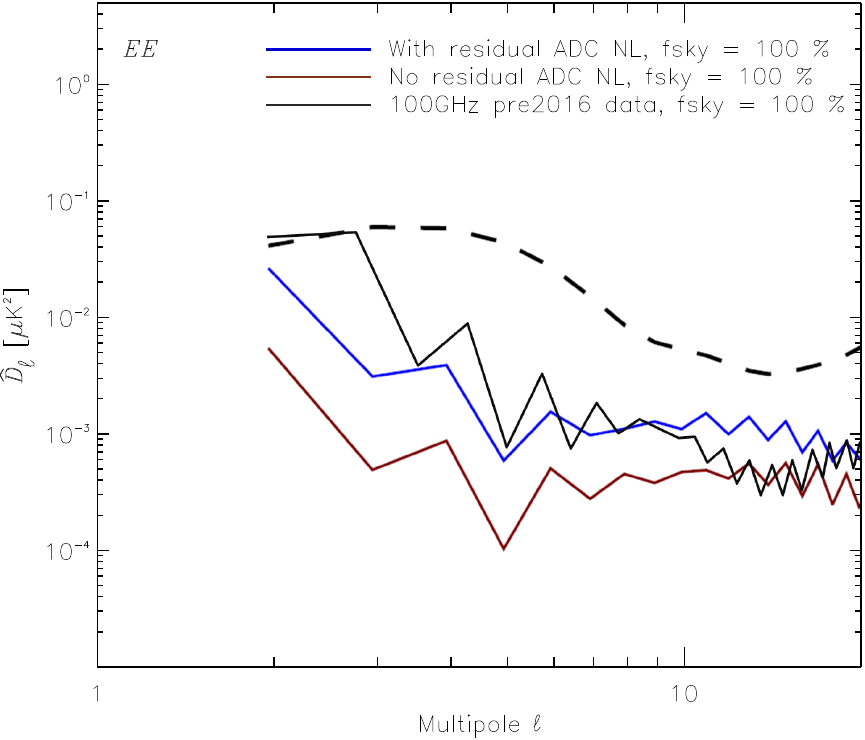} 
\caption{Simulation of the template-fitting tests for temperature-to-polarization leakage in the 100-GHz maps, with (blue) and without (red) ADC-nonlinearity dipole distortion. The black curve shows the result of the leakage fit in the HFI pre-2016 data, which is comparable to the simulation with the ADC nonlinearities. The dark red curve is lower by a factor of 5~or more, showing that dipole distortions due to ADC nonlinearity are a significant contributor to the leakage.  The black dashed line corresponds to the fiducial model F-$EE$ with $\tau = 0.066$.}
\label{fig:leak_sim_dip_dist}
\end{figure}

Figure~\ref{fig:leak_sim_dip_dist} shows that auto-spectra exhibit a leakage in the pre-2016 100-GHz data (black line), comparable to the simulations with (blue line) full ADC nonlinearity effects. The red line, with the ADC dipole distortion effect removed, is lower by a factor of 5 or more.
This demonstrates that the detected leakage at 100\,GHz contains a potentially significant, and possibly dominant, spurious detection due to a degeneracy between the dipole distortion and the leakage templates. This is in line with the low level of the leakage and the level of the dipole distortion discussed in Sect.~\ref{sec:adc}. This can only be confirmed through E2E simulations and comparison with data residuals from an appropriate null test. The remaining systematic effects are at a level of $8\times 10^{-4} \muK^2$ for $\ell>3$. The quadrupole systematic term is an order of magnitude larger. As discussed before, quadrupole and octopole systematic effects dominate in Fig.~\ref{fig:plmapresiduspectre}. The residuals of the ADC nonlinearity dipole distortion at $\ell >3$ are small with respect to the F-$EE$\footnote{We define F-$xx$ (F-$TT$, F-$EE$, F-$BB$) as the CMB fiducial power spectra based on \Planck\ cosmological parameters with $\tau=0.066$ and $r=0.11$ \citep{planck2014-a15}.} signal.

\begin{figure}[htbp!]
\includegraphics{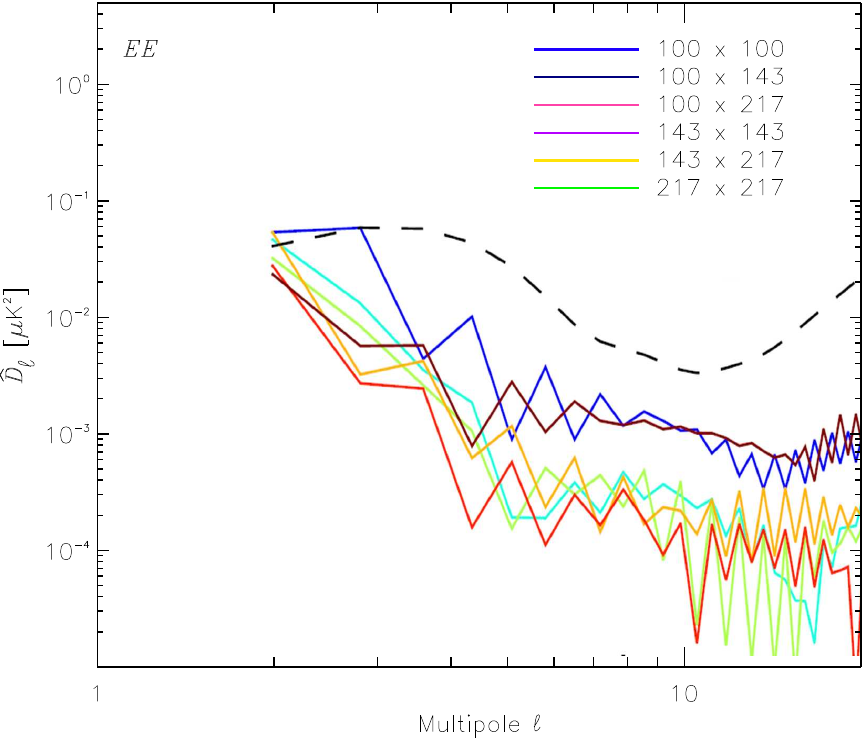} 
\caption{$EE$ auto- and cross-spectra of the global fit test of the temperature-to-polarization leakage, for 100, 143, and 217\,GHz. Levels are similar to that for 100\,GHz, for which Fig.~\ref{fig:leak_sim_dip_dist} shows that this is dominated by the ADC-nonlinearity dipole distortion. The dashed black line corresponds to the fiducial model F-$EE$ with $\tau = 0.066$.}
\label{fig:leak_dip_dist}
\end{figure}

\begin{figure*}[htbp!]
\includegraphics[width=\textwidth]{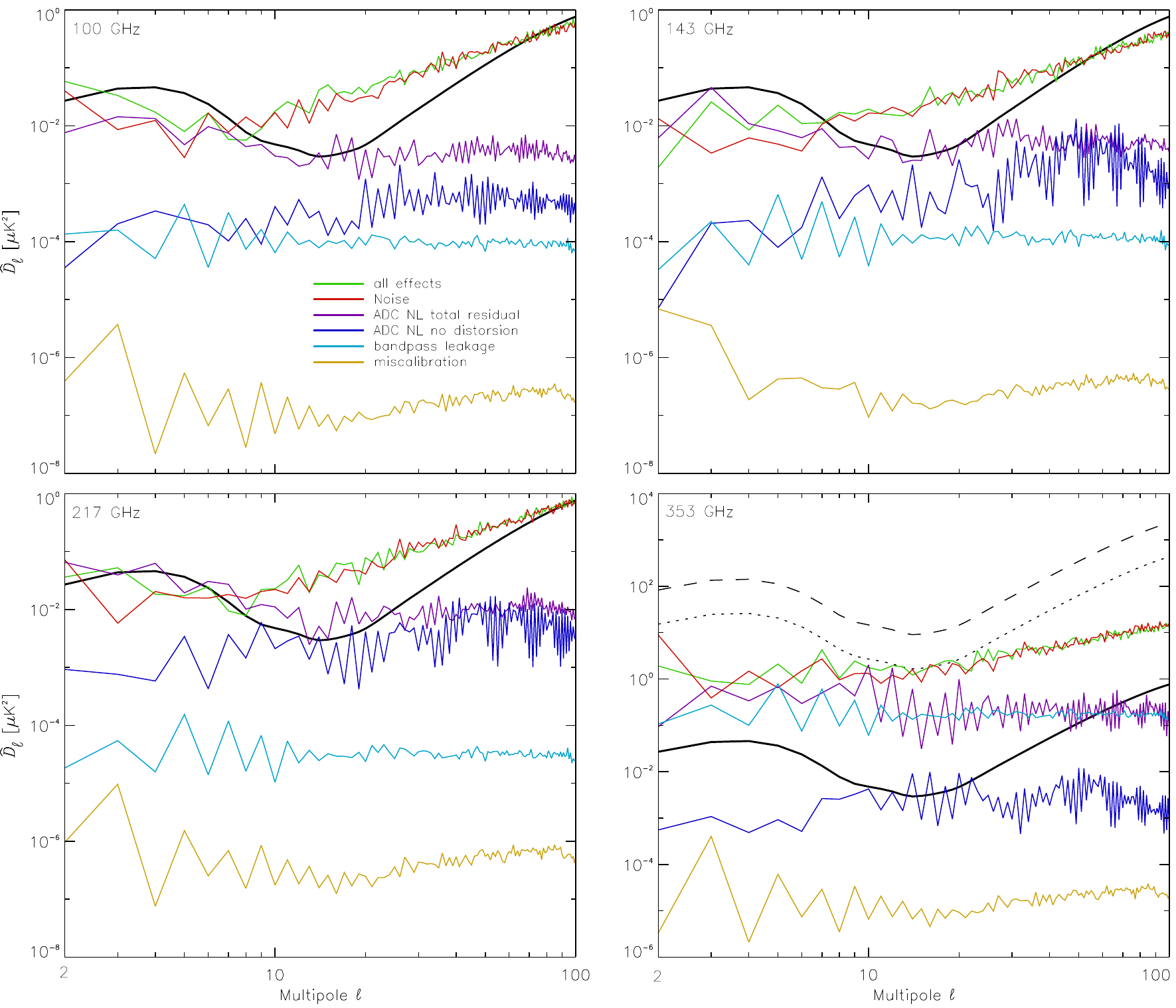} 
\caption{Residual $EE$ auto-power spectra of systematic effects from the HFI pre-2016 E2E simulations computed on 50\,\% of the sky (colours specified in the top left panel apply to all panels). The purple line (ADC NL total residual) shows the sum of all effects associated with ADC nonlinearity. The dark blue line (ADC NL no distortion) shows the level without the dominant dipole distortion. The plots show also the F-$EE$ model (black curves). The 100-GHz and 143-GHz model scaled to 353\,GHz with a dust SED is shown as dashed and dotted lines,
respectively.}
\label{fig:DIFFMERGE}
\end{figure*}

Figure~\ref{fig:leak_dip_dist} shows results for 100, 143, and 217-GHz data. All spectra show a rise similar to the one seen for the blue line in Fig.~\ref{fig:leak_sim_dip_dist}, which simulates the ADC nonlinearity dipole distortion at  100\,GHz. The higher frequencies also exhibit negligible temperature-to-polarization leakage, but have comparable ADC nonlinearity dipole distortions. This is a strong indication that, at these frequencies, the dominant residual systematic is also due to ADC nonlinearity acting on the dipole, which is not removed from the maps, in agreement with Fig.~\ref{fig:DIFFMERGE}.  This effect has to be accounted for in the likelihood before science results can be extracted from the pre-2016 maps. The temperature-to-polarization leakage is at a level of $10^{-3}\mu$K$^2$ or lower at $\ell>4$.  The dominant systematic at 353\,GHz, in contrast, is calibration uncertainty (see Sect.~\ref{INTER_DETECT_CALIB}).

\subsection{Null tests}
\label{sec:nulltests_powerspectra}

In this section we use null tests on power spectra, together with our understanding of the simulated systematic effects discussed in Appendix~\ref{sroll}, to demonstrate that we have identified all detectable systematic effects. In Fig.~\ref{fig:DIFFMERGE}, the green lines show the total of noise and systematic effects. Noise (red lines) dominates at multipoles $\ell>10$ for all frequencies. For the CMB channels (100, 143, and 217\,GHz) the main systematic is the ADC-nonlinearity dipole distortion (purple lines), which is not removed in the mapmaking; for this systematic, it is not the residual but the effect itself that remains in the maps. The residuals from other effects of ADC nonlinearities are smaller (dark blue lines), and the bandpass leakage residual (light blue) is smaller still.  Frequency-band intercalibration residuals (orange) are even lower, as are other systematic effects that are not shown.

For the 353-GHz channel, which is used only to clean foreground dust, the simulated systematic effects at low multipoles are all negligible with respect to the F-$EE$ model level scaled by the foreground correction coefficients for 100 and 143\,GHz (dashed and dotted black). Although the inter-survey calibration difference shown in Fig.~\ref{fig:gainvar} is greater for 353\,GHz than for the CMB channels, this does not affect the present results.

We perform null tests on four different data splits. The first is the odd$\,-\,$even survey differences. These map differences test residuals associated with the scanning direction and far sidelobes, short and long time constants in the bolometers, and beam asymmetry. The only clear evidence for any residual systematic effects associated with scanning direction is the odd$\,-\,$even Solar dipole amplitude oscillation displayed in Fig.\ref{fig:gainvar}. This is above the noise level only at 353\,GHz (Sects.~\ref{sec:timeresponse} and \ref{INTER_DETECT_CALIB}). Other scanning-direction systematic effects are almost entirely eliminated by the {\tt SRoll} mapmaking algorithm, and are not discussed further in this section. 

The other three data splits used for null tests are by ``detset'' \citep{planck2014-a08}, by the first and second halves of each stable pointing period \citep{planck2014-a08}, and by the first and second halves of the mission \citep{planck2014-a09}. In each case, we compute $C_\ell$ spectra of the difference between the maps made with the two subsets of the data. The spectra are rescaled to the full data set. The resulting spectra contain noise and systematic errors, and can be compared with the FFP8 simulations \citep{planck2014-a14} of the \Planck\ sky signal and TOI noise only. 

Detset differences are sensitive to errors in detector parameters, including polarization angle, cross-polar leakage, detector-mismatch leakage, far sidelobes, and time response. 

Half-ring differences are sensitive to detector noise at levels predicted by the analysis of TOI (Sect.~\ref{sec:detnoise}). All systematic errors cancel that are associated with different detector properties, drifts on timescales longer than 30\,min, and leakage patterns associated with the scanning strategy.\footnote{A small correlated noise component induced by the deglitching procedure and the TTF corrections leads to non-white noise spectra (as in Fig.~\ref{fig:A08meannoisescaled}) different from the noise levels predicted from the TOI analysis \citep{planck2014-a08, planck2014-a09}. The differences are significant only at high multipoles.}

Half-mission differences are sensitive to long-term time drifts, especially those related to the position of the signal on the ADC and related changes of the 4-K lines affecting the modulated signal.

\begin{figure*}[htbp!]
\includegraphics[width=\textwidth]{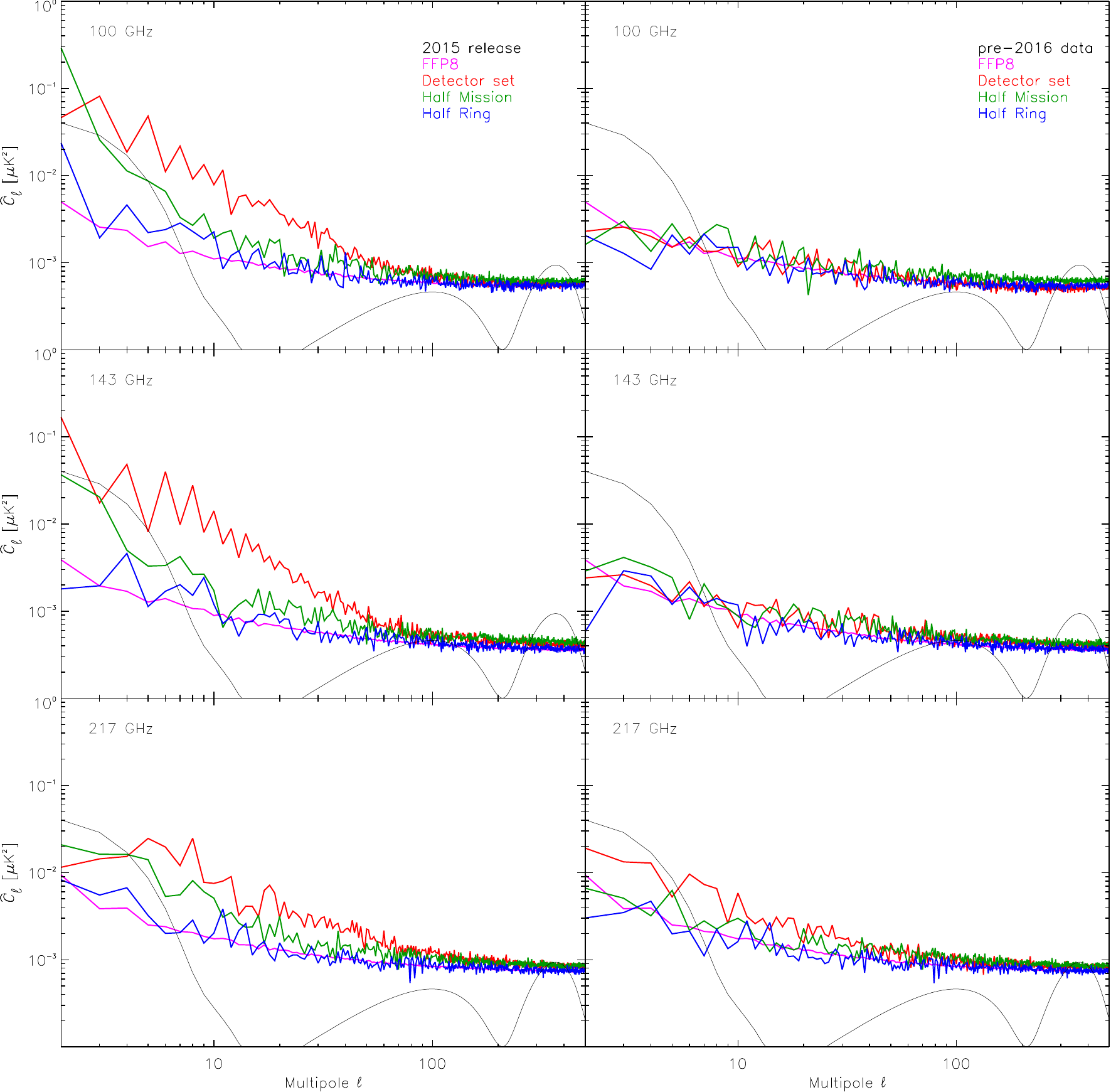} 
\caption{$EE$ auto-power-spectra of detector-set, half-mission, and half-ring difference maps for 100, 143, and 217\,GHz. $C_{\ell}$ rather than $D_{\ell}$ is plotted here to emphasize low multipoles. Results from the \Planck\ 2015 data release are on the left; results from the pre-2016 maps described in this paper are on the right. Colour-coding is the same for all frequencies. We also show for reference an average of FFP8 simulations boosted by 20\,\% to fit the half-ring null tests. The black curves show the F-$EE$ model.}
\label{fig:DX11_RD12_comp}
\end{figure*}

\begin{figure*}[htbp!]
\includegraphics[width=177mm]{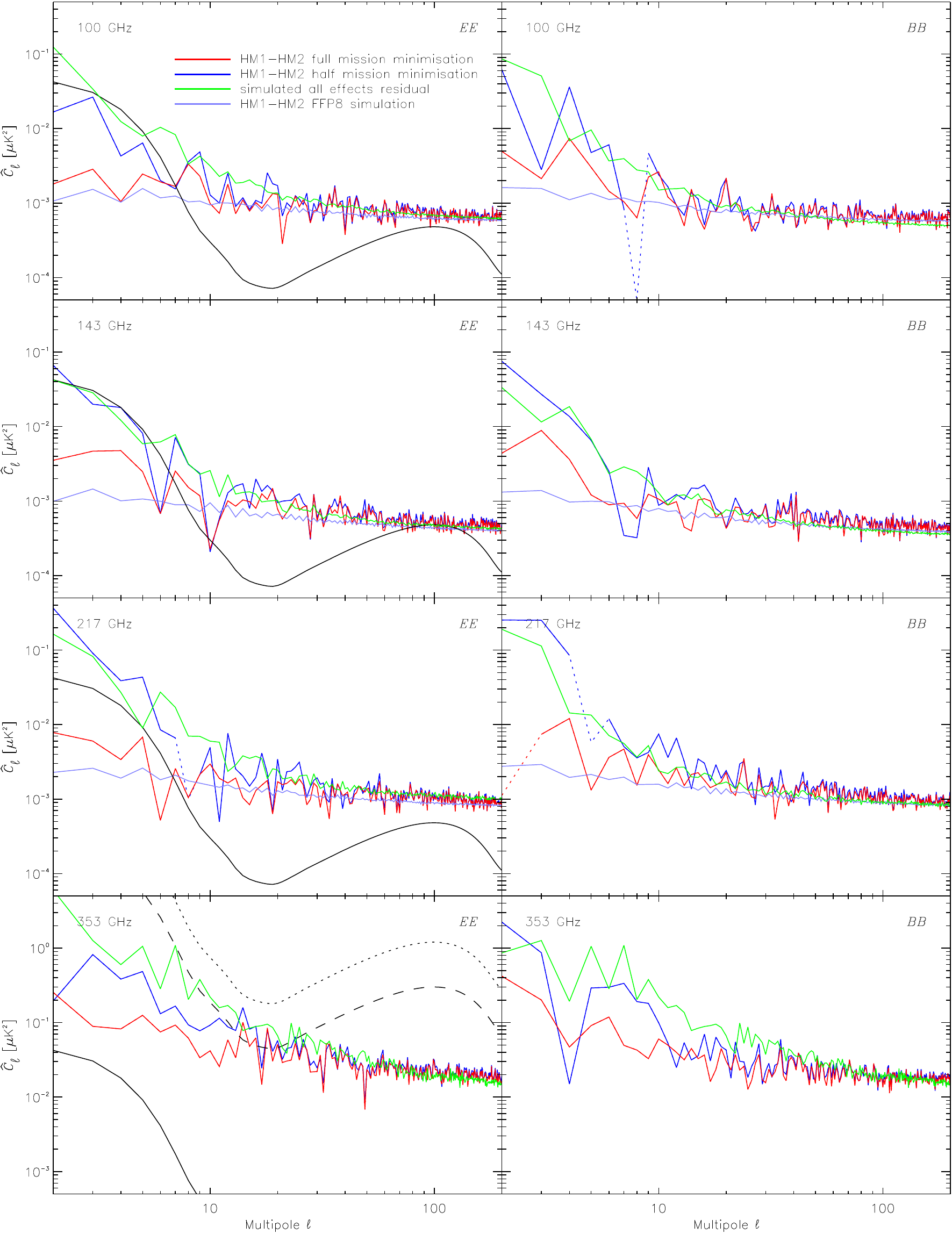} 
\caption{$EE$ and $BB$ cross-power spectra of the residual effect computed from null tests between half-mission maps based on full mission minimization (red curve) or independent minimization for each half mission (blue curve). This second approach clearly shows a systematic effect. The sum of all systematic effects, dominated by the ADC-nonlinearity dipole distortion shown in green in Fig.~\ref{fig:DIFFMERGE}, is at the same level as the simulated null test with independent minimization. The FFP8 power spectrum is given again for reference.}
\label{fig:plclhm}
\end{figure*}

\begin{figure*}[htbp!]
\includegraphics[width=177mm]{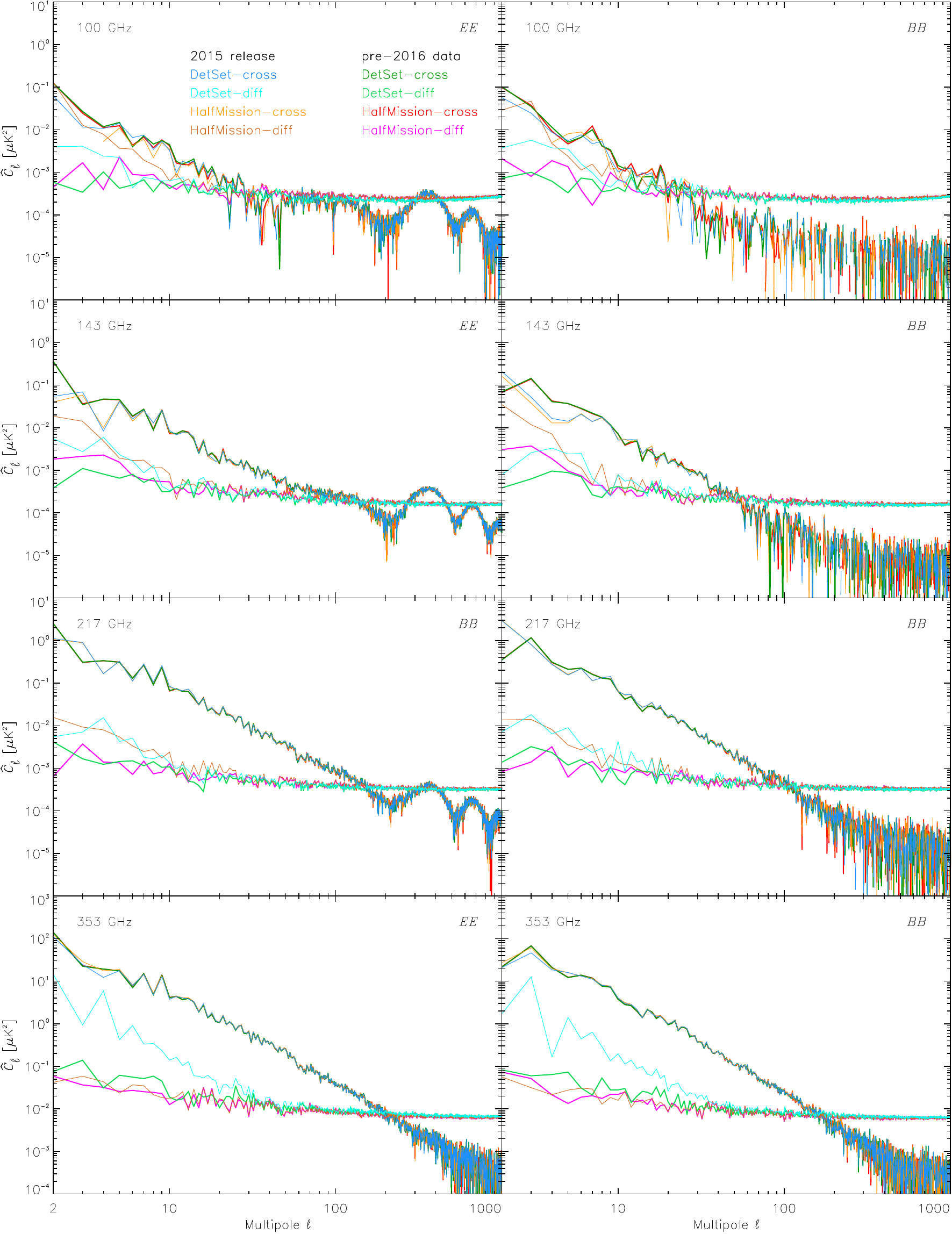} 
\caption{$EE$ and $BB$ spectra of the 2015 maps and the pre-2016 maps used in this work at 100, 143, 217, and 353\,GHz. The cross-spectra of detset and half-mission maps and the auto-spectra of the detset and half-mission difference maps are shown. The maps are masked so that 43\,\% of the sky is used.}
\label{fig:HFI_spectra}
\end{figure*}

The pre-2016 E2E simulations can be used to distinguish whether systematic effects are comparable to the level of the base \LCDM\ $EE$ spectrum and could therefore affect the $\tau$ determination, or whether the systematic effects are negligible or accurately corrected by {\tt SRoll}.

Figure~\ref{fig:DX11_RD12_comp} compares $EE$ auto-spectra of detset, half-mission, and half-ring null-test difference maps for the pre-2016 data and the 2015 release data. The half-ring null test results (blue lines) agree with FFP8 as expected. For detset (red lines) and half mission (green lines) null tests, the 2015 data show large excesses over the FFP8 simulation up to $\ell\,=\,100$. In contrast, the HFI pre-2016 data detset differences for 100 and 143\,GHz are in good agreement with the FFP8 reference simulation. This is no surprise, as systematics detected by this test have been shown to be small. At 217\,GHz, the detset test is not yet at the level of FFP8.

For the half-mission null tests, the analysis of systematic effects shows that the ADC-nonlinearity dipole distortion, which has not been removed, dominates, and should leave an observable excess at low multipoles in this test, which is not however seen. Thus this null test does not agree with the systematic analysis. A possible explanation is that the destriping is done on the full mission and applied to the two halves of the mission. The correlation thereby introduced in the two halves could lead to an underestimate of the residual seen in the null test. We checked this by constructing a set of maps in which the destriping is done independently for each half-mission. Figure~\ref{fig:plclhm} shows the results of this check. The independent destriper for the two half-missions (blue lines) shows a systematic effect at all frequencies in the half-mission null test that is not seen for the full mission minimization (red lines). The separated minimization can also be compared to the sum of all simulated effects (green lines); it shows the expected behaviour for the 100 to 217-GHz bands. We conclude that the uncorrected ADC-nonlinearity dipole distortion accounts for most of the systematic detected by this new half-mission null test. This, of course, does not imply that the maps should be constructed using the separate half-mission minimizations, because these use fewer redundancies than the full mission ones, and are therefore less powerful. 

At 353\,GHz, the null test is below the sum of all systematic effects between multipoles 2 and 50, and is not catching all systematic effects. The calibration and transfer functions for 353\,GHz are not captured by the half-mission null test; this can plausibly explain the difference.

We have shown that the destriping of the two half-missions should be done independently in order to detect ADC nonlinearity residuals. The 2016 data release will therefore include half-mission {\tt SRoll} fits of parameters.

Figure~\ref{fig:HFI_spectra} shows $EE$ and $BB$ spectra of the difference maps and cross-spectra (signal) over 43\,\% of high latitude sky for both the 2015 HFI maps and for the HFI pre-2016 polarization maps described in this paper, and shows the relative improvements made by the {\tt SRoll} processing. Differences between 2015 and pre-2016 power spectra are not particularly sensitive to the polarization mask.
The cross-spectra show the total signal level, dominated by polarized dust emission at $\ell \la 200$ and, in $EE$, by the CMB at $\ell \ga 200$. This allows a direct comparison of the signal with noise plus systematic effects.

In Fig.~\ref{fig:HFI_spectra}, the 353-GHz detset null test for the 2015 data (blue line) at $3\le\ell\le55$ is 30 times larger than the FFP8 noise, and is at a level larger than 10\,\% of the dust foreground spectrum. In the 2015 data release, systematic effects in the 353-GHz maps constitute the main uncertainty in the removal of dust emission from 100 and 143\,GHz at low multipoles, dominating over statistical uncertainty in the dust removal coefficient (around 3\,\%, see Sect.~\ref{sec:compsep}). In the pre-2016 data detset differences (green line), the systematic effects are much lower, but not yet at the TOI noise level.

In summary, all known systematic residuals have been seen in at least one null test at the expected level. Conversely, there is no excess over noise seen in a null test that is not accounted for by a known systematic. This important conclusion fulfills the goal of this section. However, one systematic effect has not been corrected at all, namely the ADC-induced dipole distortion, because it requires a better ADC model that can be applied at the TOI or ring level simultaneously with the correction of other systematic effects. This will be done in the next generation of data.

\section{LFI low-$\boldsymbol{\ell}$ polarization characterization}
\label{sec:LFI}

The measurement of $\tau$ for the \Planck\ 2015 release was based on the LFI 70-GHz maps, cleaned of synchrotron and dust emission with 30-GHz and 353-GHz templates, respectively \citep{planck2014-a15}. In this paper we use the 70, 100, and 143-GHz polarization maps to calculate $70\times100$ and $70\times143$ cross-spectra (see Sect.~\ref{sec:analysis}). Since the LFI and HFI instruments are based on very different technologies, the systematic effects they are subject to are largely independent. In particular, the dominant systematic effects in the two instruments (gain uncertainties for LFI and residual ADC effects for HFI) are not expected to be correlated. Therefore LFI\,$\times$\,HFI cross-spectra provide a cross-check on the impact of certain systematic effects in the estimate of $\tau$. There can be common mode systematics as well as chance correlations, however, so the cross-spectra cannot be assumed to be perfectly free of systematic effects.

\begin{figure}[htbp!]
\includegraphics[width=\columnwidth]{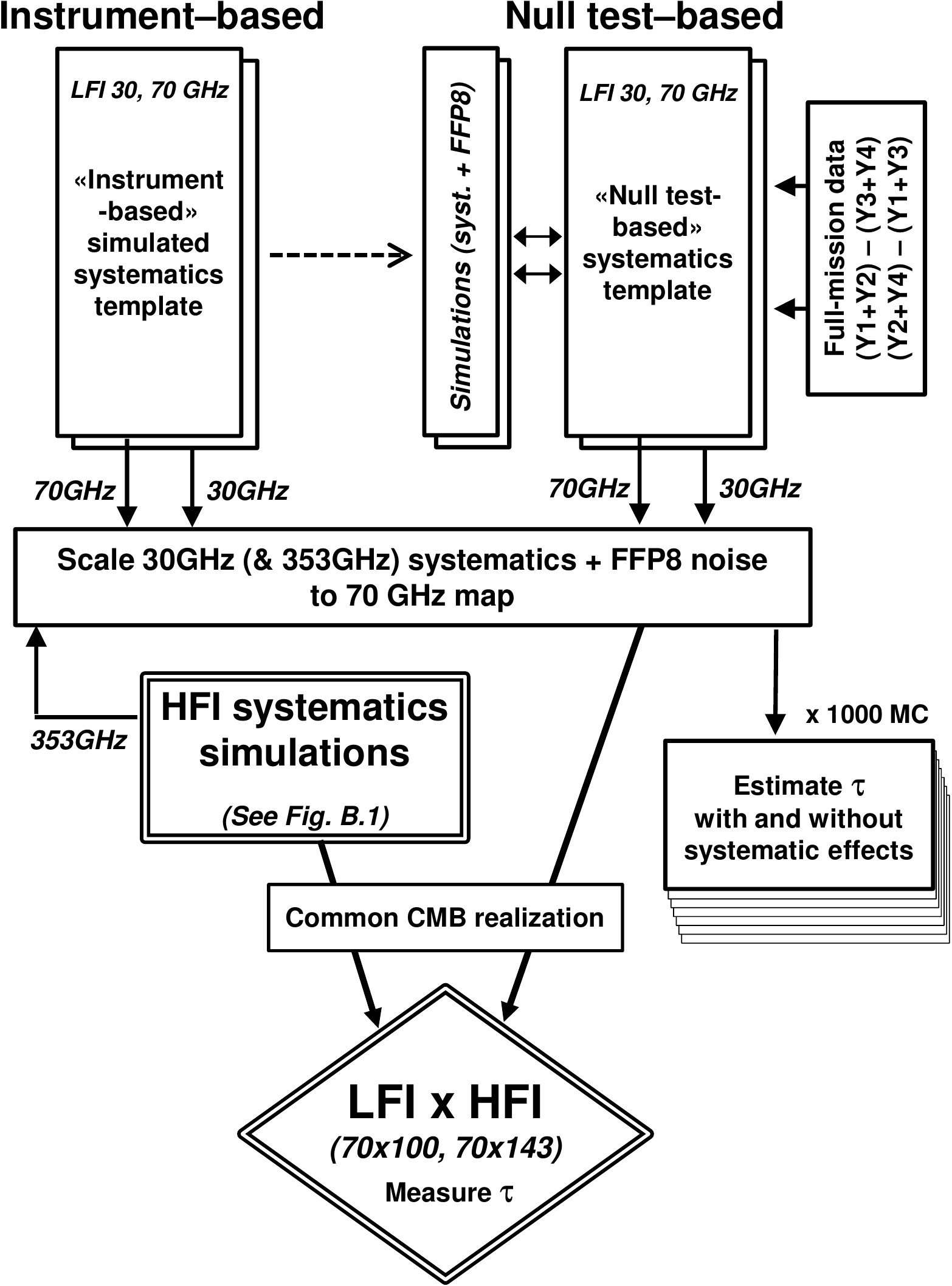} 
\caption{Schematic of the simulation plan to characterize the LFI polarization data at low multipoles. Both the instrument-based and null-test-based strategies are represented (see text). The lower left part of the diagram outlines the cross-spectrum simulation analysis involving LFI and HFI data, including systematic effects (see Sect.~\ref{sec:analysis}).}
\label{fig:LFIlowEllSysScheme} 
\end{figure}

In this analysis we use LFI data from the 2015 release. A detailed discussion of the systematic effects in those data is given in \cite{planck2014-a04}. To assess the suitability of the LFI data for low-$\ell$ polarization analysis, we analyse residual systematic effects in two ways, first using our model of all {\it known} instrumental effects (the ``instrument-based'' approach; \citealt{planck2014-a04}), and second using null maps of measured data as a representation of residual systematic effects (the ``null-test-based'' approach). In each case, we evaluate the impact of systematic effects on the extraction of $\tau$ by propagating them through foreground removal, power spectrum estimation, and parameter extraction (Fig.~\ref{fig:LFIlowEllSysScheme}).
We then use our ``instrument-based'' simulations to support a cross-spectrum analysis between the LFI 70-GHz channel and the HFI 100- and 143-GHz channels. A summary of the effects that are most relevant for the present analysis is given in Appendix~\ref{annex_LFI}.

\subsection{Instrument-based approach}
\label{sec:instrument-based}

Following \citet{planck2014-a04}, we produce a map of all systematic effects (the ``systematics template'') at 70\,GHz and add this to realizations of the full-mission noise and the CMB from FFP8. To quantify the impact of systematic effects in the foreground removal process, we use the corresponding systematic effects templates at 30\,GHz and 353\,GHz, and scale them to 70\,GHz with spectral indexes $\alpha_{30/70} = 0.063$ and $\beta_{353/70} = 0.0077$ \citep{planck2014-a13}. This is equivalent to assuming that our cleaning procedure leaves no synchrotron or dust contamination in the final 70-GHz map, while we evaluate the impact of the rescaled noise and systematic effects. We then extract the power spectra for the temperature and polarization components at $\ell < 30$. The spectra are calculated over the same sky region used to derive $\tau$ in \citet{planck2014-a13}.

To calculate the bias introduced by systematic effects on $\tau$, $r$, and $\log(A_\mathrm{s})$, we compare 1000 FFP8 realizations of the polarized CMB (for a fiducial value $\tau= 0.065$), plus white and $1/f$ noise including systematic effects, with 1000 similar simulations containing only CMB and noise. For each realization, we calculate the marginalized distributions for each of the three parameters $X$ ($=\tau$, $r$, or $A_{\rm s}$), and calculate the differences $\Delta X = X_{\rm syst.} - X_{\rm no syst.}$, which represent the bias introduced in the estimates of parameter $X$ by the combination of all systematic effects.

\begin{figure}[htbp!]
\includegraphics[width=\columnwidth]{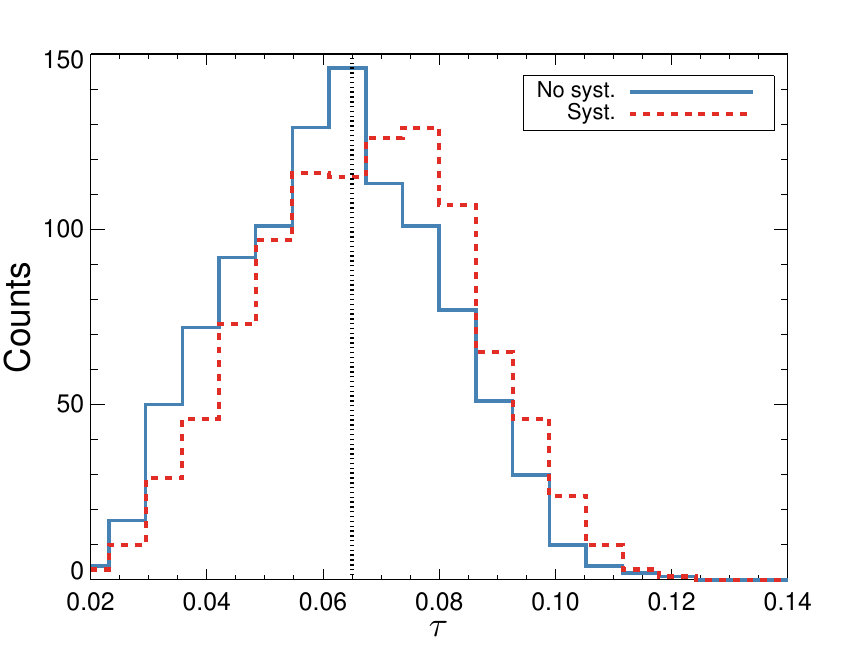} 
\caption{\label{fig:LFITauBias} Bias on $\tau$ due to systematic effects, specifically showing the distribution of $\tau$ with (red/dashed) and without (blue/solid) the systematics template. The vertical dotted line shows the input value to the simulation, $\tau = 0.065$.}
\label{fig:fig1}
\end{figure}

For the optical depth, we find a mean bias $\langle \Delta \tau \rangle = 0.005$ (Fig.~\ref{fig:LFITauBias}), or approximately 0.25 times the standard deviation of the value of $\tau$ measured by LFI \citep{planck2014-a15}. The positive measured bias may suggest a slight over-estimate of the measured value of $\tau$ caused by systematics. We also tested the sensitivity of the measured bias with the fiducial value of $\tau$. We repeated the analysis replacing the FFP8 CMB simulations with a set of realizations drawn from a cosmological model with $\tau = 0.05$ (all other parameters were fixed to the FFP8 values) and find no measurable difference from the previous case: $\langle \Delta \tau \rangle = 0.005$.

In the above analysis, the values of the scaling coefficients $\alpha_{30/70}$ and $\beta_{353/70}$ were fixed at the values measured in the real data, as appropriate in quantifying the impact of residual systematic effects in the released LFI maps and likelihood.  For a more general assessment of the component-separation and parameter-estimation procedures in the presence of systematic effects, these quantities should be re-estimated for each individual realization \citep[e.g., section~2.5 of][]{planck2014-a15}.

\subsection{Null-test-based approach}
\label{sec:null-test-based}

\begin{figure*}[htbp!]
\includegraphics[width=\textwidth]{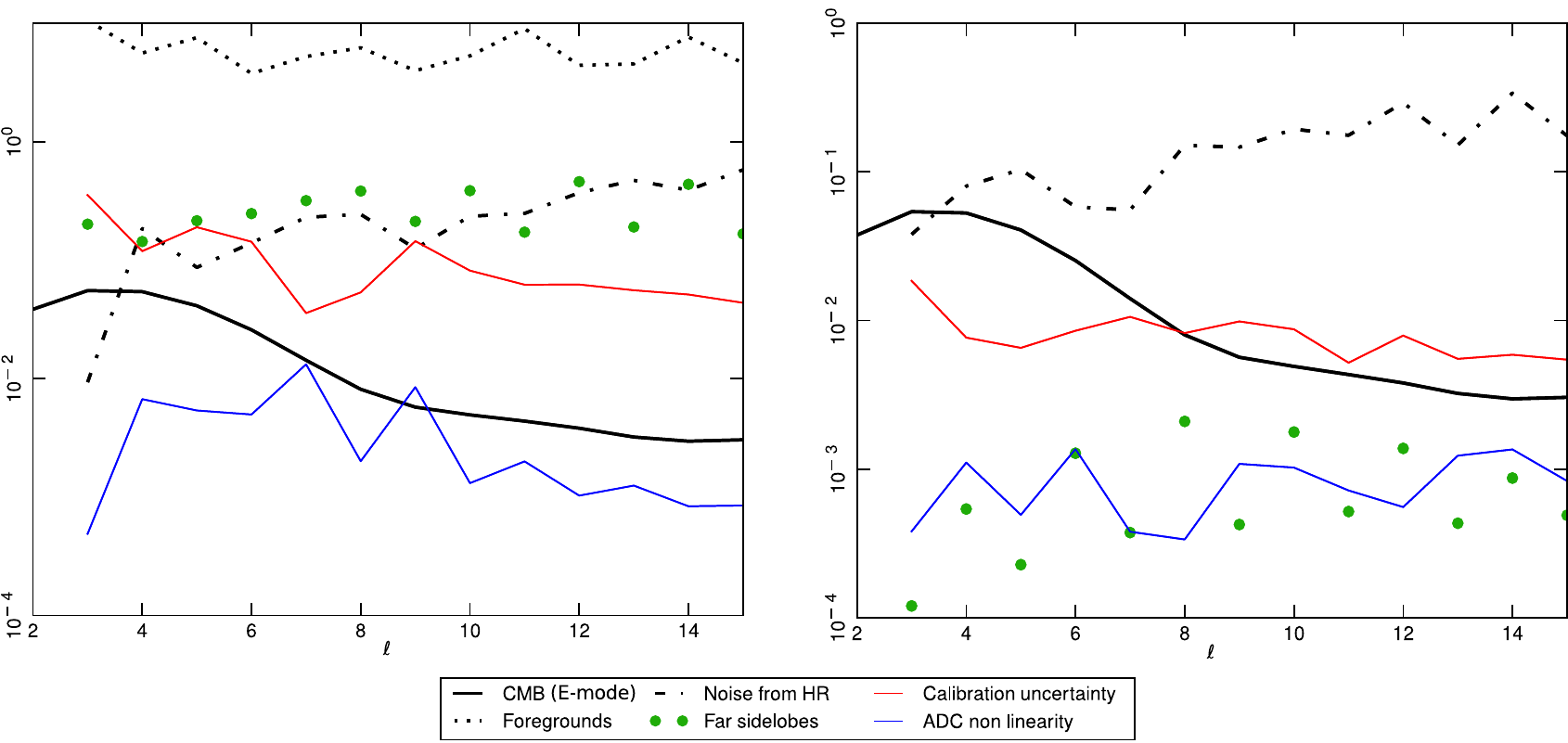} 
\caption{Power spectra ($D_\ell$ in $\mu$K$^2$) of systematic effects at 30\,GHz (left) and 70\,GHz (right), with each effect coded as indicated in the legend.}
\label{fig:lowell_spectrum_all_effects} 
\end{figure*}

The above analysis only probes systematic effects that are known a priori, and is limited by the accuracy of our instrument model. Furthermore, Fig.~\ref{fig:lowell_spectrum_all_effects} shows that the amplitude of systematic residuals at large scales is comparable to the noise at 30\,GHz, and somewhat lower than the noise at 70\,GHz. Therefore tests are also needed that do not rely on a model for the systematic effects. For this purpose, we use null maps of measurement data as ``templates'' of the LFI systematic effects, add to them simulated CMB realizations, and extract the value of $\tau$. This null-test-based approach has the advantage of using real data, instead of simulations, but it assumes that difference maps contain a level of contamination that is representative of that in the full maps. If systematic effects are correlated between Surveys or years, however, under- or over-estimates of some effects could result.

\begin{figure}[htbp!]
\includegraphics[width=\columnwidth]{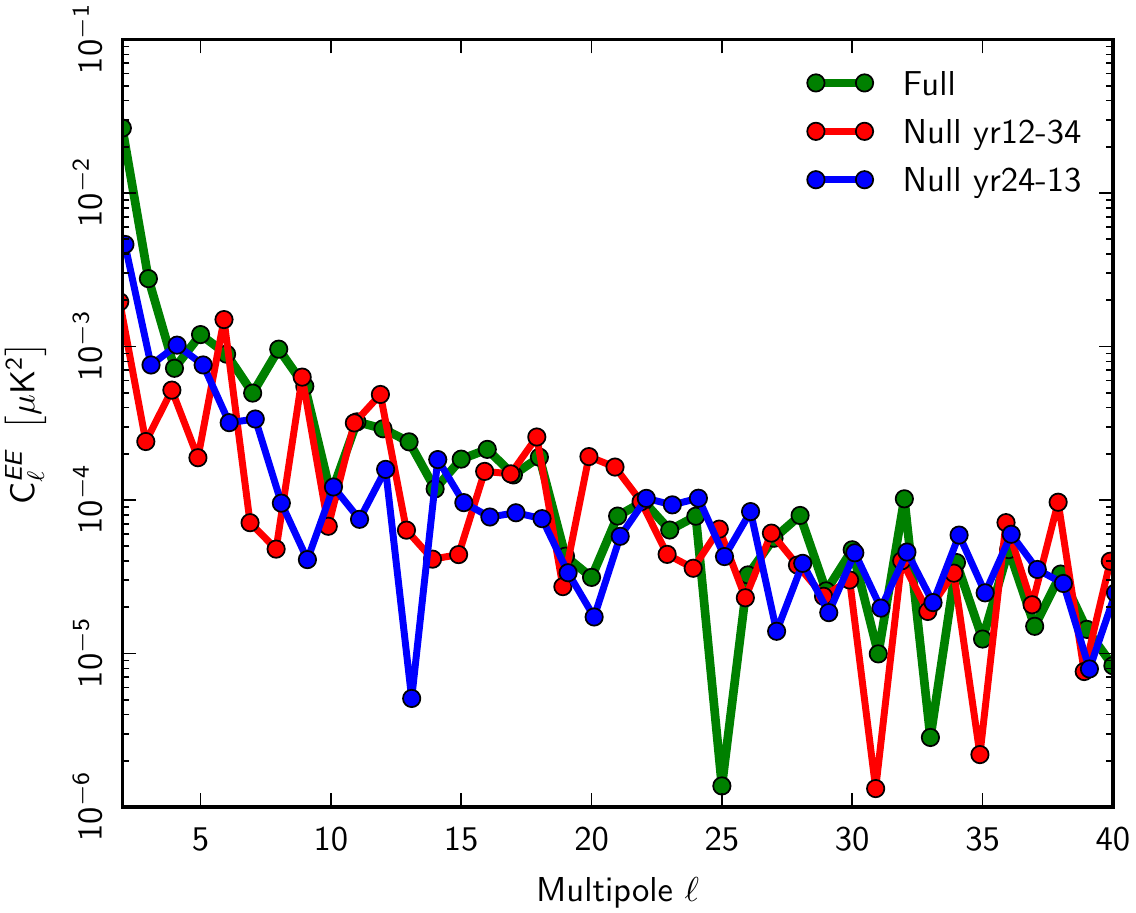} 
\caption{\label{fig:NULLsyst_sims} Power spectra for 70\,GHz of systematic effects from our ``instrument-based'' systematics simulations (with no noise) for: the full 4-year mission in green; the map difference
$(\mathrm{yr1} + \mathrm{yr2}) - (\mathrm{yr3} + \mathrm{yr4})$ in red;
and the map difference
$(\mathrm{yr2} + \mathrm{yr4}) - (\mathrm{yr1} + \mathrm{yr3})$ in blue.}
\end{figure}

We test this assumption using our ``instrument-based'' systematics model, comparing the power spectrum of the systematic effects predicted by our model for the full mission with the power spectra predicted for two combinations of half-mission null maps. Figure~\ref{fig:NULLsyst_sims} shows results for the 70-GHz channel. At $\ell<15$, the full-mission spectrum is higher on average by a factor of about 2.5 than the spectra of the difference maps, likely due to partial cancellation of common modes. However, the comparison indicates a reasonable consistency of the amplitudes of the three power spectra, within the observed scatter, at a level $\approx 10^{-3}$ for $\ell >2$. This indicates that applying the null-test-based approach to the real data provides a useful test of LFI data quality.

Next, we compute the power spectra for the full mission and for the half-mission differences using the measurement data.
Residuals in year-map differences (e.g., year 1$\,-\,$year 2) are more consistent with noise than residuals in Survey differences, as expected from the increased efficiency of the destriping algorithm \citep{planck2014-a07}; however, we find only marginal improvement when using 2-year map differences compared to combinations of single-year maps. Thus we adopt the latter, which allows us to evaluate the two half-mission combinations
\begin{align}
D_{\rm 12-34}\equiv& \,\, \frac{(\mathrm{yr1} + \mathrm{yr2}) - (\mathrm{yr3} + \mathrm{yr4})}{w8}, \\
D_{24-13}\equiv& \,\, \frac{(\mathrm{yr2} + \mathrm{yr4}) - (\mathrm{yr1} + \mathrm{yr3})}{w8},
\end{align}
where 
\begin{equation}
w8=\sqrt{N_{\rm hits}^{\rm tot}\times(1/N_{\rm hits}^{\rm yr1} + 1/N_{\rm hits}^{\rm yr2} + 1/N_{\rm hits}^{\rm yr3} + 1/N_{\rm hits}^{\rm yr4})}
\end{equation}
is a weight based on the number of hits per pixel, $N_{\rm hits}$, needed to equalize the noise in the two maps. We apply the same mask used for extraction of $\tau$ in the instrument-based simulations. In order to correct for the mask mode-coupling and the polarization leakage effect we used 
{\tt CrossSpect}, a {\tt MASTER}-like \citep{2002ApJ...567....2H} power spectrum estimator. 

\begin{figure}[htbp!]
\includegraphics[width=\columnwidth]{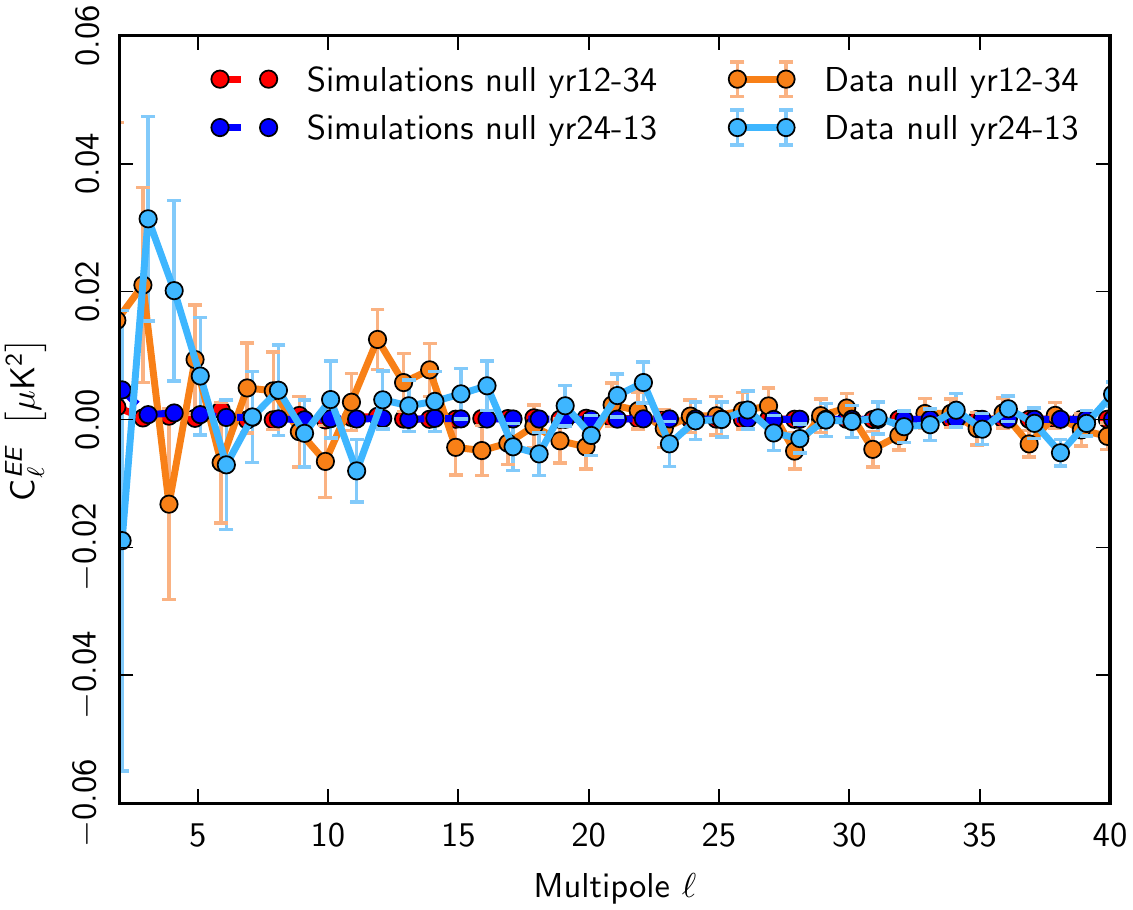} 
\caption{\label{fig:NULLsyst_data} Power spectrum of the measured 70-GHz null maps for the differences $D_{\rm 12-34}$ (orange) and $D_{24-13}$ (light blue). The scatter is dominated by the measurement noise. The error bars are computed from 1000 Monte Carlo simulations from FFP8 noise. Negative values are due to the fact that the noise bias has been removed. For comparison we also plot the systematics-only spectra (red and blue). }
\end{figure}

Figure~\ref{fig:NULLsyst_data} shows the 70-GHz auto-spectra of the difference maps $D_{\rm 12-34}$ and $D_{\rm 24-13}$ using real data, after removal of the noise bias. For comparison, in the same figure we also show the same spectra of Fig.~\ref{fig:NULLsyst_sims} from simulated systematic effects (note the different scale). As expected, the scatter for the data is much larger than the scatter for the models of systematic effects. To verify whether the scatter observed in the data is consistent with instrument noise, we compute statistical error bars at each multipole from 1000 realizations of FFP8 noise. Over the multipole range $2 \leq \ell \leq 29$, we find reduced $\chi^2$ values of 0.94 for $D_{\rm 12-34}$ (${\rm PTE}=0.55$), and 1.06 for $D_{\rm 24-13}$ (${\rm PTE}=0.38$), accounting for $\ell$-to-$\ell$ correlations via the empirical noise $C_\ell$ covariance matrix. Considering only the lowest multipoles, $2 \leq \ell \leq 9$, we find reduced $\chi^2$ values of 0.61 (${\rm PTE}=0.77$) and 1.25 (${\rm PTE}=0.26$), respectively.

The fact that the scatter is consistent with FFP8 noise confirms that the amplitude of systematic effects, while not negligible, is smaller than white and $1/f$ noise, even at large scales. This level of systematic effects could still lead to small biases in the cosmological parameters, so we perform an analysis of the recovery of $\tau$ in the presence of systematic effects also for the null-test-based templates. Since these are based on a single realization of the real data, we cannot use the approach of Sect.~\ref{sec:instrument-based} and marginalize over the CMB and noise realizations to obtain the average systematic impact on parameters. Rather, we proceed similarly to what was done in the above analysis of null-map power spectra. We start by calculating the $D_{\rm 12-34}$ and $D_{24-13}$ yearly differences for the 30 and 70\,GHz ``instrument-based'' systematic templates discussed above. We then generate two sets of 1000 CMB+noise and CMB+noise+simulated null maps, and analyse them as in the previous section. The resulting average bias on $\tau$ is $\approx 0.002$ and $\approx 0.003$, for the simulated $D_{\rm 12-34}$ and the $D_{24-13}$ combinations, respectively. In agreement with the power spectrum results of Fig.~\ref{fig:NULLsyst_sims}, we find that the null templates give rise to a lower bias than the full mission systematics templates.

\begin{figure}[htbp!]
\includegraphics[width=\columnwidth]{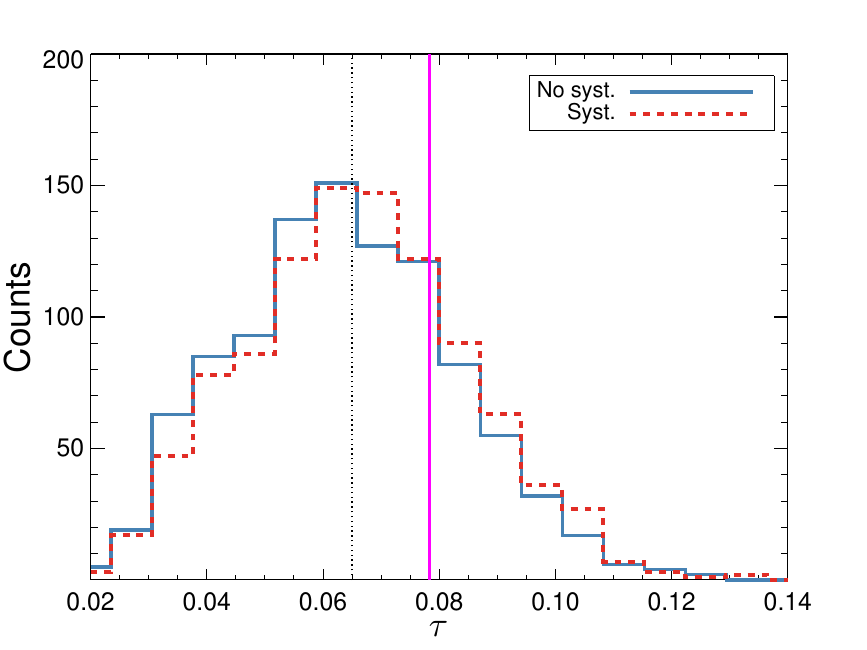} 
\includegraphics[width=\columnwidth]{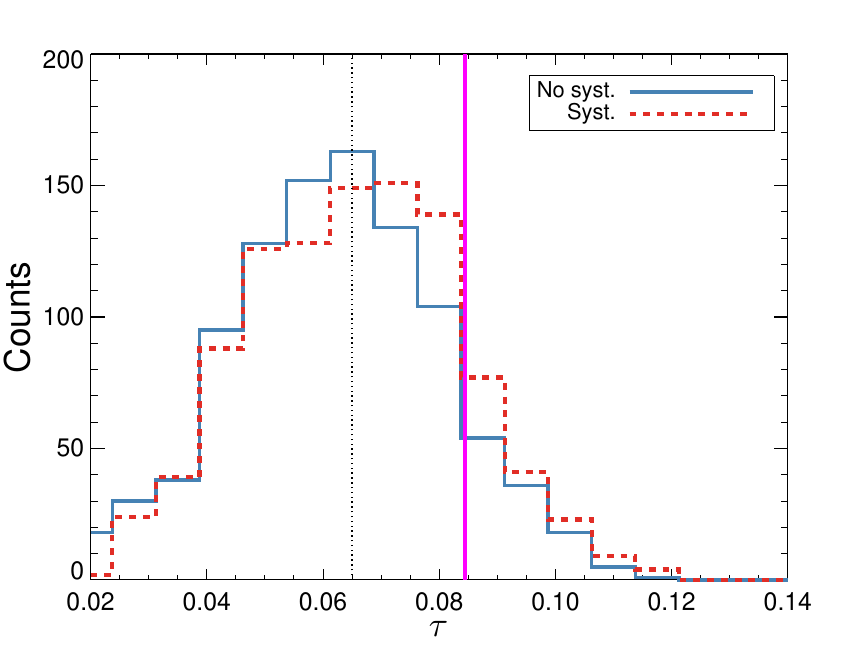}
\caption{\label{fig:top_down_params} Bias on $\tau$ due for the null-test-based systematics templates. The blue line shows the distribution of $\tau$ for the reference CMB+noise simulations, the red line shows the resulting distribution when including also the null systematic maps built from the ``instrument-based'' simulations. The vertical magenta line shows the value of $\tau$ estimated for the single CMB realization to which the null maps from actual data, as opposed to the null maps from the instrumental simulations, were added. The dotted line shows the input value to the simulation, $\tau = 0.065$. {\it Top\/}: $D_{\rm 12-34}$ null combination. {\it Bottom\/}: $D_{24-13}$ null combination.}
\end{figure}

\begin{figure*}[!ht]
\includegraphics[width=\textwidth]{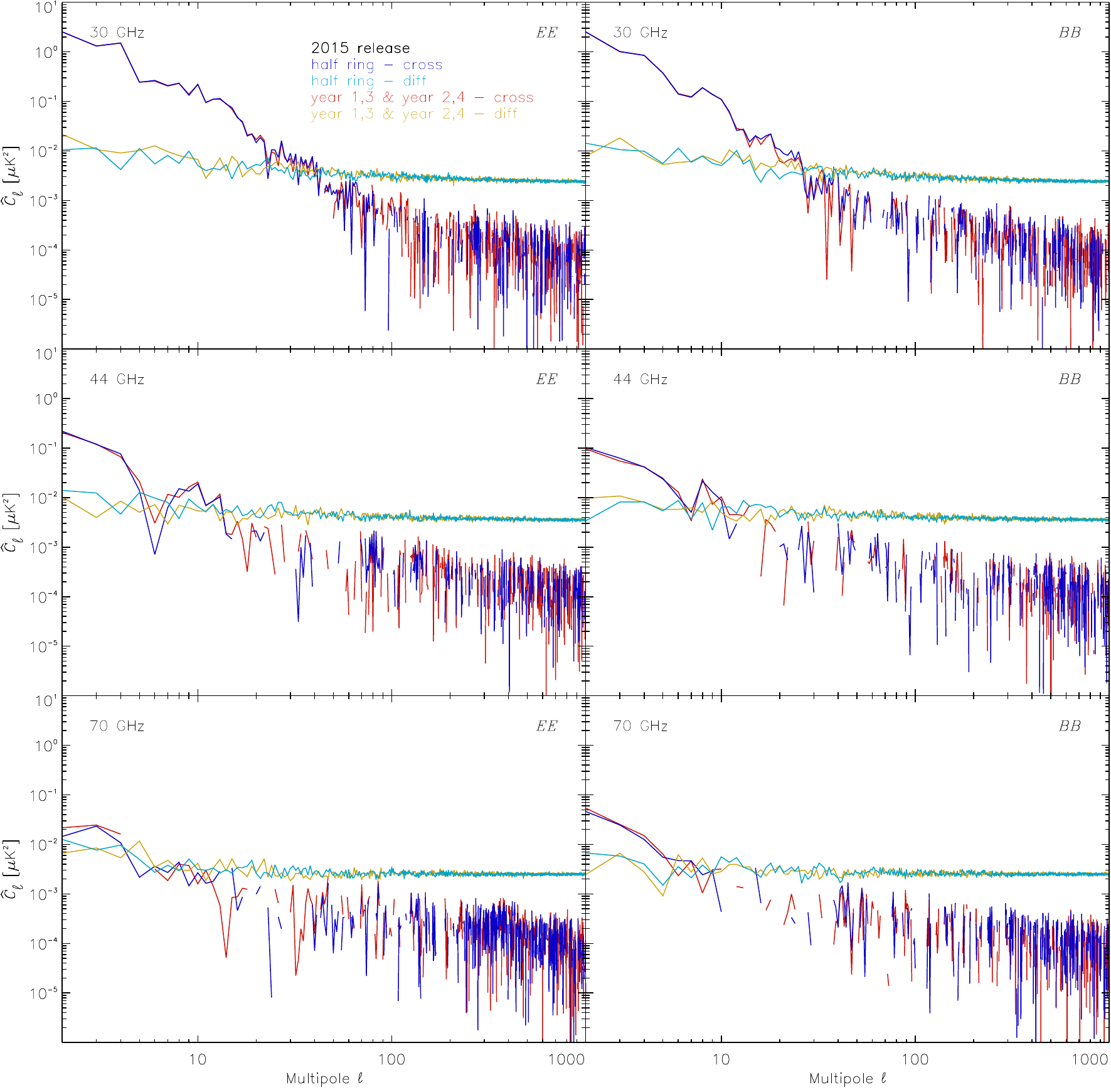} 
\caption{$EE$ and $BB$ differences and cross-spectra of the 2015 maps at 30, 44, and 70\,GHz, for different data splits. These maps are masked so that 43\,\% of the sky is used. This figure show similar plots to Fig.~\ref{fig:HFI_spectra} for LFI frequencies.}
\label{fig:LFI_spectra}
\end{figure*}

We then add a single CMB realization to the null-test-based templates, and estimate parameters for these maps as well. Fig.~\ref{fig:top_down_params} shows that the resulting estimates are well within the distributions. Note also that the bias in the case of $D_{\rm 12-34}$ is somewhat lower than in the case of $D_{24-13}$, consistent with the behaviour expected from simulations (Fig.~\ref{fig:NULLsyst_sims}). These results suggest that the instrument-based and null-test-based null maps have a consistent impact on power spectra and parameters, and provide support for the use of the instrument-based simulations in the cross-instrument analysis.

Figure~\ref{fig:LFI_spectra} shows $EE$ and $BB$ pseudo-power-spectra from the 2015 LFI data release used in this paper, showing the difference maps and cross-spectra (signal) over 43\,\% of the sky. Foregrounds are clearly visible in the cross-spectra only at $\ell<30$ at 30\,GHz and $\ell<5$ at 70\,GHz. As will be shown in Sect.~\ref{sec:compsep}, although the removal of the modest synchrotron foreground component at 70\,GHz may not be extremely accurate, its final uncertainty is well within the noise.

\section{Interfrequency calibration, component separation, and power spectra}
\label{sec:Tests}

The low-multipole polarization signals measured by \Planck\ include Galactic foregrounds.  In this section, we describe the combination of \Planck\ frequency bands that we use to separate the CMB signal from these foregrounds. First we revisit dipole residuals in the single-frequency maps to establish the level of precision of the inter-calibration error between the frequency bands of \Planck\ in both instruments. Then, we use the {\tt SMICA} component-separation code to show that the polarization signal at large angular scales consists mostly of CMB and two Galactic foreground components.  A third component does not improve the fit, and we show that a simple internal linear combination (ILC) method is sufficient to clean the CMB-dominated bands by regressing synchrotron with 30\,GHz and dust with 353\,GHz.  Finally, we describe the angular power spectra at low multipoles using two estimators and subtracting the foreground components.

\subsection{Interfrequency calibration}
\label{sec:interfrequencycal}

For the 44, 70, 100, 143, 217, and 353\,GHz channels, the photometric calibration is based on the ``orbital dipole,'' i.e., the modulation of the Solar dipole induced by the orbital motion of the satellite around the Solar System barycentre (see details in Appendix~\ref{sroll}).  The ``Solar dipole,'' induced by the motion of the Solar System barycentre with respect to the CMB, is stronger by an order of magnitude, and can be measured after masking the Galactic plane and removing foreground emission outside the mask.  If the uncertainties associated with foreground removal are negligible, the amplitude of the Solar dipole provides an excellent check on the relative calibration between different frequencies. 

Most of the higher-order terms discussed in \cite{Notari:2015kla} affect all CMB dipoles in the same way, and thus do not induce errors in the relative calibration of the Solar dipole with respect to the orbital one but can contribute small factors to the frequency calibration differencies.

The frequency-dependent kinematic-dipole-induced second-order quadrupole \citep{KamionKnox} has amplitude $2.0\times10^{-4} \mu$K at 100\,GHz, $3.9\times10^{-4} \mu$K at 143\,GHz, $8.2\times10^{-4} \mu$K at 217\,GHz, and $1.7\times10^{-3}\mu$K at 353\,GHz.  These corrections have a negligible impact on the calibration of the CMB channels.

Table~\ref{tab:dipole_direction_RD12} summarizes the HFI measurements of the Solar dipole from 100 to 353\,GHz.
\begin{table*}[tb]
\newdimen\tblskip \tblskip=5pt
\caption{Solar dipole amplitude and direction by frequency, as well as for the average (AVG) of the 100 and 143-GHz maps. Here dust emission is removed using the CMB-free 857-GHz map as a template.  Fits are performed using three different masks based on the $I_{857}$ temperature, corresponding to 37\,\%, 50\,\%, and 58\,\% of the sky.  Uncertainties are purely statistical.}
\label{tab:dipole_direction_RD12}
\vskip -6mm
\footnotesize
\setbox\tablebox=\vbox{
 \newdimen\digitwidth
 \setbox0=\hbox{\rm 0}
 \digitwidth=\wd0
 \catcode`*=\active
 \def*{\kern\digitwidth}
 \newdimen\signwidth
 \setbox0=\hbox{+}
 \signwidth=\wd0
 \catcode`!=\active
 \def!{\kern\signwidth}
\halign{\hbox to 2.0cm{#\leaderfil}\tabskip 1em&
  \hfil#\hfil\tabskip 1em&
  \hfil$#$\hfil\tabskip 1.5em&
   \hfil$#$\hfil\tabskip 1.5em&
    \hfil$#$\hfil\tabskip 0em\cr
\noalign{\doubleline}
\omit\hfil $\nu$\hfil&$I_{857}$ threshold&A&l&b\cr
\omit\hfil [GHz]\hfil&[M\,Jy\,sr\mo]&[\muK]&$[deg]$&$[deg]$\cr
\noalign{\vskip 3pt\hrule\vskip 5pt}
100&   2& 3361.25 \pm 0.06& 263.937 \pm 0.002& 48.2647 \pm 0.0008\cr
\omit& 3& 3361.22 \pm 0.05& 263.937 \pm 0.002& 48.2644 \pm 0.0006\cr
\omit& 4& 3361.46 \pm 0.04& 263.942 \pm 0.002& 48.2634 \pm 0.0006\cr
\noalign{\vskip 4pt}
143&   2& 3362.85 \pm 0.04& 263.913 \pm 0.001& 48.2629 \pm 0.0004\cr
\omit& 3& 3362.46 \pm 0.03& 263.910 \pm 0.001& 48.2647 \pm 0.0004\cr
\omit& 4& 3362.15 \pm 0.02& 263.914 \pm 0.001& 48.2664 \pm 0.0003\cr
\noalign{\vskip 4pt}
217&   2& 3366.56 \pm 0.06& 263.852 \pm 0.002& 48.2645 \pm 0.0008\cr
\omit& 3& 3365.37 \pm 0.05& 263.840 \pm 0.002& 48.2713 \pm 0.0007\cr
\omit& 4& 3364.38 \pm 0.04& 263.846 \pm 0.001& 48.2765 \pm 0.0006\cr
\noalign{\vskip 4pt}
353&   2& 3364.19 \pm 0.26& 263.385 \pm 0.009& 48.3191 \pm 0.0035\cr
\omit& 3& 3358.51 \pm 0.20& 263.399 \pm 0.007& 48.3710 \pm 0.0030\cr
\omit& 4& 3352.99 \pm 0.18& 263.452 \pm 0.006& 48.4207 \pm 0.0028\cr
\noalign{\vskip 4pt}
545&   2& 3398.72 \pm 1.99& 255.890 \pm 0.070& 48.1571 \pm 0.0281\cr
\omit& 3& 3350.02 \pm 1.56& 255.878 \pm 0.058& 48.6087 \pm 0.0245\cr
\omit& 4& 3292.11 \pm 1.38& 256.181 \pm 0.052& 49.1146 \pm 0.0232\cr
\noalign{\vskip 4pt}
AVG&   2& 3362.05 \pm 0.04& 263.925 \pm 0.001& 48.2641 \pm 0.0004\cr
\omit& 3& 3361.84 \pm 0.03& 263.924 \pm 0.001& 48.2650 \pm 0.0004\cr
\omit& 4& 3361.80 \pm 0.02& 263.928 \pm 0.001& 48.2652 \pm 0.0003\cr
\noalign{\vskip 3pt\hrule\vskip 5pt}}}
\endPlancktablewide
\end{table*}
At 100\,GHz, the direction and amplitude of the Solar dipole remain within two standard deviations as the sky fraction changes from 37\,\% to 58\,\%.  At 143\,GHz, the direction is the same as at 100\,\GHz\, within $2\,\sigma$ of the statistical noise. Higher frequencies show drifts in direction with changing sky fraction that are larger. The amplitude also drifts when the sky fraction is reduced.  

\begin{figure}[htbp!]
\includegraphics[width=\columnwidth]{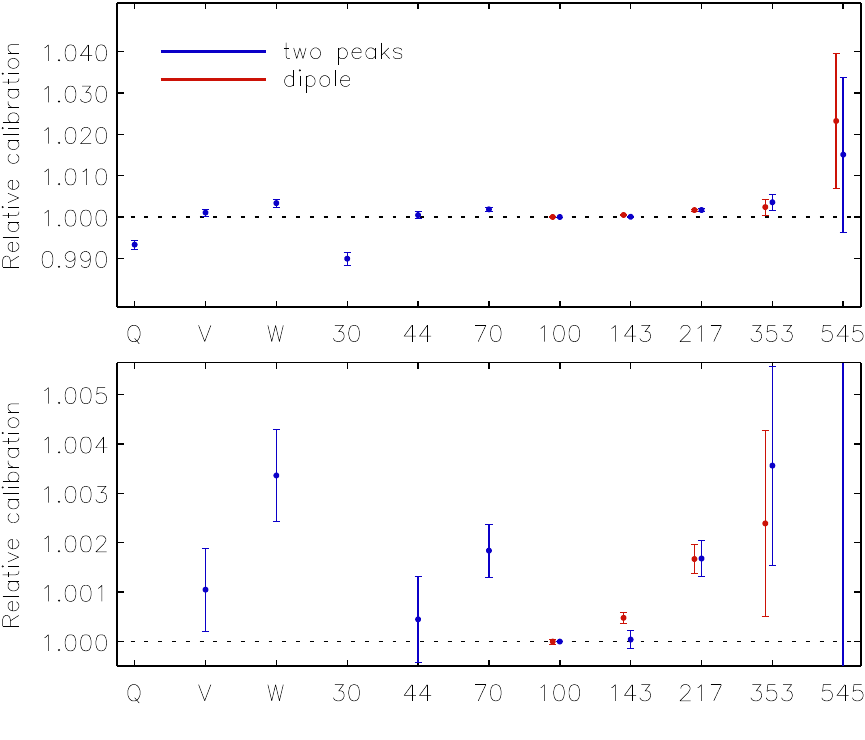} 
\caption{Relative calibration based on measurement of the first two acoustic peaks (blue) and on the Solar dipole (red). The bottom panel is a zoom of the top panel.  We use 100\,GHz as the reference frequency for the dipole calibration method.  In this plot, \Planck\ 30-GHz and WMAP Q-band data may be affected by their relatively low angular resolution.}
\label{fig:intercalib}
\end{figure}

The excellent stability of the Solar dipole amplitude and direction with changes in the sky fraction implies that the orbital dipole calibration at each frequency is better than 0.1\,\% for \Planck\ CMB frequencies and WMAP W-band, as shown in Fig.~\ref{fig:intercalib} and Table~\ref{tab:dipole}, which give the properties of the Solar dipoles measured at \Planck\ frequencies and in the WMAP bands.

Table~\ref{tab:dipole} gives the Solar dipole and the relative amplitude as measured by \Planck\ from 44 to 545\,GHz, plus that of WMAP at 94\,GHz, as well as the relative calibration determined from the first and second peak amplitude of the CMB spectrum.  The relative amplitudes determined both ways are plotted in Fig.~\ref{fig:intercalib}.  The relative calibration determined from the first two peaks (specifically, over $110\le\ell\le500$) of the CMB power spectrum, is very close to that determined from the Solar dipole.  The difference in calibration between the Solar dipole and the first two CMB peaks gives an upper limit on the residual transfer function error relative to 100\,GHz, between $\ell=1$ and $\ell \approx 300$, also shown in Table~\ref{tab:dipole}.

\begin{table*}[tb]
\newdimen\tblskip \tblskip=5pt
\caption{Relative calibration between the Solar dipole and the first and second peaks ($\ell=110$--500) in the CMB power spectrum with respect to 100\,GHz. The ``Transfer function'' in column~7 is the difference between the relative responses to the dipole and to the CMB first acoustic peak ($\ell=2$--300).}
\label{tab:dipole}
\vskip -6mm
\footnotesize
\setbox\tablebox=\vbox{
 \newdimen\digitwidth
 \setbox0=\hbox{\rm 0}
 \digitwidth=\wd0
 \catcode`*=\active
 \def*{\kern\digitwidth}
 \newdimen\dpwidth
 \setbox0=\hbox{.}
 \dpwidth=\wd0
 \catcode`|=\active
 \def|{\kern\dpwidth}
 \newdimen\signwidth
 \setbox0=\hbox{+}
 \signwidth=\wd0
 \catcode`!=\active
 \def!{\kern\signwidth}
\halign{\hbox to 2.5cm{#\leaderfil}\tabskip 2em&
 \hfil#\hfil\tabskip 0.5em&
 \hfil#\hfil&
 \hfil#\hfil&
 \hfil#\hfil\tabskip 2em&
 \hfil#\hfil\tabskip 2em&
 \hfil#\hfil\tabskip 0em\cr
\noalign{\doubleline}
\omit&\multispan4\hfil Solar dipole\hfil&First and second peaks\cr %
\noalign{\vskip -3pt}
\omit&\multispan4\hrulefill&\hrulefill&Transfer function\cr
\noalign{\vskip 1pt}
\omit\hfil Frequency\hfil&Amplitude&$l$&$b$&Rel. amplitude&Rel. amplitude&$\Delta\ell=2$--$300$\cr
\noalign{\vskip 3pt} 
\omit\hfil [GHz]\hfil&[\muK]&[deg]&[deg]&[\%]&[\%]&[\%]\cr
\noalign{\vskip 3pt\hrule\vskip 5pt}
\noalign{\vskip 2pt}
*44&         \dots&  \dots&  \dots&  \dots& 0.05&  \dots\cr	
*70&       3363.1*&263.97*&48.26**&  !0.06& 0.18&  !0.13\cr
*94 (WMAP)&3355|**&  \dots&  \dots&$-$0.19& 0.34&  !0.52\cr
100&       3361.25&263.937&48.2647& !0.00&   Ref.&   !Ref.\cr
143&       3362.85&263.913&48.2629&  !0.05& 0.00&  $-$0.04\cr
217&       3366.56&263.852&48.2645&  !0;16& 0.17&  !0.01\cr
353&       3364.19&263.385&48.3191&  !0.09& 0.36&  !0.27\cr
\noalign{\vskip 3pt\hrule\vskip 5pt}}}
\endPlancktablewide
\end{table*}

The 545-GHz channel is difficult to calibrate on the orbital dipole because the orbital dipole is so weak with respect to the dust emission.  Instead, the 545-GHz channel was calibrated on Uranus and Neptune \citep{planck2014-a09}.  Nevertheless, the (much stronger) Solar dipole and acoustic peak amplitudes {\it can\/} be compared with those of the CMB channels, once again after cleaning dust using the 857-GHz template.  The result is that the planet calibrations agree with the CMB calibrations within 1.5 or 2.5\,\% (for 545 and 847\,GHz, respectively).  The absolute calibration uncertainty using planets was estimated at 5\,\%, and is now shown to be within 2\,\% of the photometric CMB calibration. Since the {\it Herschel\/} observatory was calibrated with the same planet models, inter-calibration between {\it Herschel\/} and \Planck\ is also at the $2\,\%$ level \citep{bertincourt2016}.

In summary, absolute calibration of the 100-GHz channel on the orbital dipole is insensitive to sky fraction (and thus not affected by Galactic foreground removal).  Based on that absolute calibration, we use the Solar dipole to compare the calibration of the other \Planck\ CMB channels and WMAP W-band.  We also compare the calibration of the CMB channels using the CMB anisotropies in the multipole range 110--500. There is evidence of some potential orbital dipole calibration errors due to the dust foreground at frequencies above 100\,GHz, but also of very stable behavior of the transfer functions with \Planck\ frequencies from 70 to 217\,GHz. At 353\,GHz, we still see a 0.3\,\% discrepancy between various different calibrations, and for WMAP W band a 0.5\,\% difference. Whether the origin of these discrepancies is in the calibration itself or is a result of transfer function errors cannot be determined by this analysis.

\subsection{Diffuse component separation}
\label{sec:compsep}

In this section we discuss the effects of polarized diffuse component separation on the CMB spectrum (point sources can be ignored for low-multipole work).  To begin, we use the (blind) {\tt SMICA} code \citep{planck2013-p06}, with no assumptions about the number and properties of the foreground components, to establish the number and properties of the foreground components that must be removed, and the fraction of sky $f_{\rm sky}$ to use. It turns out that two polarized foreground components are enough for polarization, i.e., including a third foreground component does not improve the fit.  Not surprisingly, the two components are easily identified with dust and synchrotron emission.  Changing $f_{\rm sky}$ from 0.4 to 0.5 makes essentially no difference in the resulting CMB power spectrum, but changing to 0.6 or 0.7 makes a substantial difference. This leads us to adopt $f_{\rm sky} = 0.5$.

\begin{figure}[htbp!]
\includegraphics[width=\columnwidth]{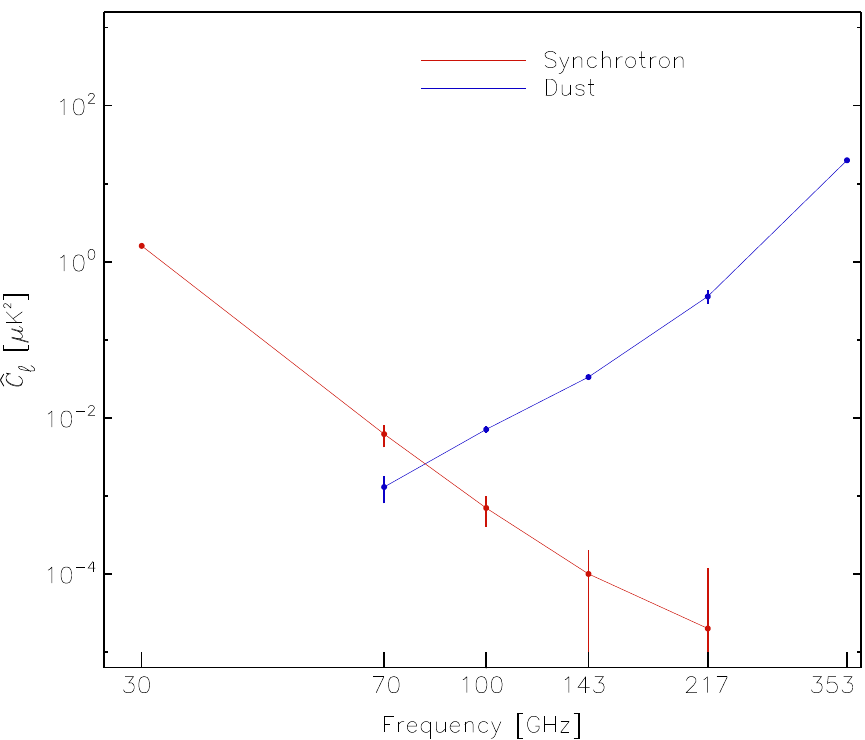} 
\caption{Dust correction to $C_{\ell}$ in $\mu$K$^2$, using the 353-GHz channel as a template (in blue), along with the synchrotron correction to $C_{\ell}$ in $\mu$K$^2$ using the 30-GHz channel as a template (in red).  These are plotted for the 70--217\,GHz channels as a function of frequency. Note that the points for synchrotron at 143 and 217\,GHz are very low and can only be considered as upper limits.}
\label{fig:ILCCOEFRD12}
\end{figure}

The two likelihood methods described in Sect.~\ref{sec:analysis} use internal linear combination (ILC) or template fiting to remove synchrotron and dust emission.  The 30-GHz map is used as a template for synchrotron emission that is uncorrelated with dust, and the 353-GHz map is used as a template for dust emission.  Table~\ref{tab:coeffs} shows the ``projection coefficients'' used by the different likelihoods.  Table~\ref{tab:coeffs} also gives the {\tt SMICA} projection coefficients for two blind components, and shows that the diffuse-component-separation procedure is very stable between the different approaches for 100 and 143\,\GHz. At 70\,GHz, the dispersion is larger, but if no synchrotron component were explicitly removed, the synchrotron that is correlated with dust would still be taken care of; additionally, the uncorrelated part would have no effect to first order on the $100\times143$\,GHz cross-spectra, because the synchrotron signal is very weak at 143\,GHz, as shown in Fig.~\ref{fig:ILCCOEFRD12}.

\begin{table*}[tb] 
\newdimen\tblskip \tblskip=5pt
\caption{Projection coefficients used by the two likelihood methods described in Sect.~\ref{sec:analysis}, as well as {\tt SMICA}, for removal of dust and synchrotron emission using the 353- and 30-GHz maps, respectively. In all these cases, $f_{\rm sky} = 0.5$.}
\label{tab:coeffs}
\vskip -6mm
\footnotesize
\setbox\tablebox=\vbox{
\newdimen\digitwidth
\setbox0=\hbox{\rm 0}
\digitwidth=\wd0
\catcode`*=\active
\def*{\kern\digitwidth}
\newdimen\signwidth
\setbox0=\hbox{.}
\signwidth=\wd0
\catcode`!=\active
\def!{\kern\signwidth}
\halign{\hbox to 1.5 cm{#\leaderfil}\tabskip 1.5em&
\hfil#\hfil\tabskip 0.5em&
\hfil#\hfil&
\hfil#\hfil&
\hfil#\hfil&
\hfil#\hfil&
\hfil#\hfil\tabskip 1.5em&
\hfil#\hfil\tabskip 0.5em&
\hfil#\hfil&
\hfil#\hfil&
\hfil#\hfil&
\hfil#\hfil&
\hfil#\hfil\tabskip 0pt\cr
\noalign{\doubleline}
\omit&\multispan5\hfil Dust\hfil&\multispan5\hfil Synchrotron\hfil\cr
\noalign{\vskip -3pt}
\omit&\multispan5\hrulefill&\multispan5\hrulefill\cr
\omit\hfil Frequency\hfil&{\tt Lollipop}&{\tt SimBaL}&{\tt SMICA}&mean&error&{\tt Lollipop}&{\tt SimBaL}&{\tt SMICA}&mean&error\cr
\noalign{\vskip 3pt}
\omit\hfil [GHz]\hfil\cr
\noalign{\vskip 3pt\hrule\vskip 5pt}
\noalign{\vskip 2pt}
*70&0.0084&0.0060&0.0095&0.0080&16.7\,\%&0.0679&0.0520&0.0621&0.0621&16\,\%\cr
100&0.0183&0.0193&0.0183&0.0186&*4.4\,\%&0.0390&0.0200&0.0210&0.0267&20\,\%\cr
143&0.0399&0.0427&0.0399&0.0408&*2.3\,\%&0.0199&0.0076&0.0080&0.0118&45\,\%\cr
\noalign{\vskip 3pt\hrule\vskip 5pt}}}
\endPlancktable
\end{table*}

Table~\ref{tab:comsep} gives the average value of the power spectrum removed for each foreground at $\ell =4$, at the peak of the $EE$ reionization feature, together with the associated uncertainties computed with the ILC method. The uncertainties are always lower than the F-$EE$ signal, by more than an order of magnitude.  Figure~\ref{fig:ILCCOEFRD12} shows this graphically.  

At the two \Planck\ frequencies with the lowest noise, 100 and 143\,GHz, the level of synchrotron emission is lower than that of dust by factors of 10 and 300, respectively.  For 100\,GHz, not removing the synchrotron would introduce a bias of less than 6\,\% of F-$EE$. 
The 70-GHz spectrum is limited by noise rather than the large uncertainty (16\,\%) in the synchrotron foreground removal, even though this is the largest foreground removal uncertainty in the \Planck\ CMB channels for the reionization peak.

\begin{table}[tb]
\newdimen\tblskip \tblskip=5pt
\caption{For each CMB frequency, the amplitude of the power spectrum $D_\ell$ at $\ell=4$ that is removed for the dust and synchrotron foregrounds. These are computed from the 353 and 30-GHz power spectra at $\ell=4$; uncertainties are scaled appropriately by the projection coefficients and relative errors.}
\label{tab:comsep}
\vskip -6mm
\footnotesize
\setbox\tablebox=\vbox{
 \newdimen\digitwidth 
 \setbox0=\hbox{\rm 0}
 \digitwidth=\wd0
 \catcode`*=\active
 \def*{\kern\digitwidth}
  \newdimen\signwidth
  \setbox0=\hbox{.}
  \signwidth=\wd0
  \catcode`!=\active
  \def!{\kern\signwidth}
\halign{\hbox to 2.0 cm{#\leaderfil}\tabskip 1em&
    \hfil#\hfil&
    \hfil#\hfil&
    \hfil#\hfil&
    \hfil#\hfil\tabskip 0em\cr
\noalign{\doubleline}
\omit&\multispan2\hfil Dust\hfil&\multispan2\hfil Synchrotron\hfil\cr %
\noalign{\vskip -3pt}
\omit&\multispan2\hrulefill&\multispan2\hrulefill\cr
\omit\hfil Frequency\hfil& Mean& Uncertainty& Mean& Uncertainty\cr
\noalign{\vskip 3pt} 
\omit\hfil [GHz]\hfil&[$\mu$K$^2$]&[$\mu$K$^2$]&[$\mu$K$^2$]&[$\mu$K$^2$]\cr
\noalign{\vskip 3pt\hrule\vskip 5pt}
\noalign{\vskip 2pt}
*70&0.0041&0.0010&0.019*&0.005*\cr
100&0.0227&0.0020&0.0036&0.0011\cr
143&0.106*&0.0052&0.0007&0.0004\cr
\noalign{\vskip 3pt\hrule\vskip 5pt}}}
\endPlancktable
\end{table}

The spatial variation of the dust spectral energy distribution at high latitude has been analysed on 400\,deg$^2$ patches, and is smaller than the ILC uncertainty \citep{planck2014-a32}.
 
\begin{figure}[htbp!]
\includegraphics[width=\columnwidth]{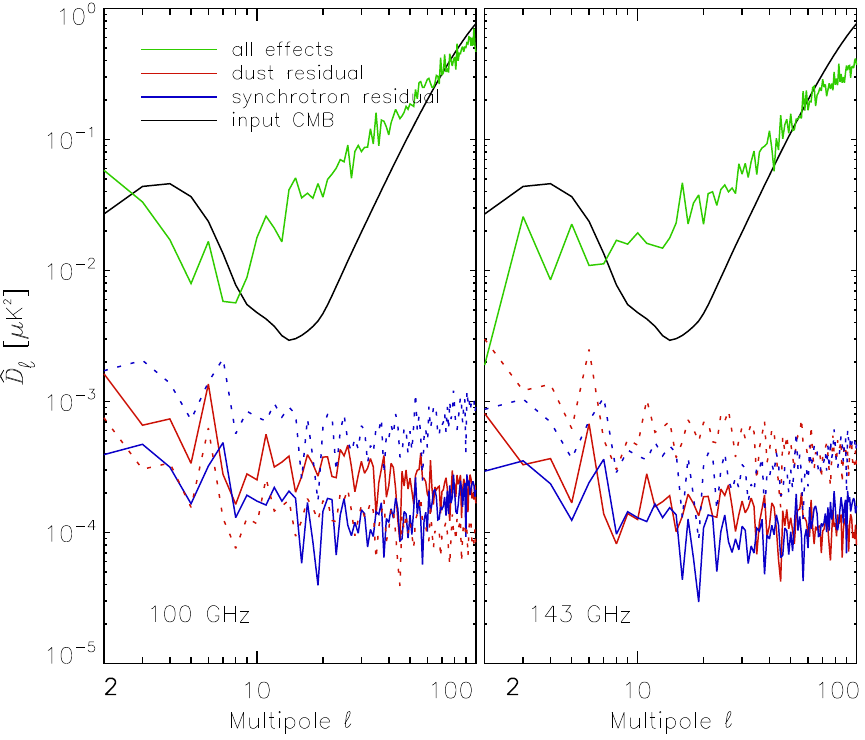} 
\caption{Residual errors in $EE$ from component separation, estimated from the scatter of the component-separation coefficients. The fiducial $EE$  spectrum and the noise plus systematics residuals (green line) are also shown.}
\label{fig:plcompsep}
\end{figure}

Figure~\ref{fig:plcompsep} shows the level of residual foregrounds. It is clear that the component-separation errors propagated to the power spectra are much lower than the sum of all systematic effects, and have a negligible impact on the determination of $\tau$.

\subsection{Power spectra}
\label{sec:powerspectra}
Likelihood analyses for CMB cosmological parameters were originally developed for CMB experiments that were dominated by CMB signal and detector noise, in contrast to more recent experiments with much lower noise in which residuals from instrumental systematic effects and foreground subtraction dominate over the detector noise, especially at low multipoles. Such analysis approaches rely on pixel-pixel covariance matrices coming from a large number of simulations of CMB and noise, which are assumed to be two statistically independent Gaussian fields for the purposes of these likelihood codes. Pixel-based likelihoods can also account for noise correlations and masking of the sky.
In the \Planck\ likelihoods, the ``noise'' description combines the detector and readout chain white noise, some correlated fraction due to $1/f$ noise, and residuals of systematic effects that have statistical properties that are not well understood. These systematic effect residuals thus need to be brought to levels much lower than the Gaussian noise.

\subsubsection{Cross-spectra at very low multipoles}
\label{sec:crossspectra}

\begin{figure}[htbp!]
\includegraphics[width=\columnwidth]{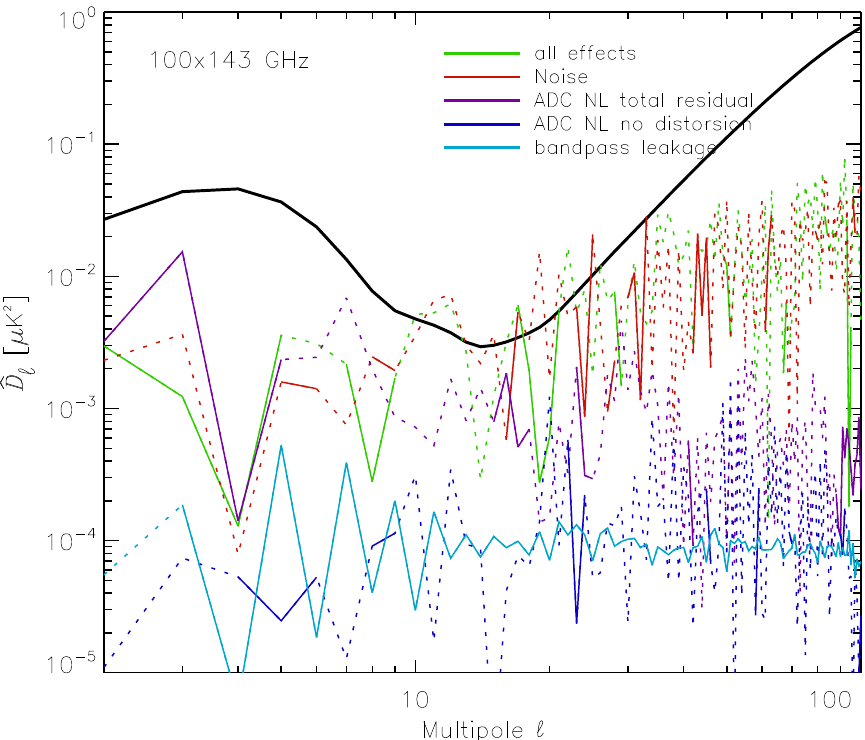} 
\caption{Similarly to Fig.~\ref{fig:DIFFMERGE}, residual $EE$ cross-power spectra of systematic effects from the HFI pre-2016 E2E simulations computed on 50\,\% of the sky are shown for $100 \times 143$. Dotted lines are used to join two multipoles when one is negative.}
\label{fig:domergemapcross}
\end{figure}

\begin{table*}[!ht]
\newdimen\tblskip \tblskip=5pt
\caption{$\langle D_{\ell}\rangle_{\rm rms}$ over $2\le\ell\le8$ for auto-spectra simulations of the ADC-induced dipole distortion and noise at 100 and 143\,GHz (Fig.~\ref{fig:DIFFMERGE}), together with simulations of $100\times143$ $EE$ cross-spectra (Fig.~\ref{fig:domergemapcross}) and data $BB$ cross-spectra (bottom panel of Fig.~\ref{fig:100143}) tracing final residuals and noise (signal is negligible).}
\label{tab:dell}
\vskip -6mm
\footnotesize
\setbox\tablebox=\vbox{
\newdimen\digitwidth
\setbox0=\hbox{\rm 0}
\digitwidth=\wd0
\catcode`*=\active
\def*{\kern\digitwidth}
\newdimen\signwidth
\setbox0=\hbox{.}
\signwidth=\wd0
\catcode`!=\active
\def!{\kern\signwidth}
\halign{\hbox to 3.5 cm{#\leaderfil}\tabskip 2em&
        \hfil#\hfil&
        \hfil#\hfil&
        \hfil#\hfil&
        \hfil#\hfil&
        \hfil#\hfil&
        \hfil#\hfil&
        \hfil#\hfil&
        \hfil#\hfil&
        \hfil#\hfil&
        \hfil#\hfil&
        \hfil#\hfil&
        \hfil#\hfil\tabskip 1em\cr
\noalign{\doubleline}
\noalign{\vskip 3pt}
\omit&\multispan3\hfil\sc Simulated auto-spectra\hfil&\sc Data $BB$\cr
\noalign{\vskip -3pt}
\omit&\multispan3\hrulefill&\hrulefill\cr
\omit&100&143&$100\times143$&$100\times143$\cr
\noalign{\vskip 3pt}
\omit\hfil &[10$^{-3}\mu$K$^2$]&[10$^{-3}\mu$K$^2$]&[10$^{-3}\mu$K$^2$]&[10$^{-3}\mu$K$^2$]\cr
\noalign{\vskip 3pt\hrule\vskip 5pt}
\noalign{\vskip 2pt}
ADC dipole distortion&*9&13&6.2&\cr
\noalign{\vskip 4pt}
Noise&                         15&*8&2.0\cr
\noalign{\vskip -18pt}
\omit&&&&3.0\cr
\noalign{\vskip 11pt}
\noalign{\vskip 3pt\hrule\vskip 5pt}}}
\endPlancktable
\end{table*}
Figure~\ref{fig:plmapresiduspectre} shows that systematic effects in the 100-GHz channel are below $10^{-3}\,\mu{\rm K}^2$ for $4<\ell<100$.  At $\ell = 2$ and 3, the ADC-induced dipole distortion dominates the 
systematic effects in simulated auto-spectra (Fig.~\ref{fig:DIFFMERGE}).  Figure~\ref{fig:domergemapcross} shows that the total level of simulated systematic effects (purple line) in the $100 \times 143$ cross-spectrum is significantly reduced compared to the auto-spectra of both frequencies (Fig.~\ref{fig:DIFFMERGE}) for most 
multipoles relevant for the reionization feature. In the first two columns of Table~\ref{tab:dell} we give $\langle D_{\ell}\rangle_{\rm rms}$ for $2\le\ell\le8$ for the auto-spectrum of the ADC-induced dipole distortion and noise from simulations (Fig.~\ref{fig:DIFFMERGE}). The third column gives the same $\langle D_{\ell}\rangle_{\rm rms}$ for the simulated ADC induced dipole distortion $EE$ cross-spectra $100 \times 143$
(Fig.~\ref{fig:domergemapcross}) tracing the final residuals. The fourth column gives this same quantity for the $BB$ data cross-spectrum (bottom panel of Fig.~\ref{fig:100143}) for noise and residual systematics after removal of the average of the simulation of systematic effects (where the expected signal is very small) for QML2.

The steps in the reduction of the ADC residuals from the simulation auto-spectra ($9 \times 10^{-3}$ and $13\times 10^{-3}\,\mu{\rm K}^2$) to the cross-power spectrum ($6\times 10^{-3}\,\mu{\rm K}^2$) and finally the QML $BB$ data cross-spectrum of residual systematics and noise ($3\times 10^{-3}\,\mu{\rm K}^2$) lead us to only a small excess with respect to the expected noise ($3.0\times 10^{-3}$ to $2.0\times 10^{-3}\,\mu{\rm K}^2$).

Cross-spectra calculated from two different detsets within a single frequency, for both 100 and 143\,GHz, have a higher level of residual systematics than those from cross-frequency detsets, as discussed below.
As explored in Sect.~\ref{sec:compsep}, $100 \times 143$ $EE$ cross-spectra can be cleaned of dust and of the correlated synchrotron fraction (60\,\%) with the 353-GHz template, and have negligible synchrotron power remaining, because synchrotron emission is so weak at 143\,GHz.  We also described in Sect.~\ref{sec:compsep} why we adopted $f_{\rm sky} = 0.5$. For these reasons, we choose the full frequency $100\times143$ cross-spectra and $f_{\rm sky} = 0.5$ as our baseline.

\subsubsection{Estimators}
\label{sec:estimators}
We use both pseudo-$C_\ell$ (PCL) and quadratic maximum likelihood (QML) estimators to extract cross-spectra from maps. 
Two versions of the PCL cross-spectra were produced, using {\tt Xpol} \citep{2005MNRAS.358..833T} and {\tt Spice}, a MASTER-like code \citep{2002ApJ...567....2H}.

QML estimators have been widely discussed in the literature \citep[e.g.,][]{Tegmark:1996,Bond:1998,Efstathiou:2004,Efstathiou:2006}. QML auto-spectra are close to optimal, but must be corrected for noise bias. The removal of this bias requires an accurate estimate of the pixel noise covariance matrix $N_{ij}$. In analogy with QML auto-spectra, it is straightforward to define a quadratic cross-spectrum estimator that is unbiased. The resulting cross-spectrum estimator will not have minimum variance, but nevertheless we retain the nomenclature ``QML'' to distinguish it from the PCL estimators described above.

For two maps ``$a$'' and ``$b$,'' the cross-spectrum estimate is defined as
\begin{equation}
y^{ab} _{\ell r} = x^{a}_i x^{b}_j {\tens{E}}^{ab}_{\ell r, ij}\, ,
\label{QML1}
\end{equation}
where $i$ and $j$ are pixel numbers, and $\ell$ and $r$ form a paired
index with $\ell$ denoting multipole number and $r$
denoting spectrum type (e.g.\ $TT$, $TE$, $EE$ or $BB$). The matrix $\tens{E}$ here is
\begin{equation}
 \tens{E}^{ab}_{\ell r} = 
{1 \over 2}(\check{\tens{C}}^a)^{-1} {\partial \tens{C} \over \partial
  C_{\ell r}} (\check{\tens{C}}^b)^{-1},  \label{QML2}
\end{equation}
where the covariance matrix $\tens{C}=\tens{S}+\tens{N}$ and $\check{\tens{C}}$ is a ``reshaped'' covariance matrix of the form
\begin{equation}
 \check{\tens{C}} 
 =  {\left ( \begin{array}{ccc}
        \tens{C}^{TT}&  0&  0  \\
        0& \tens{C}^{QQ}& \tens{C}^{QU} \\
        0& \tens{C}^{UQ}& \tens{C}^{UU} 
       \end{array} \right ) },  \label{QML3}
\end{equation}
that does not mix temperature estimates with polarization estimates.  Provided that the noise between maps $a$ and $b$ is uncorrelated, the expectation value of Eq.~(\ref{QML1}) is 
\begin{equation}
 \langle y_{\ell r} \rangle 
 =    \check{\tens{F}}^{ab}_{\ell r \ell^\prime r^\prime}
 C_{\ell^\prime r^\prime} , \label{QML4}
\end{equation}
where
\begin{equation}
 \check{\tens{F}}^{ab}_{\ell r \ell^\prime r^\prime} =  {1 \over 2}
{\rm Tr} \left [  {\partial \tens{C} \over \partial C_{\ell^\prime r^\prime}}  (\check{\tens{C}}^{a})^{-1}
{\partial \tens{C} \over \partial C_{\ell r}}  (\check{\tens{C}}^{b})^{-1} \right ].
\label{QML5}
\end{equation}
Equation~(\ref{QML4}) can be inverted to give a deconvolved estimator of $C_{\ell
  r}$. The variance of the cross-spectrum estimator can be written in terms of the matrices $\tens{S}$, $\tens{N}$, and $\tens{E}$ as
\begin{eqnarray}
 \langle y^{ab}_{\ell}y^{ab}_{\ell^\prime} \rangle - \langle y^{ab}
 _{\ell r} \rangle \langle 
y^{ab} _{\ell^\prime r^\prime} \rangle &=& \left[2 \tens{S}_{ip} \tens{S}_{jq} 
+ (\tens{N}^{a}_{ip} + \tens{N}^{b}_{ip}) \tens{S}_{jq}\right.\cr
& & \left. \qquad\qquad +\, \tens{N}^{a}_{ip} \tens{N}^{b}_{jq}\right]
\tens{E}^{ab}_{\ell r, ij} \tens{E}^{ab}_{\ell^\prime r^\prime pq} \,.
\label{QML6}
\end{eqnarray}
Inaccurate determinations of the noise covariance matrices $\tens{N}^{a}$ will therefore not bias the power spectrum estimates, but will lead to inaccurate estimates of the variance, via Eq.~(\ref{QML6}).

In Sects.~\ref{sec:TOIsqualitytests} and \ref{sec:hfimaps} we discussed systematic effects in full resolution maps. Most residual systematic effects are negligible above $\ell=50$.  For science analysis at $\ell < 20$, we degrade the resolution to {\tt HEALPix} \Nside=16 with the following smoothing: 
\begin{equation}
f(\ell) = \left\{  
		\begin{array} {ll}            1  & \ell \le \Nside; \\
            {1 \over 2} \left ( 1 + \cos \left ( {(\ell - \Nside) \pi \over 2 \Nside } \right ) \right ) & \Nside < \ell \le 3 \ \Nside \, ; \\
            0 & \ell > 3 \ \Nside.
		\end{array}
\right. 
\end{equation}
This does not affect the analysis of systematic effects.

\subsubsection{Bias estimate}

\begin{figure}[htbp!]
\includegraphics[width=\columnwidth]{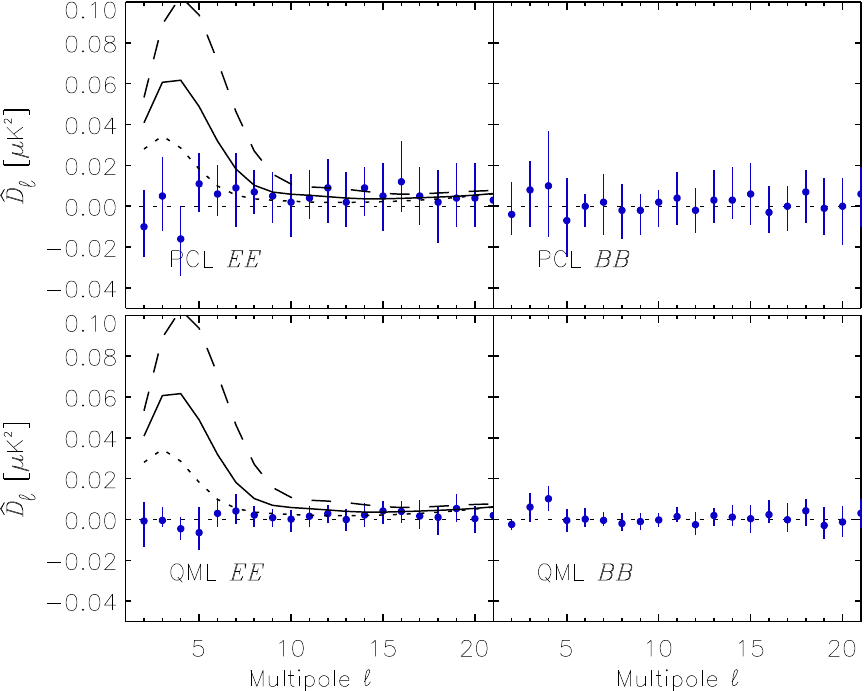} 
\caption{$100\times 143$ cross-spectra of noise and systematics from the average of HFPS1, HFPS2, and HFPS3 simulations, calculated using both PCL and QML estimators (with the covariance matrix from HFPS2). Theoretical spectra are plotted (in black) for $\tau =0.05$, 0.07, and 0.09.}
\label{fig:plspectresimu}
\end{figure}

To estimate the bias induced by residual systematic effects at the power spectrum level, we average simulations HFPS1, HFPS2, and HFPS3  (Sect.~\ref{sec:intro2e2e}) and calculate cross-spectra on the average data with both estimators (Fig.~\ref{fig:plspectresimu}). These averages are then removed from the data.  Note that the biases and their uncertainties are relatively small compared to the theoretical spectra shown in Fig.~\ref{fig:DIFFMERGE}.  This is due to the use of cross-spectra, as mentioned above. 

\subsubsection{Building the cross-spectra}
\label{sec:buildingcrossspectra}

For QML estimators, full mission pixel-pixel noise covariance matrices at \Nside=16 were produced from the 2015 FFP8 simulations,  which capture some aspects of correlations via pixel hit counts and the \Planck\ scanning pattern, but which do not include errors from instrumental systematic effects. To account for instrumental systematic effects, a covariance matrix is derived from the HFPSs. The signal covariance matrix assumes the 2015 base \LCDM\ model parameters with $\tau=0.07$.  After adding CMB signal, this matrix can be inverted. We used the HFPS1 (83 realizations) and HFPS2 (100 realizations) simulations. One set is used to build the pixel covariance matrix.  We compute the QML spectra using either the same set or the other one, to test the effect of overfitting and noise bias introduced by using the same set when the number of simulations is small.  This bias is demonstrated in Appendix~\ref{simbal}, which shows that using the same set of simulations produces significant effects in the dispersion of the QML simulated spectra, as well as more discrepant PTEs.  These biases decrease as the number of simulations increases.  As an extra check, we swap the two independent sets and find fully compatible results.  
\begin{figure}[htbp!]
\includegraphics[width=\columnwidth]{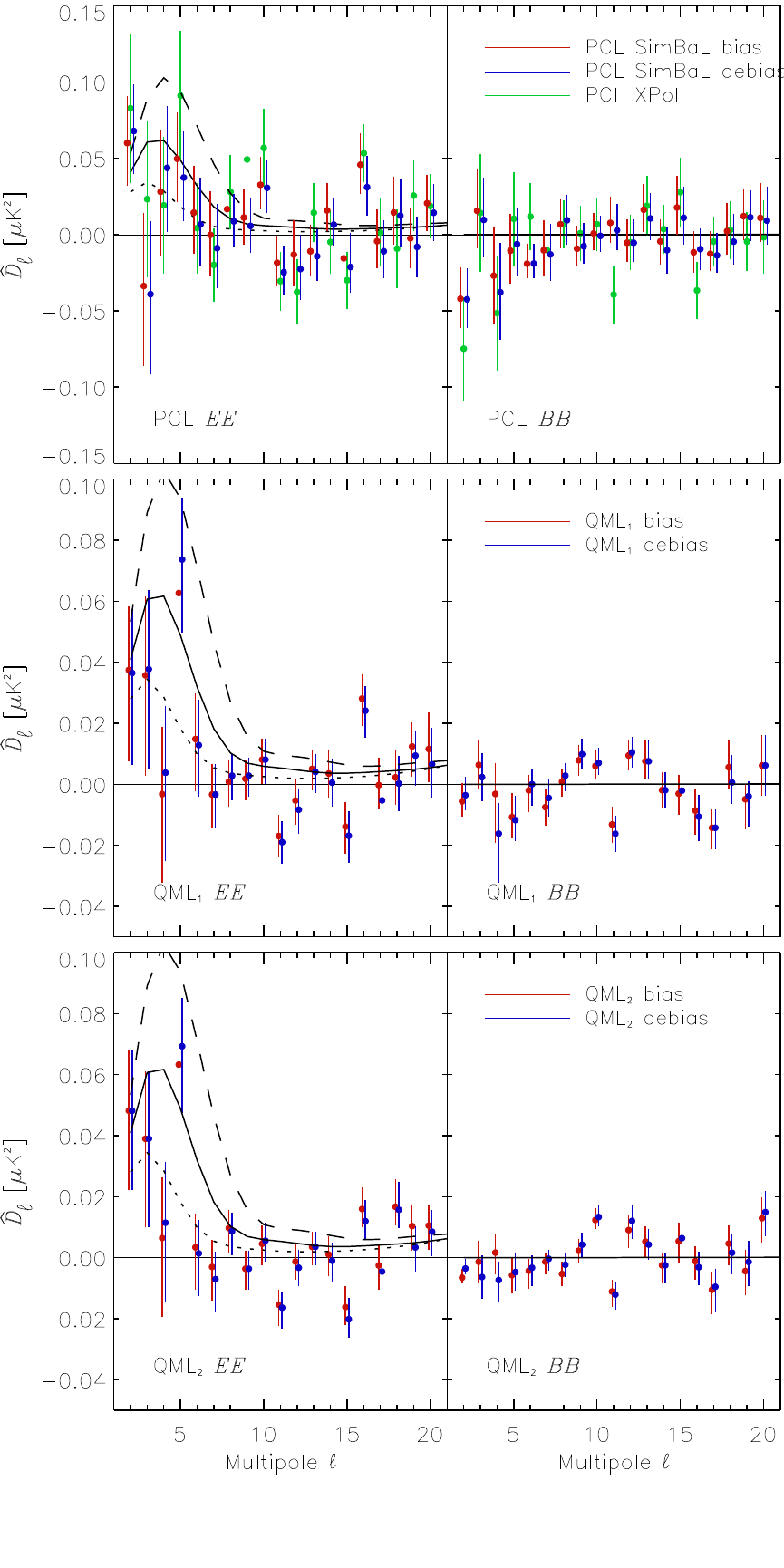} 
\caption{$100 \times 143$ cross-spectra used in this paper. {\it Top\/}: Debiased {\tt XPol} PCL (green), and the biased (red) and debiased (blue) {\tt SimBaL PCL} spectra.  {\it Middle\/}: QML power spectra (red biased, blue debiased) for HFPS1.  {\it Bottom\/}: Same for HFPS2. Model spectra for $\tau$=0.050, 0.070, and 0.090 are displayed in black (dashed) lines.}
\label{fig:100143}
\end{figure}

The $100\times143$ cross-spectra for science analysis are displayed in Fig.~\ref{fig:100143}. Debiasing is shown for the {\tt SimBaL} pseudo-$C_{\ell}$ spectra, and is significant only at very low multipoles. The debiased {\tt Xpol} and {\tt SimBaL} PCL spectra are fully consistent. 
 
The QML power spectra calculated from HFPS1 and HFPS2 differ only slightly. A quantitative estimate of the implications when propagated all the way to $\tau$ is given in Appendix~\ref{simbal}. The $BB$ power spectra from both codes show negligible signal, as expected, and can be used as an estimate of the noise and an upper limit on residual bias. 

The QML estimator produces more local estimates, i.e., with less covariance between multipoles, and better isolation of $E$- and $B$-modes. The latter is clearly seen in Fig.~\ref{fig:100143}.  We also expect the QML approach to give more optimal estimates than the PCL approach, and indeed the QML error bars are found to be significantly smaller than the PCL ones. In both cases, the impact of debiasing the spectra for the ADC nonlinearity distortion is small compared to the uncertainties.

\begin{figure}[htbp!]
\includegraphics[width=\columnwidth]{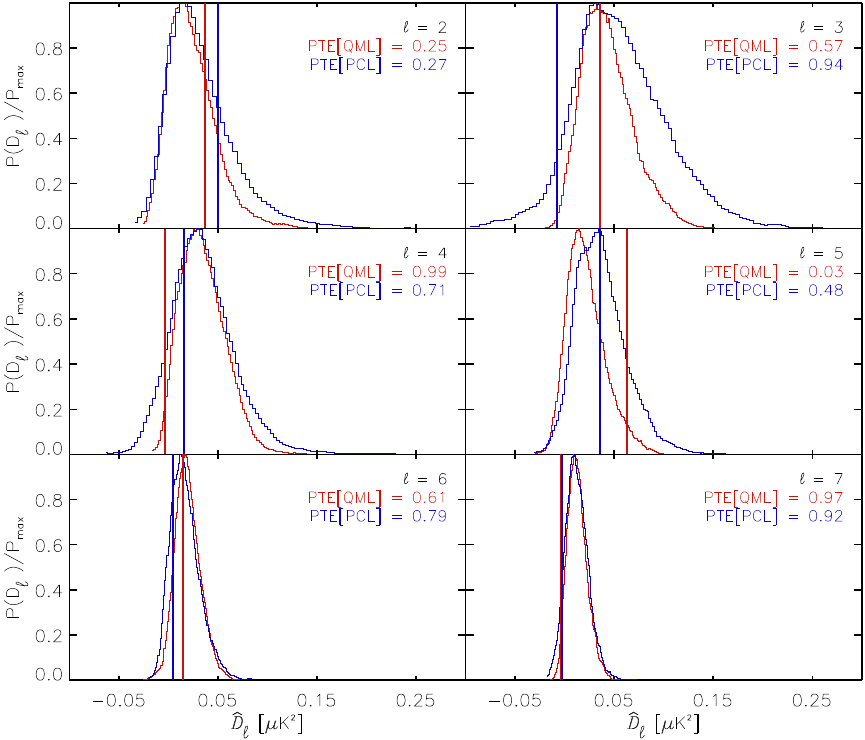} 
\caption{Expected $EE$ cross-spectrum statistical distributions for $2\le\ell\le7$, computed from the HFPS1 simulations, with a fiducial $\tau=0.055$. The blue (PCL) and red (QML) vertical lines show the observed values. The probability to exceed (PTE) is given in each panel.}
\label{fig:jmdmilan2}
\end{figure}

\begin{figure}[htbp!]
\includegraphics[width=\columnwidth]{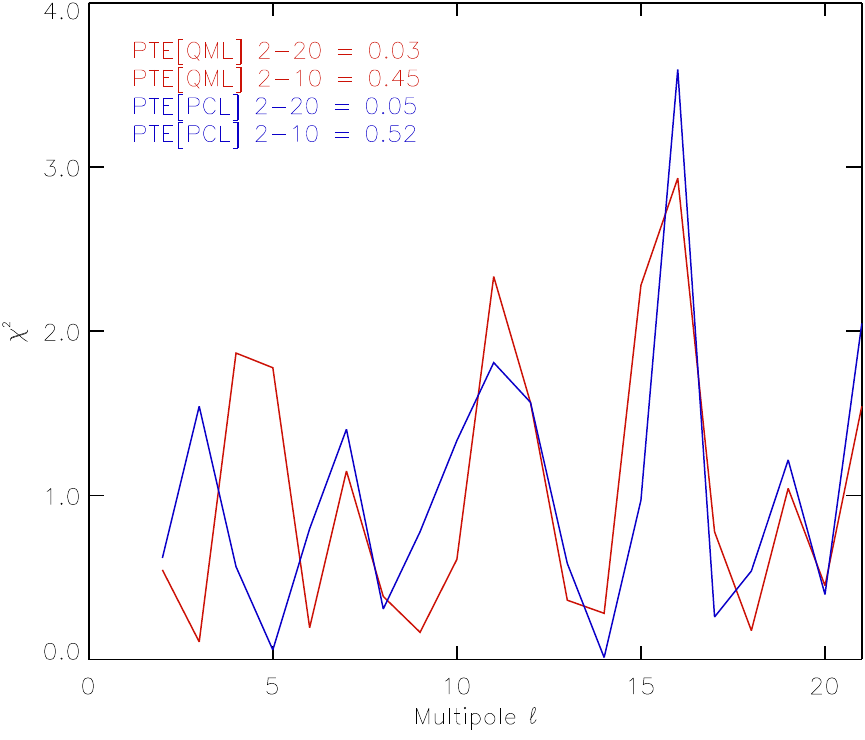} 
\caption{PTE expressed as equivalent $\chi^{2}$ for both the QML and PCL estimators, derived from the statistic shown in Fig.~\ref{fig:jmdmilan2}.}
\label{fig:specpte}
\end{figure}

Fig.~\ref{fig:jmdmilan2} compares statistical distributions of expected values of the $EE$ cross-spectra derived from the HFPSs to observed values (vertical lines), and gives the associated probability-to-exceed (PTE) those values.  For $2\le\ell\le5$, we see non-Gaussian and asymetrical probability distributions coming from systematic residuals and cosmic variance. Figure~\ref{fig:specpte} shows the PTE expressed as the equivalent $\chi^{2}$ of a Gaussian distribution, for a broader range of multipoles. The PCL and QML estimators are largely in agreement. There is a significant outliers : $\ell=16$ which is bad for both estimators. The PTEs are similar, demonstrating that the results from both procedures are in agreement with the expected statistics. Since the QML estimate gives significantly smaller error bars (Fig.~\ref{fig:jmdmilan2}), we use it for the final results.

\subsubsection{$TE$ cross-spectra}
\begin{figure}[htbp!]
\includegraphics[width=\columnwidth]{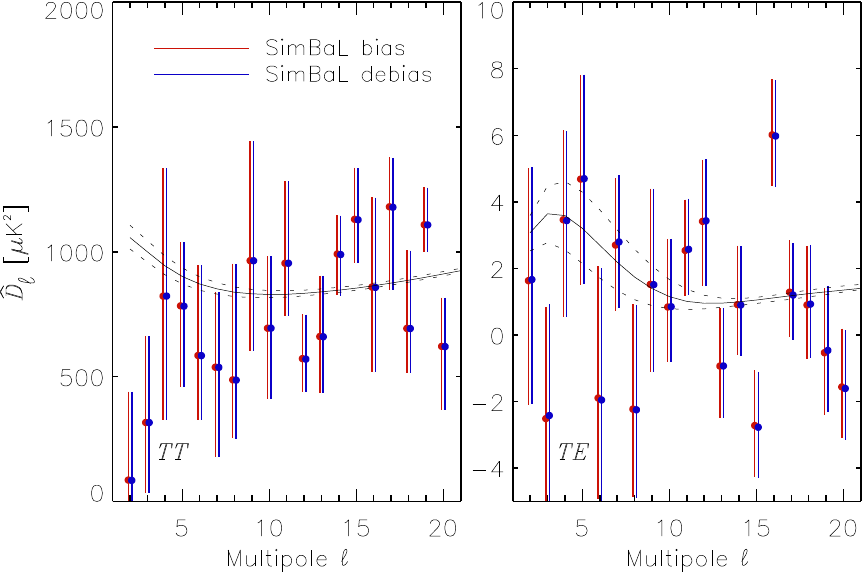} 
\caption{$TT$ and $TE$ $100\times143$ cross-spectra, plotted for the {\tt SimBaL} results with and without the bias correction. The black lines shows the fiducial spectra for $\tau=0.05$, 0.07, and 0.09.}
\label{fig:spectraTE}
\end{figure}

The $TE$ cross-correlation spectrum can in principle also contribute to the determination of $\tau$, and has been used when the $EE$ spectra were either too noisy (e.g., in early WMAP results) or limited by systematic effects (in earlier \Planck\ data releases). Figure~\ref{fig:spectraTE} shows that at present the $TE$ spectrum is compatible with the range of $\tau$ values allowed by $EE$. Nevertheless the uncertainties are such that $TE$ cannot bring any significant improvement to the determination of $\tau$ from $EE$, partly due to the large cosmic variance of the temperature signal at very low multipoles \citep[see also figure~3 in][]{planck2014-a25}. Furthermore, to include $TE$ fully would require a more comprehensive analysis of component separation for the large-scale temperature map, which has not yet been done.

\section{Likelihoods for $\boldsymbol{\tau}$}
\label{sec:analysis}

\subsection{Description of different likelihoods}

In this section we estimate the reionization optical depth $\tau$, using PCL and QML estimators of $EE$ cross-spectra between the two best \Planck\ channels, 100 and 143\,GHz, at very low multipoles. The $\tau$ parameter is strongly degenerate with the amplitude of the primordial spectrum $A_{\rm s}$. The $TT$ power spectrum constrains the combination $A_{\rm s}e^{-2\tau}$ at the sub-percent level, $10^9\,A_{\rm s}e^{-2\tau} = 1.875 \pm 0.014$ \citep{planck2014-a15}.  The $\tau$--$A_{\rm s}$ degeneracy can be broken using \Planck\ $TT$ data along with CMB lensing, or by combining with external data on large-scale structure, both of which constrain $A_{\rm s}$.  However, the amplitude of the $EE$ reionization feature depends quadratically on $\tau$, with very little dependence on the other parameters of the $\Lambda$CDM model.  A 10\,\% constraint on $\tau$ from $EE$ data constrains $A_{\rm s}$ at the 1\,\% level ($\delta A_{\rm s} / A_{\rm s} \approx 2\,\delta \tau$ for $\tau \approx 0.06$), and can affect some of the tensions within the cosmological parameters. 

We use the following two likelihoods with the cross-spectra from Sect.~\ref{sec:estimators} to estimate $\tau$.
\begin{itemize}

\item {\tt Lollipop} \citep{Mangilli01112015} is based on a modification of the Hamimeche and Lewis approach \citep{2008PhRvD..77j3013H} for cross-spectra at low multipoles. The offset is proportional to an effective noise term $o_\ell =N_\ell^{\rm eff} \approx \sqrt{(2 \ell+1)}\, \Delta C_\ell$, derived from the HFPS set from which the covariance matrix is also computed.   We use this likelihood only with PCL spectra in this paper \citep[see also][]{planck2014-a25}.

\item {\tt SimBaL} is a likelihood code based on simulations, and targeted at estimating $\tau$, as described in Appendix~\ref{simbal}.  We apply it to both PCL and QML spectra.  For QML spectra, we use either two independent subsets of HFPSs to determine the spectra and the covariance matrix, or (to limit the bias due to using the same small set of simulations) we use the full set, but with only a few eigenmodes to describe the systematic effects.

\end{itemize}

In the {\tt Lollipop} likelihood, the bias is subtracted directly from the power spectra, while {\tt SimBaL} corrects for the bias by directly using the statistics of the simulations.   

For QML, we define three versions of {\tt SimBal}.  {\tt SimBaL1} and {\tt SimBaL2} use covariance matrices determined from the simulation sets HFPS1 and HFPS2, respectively, while {\tt SimBaL3} uses the covariance matrix determined from all 283 simulations (HFPS1, HFPS2, and HFPS3) and only four eigenmodes.  These three versions of the {\tt SimBal} likelihood deal in two different ways with the difficulty of the limited number of simulations (as discussed in Appendix~\ref{simbal}). 

The two likelihoods sample $\tau$ in the range 0.01--0.15 with a step $ \Delta \tau= 0.001$, and with all other parameters, except $A_{\rm s}$, fixed to the \Planck\ 2015 best-fit values.  We keep $A_{\rm s}\,e^{-2\tau}$ at the fixed value from \Planck\ 2015 (since it is tightly constrained by the higher multipoles). It was shown in Sect.~\ref{sec:adc} that the main systematic effect left in the maps is small, being significant only for $\ell=2$ and 3, and can be simulated and removed (see \ref{sec:debias}). Figure~\ref{fig:pltestlmin} shows that taking $\ell_{\rm min}$ to be 2, 3, or 4 shifts the $\tau$ posterior distributions by less than $4 \times 10^{-3}$, confirming that the final removal of this last systematic effect (by the subtraction of the simulated effect, combined with the use of QML cross-spectra) is very good. In Sect.~\ref{sec:debias}, we test the robustness of our analysis with respect to the removal of residual systematics, and with respect to the two estimators of the power spectra described in Sect.~\ref{sec:powerspectra}. 
We also test the consistency of our results by using cross-spectra between
HFI and LFI, specifically $70\times100$ and $70\times143$.

\begin{figure}[htbp!]
\includegraphics[width=\columnwidth]{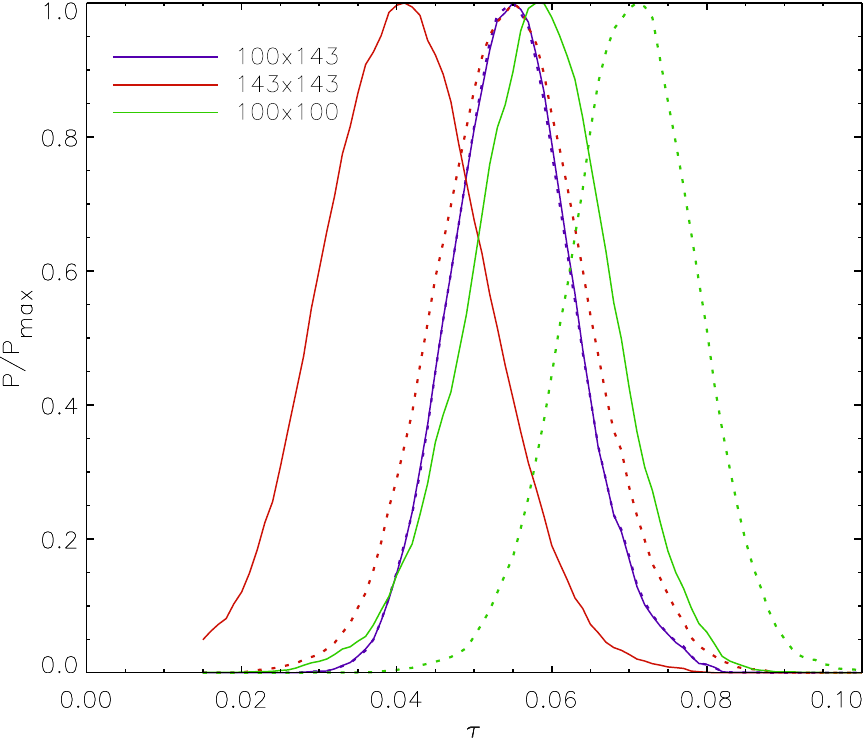} 
\caption{$\tau$ posteriors computed by {\tt SimBal} from the 100 and 143-GHz QML cross- and auto-spectra, both with (solid curves) and without (dashed curves) debiasing by the mean of the simulated power spectra.  While the $100\times143$ cross-power spectrum is not affected by debiasing, the $100\times100$ and $143\times143$ auto-power spectra, which are still dominated by the systematic effects, are more affected. Nevertheless all debiased $\tau$ posteriors are consistent.}
\label{fig:plother}
\end{figure}

\subsection{$\tau$ determination from \textit{EE} cross-spectra only}
\label{sec:debias}
Figure~\ref{fig:plother} shows $\tau$ posteriors computed by {\tt SimBal} from the 100 and 143-GHz QML cross- and auto-spectra, both with and without debiasing.  Debiasing makes essentially no difference for the $100\times143$ cross-spectrum (blue full and dotted lines); however it makes a significant difference in the auto-spectra, as expected, because the use of QML cross-spectra removes part of the systematics.  In the following, we always use debiased $EE$ $100\times143$ cross-spectra to extract $\tau$.

\begin{figure}[htbp!]
\includegraphics[width=\columnwidth]{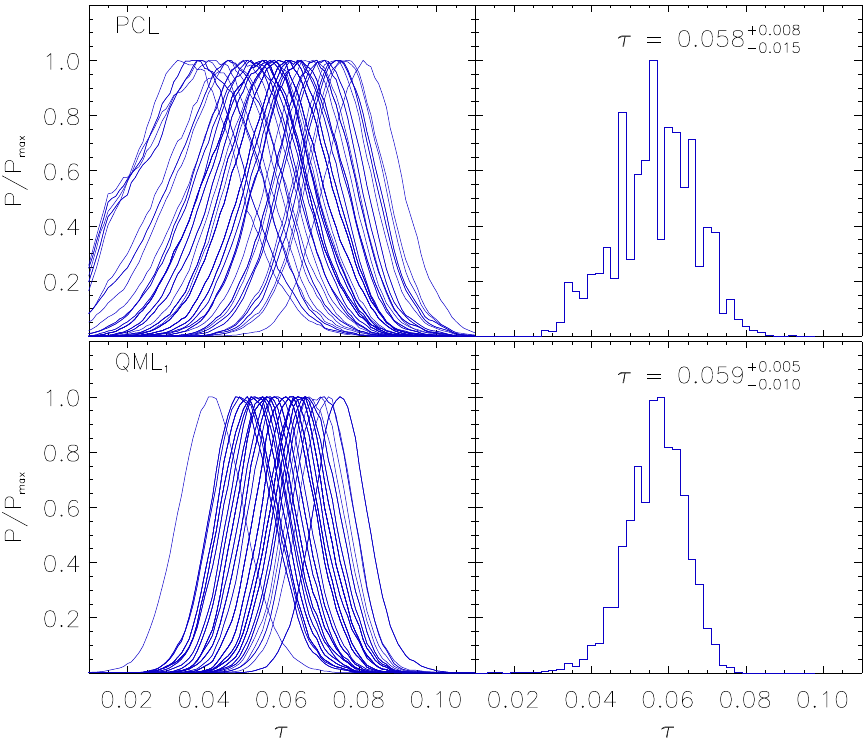} 
\caption{Left column: simulation of {\tt SimBaL} posterior distributions using the HFPS sets for noise and systematic effects, with the fiducial value of $\tau=0.060$. Right column: distribution of the $\tau$ peak value from the left column. The top row is for the PCL estimator and HFPS1 simulations, and the bottom row for the QML estimator and HFPS2 simulations.}
\label{fig:gpemilan10}
\end{figure}

The accuracy of the {\tt SimBal} likelihood in recovering $\tau$ is tested using the HFPS set and shown in Fig.~\ref{fig:gpemilan10} for both QML and PCL spectra.  The input value of $\tau = 0.06$ is recovered accurately as $\tau = 0.059^{+0.005}_{-0.010}$ (QML, 83 realizations in HFPS1, with pixel-pixel covariance matrix from 100 realizations in HFPS2; see Appendix~\ref{simbal}) and $\tau = 0.058^{+0.008}_{-0.015}$ (PCL, 83 realizations in HFPS1).  This demonstrates that our method does not remove signal and provides an essentially unbiased estimator for $\tau$.  The QML estimator produces smaller error bars, as expected, with a narrower and more symmetrical distribution.

\begin{figure}[htbp!]
\includegraphics[width=\columnwidth]{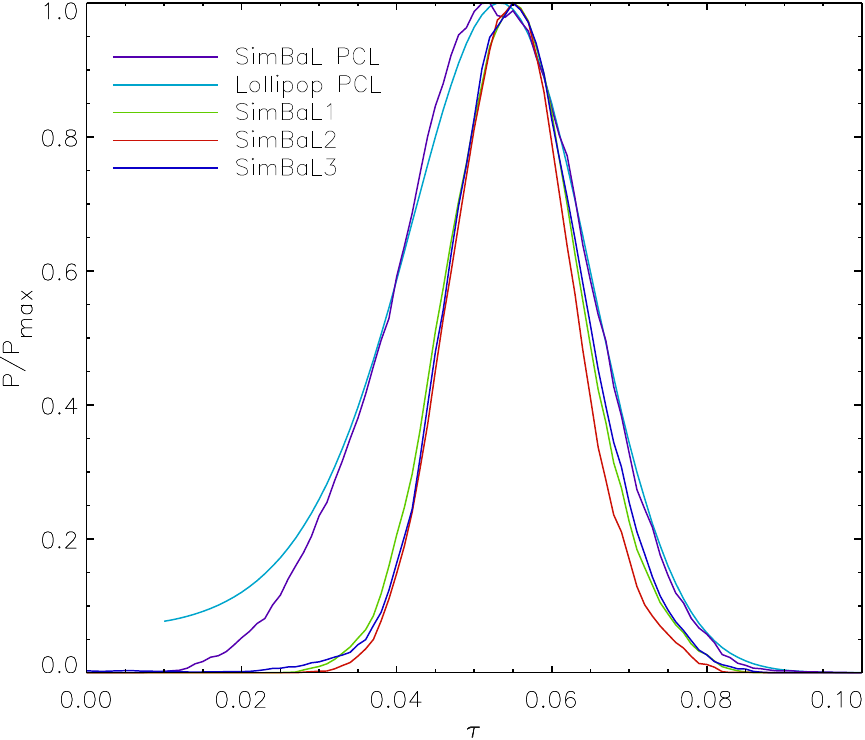} 
\caption{$\tau$ posteriors obtained with the two likelihood methods ({\tt Lollipop} and {\tt SimBal}) for the $100 \times 143$ QML and PCL cross-spectra estimators, colour coded as indicated in the legend.}
\label{fig:biasdebias}
\end{figure}

Figure~\ref{fig:biasdebias} shows $\tau$ posteriors for the data computed by {\tt Lollipop} and {\tt SimBal} from $100\times143$ PCL and QML cross-spectra.  For the PCL spectra, results from {\tt Lollipop} and {\tt SimBal} are fully consistent. The asymmetry of the posterior distribution at low $\tau$ is smaller for {\tt SimBaL} than for {\tt Lollipop}, due to {\tt SimBaL}'s better handling of the statistics of the $C_\ell$s (see Appendix~\ref{simbal}).  For QML, the three {\tt SimBal} estimates give consistent posterior distributions, with significantly narrower width than the PCL results.  They therefore provide our tightest constraints on $\tau$. The near-coincidence of the high-$\tau$ tails of the distributions would imply essentially identical upper limits on $\tau$ for the PCL and QML approaches, if the posteriors were used to provide upper limits on the reionization redshift.  However, the posteriors with QML cross-spectra show a clear detection of $\tau$ at the level of $3.5\,\sigma$, as discussed quantitatively in Appendix~\ref{simbal}.

\begin{figure}[htbp!]\includegraphics[width=\columnwidth]{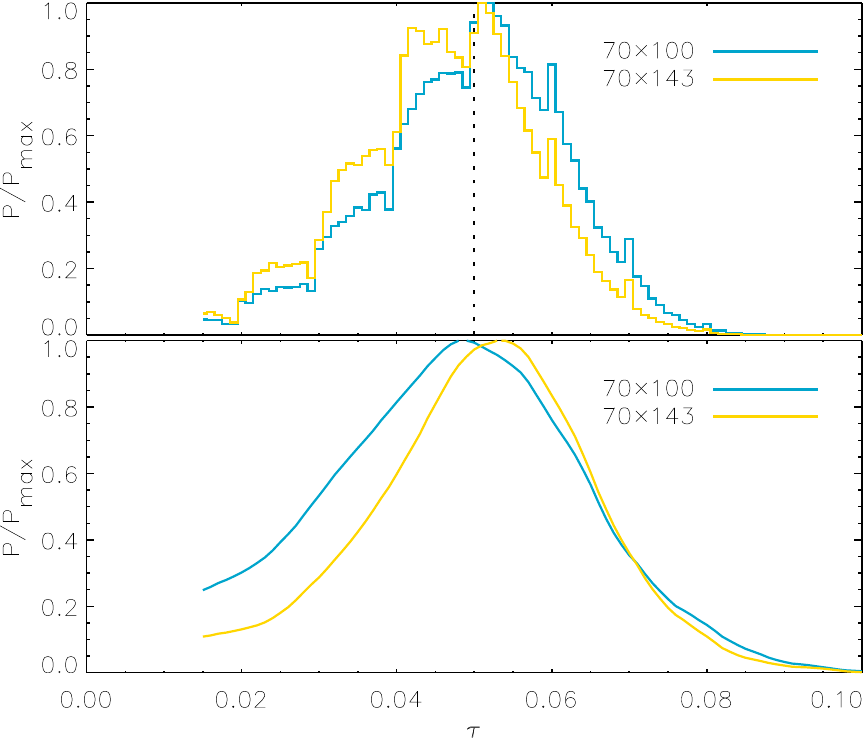} 
\caption{Posterior distributions of $\tau$ calculated by {\tt SimBaL}  from LFI\,$\times$\,HFI cross-spectra. {\it Top\/}: Results from simulations using 10 LFI systematic effect realizations, 83 HFPS1 realizations, and 100 CMB realizations with input reionization parameter $\tau=0.05$.  {\it Bottom\/}: Results from data, specifically the HFI pre-2016 maps at 100 and 143\,GHz and the LFI 70-GHz 2015 released maps \citep{planck2014-a07}.  Results from $70\times100$ and $70\times143$ are consistent.}
\label{fig:simballfi}
\end{figure}

To test how efficiently cross-spectra between the two \Planck\ instruments suppress uncorrelated systematics, we used the HFI HFPS1 and the LFI end-to-end simulations of low-multipole systematic effects (83 for HFI, 10 for LFI).  In the LFI 70-GHz channel maps, residuals from calibration uncertainties dominate systematic effects.  They are at the same level as the dominant HFI systematic, but are lower than the noise (as shown in Sect.~\ref{sec:LFI}).  To each of these simulations of systematic errors we add 100 CMB simulations, with an input value of $\tau=0.05$. We then calculate $70\times100$ and $70\times143$ PCL cross-spectra and run the {\tt SimBaL} likelihood.

Figure~\ref{fig:simballfi} (top) shows results from simulations. The peak values are not significantly biased with respect to the input value of $\tau=0.05$. This indicates that the cross-correlation between LFI and HFI frequencies removes the simulated residual systematic effects rather well for both instruments.  The distributions are asymetric and not very smooth, because of the small number of LFI simulations. They show full width at half maximum 1.5 to 2 times larger than that obtained for the QML estimate using the HFI frequencies alone.  
Figure~\ref{fig:simballfi} (bottom) shows results from the data, specifically the HFI pre-2016 maps at 100 and 143\,GHz, and the LFI 70-GHz 2015 released maps \citep{planck2014-a07}.  As in the simulations, $\tau$ is extracted from the $70\times100$ and $70\times143$ PCL cross-spectra using {\tt SimBal}.  The peak values and 68\,\% upper and lower limits are
\begin{eqnarray}
\tau&=&0.049_{-0.019}^{+0.015} \ \textrm{for the 70$\times$100 cross-spectra}, \nonumber \\
\tau&=&0.053_{-0.016}^{+0.012} \ \textrm{for the 70$\times$143 cross-spectra}.\nonumber
\end{eqnarray}
Very low values are not excluded by these combinations of data, but the peak values are compatible with the baseline results from the HFI $100 \times 143$ cross-spectrum discussed above.

\subsection{Summary of results}
Table~\ref{tab:tau} summarizes the results on the posterior distributions of $\tau$ based on the HFI $100\times 143$ $EE$ cross-spectra PCL and QML estimators shown in Fig.~\ref{fig:biasdebias}.
\begin{table}[tb]
\newdimen\tblskip \tblskip=5pt
\caption{Peak values, 68\,\% upper and lower limits, together with 95\,\% upper limits, for the two likelihood methods ({\tt SimBaL} and {\tt Lollipop}) and the two cross-spectra estimators (PCL and QML).}
\label{tab:tau}
\vskip -6mm
\footnotesize
\setbox\tablebox=\vbox{
\newdimen\digitwidth
\setbox0=\hbox{\rm 0}
\digitwidth=\wd0
\catcode`*=\active
\def*{\kern\digitwidth}
\newdimen\signwidth
\setbox0=\hbox{.}
\signwidth=\wd0
\catcode`!=\active
\def!{\kern\signwidth}
\halign{\hbox to 2cm{#\leaderfil}\tabskip 1em&
\hfil#\hfil&
\hfil#\hfil&
\hfil#\hfil&
\hfil#\hfil\tabskip 0em\cr
\noalign{\doubleline}
\omit&\multispan2\hfil PCL\hfil&\multispan2\hfil QML\hfil\cr %
\noalign{\vskip -3pt}
\omit&\multispan2\hrulefill&\multispan2\hrulefill\cr
\omit\hfil Method\hfil&peak $\pm 1\,\sigma$&peak $+ 2\,\sigma$&peak $\pm 1\,\sigma$&peak $+ 2\,\sigma$\cr
\noalign{\vskip 3pt\hrule\vskip 5pt}
\noalign{\vskip 2pt}
{\tt Lollipop}&$0.053_{-0.016}^{+0.011}$&0.075&\dots&\dots\cr
\noalign{\vskip 4pt}
{\tt SimBaL1}&$0.052_{-0.014}^{+0.011}$&0.076&$0.055_{-0.009}^{+0.009}$&0.073\cr
\noalign{\vskip 4pt}
{\tt SimBaL2}&\dots&\dots&$0.055_{-0.008}^{+0.008}$&0.071\cr
\noalign{\vskip 4pt}
{\tt SimBaL3}&\dots&\dots&$0.055_{-0.008}^{+0.009}$&0.073\cr
\noalign{\vskip 3pt\hrule\vskip 5pt}}}
\endPlancktable
\end{table}

The QML results give a detection of $\tau$ at more than $3.5\,\sigma$ (see Appendix~\ref{simbal}), with the smallest uncertainties obtained so far from CMB data.  For the same likelihood method and sky fraction, a cosmic-variance-limited measurement would have an uncertainty of 0.006.  The peak values obtained with the PCL and QML methods agree to within $0.2\,\sigma$.  Cross-spectra between the two \Planck\ instruments ($70\times100$ and $70\times143$) also give compatible results, but with larger uncertainties.

\begin{figure*}[htbp!]
\includegraphics[width=\textwidth]{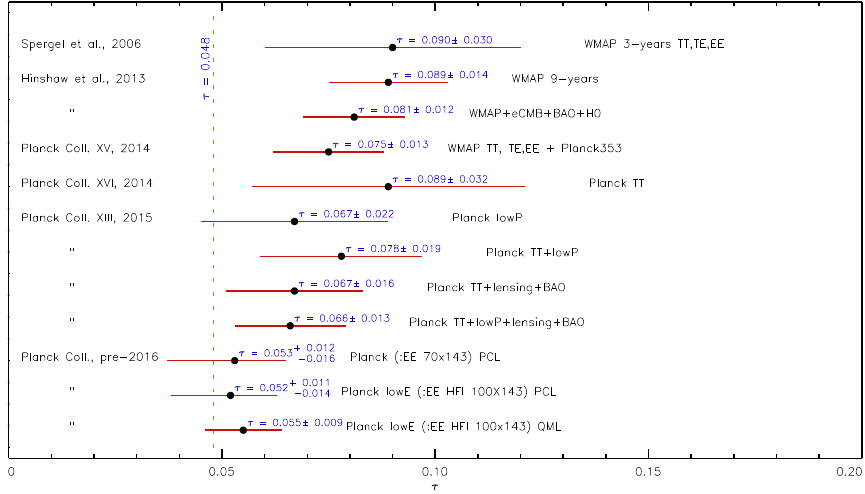} 
\caption{History of $\tau$ determination with WMAP and \Planck. We have omitted the first WMAP determination ($\tau=0.17\pm0.04$, \citealt{bennett2003a}), which was based on $TE$ alone.}
\label{fig:taustory}
\end{figure*}

The most stringent results in Table~\ref{tab:tau} are obtained with the {\tt SimBal} likelihood from $100 \times 143$ QML cross-spectra.  Taking a conservative uncertainty between the three QML results, we obtain
\begin{equation}
\tau=0.055 \pm 0.009,\qquad \qquad \lowE\,.
\label{eq:resulttau}
\end{equation}
We will refer to this as the ``\lowE'' data set and likelihood.\footnote{Since it is based on HFI $EE$ modes,
and in distinction to ``lowP'' that we used in the 2015 release, based on LFI low-$\ell$ multipoles.}  There has been a significant decrease in the peak value of $\tau$ since its first determination from CMB $TE$ measurements in 2003 and subsequent refinement using $EE$ measurements from 2006.  Figure~\ref{fig:taustory} shows the history of $\tau$ estimates.

Reionization history models based on astrophysical observations of high redshift sources predict asymptotic values of $\tau$ at high redshift in the range
0.048--0.055 \citep[figure~7 of][]{2016MNRAS.455.2101M} or 0.05--0.07 \citep[figure~2 of][]{2015ApJ...802L..19R}. Our results are fully consistent with these expectations. For the first time, the upper limit on $\tau$ derived from CMB $EE$ observations gives meaningful limits on how such models can be extrapolated to redshifts larger than $10$.

\section{Implications of a lower value of $\boldsymbol{\tau}$ for cosmological parameters}
\label{sec:cosmo}

The first \Planck\ results on polarization at low multipoles \citep{planck2014-a13, planck2014-a15}, based on the LFI maps, gave a lower value of $\tau$ than that given in the 9-year WMAP analysis \citep{hinshaw2012}.  We also showed that cleaning the WMAP polarization maps for polarized dust emission using the HFI 353-GHz maps led to a reduction in $\tau$, consistent with the results from the LFI-based analysis. The results presented in this paper lead to even lower values of $\tau$. 

The amplitude of the reionization bump in the $EE$ power spectrum at low multipoles scales approximately as $\tau^2$. Low values of $\tau$ are therefore difficult to measure from large-scale polarization measurements of the CMB. As this paper demonstrates, exquisite control of systematic errors, polarized foregrounds, and instrument noise are required in order to measure the small polarization signal induced by cosmic reionization. The main results presented in this paper are based on the $100\times 143$ $EE$ cross-spectrum, summarized in Fig.~\ref{fig:biasdebias} and Table~\ref{tab:tau}. The posteriors for $\tau$ in Fig.~\ref{fig:biasdebias} show clear narrow peaks, with maxima at $\tau= 0.055$, indicating detection of an $EE$ reionization feature. Furthermore, in Table~\ref{tab:tau} we find a 95\,\% upper limit of $\tau < 0.072$. This limit is consistent with the Planck+LowP LFI results, the \Planck\ dust-cleaned WMAP results reported in \cite{planck2014-a15}, and the LFI\,$\times$\,HFI results summarized in Fig.~\ref{fig:simballfi}.

Current astrophysical observations exclude very low values of $\tau$. If the Universe is abruptly reionized at redshift $z_{\rm re}$, the optical depth caused by Thomson scattering is \citep[e.g.,][]{ShullV2008}
\begin{equation}
\tau = {2c\sigma_{\rm T} (1-Y_{\rm P}) \over m_{\rm p}} {\Omega_{\rm b} \over \Omega_{\rm m}} {H_0 \over 8 \pi G} \left\{ \left[\Omega_{\rm m} (1+z_{\rm re})^3 + \Omega_\Lambda \right]^{1/2} - 1 \right\} \label{Sec7_1}
\end{equation}
for the base \LCDM\ model with a helium abundance by mass of $Y_{\rm P}$ (assuming that the helium remains neutral). The Gunn-Peterson test \citep{Gunn:1965, Fan:2006} provides strong astrophysical evidence that the intergalactic medium was highly ionized by a redshift of $z=6.5$. For the \Planck\ 2015 base \LCDM\ parameters, Eq.~(\ref{Sec7_1}) gives $\tau=0.039$ for $z_{\rm re}=6.5$. This conservative {\it astrophysical} constraint on $\tau$, which eliminates the low $\tau$ regions of the posteriors shown in Fig.~\ref{fig:biasdebias}, is in excellent agreement with our current constraint.  The result also stands in good agreement with the PlanckTT+lensing+BAO constraints reported in \cite{planck2014-a15}, $\tau = 0.067 \pm 0.016$, which makes no use of CMB polarization at low multipoles.

\begin{figure}[htbp!]
\includegraphics[width=\columnwidth]{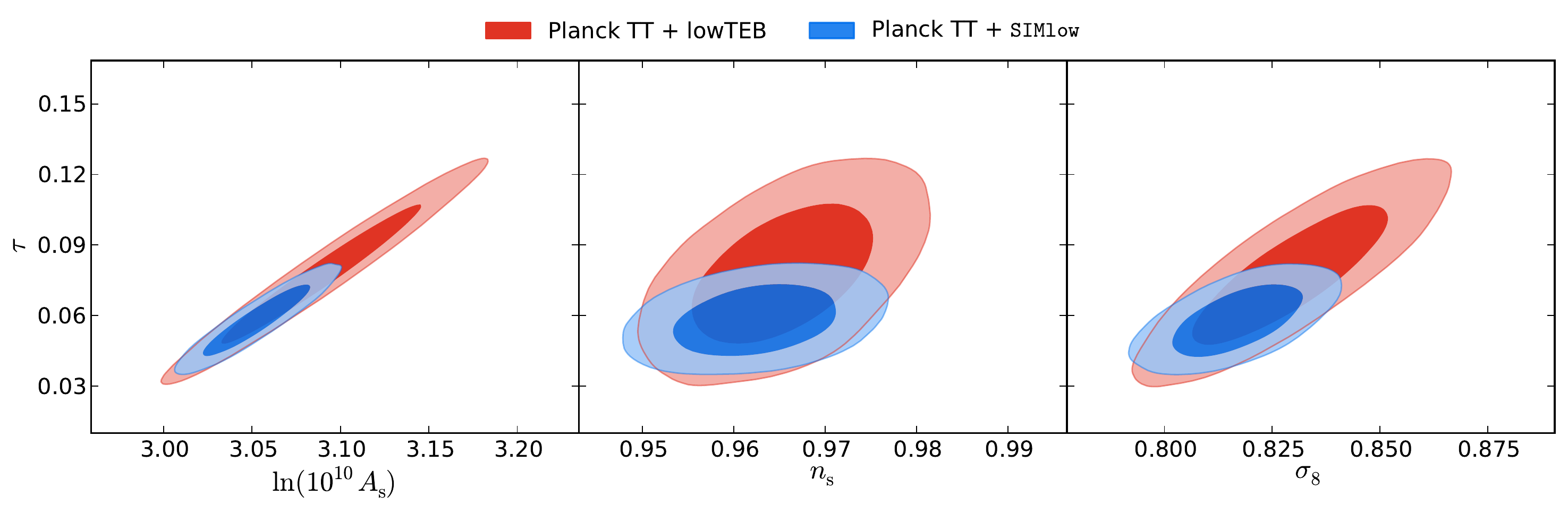} 
\caption{Parameter constraints for the base \LCDM\ cosmology, illustrating the $\tau$--$\ns$ degeneracy and the impact of replacing the LFI-based lowP likelihood used in the 2015 \Planck\ papers with the HFI-based {\tt SimLow} likelihood discussed here. The values of $\tau$ and $\sigma_8$ shift downwards.}
\label{fig:tau_degeneracies2}
\end{figure}

Of course the true value of $\tau$ is important in understanding the formation of the first stars and the process of reionization, but it also has an impact on cosmological parameters that are of more fundamental significance. This is illustrated in Fig.~\ref{fig:tau_degeneracies2}. The principal degeneracies are between $\tau$ and $A_{\rm s}$ and between $\tau$ and $\sigma_8$. However, $\tau$ is also correlated with the spectral index of the fluctuation spectrum, $n_{\rm s}$, which is set by early Universe physics. As we will see below, improving the constraint on $\tau$ helps break this particular degeneracy. 

\begin{table*}
\begingroup
\caption{Parameter constraints for the base \LCDM\ cosmology \citep[as defined in][]{planck2013-p11}, illustrating the impact of replacing the LFI-based lowP likelihood (used in the 2015 \Planck\ papers) with the HFI-based {\tt SimLow} likelihood discussed in the text. We also present here the change when including the high-$\ell$ polarization.}
\label{tab:cosmo:res}
\openup 5pt
\newdimen\tblskip \tblskip=5pt
\nointerlineskip
\vskip -3mm
\scriptsize
\setbox\tablebox=\vbox{
    \newdimen\digitwidth
    \setbox0=\hbox{\rm 0}
    \digitwidth=\wd0
    \catcode`"=\active
    \def"{\kern\digitwidth}
    \newdimen\signwidth
    \setbox0=\hbox{+}
    \signwidth=\wd0
    \catcode`!=\active
    \def!{\kern\signwidth}
\halign{
     \hbox to 0.9in{$#$\leaderfil}\tabskip=1.5em&
     \hfil$#$\hfil&
     \hfil$#$\hfil&\hfil$#$\hfil&
     \hfil$#$\hfil\tabskip=0pt\cr
\noalign{\doubleline}
\multispan1\hfil \hfil&\multispan1\hfil PlanckTT+lowP\hfil&\multispan1\hfil PlanckTT+SIMlow\hfil&\multispan1\hfil PlanckTTTEEE+lowP\hfil&\multispan1\hfil PlanckTTTEEE+SIMlow\hfil\cr
\noalign{\vskip -3pt}
\omit\hfil Parameter\hfil&\omit\hfil 68\,\% limits\hfil&\omit\hfil 68\,\% limits\hfil&\omit\hfil 68\,\% limits\hfil&\omit\hfil 68\,\% limits\hfil\cr
\noalign{\vskip 3pt\hrule\vskip 5pt}
\Omega_{\mathrm{b}} h^2&0.02222\pm 0.00023&0.02214\pm 0.00022&0.02225\pm 0.00016&0.02218\pm 0.00015\cr
\Omega_{\mathrm{c}} h^2&0.1197\pm 0.0022&0.1207\pm 0.0021&0.1198\pm 0.0015&0.1205\pm 0.0014\cr
100\theta_{\mathrm{MC}}&1.04085\pm 0.00047&1.04075\pm 0.00047&1.04077\pm 0.00032&1.04069\pm 0.00031\cr
\tau&0.078\pm 0.019&0.0581\pm 0.0094&0.079\pm 0.017&0.0596\pm 0.0089\cr
\ln(10^{10} A_\mathrm{s})&3.089\pm 0.036&3.053\pm 0.019&3.094\pm 0.034&3.056\pm 0.018\cr
n_\mathrm{s}&0.9655\pm 0.0062&0.9624\pm 0.0057&0.9645\pm 0.0049&0.9619\pm 0.0045\cr
H_0&67.31\pm 0.96&66.88\pm 0.91&67.27\pm 0.66&66.93\pm 0.62\cr
\Omega_{\mathrm{m}}&0.315\pm 0.013&0.321\pm 0.013&0.3156\pm 0.0091&0.3202\pm 0.0087\cr
\sigma_8&0.829\pm 0.014&0.8167\pm 0.0095&0.831\pm 0.013&0.8174\pm 0.0081\cr
\sigma_8 \Omega_{\mathrm{m}}^{0.5}&0.466\pm 0.013&0.463\pm 0.013&0.4668\pm 0.0098&0.4625\pm 0.0091\cr
\sigma_8 \Omega_{\mathrm{m}}^{0.25}&0.621\pm 0.013&0.615\pm 0.012&0.623\pm 0.011&0.6148\pm 0.0086\cr
z_{\mathrm{re}}&9.89^{1.8}_{-1.6}&8.11\pm 0.93&10.0^{1.7}_{-1.5}&8.24\pm 0.88\cr
10^9 A_{\mathrm{s}} e^{-2\tau}&1.880\pm 0.014&1.885\pm 0.014&1.882\pm 0.012&1.886\pm 0.012\cr
\mathrm{Age}/\mathrm{Gyr}&13.813\pm 0.038&13.829\pm 0.036&13.813\pm 0.026&13.826\pm 0.025\cr
\noalign{\vskip 5pt\hrule\vskip 3pt}
} 
} 
\endPlancktable
\endgroup
\end{table*}

To assess the impact of our new polarization data on the cosmological parameters, we have constructed a simplified, low-multipole likelihood based on simulations that include systematic effects.  This is {\tt SimLow}, based on the $100\times143$ $EE$ spectrum between $\ell=2$ and 20. {\tt SimLow} is a simulation-based likelihood that gives the posterior distribution of fiducial $C_{\ell}$s for each multipole; it is fully described in Appendix~\ref{sec:simlow}.  To explore the effects on cosmological parameters, we combine {\tt SimLow} with the PlanckTT likelihood (i.e., {\tt Commander} at $\ell<30$, and {\tt Plik} at higher $\ell$).

In the \Planck\ 2015 parameter analysis, we used an LFI-based polarization likelihood (referred to as ``lowP'').  The lowP likelihood gives
\begin{equation}
             \tau = 0.067 \pm 0.022, \qquad \qquad \rm{lowP} \ {\rm only},            \label{eq:Sec7_2}
\end{equation}
on its own, with higher multipoles only used to fix $A_{\rm s}e^{-2\tau}$ to the 2015 base \LCDM\ best fit.
This constraint is compatible with the HFI results presented in Sect.~\ref{sec:analysis}, although it is statistically weaker. Combining the LFI polarization likelihood with the high multipole PlanckTT data, for base \LCDM\ we obtain 
\begin{equation}
\tau = 0.078 \pm 0.019, \qquad \qquad \rm{PlanckTT{+}lowP}.
\label{eq:Sec7_3}
\end{equation}
Adding the PlanckTT likelihood drives the value of $\tau$ upwards by about $0.5\,\sigma$. The $A_{\rm s} e^{-2\tau}$ degeneracy at intermediate multipoles ($\ell<1500$) is broken by the lensing effect seen in the higher part of the spectrum. 

However, the $\ell\ga1000$ part of the \Planck\ spectrum is characterized by peaks that are slightly broader and smoother than what the \LCDM\ model predicts. The high-multipole peak smoothing is compatible with a slightly stronger lensing amplitude, and translates into a roughly $2\,\sigma$-high phenomenological parameter $\Alens$ value. The $A_L^{\Phi \Phi}=0.95\pm0.04$ value derived from the lensing power spectrum \citep{planck2014-a15} supports that this would just be a statistical fluctuation, rather than a peculiar feature of the lensing power spectrum itself. Nevertheless, the preference for a larger lensing amplitude at high multipoles pushes the normalization and the optical depth values up.
The lowP likelihood was not statistically powerful enough to counteract this trend, and so in the PlanckTT+lowP analysis $\tau$ is driven upwards compared to Eq.~(\ref{eq:Sec7_2}). This effect is discussed at length in \citet{planck2013-p11} and \citet{planck2014-a15}.

Adding the \Planck\ lensing measurements, which are compatible with lower values of $\As$, drives $\tau$ down again, close to the original lowP value:
\begin{equation}
\tau = 0.066 \pm 0.016, \qquad \qquad \rm{PlanckTT{+}lowP{+}lensing}.
\label{eq:Sec7_4}
\end{equation}
These shifts, and in fact the low-multipole power deficit, are not of sufficiently high significance to suggest new physics. Moreover, the fact that adding \Planck\ lensing causes $\tau$ to shift downwards suggests that $\tau$ is lower than the value in Eq.~(\ref{eq:Sec7_2}). Indeed {\tt SimLow} alone gives the following constraint on $\tau$:
\begin{equation}
\tau = 0.055 \pm 0.009, \qquad \qquad {\tt SimLow}.
\label{eq:tau:simlow}
\end{equation}

We can anticipate what will happen if we replace the lowP likelihood with a statistically more powerful polarization likelihood favouring a low value of $\tau$ --- the main effect will be to shift $\sigma_8$ towards lower values, with a proportionately smaller shift of $n_{\rm s}$ also to smaller values. Furthermore this would not be consistent with solving the high-multipole peak smoothing through an underestimate of the effect of lensing.
In fact, adding the \Planck\ lensing measurement to {\tt SimLow} and PlanckTT has a small impact on the value of $\tau$, giving $\tau=0.057\pm 0.0092$.

Figure~\ref{fig:tau_degeneracies2} compares the {\tt SimLow} and lowP parameter constraints on $\tau$, $\sigma_8$, and $\ns$, while Table~\ref{tab:cosmo:res} gives numerical results for parameters of the base \LCDM\ model. The tighter constraint on $\tau$ brought by {\tt SimLow} reduces the correlation between $\ns$ and $\tau$, and leads to slightly tighter bounds on $\ns$. However, larger parameter changes are seen for $\tau$ and $A_s$, each changing by about $1\,\sigma$. We specifically find
\begin{equation}
\tau = {0.058 \pm 0.009}, \qquad \qquad \rm{PlanckTT+}{\tt SimLow},
\label{eq:Sec7_3}
\end{equation} 
in excellent agreement with the result from {\tt SimLow} alone. The present day amplitude of the fluctuations, $\sigma_8$, decreases by about $1\, \sigma$ and its uncertainty shrinks by about $33\,\%$. This shift goes in the right direction to reduce the tensions with cluster abundance and weak galaxy lensing, as discussed in \cite{planck2014-a15}, although it is not yet sufficient to remove them entirely.

Changes in most of the other cosmological parameters are small, with deviations being less than $0.5\,\sigma$.  The largest deviation is in the spectral index $\ns$, due to its partial correlation with $\tau$.  This slight decrease in $n_{\rm s}$ means that we can now reject the scale-invariant spectrum at the $6.7\,\sigma$ level ($8.7\,\sigma$ when using the high-$\ell$ polarization data). The Hubble constant $H_0$ decreases by $0.4\,\sigma$; within the framework of the base \LCDM \ model, this increases the tension with some recent direct local determinations of $H_0$ \citep{2016arXiv160401424R} to around $3.2\,\sigma$.

The use of {\tt SimLow} in place of lowP has little effect on most of the usual extentions to the \LCDM\ model, as can be seen in Table~\ref{tab:cosmo:ext}. The number of relativistic species, for example, remains compatible with 3. The phenomenological $A_{\rm L}$ parameter is essentially unchanged, and still discrepant at roughly the $2\,\sigma$ level from its expected value of 1. Additionally, the running of the spectral index is constrained to be even closer to zero. 

Due to the lowering of the normalization, the CMB constraint on neutrino masses improves from $\sum m_\nu < 0.72\,{\rm eV}$ (PlanckTT+lowP) to $\sum m_\nu < 0.59\,{\rm eV}$ (PlanckTT+SimLow) and $\sum m_\nu < 0.34$\,eV (PlanckTTTEEE+SimLow). When adding BAO information \citep[see][for details]{planck2014-a15}, these constraints improve further to $m_{\nu}<0.17$\,eV (PlanckTT+SimLow+lensing+BAO) and $m_{\nu}=0.14$\,eV (PlanckTTTEEE+SimLow+lensing)

\begin{table*}
\begin{center}
\begingroup
\openup 5pt
\newdimen\tblskip \tblskip=5pt
\nointerlineskip
\vskip -3mm
\setbox\tablebox=\vbox{
    \newdimen\digitwidth
    \setbox0=\hbox{\rm 0}
    \digitwidth=\wd0
    \catcode`"=\active
    \def"{\kern\digitwidth}
    \newdimen\signwidth
    \setbox0=\hbox{+}
    \signwidth=\wd0
    \catcode`!=\active
    \def!{\kern\signwidth}
\halign{
    \hbox to 0.9in{$#$\leaderfil}\tabskip=1.5em&
        \hfil$#$\hfil&
        \hfil$#$\hfil&
        \hfil$#$\hfil&
        \hfil$#$\hfil\tabskip=0pt\cr
\noalign{\doubleline}
\multispan1\hfil \hfil&\multispan1\hfil PlanckTT+lowP\hfil&\multispan1\hfil PlanckTT+SIMlow\hfil&\multispan1\hfil PlanckTTTEEE+lowP\hfil&\multispan1\hfil PlanckTTTEEE+SIMlow\hfil\cr 
\noalign{\vskip -3pt}
\omit\hfil Parameter\hfil&\omit\hfil 95\,\% limits\hfil&\omit\hfil 95\,\% limits\hfil&\omit\hfil 95\,\% limits\hfil&\omit\hfil 95\,\% limits\hfil\cr
\noalign{\vskip 3pt\hrule\vskip 5pt}
\Omega_K& -0.052^{+0.049}_{-0.055}& -0.053^{+0.044}_{-0.046}& -0.040^{+0.038}_{-0.041}& -0.039^{+0.032}_{-0.034}\cr
\Sigma m_\nu\,[\mathrm{eV}]& < 0.715& < 0.585& < 0.492& < 0.340\cr
N_{\mathrm{eff}}& 3.13^{+0.64}_{-0.63}& 2.97^{+0.58}_{-0.53}& 2.99^{+0.41}_{-0.39}& 2.91^{+0.39}_{-0.37}\cr
Y_{\mathrm{P}}& 0.252^{+0.041}_{-0.042}& 0.242^{+0.039}_{-0.040}& 0.250^{+0.026}_{-0.027}& 0.244^{+0.026}_{-0.026}\cr
\mathrm{d}n_{\mathrm{s}}/\mathrm{d}\ln k& -0.008^{+0.016}_{-0.016}& -0.004^{+0.015}_{-0.015}& -0.006^{+0.014}_{-0.014}& -0.003^{+0.014}_{-0.013}\cr
r_{0.002}& < 0.103& < 0.111& < 0.0987& < 0.111\cr
w& -1.54^{+0.62}_{-0.50}& -1.57^{+0.61}_{-0.49}& -1.55^{+0.58}_{-0.48}& -1.59^{+0.58}_{-0.46}\cr
A_{\mathrm{L}}& 1.22^{+0.21}_{-0.20}& 1.23^{+0.20}_{-0.18}& 1.15^{+0.16}_{-0.15}& 1.15^{+0.13}_{-0.12}\cr
\noalign{\vskip 5pt\hrule\vskip 3pt}
} 
} 
\endPlancktable
\endgroup
\caption{Constraints on 1-parameter extensions of the base \lcdm\ model obtained using the \planckTTonly\ likelihood in combination with lowP and {\tt SimLow}. Note that contrary to Table~\ref{tab:cosmo:res}, where all the parameters correspond to a single cosmological fit, the results in each line here are obtained from a separate \lcdm+extension fit. Uncertainties are at $95\,\%$CL.}
\label{tab:cosmo:ext}
\end{center}
\vskip -0.7cm
\end{table*}

The HFI polarization measurements presented in this paper reduce some of the tensions for some of the cosmological parameters, but have no substantial impact on the qualitative conclusions of the \Planck\ 2015 cosmology papers.  In particular, the \LCDM\ model remains an excellent description of the current data.

Further interpretation of these results in terms of reionization models can be found in the companion paper \citet{planck2014-a25}.  A more complete quantitative description of all the cosmological consequences of the new HFI-based constraints, along with a complete reanalysis of the high-$\ell$ polarization using these new data, will be performed in a forthcoming \Planck\ release.

\section{Conclusions}

This paper presents a value of the parameter $\tau$ derived solely from low-multipole $EE$ polarization, which is both the most accurate to date and the lowest central value.  It depends on the high-$\ell$ multipoles of the $TT$ power spectrum only through the constraint required to fix $A_{\rm s}\,e^{-2\tau}$.

Measurements of polarized CMB anisotropies on these large ($>10\deg$) scales require all-sky coverage and very low noise. Only space experiments, with great stability, multiple redundancies, and enough frequencies to remove Galactic foregrounds, can achieve all of these simultaneously. The WMAP satellite was the first to reach most of these goals, with two telescopes to directly measure very large scales, passively cooled detectors, and nine years of observations; however, its lack of high frequencies did not allow a good enough dust foreground subtraction to be carried out.  The \Planck\ mission achieved much lower noise, especially with the 100\,mK bolometers of the HFI.  Scanning the sky in nearly great circles at 1\,rpm for 2.5\,years, \Planck\ HFI required extreme stability and many levels of redundancy to measure the polarized CMB on the largest scales.  The challenge was such that, for the first two cosmological data and science releases by the \Planck\ team, there were still obvious signs of poorly-understood systematic effects, which prevented the use of large-scale HFI polarization data.

At 545 and 857\,GHz, calibration is not as accurate as in the CMB-calibrated channels (70--353\,GHz), but polarized dust emission can be removed well enough that $\tau$ is unaffected by dust residuals.  Furthermore, most of the synchrotron foreground that is uncorrelated with dust is also mostly removed when using only $100\times143$ cross-spectra. At CMB frequencies, polarized systematic effects have now been understood and modelled.   One systematic effect is not yet corrected in the maps, but is removed as a small correction to the power spectra using simulations.   Finally the use of the $100\times143$ QML cross-spectrum reduces even further the impact of the remaining systematic effects at CMB frequencies.  These advances enable a measurement of $\tau$ to be made using only the low $EE$ multipoles from HFI, along with the high-multipole $TT$ constraint on $A_{\rm s}\,e^{-2\tau}$.  The path is now clear to the final \Planck\ release in 2016, which will provide full-sky, polarized, CMB maps that are reliable on all scales.

This $\tau$ determination fully reconciles the CMB results with other astrophysical measurements of reionization from sources at high redshift. It also gives constraints on the level of reionization at redshifts beyond that of the most distant sources ($z\ga10$). 

Large-scale CMB polarization measurements ultimately contain more information on the reionization period than just the value of $\tau$, and can also constrain turbulence in interstellar magnetic fields and $B$-mode polarization due to primordial gravitational waves, hence the physics of inflation and the early Universe. This last objective requires not only the knowledge of $\tau$ that is now achievable, but also data that are fully noise-limited at all frequencies.

\begin{acknowledgements}
The Planck Collaboration acknowledges the support of: ESA; CNES and CNRS/INSU-IN2P3-INP (France); ASI, CNR, and INAF (Italy); NASA and DoE (USA); STFC and UKSA (UK); CSIC, MINECO, JA, and RES (Spain); Tekes, AoF, and CSC (Finland); DLR and MPG (Germany); CSA (Canada); DTU Space (Denmark); SER/SSO (Switzerland); RCN (Norway); SFI (Ireland); FCT/MCTES (Portugal); ERC and PRACE (EU). A description of the Planck Collaboration and a list of its members, indicating which technical or scientific activities they have been involved in, can be found at \href{http://www.cosmos.esa.int/web/planck/planck-collaboration}{\texttt{http://www.cosmos.esa.int/web/planck/planck-collaboration}}. 
\end{acknowledgements}

\bibliographystyle{aat}
\bibliography{Planck_bib,Lowell}

\appendix

\section{HFI Systematic Errors}
\label{TOIproc}

Here we describe tests of known systematic errors in the HFI instrument and
their effects on low-multipole polarization.

\subsection{Glitches}
\label{sec:glitches}

Glitches induced in the signal by cosmic ray (CR) hits on the
bolometers are a major source of systematic errors in
HFI and are described in detail in several earlier publications
\citep{planck2013-p03,planck2013-p03e,planck2014-a08}.
In the time-ordered information (TOI) processing software, glitches are
detected via their transient nature, and are then masked and corrected. 

Additionally, the amplitude of a long-tail glitch template, built
from stacking many events, is fitted to each detected event and the
tail is subtracted from the data. This procedure reduces the
noise power by an order of magnitude at frequencies around
0.05\,Hz.  However, uncertainties in this correction contribute to, but do not dominate, the low frequency noise of the detectors. 
The size of this effect has been evaluated at the TOI level using simulations in \cite{planck2013-p03} and \cite{planck2013-p03e}, where we have shown that it accounts for a few percent up to 30\,\% of the noise, depending on the detector, at frequencies between 0.016 and 0.1\,Hz, and is negligible at higher frequencies.

Undetected glitches that remain, below the detection threshold, 
contribute to the white noise at the level of 3 to 6\,\%.
Polarization-sensitive bolometers (PSBs) in the same module see many
coincident events, so both detected and undetected glitches are a potential
contaminant of the polarization measurements.

\begin{figure}[!ht]
\includegraphics[width=\columnwidth]{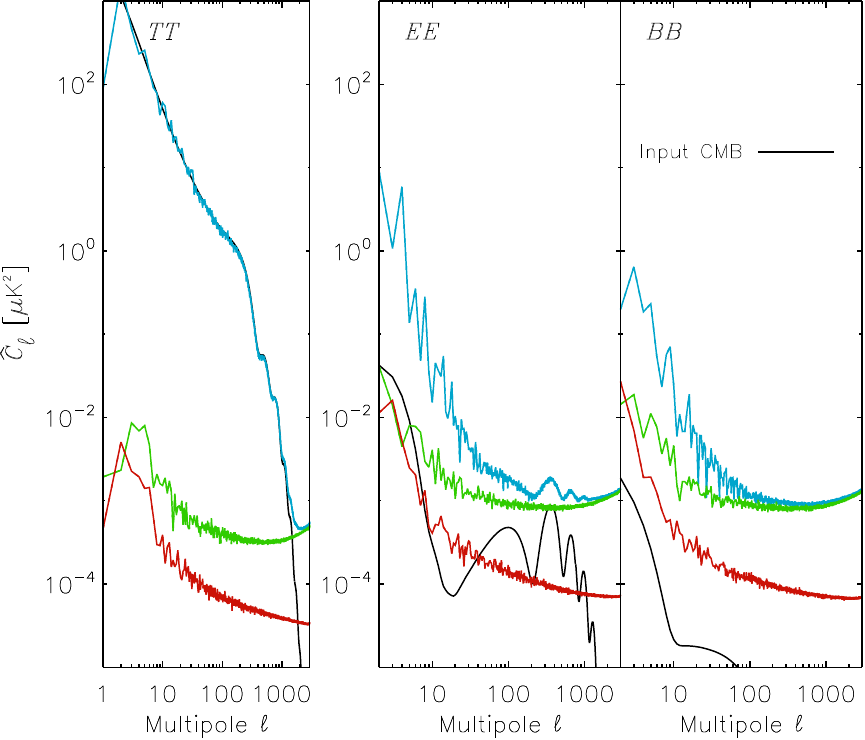} 
\caption{Estimated contribution of glitch residuals to auto-power spectra for
simulations of the 143-detset1, with the F-$EE$ model in black.  The red curve
shows the evaluation of the power spectra of the difference of maps of
simulations with and without glitch-deglitch removal. Spectra here are averaged
over four realizations. The blue curve shows the spectra after processing for
the simulated signal, with all systematic effects included. The green curve
evaluates the FFP8 noise power spectra estimated using half-ring differences,
propagated all the way to maps.}
\label{fig:nogdlitchdeglitchtteebb}
\end{figure}

To check the effect of glitches on the polarization maps and power spectra,
we have used the end-to-end simulations, which incorporate a physically
motivated model of the glitches \citep{planck2013-p03e,catalano2014}.  We have
simulated the full mission data for four detectors at 143\,GHz (the first
detset), with and without glitches. The level of glitch residuals is evaluated
by computing the power spectra of the map differences from the two sets, after
running the processing pipeline (including the glitch removal) in simulations
with glitches only, and using the same flags in both cases. The results, shown
in Fig.~\ref{fig:nogdlitchdeglitchtteebb}, indicate that, after correction,
contamination of the $E$- and $B$-mode polarization by glitches accounts for up
to $30\,\%$ of the noise in the low-$\ell$ regime ($\ell < 10$), as discussed
in Sec.~\ref{sec:detnoise}.  At low multipoles residuals are dominated by
glitch tail removal errors, while undetected glitches below the threshold give
the dominant contribution to the residuals in the white noise region at high
multipoles.

Figure~\ref{fig:nogdlitchdeglitchtteebb} presents predictions of the TOI
noise (from FFP8 simulations) and residuals of all systematic effects,
propagated to maps and
power spectra (green curve), which set the goal for null-test comparisons.
Also shown are the levels due to the glitch-deglitch residuals (i.e., residuals
after simulation of glitches and removal of the glitches in the processing
pipeline, red curve)
and the simulation of the sky signal, noise, and systematics (blue curve).

\subsection{High energy cosmic ray events}

In addition to cosmic ray transients affecting the data,
we also detect rarer, higher energy events that affect many bolometers
simultaneously.  The correlated nature of these events is of
particular concern for polarization measurements.
Very high energy cosmic rays (hundreds of GeV to TeV energies) 
interact with the material in the HFI focal plane and in the \Planck\
satellite, and create dense, energetic particle showers, after a
cascade of several hadronic interactions.
The density of these particle showers generates glitches in many
bolometers simultaneously and heats up the bolometer environment. 
The densest showers also induce a small temperature decrease in the
bolometer plate for 1 to 2\,seconds, followed by a
slow rise of the temperature for 1 to 2\,minutes, \citep{planck2013-p03e}. 
This is due to release of trapped helium on the metal and charcoal surfaces,
creating transient gas conduction between parts of the focal plane at
different temperatures (bolometer plate, dilution cooler plate, and 1.4-K box).
The two detectors of a given PSB
always show the same transient thermal behaviour in these events.
The slow common-mode temperature behaviour is removed well by subtracting
the dark bolometer baseline and does not affect the data after masking
the initial parts of the events.

After deglitching, the global effect of the high energy events is estimated to
be very weak. A very small effect has been found in polarization power spectra
when masking all events detected by 15 or more coincident glitches in the focal
plane (120\,000 events).

\subsection{Baseline removal and non-linearities}

A low frequency baseline built from the dark bolometer TOI is subtracted
from all other bolometers, removing the common effect of the
temperature fluctuation of the bolometer plate. This has no significant
residual effect on large angular scales, as shown using an end-to-end
simulation \citep[see figure~20 of][]{planck2014-a08}. The effect at low
multipoles is a small fraction of the signal ($<0.3\,\%$ for $\ell<20$ and
$2.5\,\%$ 
at $\ell=4$). This leads to levels an order of
magnitude smaller than the $EE$ fiducial model at all multipoles. 

The bolometers have a small nonlinear response
associated with the relatively large changes in 
the operating point associated with the slow drift of the temperature of 
the bolometer plate (with the instantaneous linear gain being given by the 
slope of the tangent at the operating point);
these changes in the temperature of the bolometer plate
come from Galactic cosmic ray rate 
modulation by the inhomogeneities of the solar wind, on timescales of hours
to weeks ($2$ to $10\,\mu$K), as well as the long-term solar cycle modulation
($80\,\mu$K) during the HFI mission.
The relative gain variation for a $100\,\mu$K temperature change is
2--$3\times10^{-3}$ and is taken into account as a first estimate of the
nonlinear response term on the signal. 
The second-order term is two orders of magnitude smaller for a
$100\,\mu$K temperature change (a few $10^{-5}$), a correction that is
below other effects considered here.
\cite{planck2011-1.5} predicts a $10^{-5}$ gain change 
due to the optical loading variations coming from the CMB solar dipole at
143\,GHz, the largest CMB signal. 

\begin{figure}[!ht]
\includegraphics[width=\columnwidth]{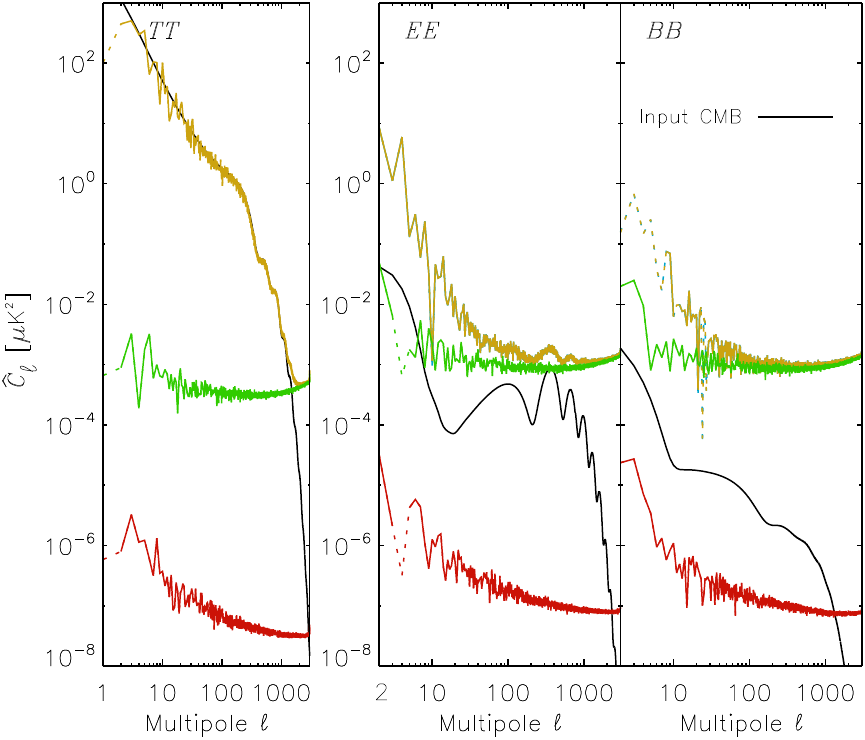} 
\caption{Simulated contribution of warm readout electronics drifts on the 
auto-power spectra of 143-detset1. The red curves show the
contribution of the electronics drift compared to the total map power (orange)
and the F-EE input to the simulations (black).
The green curve is an estimate of the noise level based on the
half-ring differences.}
\label{fig:paureucl}
\end{figure}

A second small nonlinearity is due to the changes in the temperature of the
warm electronics boxes on the satellite. The coefficients of this gain change
have been measured in ground tests and the effect has been propagated to the
power spectrum using the end-to-end simulations and the measured temperature
history of the mission. 
The results are shown in Fig.~\ref{fig:paureucl}.  The correction is less than
$10^{-4}\,\mu{\rm K}^2$ in $C_\ell$ for all multipoles. 
The uncertainty on this correction is thus negligible. 

\subsection{\HeJT\ cooler pickup}
Electrical coupling between the \HeJT\ cooler drive and the bolometer
readout (which are phase-locked) appears as very narrow lines in the power
spectrum of the timestreams (called ``4-K lines'') at harmonics of 10\,Hz.
These are removed from the TOI \citep[see, e.g.,][]{planck2014-a08}.
Residuals from this removal process can affect the angular power
spectra of the CMB, although due to the near-constant spin rate of the
\Planck\ spacecraft, the 4-K lines project to the sky at discrete angular
scales.  The lowest frequency line, seen at 10\,Hz in the TOI, projects to
$\ell\simeq600$.  Since this feature is well away from the reionization
structure in the angular power spectrum that we
consider here, we do not attempt to further clean these features from the data.

\subsection{Detector crosstalk}
Electrical coupling, or crosstalk, of the signal between the pair of detectors
in a PSB can create errors in the recovered
polarization. The ground-based calibration of HFI and the in-flight
measurements of current crosstalk put an upper limit to this coupling at the
level of $10^{-3}$ \citep{pajot2010}.
\cite{planck2011-1.5} states this limit as $-60$\,dB. Nevertheless the square
modulation of the bias current of the bolometers was kept at a constant
amplitude throughout the mission.

The voltage crosstalk is more relevant than the current crosstalk for
the coupling of the signal between 
detectors, and is estimated in-flight using cosmic ray
glitches.  While the vast majority of the long glitches are coincident
between the a and b detectors within a PSB pair, there is also a population
of coincident short glitch detections, which have a shift in time that cannot
be due to a single particle and must come from crosstalk; the time shift is on the
order of tens of milliseconds.  These phase-shifted
coincidences are then stacked to solve for a crosstalk amplitude, and
is typically in the range 1--$3\times10^{-3}$ for pairs of PSB detectors.  The
crosstalk level is consistent from a to b bolometers and from b to a,
in all cases.

Using the end-to-end simulations, we propagate the measured voltage
crosstalk between the PSBs into maps of the polarized sky using the
glitch-measured crosstalk amplitudes and 
phase shifts, and compare these to a simulated no-crosstalk
case. The main effect of the crosstalk is to bias the relative
calibration between detectors.  The {\tt SRoll} map-making algorithm
recovers the calibration correction due to crosstalk to very good
precision (as shown in Table~\ref{tab:xtalk}). 

\begin{figure}[!ht]
\includegraphics[width=\columnwidth]{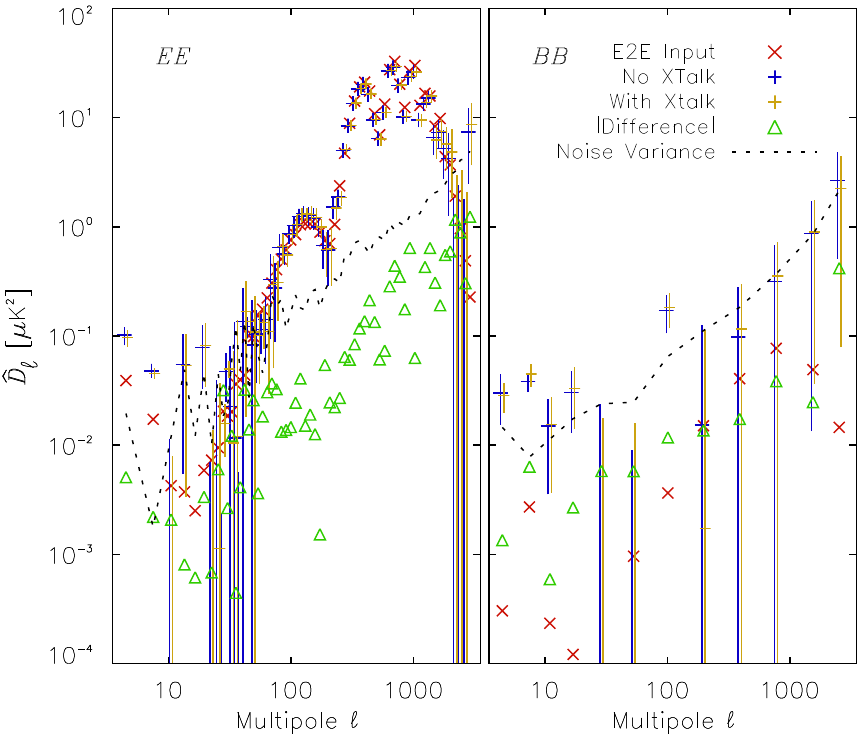} 
\caption{Auto-power spectra $EE$ (left) and $BB$ (right) of 217\,GHz end-to-end
simulations carried out to test the effects of detector crosstalk. Red crosses
indicate the F-$EE$ sky signal input to the simulation, to which a realization
drawn from the noise variance is added (dashed black).  The orange and blue
crosses show the total power in the simulations, with and without simulated
crosstalk, respectively. The absolute value of the difference (between the
simulations including crosstalk and those without crosstalk) are shown in
green.}
\label{fig:Xtalk}
\end{figure}

Figure~\ref{fig:Xtalk} shows the $EE$ and $BB$ angular power spectra
of the simulations.  Introducing detector crosstalk in the
simulation changes the recovered polarization signal by a 
small amount; the absolute value of the difference is lower than
the noise variance in the simulation. However, this might need to be corrected
in future for the detection of very low level polarized signals.

\begin{table}[tb]
\newdimen\tblskip \tblskip=5pt
\caption{Crosstalk simulation.  The calibration change solved by
{\tt SRoll} is compared to the simulated input crosstalk for different
detectors.}
\label{tab:xtalk}
\vskip -6mm
\footnotesize
\setbox\tablebox=\vbox{
 \newdimen\digitwidth
 \setbox0=\hbox{\rm 0}
 \digitwidth=\wd0
 \catcode`*=\active
 \def*{\kern\digitwidth}
  \newdimen\dpwidth
  \setbox0=\hbox{.}
  \dpwidth=\wd0
  \catcode`!=\active
  \def!{\kern\dpwidth}
\halign{\hbox to 2.2cm{#\leaderfil}\tabskip 2.5em&
    \hfil#\hfil\tabskip 1em&
    \hfil#\hfil \tabskip 0em\cr
\noalign{\doubleline}
\omit&Input&Resolved\cr
\noalign{\vskip 1pt}
\omit&crosstalk&calibration change\cr
\noalign{\vskip 3pt} 
\omit\hfil Detector\hfil&[\%]&[\%]\cr
\noalign{\vskip 3pt\hrule\vskip 5pt}
\noalign{\vskip 2pt}
217-5a&   0.224&   0.227\cr
217-5b&   0.159&   0.161\cr
217-6a&   0.112&   0.116\cr
217-6b&   0.112&   0.115\cr
217-7a&   0.447&   0.450\cr
217-7b&   0.398&   0.403\cr
217-8a&   0.200&   0.204\cr
217-8b&   0.126&   0.129\cr
\noalign{\vskip 3pt\hrule\vskip 5pt}}}
\endPlancktable
\end{table}

\subsection{Instrumental polarization}

The polarization efficiencies and angles of HFI detectors were
measured on the ground, directly on the instrument \citep{rosset2010}. The 
polarization efficiencies were determined to better than $0.2\,\%$ for
all PSBs. This error was shown to induce
an uncertainty lower than $0.1\,\%$ on the $E$-mode power spectrum.

The angles of the detectors relative to the instrument were propagated to the
sky reference frame using mechanical modeling of the telescope and
optical simulation. These errors were decomposed into two parts: a
global error of $0 \pdeg 3$, common to all detectors due to the
uncertainty on the angle of the calibrating polarizer; and an error
for each detector, dominated by systematic uncertainties in the ground
apparatus, which were estimated to be lower than $0 \pdeg 9$. Such
error levels, both for global focal plane angle and detector relative
angles, were shown to induce an uncertainty lower than 10\,\% of the cosmic
variance on the $E$-mode power spectrum in the range $\ell=2$--1000.

In order to confirm the ground calibration measurements on the sky data,
we use the $TB$ and $EB$
spectra to check for a possible global rotation of the focal plane at
100, 143, and 217\,GHz, independently. Rotation by an angle 
$\alpha$ induces $E$ and $B$ mixing, leading to spurious $TB$ and $EB$ signals
given by
\begin{eqnarray}
C_{\ell}^{TB, \mathrm{rot}} &=& C_{\ell}^{TE} \sin(2\alpha),
 \label{eq:rotation_tb} \\
C_{\ell}^{EB, \mathrm{rot}} &=& \frac{1}{2}\left(C_{\ell}^{EE}
 - C_{\ell}^{BB}\right)\sin(4\alpha)\;, \label{eq:rotation_eb}
\end{eqnarray}
where we have assumed that $C_{\ell}^{TB}$ and $C_{\ell}^{EB}$ equal zero for
the CMB.

Furthermore, beam mismatch introduces intensity-to-polarization leakage,
which may be parametrized at the spectrum level \citep{planck2014-a15} as
\begin{eqnarray}
\label{eq:beam_mismatch}
C_{\ell}^{TB, \mathrm{beam}} &= & \beta_\ell C_{\ell}^{TT} ,\\\nonumber 
C_{\ell}^{EB, \mathrm{beam}} &= & \epsilon_\ell \beta_\ell C_{\ell}^{TT} +\beta_\ell C_{\ell}^{TE},\\\nonumber 
C_{\ell}^{EE, \mathrm{beam}} &= & C_{\ell}^{EE}+ \epsilon_\ell^2C_{\ell}^{TT} + 2\epsilon_\ell C_{\ell}^{TE},\\\nonumber 
C_{\ell}^{BB, \mathrm{beam}} &= & C_{\ell}^{BB}+
  \beta_\ell^2C_{\ell}^{TT}, \\\nonumber 
C_{\ell}^{TE, \mathrm{beam}} &= & C_{\ell}^{TE} + \epsilon_\ell C_{\ell}^{TT},
\end{eqnarray}
where $\beta_l$ and $\epsilon_l$ quantify the leakage into $E$ and $B$-modes,
respectively. Assuming that differential ellipticity is the main source of beam
mismatch, $\beta_l$ and $\epsilon_l$ are dominated by $m=2$ and $m=4$
modes \citep{planck2014-a13} and can be written as
\begin{equation}
\beta_\ell = \beta_2\left(\frac{\ell}{\ell_0}\right)^2 +
\beta_4\left(\frac{\ell}{\ell_0}\right)^4 \;\;{\mathrm{and}}\;\;
\epsilon_\ell = \epsilon_2\left(\frac{\ell}{\ell_0}\right)^2
 + \epsilon_4\left(\frac{\ell}{\ell_0}\right)^4.
\label{eq:coeff}
\end{equation} 
Combining Eqs.~(\ref{eq:rotation_tb}), (\ref{eq:rotation_eb}),
(\ref{eq:beam_mismatch}), and~(\ref{eq:coeff}),
we can build the five parameter ($\alpha$, $\beta_2$, $\beta_4$, $\epsilon_2$,
$\epsilon_4$) model to be fitted to the $TB$ and $EB$ spectra at 100,
143, and 217\,GHz.

To compute the angular power spectra, we subtract dust
emission from the $Q$ and $U$ detset maps, using the 353~GHz maps
as dust templates. Masking 50\,\% of the sky as well as point sources, we use
{\tt Xpol} \citep{2005MNRAS.358..833T} to compute the cross-spectra between
the two detsets in each channel, which efficiently suppresses the noise bias.
We sample the model parameter space using a Markov chain Monte Carlo (MCMC)
algorithm from which the probability densities of the parameters may
be recovered. The angle $\alpha$ is constrained by the MCMC to the
following values (68\,\% CL, statistical only), depending on the frequency and
spectrum examined.
\begin{itemize}
\item 100\,GHz: $\alpha^{TB} = 0\pdeg01 \pm 0\pdeg21$;
 $\alpha^{EB} = 0\pdeg46 \pm 0\pdeg17$.
\item 143\,GHz: $\alpha^{TB} = 0\pdeg11 \pm 0\pdeg16$;
 $\alpha^{EB} = 0\pdeg29 \pm 0\pdeg13$.
\item 217\,GHz: $\alpha^{TB} = 0\pdeg56 \pm 0\pdeg27$;
 $\alpha^{EB} = 0\pdeg23 \pm 0\pdeg28$.
\end{itemize}
The beam mismatch parameters are not constrained by the MCMC, but they have
little impact (i.e., $\la 1\sigma$) on the values of $\alpha$ quoted above. 
Whatever the channel and spectra, the
angle is never significantly different from zero and always smaller
than the average errors on PSB orientation ($\sim 1\deg$) measured on the
ground prior to launch \citep{rosset2010}.

Comparison with the polarization orientation of the Crab nebula was discussed
in section~3.5.2 of \citep{planck2014-a35} and shown to be consistent with the
above numbers. 

We conclude that the uncertainties of the instrumental polarization parameters
as set up in the data processing pipeline have no significant effects on the
low-$\ell$ $EE$ and $BB$ cross-spectra and are also in broad agreement with 
preflight determinations.

\section{The SRoll global solution}
\label{sroll}

This appendix describes the {\tt SRoll} method for constructing
polarized sky maps from the HFI data.  The method is divided into
three steps.  The first step bins the time-ordered information 
into {\tt HEALPix} pixels per period of stable satellite pointing to
reduce the amount of data (Sect.~\ref{sec:HPRcomputation}). The second
step fits systematic effects, $1/f$ noise and calibration using
differences between measurements in the same {\tt HEALPix} pixel
(Sects.~\ref{sec:vargain} to \ref{sec:finalmin}). The third step
cleans the data using the fitted parameters, and projects the signal
to make polarized maps (Sect.~\ref{sec:projection}).  The quality of
the final maps depends on the efficiency with which the non-sky
signals can be characterized during the fitting procedure in the
second step.  After describing the method, this appendix presents
simulations in Sect.~\ref{sec:TOIsqualitytests} and tests and characterization of the efficiency with which {\tt SRoll}
can clean the signal in Sect.~\ref{sec:hfimaps}.

\subsection{Method}

\subsubsection{Systematic effects considered}
\begin{figure*}[!htbp]
\includegraphics[width=\textwidth]{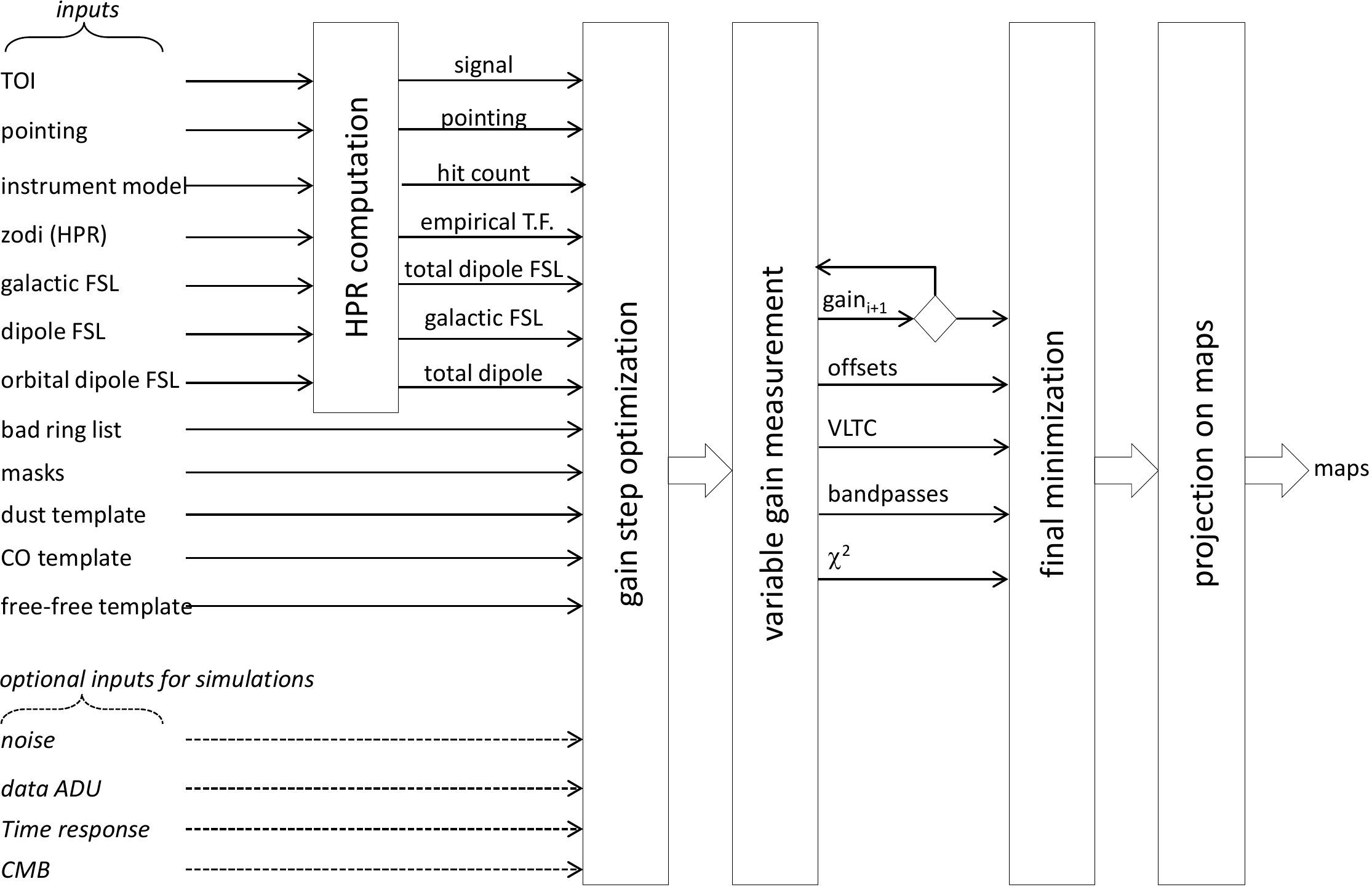} 
\caption{Overview and main functional tasks of {\tt SRoll}.}
\label{fig:srollschema}
\end{figure*}

The method corrects for baseline drifts due to $1/f$
noise in a similar way as the standard unpolarized destriping method
used for previous \Planck\ products
\citep{planck2011-1.7,planck2013-p03f,planck2014-a09}.  {\tt SRoll}
improves on that by solving the amplitudes of
templates of additional systematic effects that are not fully
corrected in the TOI processing, and leakage from temperature to
polarization.  The systematic effects considered are:
\begin{enumerate}
\item gain variation due to residual analogue-to-digital converter
nonlinearity (ADC NL);
\item pick up of Galactic emission and the orbital and solar dipoles
  in the far sidelobes of the telescope (with the far sidelobes defined
  as the response of the instrument at angles greater than $5\degr$
  from the main beam axis);
\item temperature-to-polarization leakage due to bandpass mismatch
  between bolometers, from foreground emission with SEDs differing
  from that of the CMB, namely dust, free-free, and CO;
\item temperature-to-polarization leakage due to calibration mismatch on the
  CMB dipole;
\item large-scale residual transfer functions that are not corrected
  in the deconvolution step of the standard TOI processing.
\end{enumerate}
Figure~\ref{fig:srollschema} shows an overview of the main modules of
{\tt SRoll}.

\subsubsection{Data model}
\label{subsec:srolldm}

All PSB bolometers in a given frequency band are used to solve for the sky map
at a given frequency.  The data for each bolometer (indexed with $b$) are
divided into pointing periods (indexed with $i$) and into periods of constant
gain (indexed with $k$).  The definitions of these data periods are
described in detail in Sect.~\ref{subsec:srollsamp}.  The data sample
indexed with $p$ in the equation below corresponds to a pixel within
an HPR ({\tt HEALPix} ring), not a sample in a time series.

We model the HPR data as
\begin{eqnarray}
\label{eq:datamodel}
g_{b,k}M_{b,i,p} &=&  I_p + \rho_b \left[Q_p\cos(2\phi_{b,i,p}) + U_p\sin(2\phi_{b,i,p})\right] \nonumber \\
 & &  + \sum_{h=1}^{n_{\mathrm{harm}}}\gamma_{b,h} V_{b,i,p,h} + 
\sum_{l=1}^{n_{\mathrm{comp}}}L_{b,l} C_{b,i,p,l}\nonumber \\
 & &  + D^{\mathrm{tot}}_{b,i,p} + F_{b,i,p}^{\mathrm{dip}}+ F_{b,i,p}^{\mathrm{gal}}  + O_{b,i} + g_{b,k}N_{b,i,p}.
\label{SM1}
\end{eqnarray}
For each bolometer $b$ the various terms in this equation are described below.

\begin{itemize}
\item $g_{b,k}$ is the absolute gain of bolometer $b$ during gain
  period $k$.
\item $M_{b,i,p}$ is the bolometer signal sample data in pixel $p$ of
  pointing period $i$.
\item $I_p$, $Q_p$, and $U_p$ represent the common sky maps, excluding the
  dipole, seen by all bolometers in pixel $p$.
\item $\rho_b$ is the polar efficiency, fixed at the ground measurement value;
\item $\phi_{b,i,p}$ is the detector angle of polarization compared to the
  north-south axis.
\item $V_{b,i,p,h}$ is the template of the empirical transfer function in
  bin $h$.  The amplitude of this empirical transfer function is a very small
  correction (around $10^{-3}$), so the data-built maps themselves are used to
  compute the templates using an FFT in the time domain.  The source of the
  template thus suffers from the same deconvolution errors, leading to a
  $10^{-6}$ level. On the dipole signal this induces a
  $3.4\times 10^{-3}\mu$K$_{\rm CMB}$ shift fully negligible.
  Similarly, using the data for foreground templates is more efficient than
  using a simulation model.  The templates are computed for low-order harmonics
  of the spin frequency only in a small number of bins: $h=0$ corresponds to
  the first harmonic; $h=1$ to harmonics 2 and 3; $h=2$ to harmonics from
  4 to 7, and $h=3$ to harmonics from 8 to 15.
\item $\gamma_{b,h}$ is the amplitude of the transfer function
  template for each harmonic bin $h$.
\item $C_{b,i,p,l}$ are the templates of foreground components $l$.  The
  dust component is fitted at all frequencies, the CO component is fitted at
  all frequencies except 143\,GHz, and the free-free component is fitted at
  100\,GHz only.  The templates for the components, without bandpass mismatch
  leakage, are taken initially from the \Planck\ 2013 results and then
  iterated with the maps from previous iterations. The polarization of the
  foreground is taken into consideration only for the dust.
\item $L_{b,l}$ is the amplitude of the bandpass mismatch leakage.
  For all $l$ in the range 1--$n_{\mathrm{comp}}$, we set
  $\sum_{b=1}^{n_{\mathrm{bolo}}} L_{b,l} = 0$.
\item $D^{\mathrm{tot}}_{b,i,p}$ is the total CMB dipole signal (the
  sum of Solar and orbital dipole signals) from the \Planck\ 2013
  results: $D^{\mathrm{tot}}_{b,i,p} = D^{\mathrm{orb}}_{b,i,p} +
  D^{\mathrm{sol}}_{b,i,p}$.
\item $F_{b,i,p}^{\mathrm{dip}}$ is the total dipole signal caught
 by the far sidelobes.
\item $F_{b,i,p}^{\mathrm{gal}}$ is the Galactic signal seen through
  the far sidelobes.
\item $O_{b,i}$ is the $1/f$ noise effect modelled as one offset per
  pointing period.  We set $\sum_{b=1}^{n_{\mathrm{bolo}}}
  \sum_{i=1}^{n_{\mathrm{period}}}O_{b,i}=0$, since the \Planck\ data provide
  no information on the monopole.
\item $N_{b,i,p}$ is white noise.
\end{itemize}

\subsubsection{HPR computation}
\label{sec:HPRcomputation}

The {\tt SRoll} method is based on HPRs as for the \Planck\ 2015 data
release \citep[see section~6.1 of][]{planck2014-a09}.  
This procedure reduces the amount of data, and in the
limit that the spin-axis pointing, noise, and systematic effects are
constant during a pointing period, the information loss is
negligible.

Thus, each detector's TOI for each period of stable spin-axis pointing is binned into {\tt HEALPix} pixels at \Nside=2048, compressing the data.  Furthermore, all sky-map templates fitted by {\tt SRoll} are converted from TOI to HPR (dipoles, CO maps, dust map, and free-free map). The residual transfer function templates (see Sect.~\ref{sec:TimeResponse}) are computed at the timeline level and then projected into HPRs.

\subsubsection{Empirical transfer functions}

Differential time response on long timescales remaining in the data
from the PSBs can create spurious large-scale polarization residuals.
An estimate of the time response of the HFI bolometers and
electronics, determined using planet crossings and stacked cosmic ray
glitches, have been deconvolved from the TOI
\citep{planck2013-p03c,planck2014-a08}.  However, this technique is
not sensitive to time responses on timescales longer than one second.
We refer to this additional transfer function as ``very long time constants''
(VLTCs).

Residuals of far sidelobe (FSL) contributions after removal in the
ring-making phase, together with VLTCs, induce small changes in the
phase and amplitudes of dipoles, as well as of all large-scale structures in
the maps.
 
To remove these small residuals, the HPR-maker builds transfer function
templates composed of harmonics of the spin frequency (with real part
named $R$ and imaginary part named $H$ in Fig.~\ref{fig:srollschema}).
These templates are four bands of low spin frequency harmonics of
the bolometer signal.  This provides a correction for all the residual transfer
function effects, without attributing it to a specific physical mechanism
through a physical model.

The imaginary part is easy to fit, because it is not degenerate with
the sky signal.  At 100--217\,GHz a VLTC correction had already been included 
in the \Planck\ 2015 data release to
improve the knowledge of the lowest frequency portion of the time
response \citep{planck2014-a08}. At 353\,GHz the VLTC correction had
not been deconvolved inside the time response function. After
correction in {\tt SRoll}, the residual imaginary part of the empirical transfer function is at the 
$10^{-4}$ level (see Fig.~\ref{fig:VLTCharmonicfit}), which is subdominant to other
systematic errors.

The real part of the transfer function is degenerate with the signal
in a single pointing period.  Nevertheless, the near-great circle scan
of the telescope boresight, together with the spin axis precession, ensures
that every point on the sky is crossed with at least two directions 
separated by more than $14\deg$.  The
shape of the bolometer response along the scan is fitted in
constrained regions of the sky with enough structure. For low
frequencies (100--217\,GHz), the signal-to-noise ratio of the Galactic
signal is not high enough to do this fit and
the real part is then not removed at those frequencies. The relative 
consistency of the dipole calibration and the first peak seen in
Fig.~\ref{fig:intercalib} shows that the real part of the transfer function
has a negligible difference between 100 and 143\,GHz
and may show some calibration mismatch at 217\,GHz (about $10^{-3}$).
At 353\,GHz the fit succeeds (see Fig.~\ref{fig:vltc353}), similarly to the
procedure used for 545 and 857\,GHz in \cite{planck2013-p03c}.

\subsubsection{Gain step optimization}
\label{subsec:srollsamp}

{\tt SRoll} assumes that the gain of each bolometer is constant over a
number of rings. There are $N_{\mathrm{gain}}$ time-steps that are chosen in
such a way that the total dipole gain variance is constant over each
period. If there is significant variation of the effective gain within a
defined period, then of course the method cannot account for it. This
is only a problem for the very long gain periods, which are
required when the dipole and/or Galactic signal happens to
be weak.  These periods are, by definition, characterized
by relatively small dynamical range of the signal and therefore a
correspondingly small leakage from temperature to polarization.
Furthermore, in these periods the signal samples only a small range in the ADC,
so it is unlikely to exhibit much in the way of gain variations from the ADC
NL.

\begin{figure}[!htbp]
\includegraphics[width=\columnwidth]{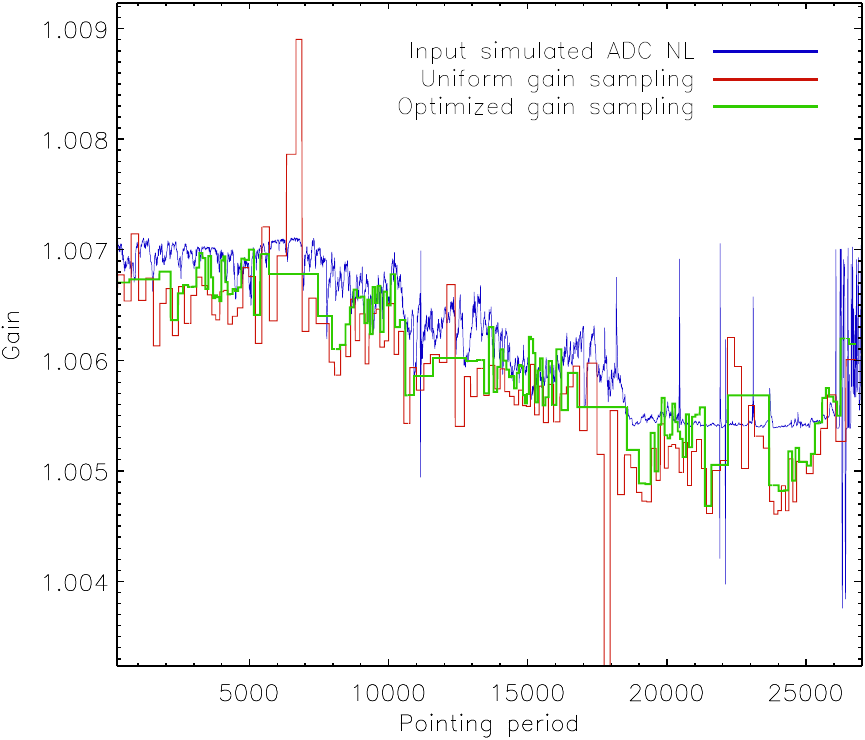} 
\caption{Gains solved by {\tt SRoll} for simulated ADC nonlinearities
  using a uniform period length for the gain (in red) and an optimized
  sampling (in green).  The blue curve shows the apparent gain variation fitted between a simulated total dipole 
and the same total dipole distorted by the simulated ADC nonlinearities. }
\label{fig:gain_sampling_res}
\end{figure}

Figure~\ref{fig:gain_sampling_res} shows the effects of the sampling
on the variable gains solved by {\tt SRoll}; a uniform period length sampling
increases the instability of the gain determination, while an optimized
sampling improves the stability of the result.

\subsubsection{Variable gain measurement}
\label{sec:vargain}

Solving for gain variability necessarily involves solving a nonlinear least squares equation. {\tt SRoll} uses an iterative scheme to solve for the gains $g_{i}$ of the bolometer at each ring $i$. At iteration $n$ we have
\begin{equation}
g_{i, n+1} = g_{i, n} + \Delta g_{i, n},
\label{eq:ge}
\end{equation}
the goal being to fit $\Delta g_{i, n}$. We can also remove the FSL effect as
\begin{equation}
g_{i,n}~M'_{i,p} = g_{i,n}~M_{i,p} - F_{i,p,b}^{\mathrm{dip}}-
F_{i,p,b}^{\mathrm{gal}}.
\label{eq:fsl}
\end{equation}
Using Eq.~(\ref{eq:ge}) and (\ref{eq:fsl}), Eq.~(\ref{eq:datamodel}) becomes
\begin{eqnarray}
g_{i,n}~M'_{i,p}  & = & (1 -  \Delta g_{i, n}) \left\{I_p +\rho  \left[Q_p\cos(2\phi_{i,p}) + U_p\sin(2\phi_{i,p})\right] \right\} \nonumber \\
&+&  D^{\mathrm{tot}}_{i,p} + \sum_{h=1}^{n_{\mathrm{harm}}}\gamma_{h} V_{i,h,p} + \sum_{l=1}^{l_{\mathrm{comp}}}L_{l,b} C_{l,p} + g_{i,n}~N_{i,p},
\label{eq:SimplifiedModel}
\end{eqnarray}
where:
\begin{itemize}
\item $i$ and $p$ are the indices of the ring $i$ and the {\tt HEALPix} pixel $p$
(it is important to remember here that the gain variations are considered stable over several rings, as described in Sect.~\ref{subsec:srollsamp} and talking about the gain at ring $i$ means the gain for the gain period $k$ that contains the ring $i$);
\item $g_{i}$ is the gain of the bolometer in ring $i$;
\item $M_{i,p}$ is the measured bolometer signal for the {\tt HEALPix} pixel
  $p$ of ring $i$;
\item $I_p$, $Q_p$, $U_p$ is the common sky map, excluding the
  dipole, seen by all bolometers in {\tt HEALPix} pixel $p$;
\item $D^{\mathrm{tot}}_{i,p}=D^{\mathrm{sol}}_p + D^{\mathrm{orb}}_{i,p}$, where $D^{\mathrm{sol}}_p$ is the Solar dipole for the {\tt HEALPix} pixel $p$ and $D^{\mathrm{orb}}_{i,p}$ is the orbital dipole at ring $i$ for the {\tt HEALPix} pixel $p$.
\end{itemize}

Based on the destripping method of \citet{keihanen2004}, for a known gain $g_{i,n}$, Eq.~(\ref{eq:SimplifiedModel}) solves for $\Delta g_{i, n}$, $\gamma_{h,b}$ and $L_{l,b}$.
Should $D^{\mathrm{tot}}_{p}$ not be degenerate with the foregrounds  or the
systematic effects, $\Delta g_{i,n}$ would be the exact difference
between the gain of the first iteration and the real gain, and the
method would converges in a single step.  Unfortunately, foregrounds
projected on a ring have a dipole component that leads to a
degeneracy between the foregrounds and the total dipole. Separating the
dipole and the orthogonal components of the signal, we write
\begin{equation}
I_p=\widetilde{I_{i,p}} + \eta~D^{\mathrm{tot}}_{i,p},
\label{eq:OrthogonalSky}
\end{equation}
where $\widetilde{I_i,p}$ is orthogonal to $D^{\mathrm{tot}}_{i,p}$ during the
period of roughly constant gain (i.e., the period $k$ as defined in
Sect.~\ref{subsec:srollsamp}).

\begin{figure}[!htbp]
\includegraphics[width=\columnwidth]{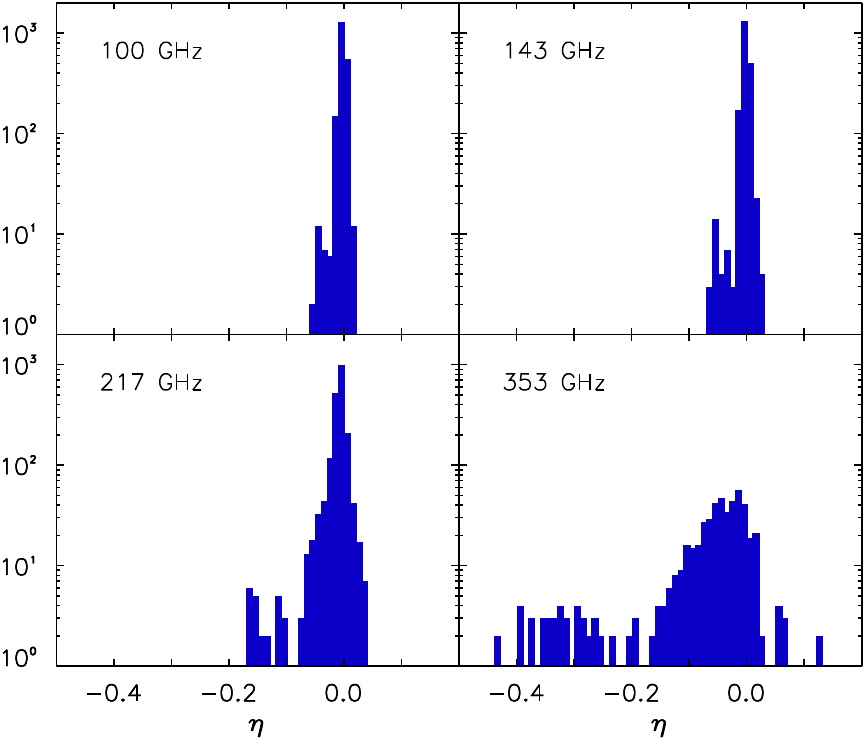} 
\caption{Histogram of the $\eta$ parameter measured on real data. For
  each gain-stable period, the total dipole is fitted on the Galactic
  signal.} 
\label{fig:plsig}
\end{figure}

Figure~\ref{fig:plsig} shows the histogram of this $\eta$
parameter. Its dispersion is explained by the variation during the
mission of the orientation of the ring with respect to the Galactic
plane. Moreover the amplitude of the dipolar component of the Galaxy
increases with frequency. Thus the magnitude of the degeneracy
increases and then the dispersion of the $\eta$ parameter increases
with frequency as well. Then, Eq.~(\ref{eq:ge}) becomes 
\begin{equation}
g_{i,n+1}= g_i + \eta~\Delta g_{i,n}.
\end{equation}
At all frequencies $|\eta|<1$ and then $g_{i,n}\rightarrow g_i$ when $n
\rightarrow \infty$. Thus the degeneracy between ${I_p}$ and
$D^{\rm tot}_{i,p}$ only impacts the convergence rate of the algorithm;
{\tt SRoll} always converges to the right gain estimation when the
data model is an accurate description of the sky signal.

Table~\ref{tab:iterations} shows how the gain convergence works in a
simulation without noise for various $\eta$ values. When $\eta=0$ the
gain is solved in one iteration. When $\eta$ is non-zero, the gain
convergence reaches $10^{-5}$ after only four iterations. Thus,
{\tt SRoll} uses four iterations.

These simulations also test the degeneracy with bandpass
leakage. They show that the degeneracy between gain determination and
bandpass leakage is small. As we will see below, more detailed 
simulations that contain noise enable us to control more precisely  the level
of accuracy that {\tt SRoll} reaches while fitting gain and other systematics.

\begin{table}[tb]
\newdimen\tblskip \tblskip=5pt
\caption{Relative error on the gain given the number of iterations for
  different foreground levels.}
\label{tab:iterations}
\vskip -6mm
\footnotesize
\setbox\tablebox=\vbox{
 \newdimen\digitwidth
 \setbox0=\hbox{\rm 0}
 \digitwidth=\wd0
 \catcode`*=\active
 \def*{\kern\digitwidth}
  \newdimen\signwidth
  \setbox0=\hbox{+}
  \signwidth=\wd0
  \catcode`!=\active
  \def!{\kern\signwidth}
\halign{\hbox to 1.5cm{#\leaderfil}\tabskip 1em&
    \hfil$#$\hfil\tabskip 1em&
    \hfil$#$\hfil&
    \hfil#\hfil\tabskip 0em\cr
\noalign{\doubleline}
\omit\hfil Iterations\hfil&\omit\hfil No foreground\hfil&\omit\hfil Low foreground\hfil&High foreground\cr
\noalign{\vskip 3pt} 
\omit\hfil \hfil & \eta =0 & \eta =-0.08 & $\eta =-0.8$\cr
\noalign{\vskip 3pt\hrule\vskip 5pt}
\noalign{\vskip 2pt}
1&4.9 \times 10^{-29}&9.6 \times 10^{-4}&$9.6 \times 10^{-3}$\cr
2& &8.9\times10^{-6}&$8.9\times10^{-4}$\cr
3& &9.9\times10^{-8}&$9.9\times10^{-5}$\cr
4& &1.2\times10^{-9}&$1.1\times10^{-5}$\cr
\noalign{\vskip 3pt\hrule\vskip 5pt}}}
\endPlancktable
\end{table}

We find that the $\eta$ variance increases when smaller gain-stable
periods are used. Figure~\ref{fig:plsig} shows that the gain-stable
period cannot be decreased at 353\,GHz, otherwise this would lead to
$|\eta|>1$. That is why we define longer periods at 353\,GHz than at
other frequencies.  
Masking pixels where the Galactic signal is degenerate with the dipole signal
also helps to obtain $|\eta|<1$.

An alternative solution could use a Galactic model to fit gain
variations on shorter timescales. This procedure would work if the
following requirements were met: first, the Galactic template must not
have noise that correlates with the data, such as using a previous map
version at the same frequency; second, the Galactic template has to have no
residual dipole remaining from a previous calibration mismatch; and thus,
{\tt SRoll} currently solves for gain variations with dipole residual
fitting.

\subsubsection{Final minimization}
\label{sec:finalmin}
During the last step of the {\tt Sroll} algorithm, the minimization is the same as in the previous
iteration, but the dipole is not removed. The minimization then
adjusts the other parameters (offsets, VLTC, and bandpasses) with the
gains as they were determined at the previous step.

\subsubsection{Projection onto maps}
\label{sec:projection}
{\tt SRoll} projects time-ordered information to pixel maps using gains and systematic effects fitted during the previous steps. From
Eq.~(\ref{eq:datamodel}) we obtain:
\begin{equation}
g_{i}~M_{i,p} - R_{i,p} = I_p+\rho_b \left[  \cos (2\phi_{_{i,p,b}})Q_p + \sin (2\phi_{_{i,p,b}})U_p \right].
\end{equation}
Thus, inside each pixel we can define
\begin{equation}
\chi^2_p = \sum_{i}\frac {\left\{\splitfrac{I_p+\rho_b \left[ \cos (2\phi_{_{i,p,b}})Q_p + \sin (2\phi_{_{i,p,b}})U_p \right] }{ - g_{i,n}~M_{i,p} + R_{i,p} }\right\}^2} {g_{i,n}^2~\sigma_{i}^2 }.
\end{equation}
A linear system is used to find the values of $I_p$, $Q_p$, and $U_p$ that
minimize this $\chi^{2}$.

\subsection{Masks}
\label{sec:srollmask}

\begin{figure}[!htbp]
\includegraphics[width=\columnwidth]{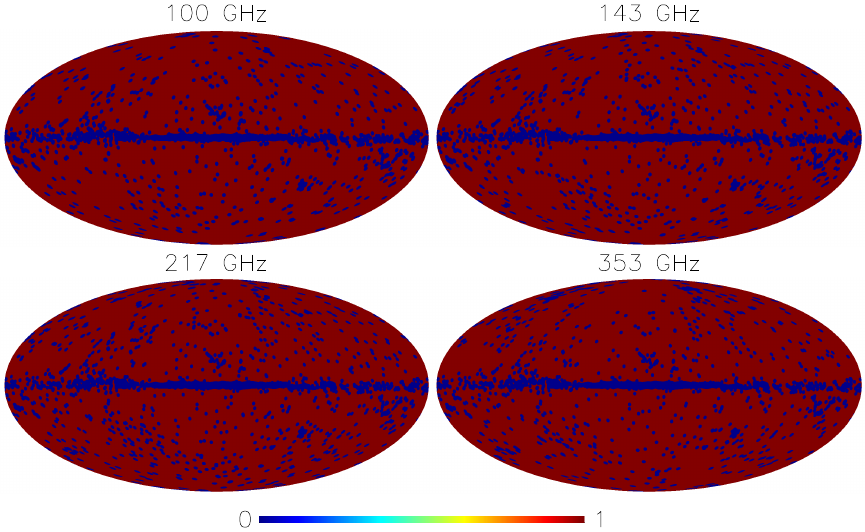} 
\caption{Masks and sky fraction left unmasked during the mapmaking
  solution.  The fractions are: $f_{\mathrm{sky}} = 0.862$ at 100\,GHz; $f_{\mathrm{sky}} = 0.856$ at 143\,GHz; $f_{\mathrm{sky}}  = 0.846$ at 217\,GHz; and $f_{\mathrm{sky}}  = 0.862$ at 353\,GHz. }
\label{fig:masksroll}
\end{figure}
The solution for the map parameters is done with the sky partly masked.  This
masking has two goals, firstly, to avoid regions of the sky with time
variable emission, and secondly to avoid regions with a strong
signal gradient. Thus, the brightest point sources are
removed (e.g., the flux of 3C273 is strong and it changes during the mission). The
Galactic plane is also masked, due to the strong signal
gradients there. Nevertheless, a relatively large sky coverage is needed to
properly solve for the bandpass mismatch extracted from the Galactic signal.
Fig.~\ref{fig:masksroll} shows the masks used and the sky fraction that is
retained (after masking) in the map solution.

\subsection{End-to-end simulations}
\label{sec:E2E}

\subsubsection{General description}
\label{sec:srollsimu}
Simulations are used to characterize the performance of {\tt Sroll}. We start from simulated TOI data sets after TOI processing, including the total CMB dipole (solar and orbital), the CMB anisotropies, Galactic and extragalactic foregrounds, and noise. These are processed to simulate all systematics relevant for mapmaking. The {\tt SRoll} mapmaking is then applied to these TOI to obtain the so-called E2E simulated frequency band $I$, $Q$, and $U$ maps and associated $TT$, $EE$, and $BB$ power spectra. For testing purposes, one can choose specific subsets of input signals, systematic effects, and processing modules.

Building a simulation data set with 100 noise and 100 CMB realizations was not feasible for this paper.
We thus test if the CMB can be simulated separately from the noise and systematics and for this purpose two sets of TOIs are simulated, the first including all the inputs, and the second including all inputs except for CMB anisotropies. Maps are made with {\tt SRoll} from both sets of TOIs.
 In the second case, the CMB anisotropies are added at the map level after the mapmaking.
The difference between these two tests shows that we can avoid including the loop over many CMB cosmological signals at the TOI level.

\begin{figure}[!htbp]
\includegraphics[width=\columnwidth]{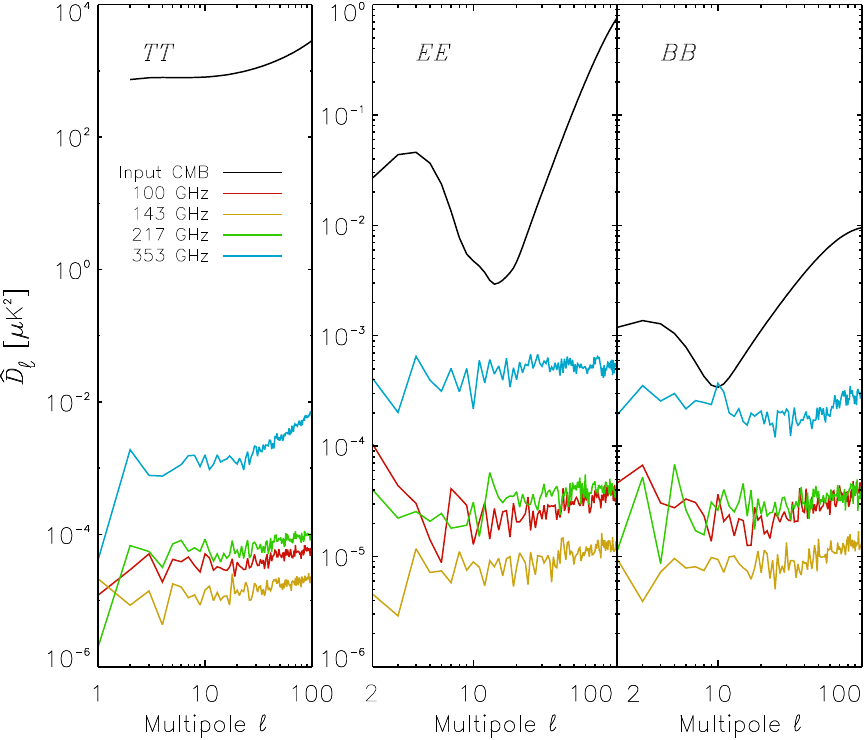} 
\caption{Difference of the power spectrum of a realization of residuals where the CMB is simulated at the TOI level and a realization where the CMB is added to the simulated residual map. In this simulation run, the $BB$ power spectrum is set to F-EE.}
\label{fig:diff_sim}
\end{figure}
Figure~\ref{fig:diff_sim} shows this difference at the power spectra level. When the CMB is simulated at the TOI level, the calibration distortion is applied to the CMB and then solved by {\tt Sroll}.
All differences are much smaller than the signal, which demonstrates that 
the CMB can indeed be added to the residual map, i.e.,
\begin{equation}
 \left [I, Q, U\right ]_{\rm sim}=\left [I, Q, U\right ]_{\rm residual}+\left [I, Q, U\right ]_{\rm CMB}.
\end{equation}
The difference is typically less than one percent of the fiducial $EE$ spectrum for the CMB channels (it is higher at 353\,GHz, but this channel is only used to remove the dust from CMB channels after multiplication by factors smaller than $10^{-2}$).

The $TT$, $EE$, and $BB$ power spectra are then computed with {\tt Spice} \citep{2001ApJ...548L.115S} from the output maps, using 50\,\% of the sky if not otherwise specified. Then these power spectra are compared with the fiducial $TT$, $EE$, and $BB$ spectra. 

We note that this also demonstrates that {\tt SRoll} does not significantly affect the CMB anisotropies in any way and that the systematic effects can be treated as additive. This is an important result, but not entirely surprising, because the dominant polarized systematics are the residual dipole leakage from calibration mismatch and the residual foreground leakage from bandpass mismatch, both of which produce spurious polarization signals that add to the CMB anisotropies. The distortions of the CMB anisotropies themselves are much lower than the noise and therefore negligible.

\subsubsection{Foregrounds}
The foreground model used in the simulations is based on the Planck Sky
Model version 1.9 \citep{delabrouille2012} in which each foreground
template is convolved with the ground-measured bandpass, leading to
realistic bandpass leakage levels.

\subsubsection{ADC nonlinearity}
\label{sec:ADCsims}
\begin{figure}[!htbp]
\includegraphics[width=\columnwidth]{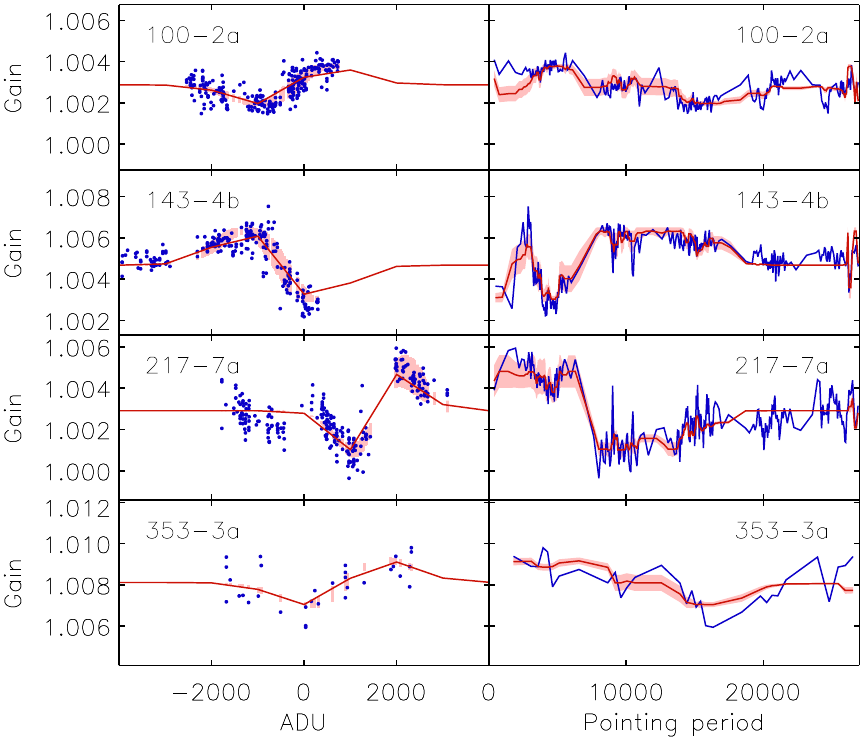} 
\caption{Evolution of the gain solved by {\tt SRoll} for four representative bolometers. The left column shows the gain as a function of the signal level in ADU, with the values determined from the data in blue. The red line is the best fit and the pink area surrounding it is the standard deviation of a four-parameter model of the residual ADC nonlinearity fitted to these data. The right column shows the gain as a function of pointing period. This shows that the model is able to reproduce the broad features of the remaining gain variation observed during the mission.}
\label{fig:ADUSIM}
\end{figure}

We build a simple four-parameter model of the ADC NL that remains after the correction at the TOI level. The TOI correction uses measurements performed during the warm phase of the mission. The most significant residual after this correction is the central discontinuity in the ADC. Our model of the residual nonlinearity is constructed as a function of the ADU value. The central discontinuity is a step, but it is smeared out by the noise in the data, so we model it as the derivative of a Gaussian. The left column of Fig.~\ref{fig:ADUSIM} shows the gain as a function of ADU value for four representative bolometers. The values determined by {\tt SRoll} from the data are shown as blue points. The red line shows the best-fit model for these gains and the pink area surrounding the line shows the standard deviation computed from 100 realizations of the uncertainties in the fit. The right column of Fig.~\ref{fig:ADUSIM} shows the measured and modelled gains as a function of pointing period, demonstrating that the model captures the broad features of the remaining gain variation observed during the mission. This model of the residual nonlinearity is used in the simulations. For each realization, we draw the parameters of the model from the fitted distribution, and the consequent nonlinearity is applied to the simulated data. The goal of the model is to simulate the distortion of the dipole that is present in the data (as discussed later in Sect.~\ref{sec:perfadc}), but not removed by {\tt SRoll}. The simulations will be used to estimate the level of these systematic effects and to take them into account in the analysis of the data.

\subsubsection{Low frequency transfer function}
Low frequency transfer functions are not simulated, but are nevertheless solved by {\tt SRoll}. The fitted parameters are thus expected to be zero. The residual low-frequency transfer function patterns after {\tt SRoll} are then interpreted as degeneracies with other simulated systematic errors.  The main aim of the resolution of this transfer function is to model the very long time constants (VLTCs) that were not characterized at the TOI level.

\subsubsection{Noise}
\label{sec:annexnoise}
The noise is assumed to be Gaussian and stationary throughout the mission  (see Sect.~\ref{sec:detnoise}). The noise is simulated by drawing a realization for each pointing period from the measured noise power spectrum of each bolometer and its readout chain. Correlations between bolometers are not included in the model.

\subsection{Validation and performance}

\subsubsection{Bandpasses}
\label{sec:srollbandpasses}

\begin{figure}[!htbp]
\includegraphics[width=\columnwidth]{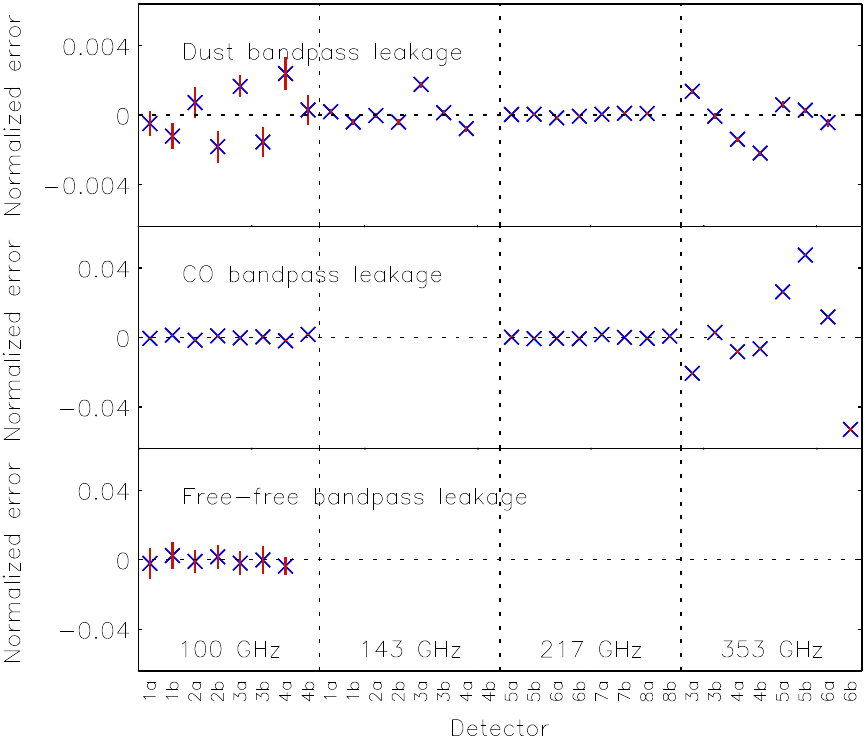} 
\caption{ Error on the recovered leakage coefficients solved by {\tt SRoll} normalized to the average of all detectors in a frequency band, on a representative set of detectors. The error bars (in red) are the statistical distribution from 20 realizations of the simulations.}
\label{fig:leaksim}
\end{figure}
Figure~\ref{fig:leaksim} shows the  error on the reconstructed leakage coefficients, normalized to the average of all detectors in a frequency band, for the three main foregrounds in the HFI channels (dust, CO, and free-free). Although the dispersion
 is in some cases larger than the statistical noise, the accuracy is better than one percent level and small enough to induce negligible effects on the gain determination and the sky maps produced by {\tt SRoll} . The excess variance of the bandpass leakage distribution at 100 and 353\,GHz is probably due to the degeneracy between the main leakage sources (CO and dust).
This is further confirmed by the extreme stability of the inter-calibration of the 100, 143, and 217\,GHz 
detectors, which have rms dispersions in the range 2--$4 \times 10^{-5}$.

Synchrotron emission is weak in the HFI channels (see Sect.~\ref{sec:compsep})
and so the synchrotron leakage will be even weaker. Using  from the ground measurements estimates of the bandpasses shows that it is negligible.

\begin{figure}[!htbp]
\includegraphics[width=\columnwidth]{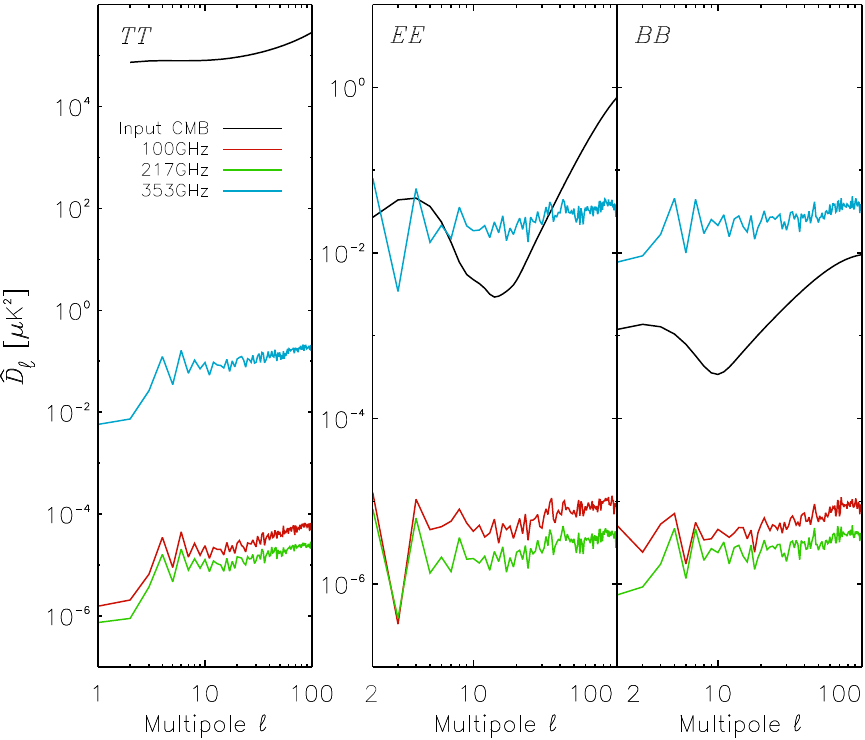} 
\caption{Residual auto-power spectra of the CO bandpass leakage residuals.}
\label{fig:DIFF_CL}
\end{figure}

\begin{figure}[!htbp]
\includegraphics[width=\columnwidth]{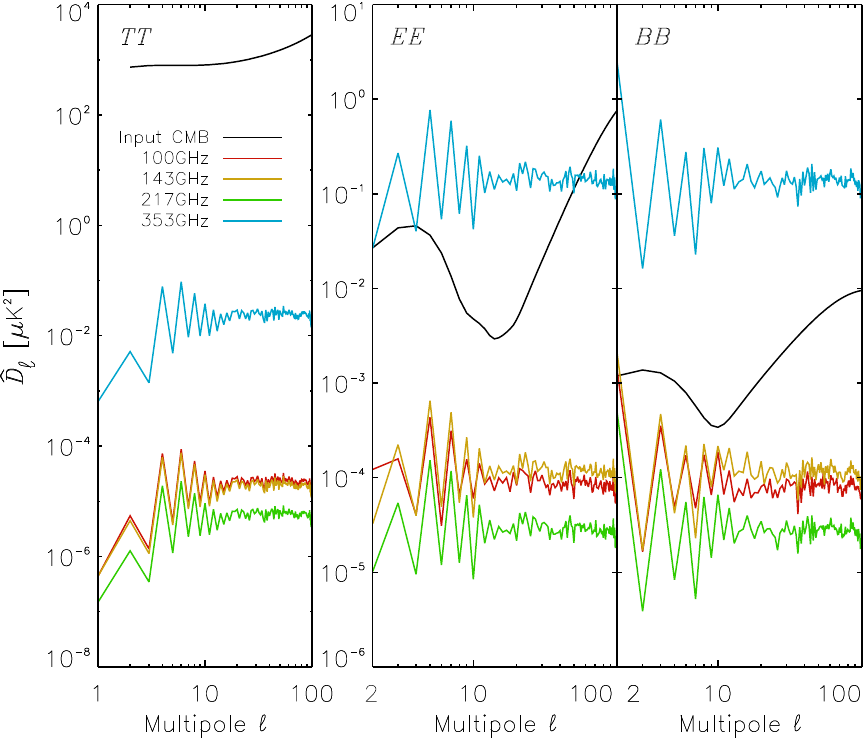} 
\caption{Residual auto-power spectra of the dust bandpass leakage residuals.}
\label{fig:DIFF_DL}
\end{figure}

\begin{figure}[!htbp]
\includegraphics[width=\columnwidth]{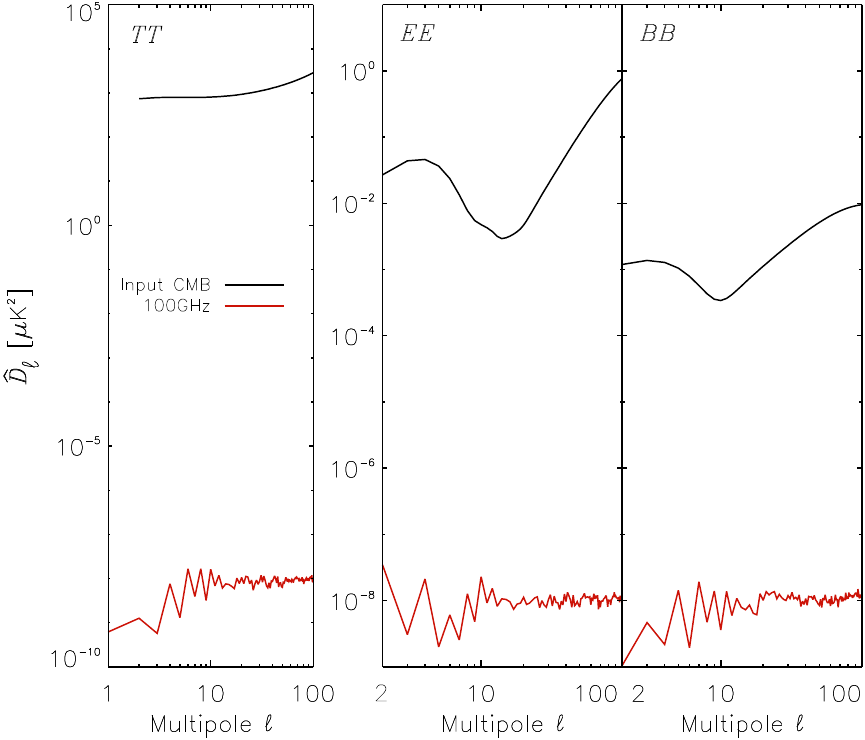} 
\caption{Residual auto-power spectra of the free-free bandpass leakage residuals.}
\label{fig:DIFF_FL}
\end{figure}

Using the reconstructed leakage coefficients, the corresponding templates, and the pointing, we build maps of the residual bandpass leakage. Figures \ref{fig:DIFF_CL}, \ref{fig:DIFF_DL}, and \ref{fig:DIFF_FL} show the power spectra of those maps for CO, dust, and free-free foregrounds, respectively. These are negligible with respect to the fiducial $EE$ spectrum at 100, 143, and 217\,GHz. The 353\,GHz channel is only used to clean the dust emission from the CMB channels, so the additional leakage induced in these channels will be scaled down by the coefficients used to do the cleaning (factors of $3.5\times 10^{-4}$, $1.7\times 10^{-3}$, and $1.6\times 10^{-2}$  at 100, 143, and 217\,GHz, respectively).

\subsubsection{ADC nonlinearities}
\label{sec:perfadc}

The full ADC NL induces dramatic distortions in the signal (as shown in Sect.~\ref{sec:adc}), which are mostly corrected at the TOI level. We investigate here the residuals after the TOI-level correction, as well as the secondary correction applied to them in the mapmaking, using the model described above in Sect.~\ref{sec:ADCsims}.

\begin{figure}[!htbp]
\includegraphics[width=\columnwidth]{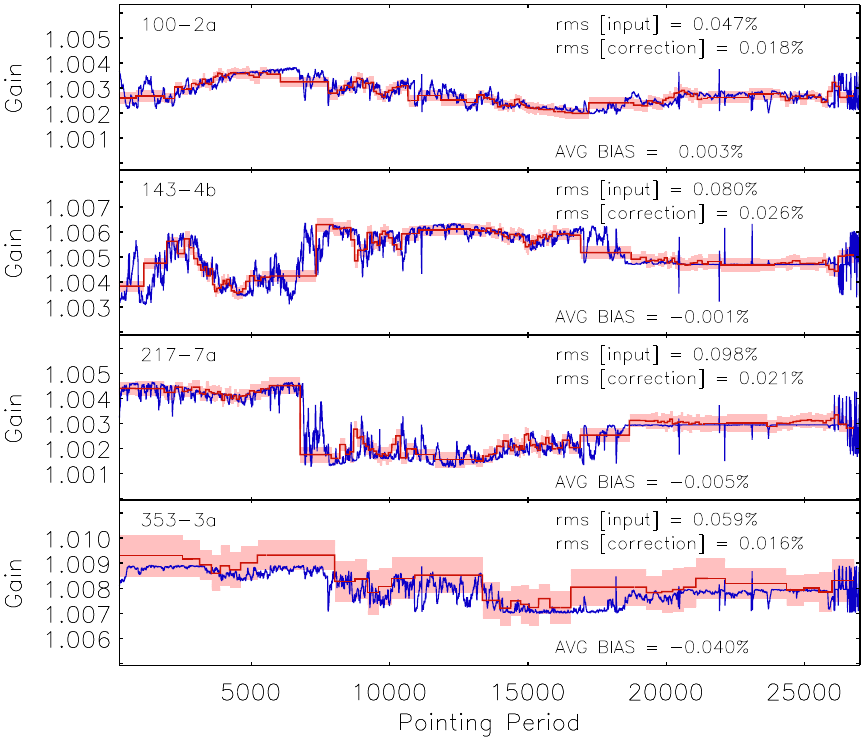} 
\caption{Gain variations solved by {\tt SRoll} (red line and band to show the variance of 20 noise simulation realizations) compared to the gain variation computed directly from the input ADC model (blue).}
\label{fig:adcgainsim}
\end{figure}
Figure~\ref{fig:adcgainsim} shows the results of ADC NL simulations for four representative bolometers. We simulate the nonlinearity using the model described in Sect.~\ref{sec:ADCsims}. The gains reconstructed by {\tt SRoll} from 20 realizations are shown with a red line (mean) and pink band (standard deviation). These can be compared to the blue line, which shows the gain variation computed directly from the input ADC model. The difference between the blue and red lines averaged over the mission gives the average bias on the gain (shown in each panel as ``AVG BIAS''), which is very small in the CMB channels (less than $10^{-4}$) and $4 \times 10^{-4}$ at 353\,GHz.

\begin{figure}[!htbp]
\includegraphics[width=\columnwidth]{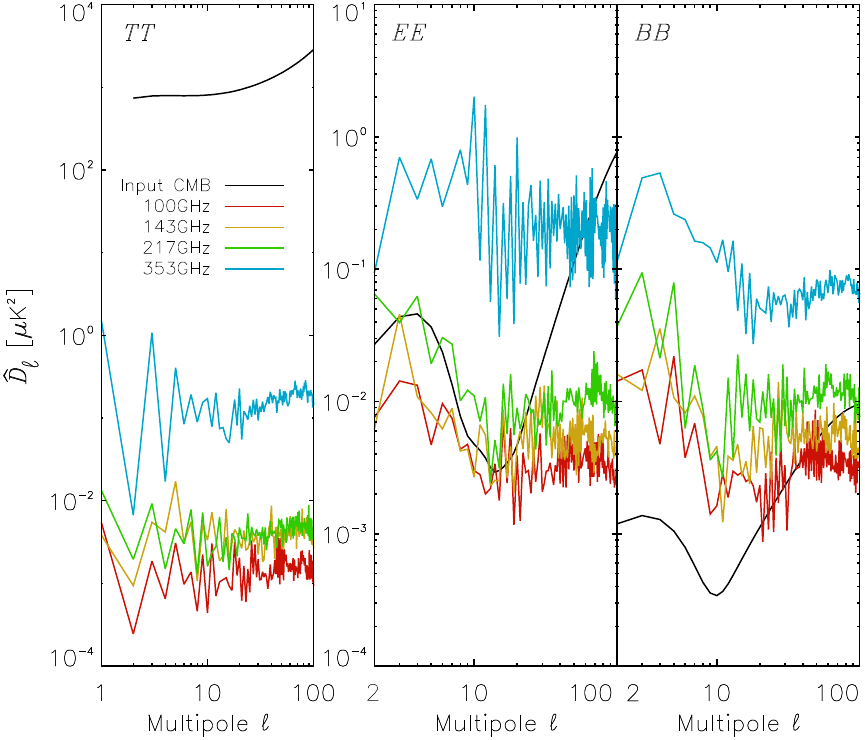} 
\caption{Residual auto-power spectrum of the ADC nonlinearity corrected by the gain variation model.}
\label{fig:DIFFGL}
\end{figure}
The result of the simulations are the residual  power spectra of the ADC NL corrected for the gain variation model, shown in Fig.~\ref{fig:DIFFGL}. This shows that these residuals in $EE$ for 100 and 143\,GHz are low for $\ell >30$, but close to the level of the fiducial $EE$ spectrum for the reionization peak at $\ell = 2$--3. At these frequencies, the apparent gain correction and inter-frequency calibration from {\tt SRoll} have been shown to be very accurate (see Fig.~\ref{fig:adcgainsim} and Table~\ref{tab:calibmismatch}).

\begin{figure}[!htbp]
\includegraphics[width=\columnwidth]{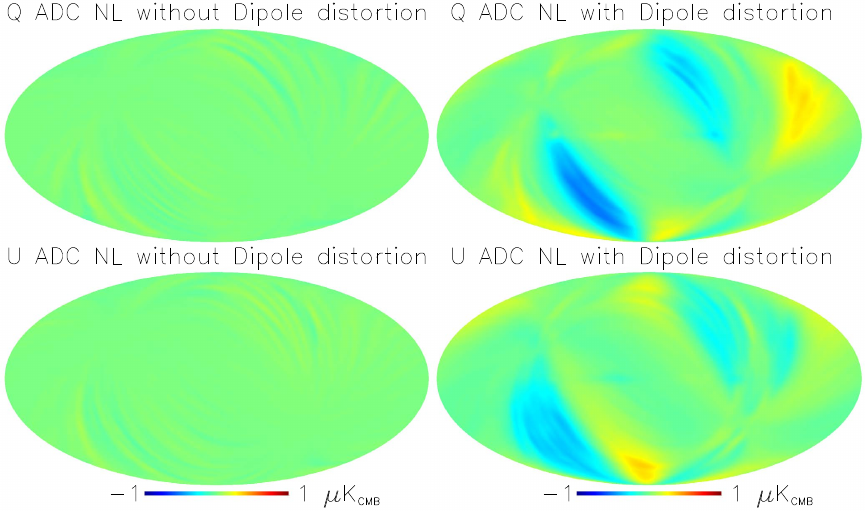} 
\caption{Simulation of residual polarization ($Q$ on the top row and $U$ on the bottom row) maps at 100\,GHz.  Maps in the left column are obtained by using a constant gain per ring, while maps in the right column are obtained when the simulations are run with the ADC NL model.}
\label{fig:adcnl}
\end{figure}
Figure~\ref{fig:adcnl} shows residual polarization maps from simulations of the 100\,GHz channel without noise. The left column shows the residuals obtained when the data are simulated using a linear model with a gain that is constant for each ring, but that varies over the mission. This is the model that {\tt SRoll} fits to the data. It is able to recover the gains almost exactly, so the residuals are at a very low level. The right column shows a simulation including the complete effect of the residual ADC nonlinearity. The additional residuals seen in this case are produced by the distortion of the dipole signal that {\tt SRoll} does not correct. The associated power spectra explains the low-$\ell$ power spectrum residuals in Fig.~\ref{fig:DIFFGL} as is demonstrated below for one specific detector.

\begin{figure}[!htbp]
\includegraphics[width=\columnwidth]{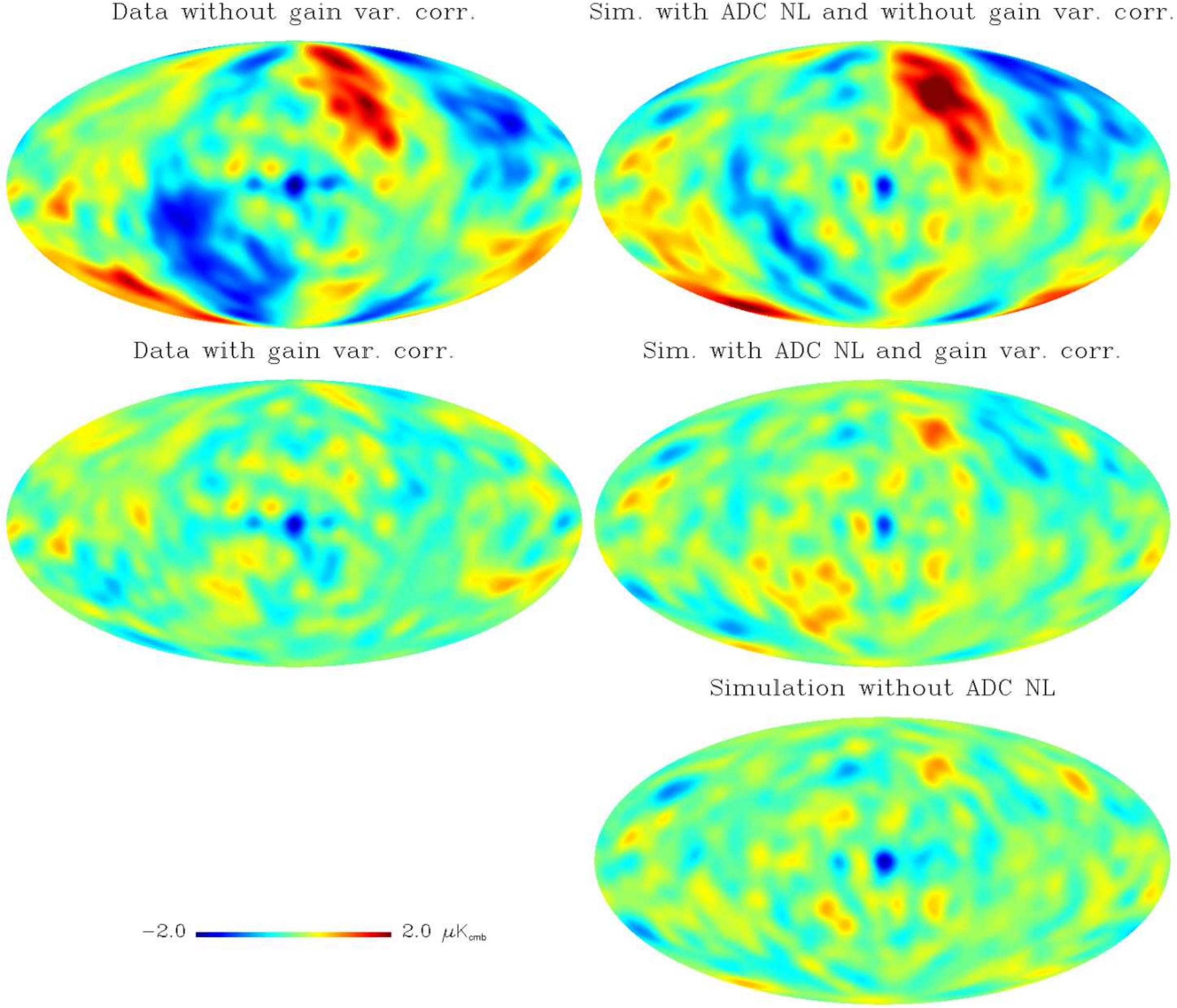} 
\caption{$Q$ maps of data (left column) and a simulation of the ADC NL effects (right column). The first row is the total effect. The second row shows the improvement brought about by the correction of the linear gain variation. The third row shows the noise and residuals from other systematics.  }
\label{fig:plmapresidumaps}
\end{figure}

Figure~\ref{fig:plmapresidumaps} shows maps of a simulation of the ADC NL (left column) compared to the data (right column) for one specific bolometer. The first row shows the maps made before correcting the ADC NL with the {\tt SRoll} variable gain model, which demonstrates that the simulations can effectively reproduce the effect seen in the data. The second row shows maps made after correction with the {\tt SRoll} variable gain model. The large-scale patterns seen in the first row are mostly removed by this correction. The dipole distortion shown in Fig.~\ref{fig:adcnl} is still present, but it is partly hidden by the noise. The third row shows the map of the simulation when the ADC NL is not introduced, and contains only noise. The dipole distortion is not present in this map. This is a clear indication that the only residual remaining after the {\tt SRoll} mapmaking is small and comparable to the noise only at large scales and is the dipole distortion. A quantitative estimate of the accuracy of the simulation is made in Sect.~\ref{sec:adc} by comparing the power spectra of these maps. It shows indeed that the dipole distortion residual is lower than the noise and other residual systematic effects at $\ell < 10$, and exeed those only at $\ell = 2$--3.

We have shown that the maps made by {\tt SRoll} contain a residual ADC NL effect from the distortion of the dipole, which cannot be corrected by the variable gain model. The distortion of the dipole can nevertheless be predicted using the gain variation based model of the nonlinearity shown in Fig.~\ref{fig:ADUSIM} and described in Sect.~\ref{sec:ADCsims}.

\subsubsection{Residual complex transfer function}
\label{sec:TimeResponse}

The correction applied to the VLTC only accounts for the shifts in the dipole that it produces, which had previously allow us to use the orbital dipole for calibration. It is clear that there must be some time constants with low amplitudes affecting the data at low multipoles beside the dipoles, which lead to additional small in-scan shifts. The FSLs lead to a cross-scan shift which is cancelled to first order in the sum of odd- and even-survey data, but which leaves an amplitude effect in the cross-scan signal. Finally, the so-called baffle component of the FSLs is not correctly described by the first-order {\tt GRASP} model used to remove them. This produces an extra in-scan shift and change in the amplitude. We therefore fit for empirical transfer functions to account for these small residuals. These transfer functions are fitted for four ranges of harmonics of the spin frequency (1, 2--3, 4--7, and 8--15). The accuracy of the reconstruction of the corresponding amplitudes is estimated by using the simulations.
\begin{figure}[!htbp]
\includegraphics[width=\columnwidth]{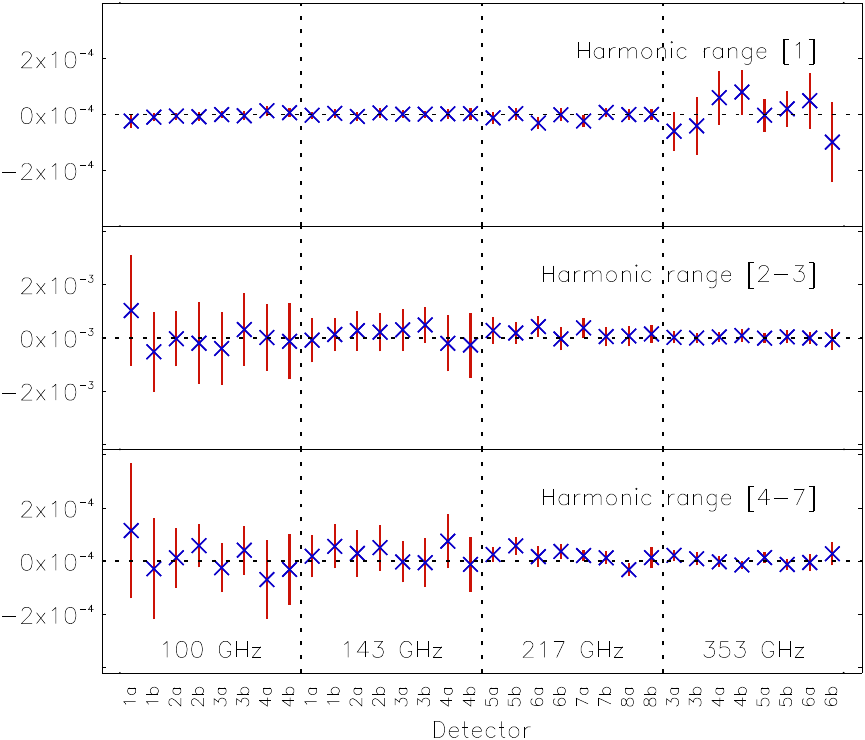} 
\caption{Low frequency complex transfer function amplitudes residuals  simulation for all bolometers 
and for three ranges of spin frequency harmonic ranges. Errors bars are computed from 20 noise simulations.}
\label{fig:plh0}
\end{figure}
Figure~\ref{fig:plh0} shows the mean and standard deviation of the reconstructed amplitudes from 20 realizations of a simulation where no extra transfer functions were included, so the true value is zero in all cases. The results are consistent with zero within the statistical uncertainties. Therefore we conclude that the degeneracies with the other systematic effects are small and that this correction for the residual transfer function will be accurate to the level of the noise.

\begin{figure}[!htbp]
\includegraphics[width=\columnwidth]{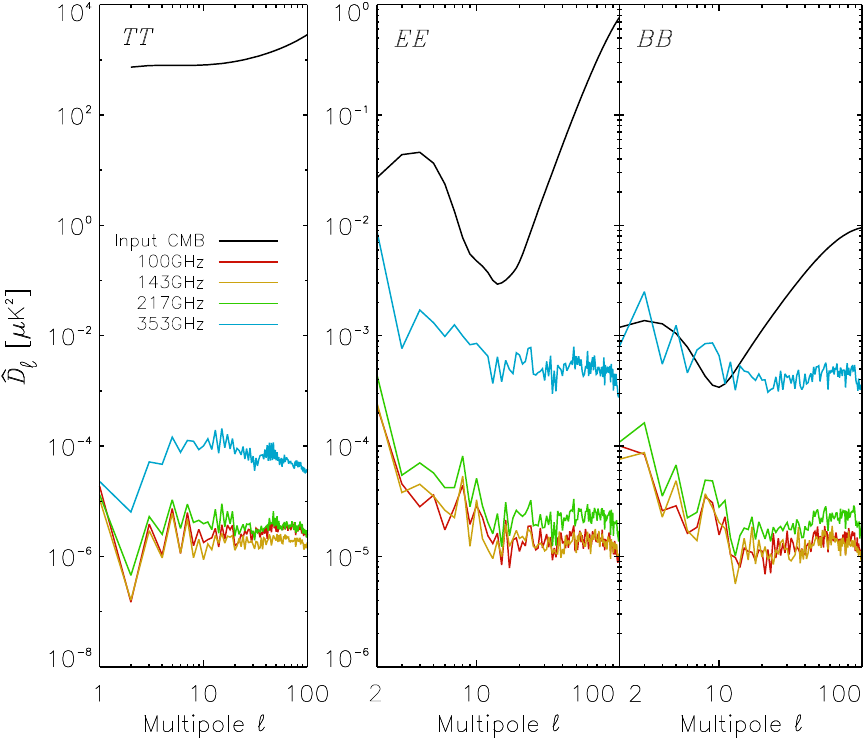} 
\caption{Simulated power spectra of the transfer function residuals compared to the fiducial CMB signal (black curve). For each bolometer, the amplitudes of the empirical transfer function were drawn from the uncertainties shown in Fig.~\ref{fig:plh0}. The residuals are negligible compared to the signal.}
\label{fig:DIFF_H0}
\end{figure}
Figure~\ref{fig:DIFF_H0} shows that the residual transfer functions in the maps lead to a negligible residual in the power spectrum for the CMB channels. The 353\,GHz channel shows an effect close to the level of the fiducial signal, but this channel is only used to clean the CMB channels, so the residuals added to those channels by this process are scaled down by the corresponding cleaning coefficients (as discussed before for other simulations).

\subsubsection{Noise}
\begin{figure}[!htbp]
\includegraphics[width=\columnwidth]{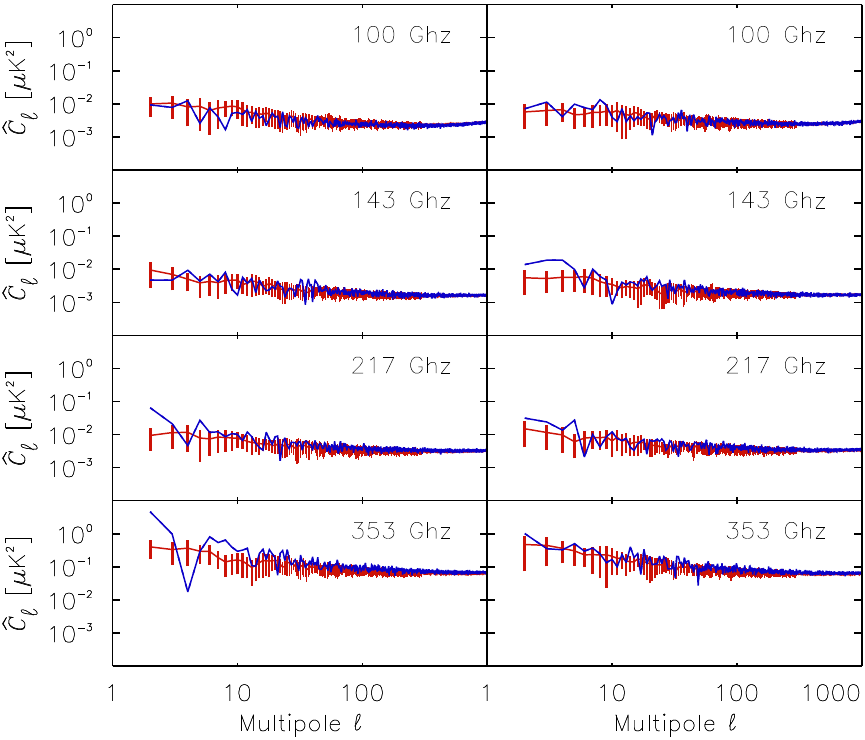} 
\caption{Power spectra of null tests on detsets (left column) and half missions (right column). The data are shown in blue and the mean and standard deviation from simulations containing noise alone are shown in red.}
\label{fig:plsimcl}
\end{figure}

Figure~\ref{fig:plsimcl} shows power spectra of null tests on detector sets and half missions. The blue lines show the data, and the red lines and error bars show the mean and standard deviation of simulations containing noise alone. Thus there is no evidence for unseen systematics in these null tests. 

\subsubsection{Calibration mismatch}

\begin{table}[tb]
\newdimen\tblskip \tblskip=5pt
\caption{Calibration mismatch from 20 simulations, in percent.}
\label{tab:calibmismatch}
\vskip -6mm
\footnotesize
\setbox\tablebox=\vbox{
 \newdimen\digitwidth
 \setbox0=\hbox{\rm 0}
 \digitwidth=\wd0
 \catcode`*=\active
 \def*{\kern\digitwidth}
  \newdimen\signwidth
  \setbox0=\hbox{+}
  \signwidth=\wd0
  \catcode`!=\active
  \def!{\kern\signwidth}
\halign{\hbox to 2.0cm{#\leaderfil}\tabskip 2em&
    \hfil$#$\hfil\tabskip 1.5em&
    \hfil$#$\hfil&
    \hfil#\hfil\tabskip 0em\cr
\noalign{\doubleline}
\omit\hfil Frequency\hfil&\omit\hfil Min\hfil&\omit\hfil Max\hfil&rms\cr
\noalign{\vskip 3pt} 
\omit\hfil [GHz]\hfil&[\%]&[\%]&[\%]\cr
\noalign{\vskip 3pt\hrule\vskip 5pt}
\noalign{\vskip 2pt}
100&-0.008&!0.010&0.003\cr
143&-0.009&-0.001&0.003\cr
217&-0.010&!0.005&0.004\cr
353&-0.078&!0.030&0.010\cr
\noalign{\vskip 3pt\hrule\vskip 5pt}}}
\endPlancktable
\end{table}

\begin{figure}[!htbp]
\includegraphics[width=\columnwidth]{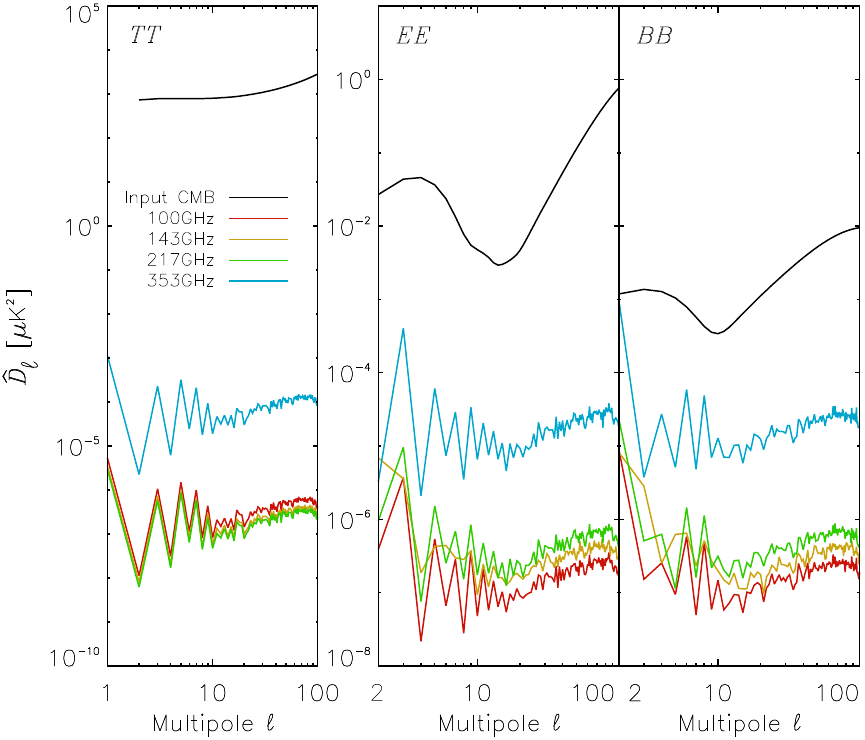} 
\caption{Auto-power spectra induced by the residual calibration mismatch in the maps. The residuals in the CMB channels are very low compared to the F-$EE$ signal.}
\label{fig:DIFF_GL}
\end{figure}

The difference between the mean of the input variable gain and the mean of the variable gain solved for by {\tt SRoll}, gives an estimate of the calibration mismatch in the final maps. Table~\ref{tab:calibmismatch} shows the minimum, maximum, and rms of the detector calibration in each frequency band. Figure~\ref{fig:DIFF_GL} shows the propagation to the power spectra of this calibration mismatch. For CMB channels, the levels are more than four orders of magnitude below the F-$EE$ spectrum.

\subsection{Summary of systematic effects}

Figure~\ref{fig:DIFFMERGE} shows the power spectra of all systematic effects, together with the $EE$ spectrum of the fiducial CMB model. The noise dominates over all systematic effects at $\ell > 5$. Bandpass and calibration mismatch are a factor of 100 below the signal. The effect of the dipole distortions induced by the ADC NL is comparable to the noise at multipoles less than 5 in the CMB channels, and at the level of the fiducial signal at 217\,\GHz. At 353\,\GHz, the ADC-induced apparent gain variations are only a factor of a few below the noise at very low multipoles. The bandpass leakage is only a small factor lower. The noise itself is low compared to the dust signal at high latitude (more than two orders of magnitude). It is therefore far below the fiducial signal when scaled with the appropriate coefficients to clean the 100 and 143\,\GHz maps. In the CMB channels, the dipole distortion (not removed from the data) is significant and particularly strong at very low multipoles. At 353\,\GHz the systematic effects are negligible if this channel is only used to clean dust emission from the channels used for measuring $\tau$.


\section{LFI low-$\boldsymbol\ell$ polarization characterization}
\label{annex_LFI}

Here we provide a summary of the Low Frequency Instrument (LFI) systematic
effects that are most relevant for the polarization analysis at low multipoles. 
A detailed discussion of all systematic effects for the LFI 2015 release 
is given in \cite{planck2014-a04}. 



We have developed  time-dependent models for the known systematic effects in
LFI.  For each of these effects, we generate a timeline and project it into
the map domain using the same pipeline used for the data, including the
\Planck\ scanning strategy and LFI mapmaking process
We call the resulting maps ``systematics templates'' and quantify the
impact of these systematic effects by comparing their power spectra to the FFP8 simulations \citep[which reproduce the measured LFI 
noise spectra][]{planck2014-a14}, and to the recovered sky signal.

Figure~\ref{fig:lowell_spectrum_all_effects} 
summarize the results of our analysis for $\ell < 45$ (resolution
approximately 4\deg). For $EE$ polarization, the most relevant effects come
from uncertainties in the radiometer gains. ADC nonlinearity and sidelobe
residuals (at 30\,GHz) contribute at a lower level.  In the rest of this
appendix we discuss each of these three sources of systematic effects.

 
\subsection{Gain reconstruction}

The gain of a balanced differential coherent receiver may vary at the
$ 1\,\%$ level on a variety of timescales, depending primarily on changes
in the thermal environment or the bias. For each LFI radiometer, we recover
the gain changes as a function of time using the dipole signal produced
by the Doppler shift from the motion of the Solar system relative to the
CMB \citep{planck2014-a06}.  Imperfect recovery of the true radiometer gain
variation produces a systematic effect. The gain solutions are
least accurate when the dipole signal is near its minimum, typically a period
of several days, which produces effects at large angular scales. 
At multipoles $\ell<30$ for Stokes $Q$ and $U$ the rms effect is
$0.4\,\mathrm{\mu K}$ at 30\,GHz, $0.3\,\mathrm{\mu K}$
at 44\,GHz, and $0.15\,\mathrm{\mu K}$ at 70\,GHz.
These are the dominant systematic errors in the present analysis.

\subsection{ADC correction}

Deviations from linearity in ADC response impact the LFI differently than the
HFI.  Linearity in the ADC requires that the voltage step size between
successive binary outputs
is constant over the entire signal dynamic range. Deviations from
this ideal behaviour affect the calibration solution. Because the
effect induces a variation of the detector white noise that is not
matched by a corresponding variation in the absolute voltage level
(which would be expected from LFI radiometers), the anomaly can be measured
and corrected. We evaluate the residual error through simulations.
Since the radiometer gain varies coherently over periods of
days to weeks, the average output of the radiometer (and therefore the part 
of the ADC range being used) also varies on this timescale.  Imperfect
removal of ADC effects thus produces residuals
at low multipoles.  Additionally, since the effect is independent for each ADC
channel, it is not mitigated by differencing orthogonally polarized
radiometers. As a consequence, the amplitude of the residuals
are comparable for temperature and polarization, implying of course a
larger scientific impact on polarization. For the $Q$ and $U$ components at
large scales, the rms residual effect is $<0.1\,\mathrm{\mu K}$ at 30 and
44\,GHz and approximately $0.15\,\mathrm{\mu K}$ at 70\,GHz.

\subsection{Impact of sidelobes}

Telescope sidelobe response can be relevant at low multipoles, both
directly and through the calibration process; this is particularly the case
at 30\,GHz.  To deal with this, we remove a model of the sidelobe signatures
(``stray light'') in the LFI timelines by combining the
{\tt GRASP} models of the beams with the measured sky maps at each frequency
\citep{planck2014-a05}. To test the impact of imperfectly modelled sidelobes on
the calibration we ran our calibration simulation pipeline in a mode where
the stray light signal was added in the input, but not removed. This
provides a worst-case limit (100\,\% error) on the impact of the stray light
on calibration. As expected, at 44 and 70\,GHz the effects are small, 
while at 30\,GHz we find a significant effect of approximately
$1\,\mathrm{\mu K}$. While this contamination is large compared 
to the expected CMB $EE$ signal, it is clearly unimportant compared 
to the synchrotron component that the 30\,GHz channel is used to measure,
resulting in only a small effect in the 70\,GHz map. 

\section{SimBaL}
\label{simbal}

\def\lmin{{\ell_\mathrm{min}}}
\def\lmax{{\ell_\mathrm{max}}}

\def\omegaref{{\mathrm{\Omega}^\star}}
\def\omegafid{{\mathrm{\Omega}^\mathrm{fid}}}

\subsection{Description of the method}
\label{sec:simbaldescription}
Within the $\Lambda$CDM paradigm, the statistics of low multipole polarization
anisotropies depend principally on $A_{\rm s}$, $r$, and $\tau$. 
The high-$\ell$ analysis of the $TT$ power spectrum yields a highly accurate 
constraint on $A_{\rm s} e ^{-2\tau}$  (0.5\,\% ) and one may
assume negligible primordial tensors. The degeneracy between these two
parameters can then be broken by an accurate measurement of the $EE$
low-$\ell$ polarization feature, since this depends roughly on
$A_{\rm s} \tau^2$, which is linearly independent of the
high-$\ell$ constraint.  We thus develop a ``simulation-based likelihood''
code, named {\tt SimBaL}, which is focused on $\tau$ estimation, and based on
$EE$ (pseudo-) cross-power spectra $\widehat C_\ell$, using either
PCL or QML estimators.

Given our current understanding of systematic errors, in {\tt SimBaL} we
typically use a single QML cross-spectrum $\widehat C_\ell$, constructed from a
pair of maps. The \Planck\ maps, at 100 and 143\,GHz, are each ILC-cleaned
using 30 and 353\,GHz maps as foreground tracers, as described in
Sect.~\ref{sec:compsep}.
The low dimensionality of the problem renders the likelihood analysis amenable
to a scheme based on computing
$P(\tau\,|\,\widehat C_{\ell}^{\rm data},\Omega)$, where $\Omega$ stands for
the other cosmological and non-cosmological parameters. We formally have some
freedom in the choice of $\Omega$, but typically use best-fit values from the
\Planck\ cosmological parameter analysis \citep{planck2014-a15}.
More explicitly, we follow these
steps.
\begin{enumerate}
\item Compute the power spectra of simulations $\widehat C_{\ell}^{\rm sim}$
for various $\tau$ values, including noise and systematic effects.
\item Compute $P\left(\widehat C_{\ell}^{\rm sim}\mid\ell,\tau,\Omega\right)$
fitting $\ell$-by-$\ell$ a model to the simulated power spectra. This model is
needed in order to reduce noise and to extrapolate beyond the measured range,
given by the simulations (see Sect.~\ref{sec:simbalmodel} for more details).
Thus, {\tt SimBaL} (as is true for all likelihoods) needs an accurate model of
the $P(\widehat C_{\ell}\,|\,\ell,\tau,\Omega)$ distribution at a given
multipole. The {\tt SimBaL} model uses:
\begin{itemize}
\item $\ln \left(P(\widehat C_{\ell}^{\rm sim}\mid\ell,\tau,\Omega)\right)
=A(\ell)$ (where $A$ is a 3rd-order polynomial) for the central part of the
fit;
\item $\ln \left(P(\widehat C_{\ell}^{\rm sim}\mid\ell,\tau,\Omega)\right)
=B(\ell)$ (where $B$ is a 1st-order polynomial) for both tails of the fit.
\end{itemize}
\item Compute the $\hat t$ estimator, which it is closely related to $\tau$,
but easy to compute on pseudo-$C_{\ell}$ spectra:
\begin{equation}\label{eq:tp:T}
\hat t = \arg\max_\tau \sum_{\lmin}^{\lmax}
 \ln \left( P(\widehat C_\ell^{\rm data}\mid\ell,\tau,\,\Omega)\right).
\end{equation}
\item Use simulations to determine $P(\hat t\mid\tau,\Omega)$.
\item Finally deduce $P( \tau\mid\hat t,\Omega)$ from
$P(\hat t\mid\tau,\Omega)$.
\end{enumerate}

The simulation set $\widehat C_{\ell}^{\rm sim}$ is built from HFPS1 (or HFPS2)
for the noise and systematic effects, to which we add for each 100 CMB realizations.
We then use the simulations described above to construct the sampling
distributions $P( \widehat C_{\ell}^{\rm sim}\mid\ell,\tau,\Omega)$,
for a grid of $\tau$ values from $0.010$ to $0.180$, in steps of $0.001$. 
The relatively low number of noise simulations leads to some apparent features
in the distributions, particularly at low $\tau$ where the CMB contribution is
smaller; however, for a given number of simulations, this would affect the result only a range of lowest $\tau$ values.
Considering the number of simulations we have this does not affect our results for the range of $\tau$ values considered in Sect.~\ref{sec:analysis}.

\begin{figure}[!ht]
\includegraphics[width=\columnwidth]{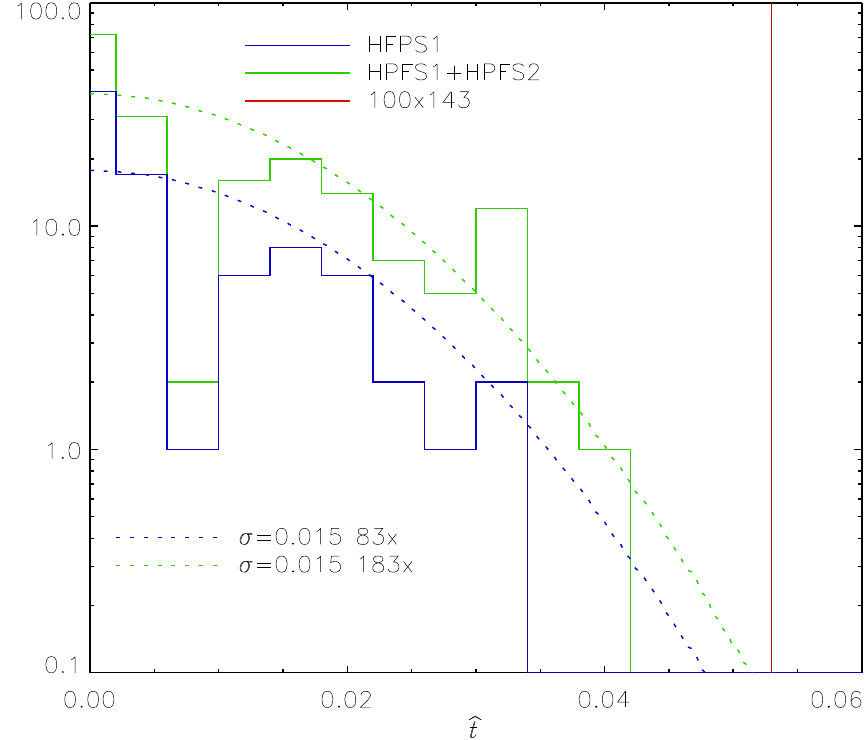}
\caption{Distribution of the {\tt SimBaL} $\hat t$ estimator from signal-free
HPFS1 and
(HFPS1+HPFS3) simulations. These two distributions (blue and green dashed
lines) are consistent with a Gaussian width $\sigma=0.015$ for the relevant numbers of simulations. The $\hat t$ value from the data (in red) is incompatible with these simulations at the $3.5\,\sigma$ level.}
\label{fig:pltestnsim}
\end{figure}

Figure~\ref{fig:pltestnsim} shows the $\hat t$ found with the data (red vertical
line) compared to the $\hat t$ distribution computed on simulations with noise and systematic effects, but without signal (green and blue curves).
Despite the limited number of simulations available, the red vertical line is
clearly well outside the histogram of signal-free simulations; this excludes
to have $\tau=0$ with a significant probability.
In order to quantify the significance of this result, we perform a
Gaussian fit to the histograms (dashed lines), showing that the $\hat t$ value
from the data is incompatible with these simulations at approximately
the $3.5\,\sigma$ level.

\subsection{Illustration of the method}
\label{sec:simbalillustration}
\begin{figure}[!ht]
\includegraphics[width=\columnwidth]{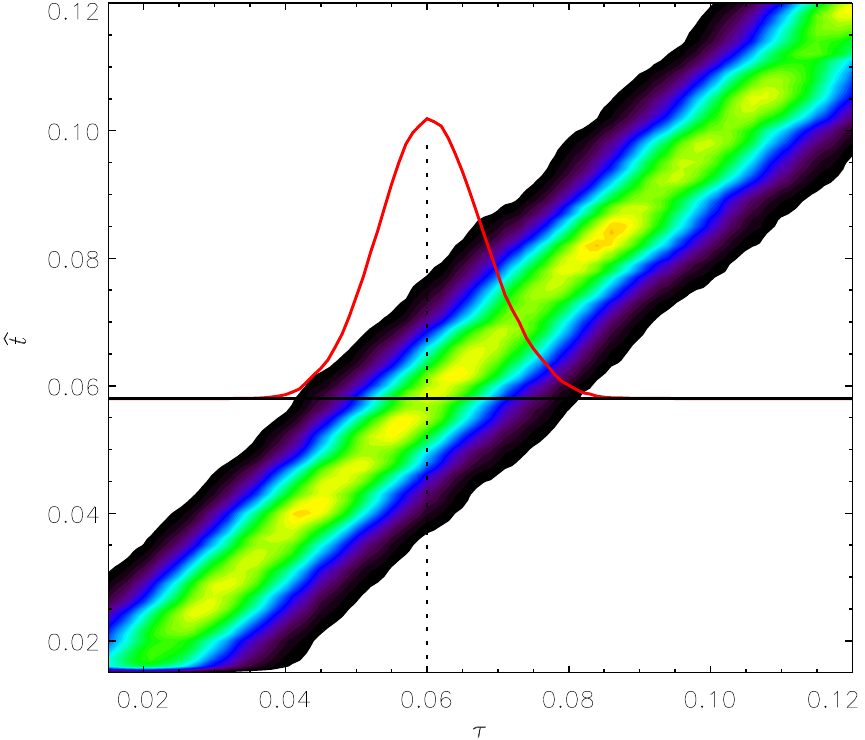} 
\includegraphics[width=\columnwidth]{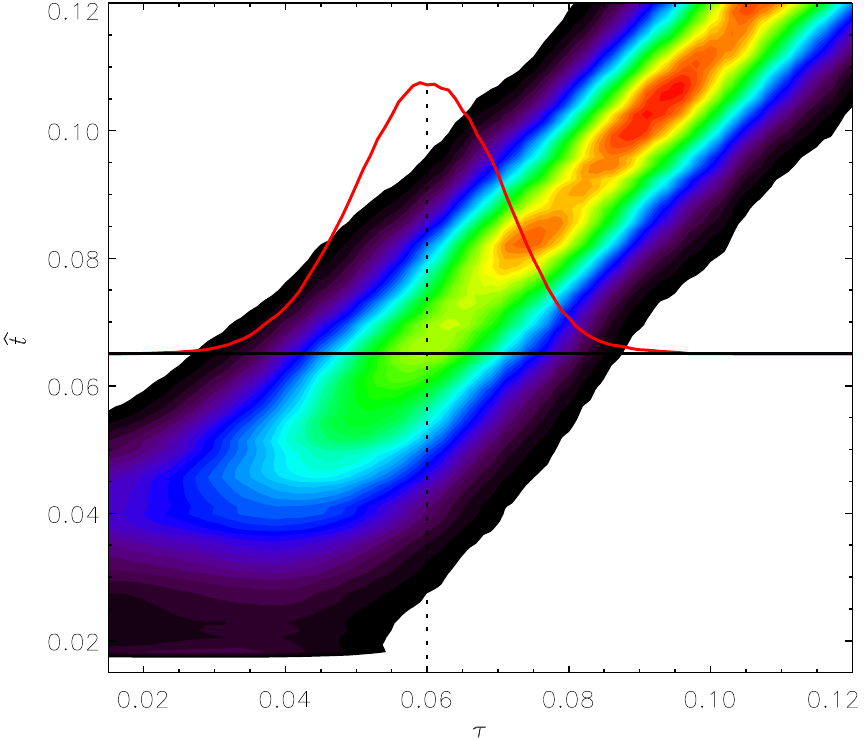} 
\caption{Distribution of the $\tau$ probability density as a function of the
$\hat t$ estimator (defined in the text) derived from the $100\times143$
cross-spectra. The top panel uses simulations containing only white noise,
while the bottom panel is for simulations with the systematic effects residuals
added. As an example of a cut through this two-dimensional plot, in the top
panel an observed value of $\hat t= 0.057$ (horizontal black line) gives a
probability distribution (red curve) that peaks at $\tau=0.060$. In the bottom
panel, the same peak value of the probability distribution is obtained with an
observed $\hat t= 0.065$.}
\label{fig:simbalsroll_rd12_distri_nobias}
\end{figure}

Figure~\ref{fig:simbalsroll_rd12_distri_nobias} shows an illustration of the
method for the $100\times143$ cross-spectra. The top panel shows, for white
noise only, the joint probability distribution in the $(\hat t, \tau)$ plane,
assuming a uniform prior on $\tau$. For a given value of $\hat t$ the red curve
shows the probability distribution of $\tau$, which peaks at a value close to the value of 
$\hat t$ by construction of the definition of $\hat t$.
The bottom panel is built using the HFPS1 simulations (with systematic effects)
and 100 CMB simulations. We note that in this case, for low values of $\tau$,
the maxima of the probability distributions shifts slightly from the diagonal,
because of the inclusion of the systematic effects. In the top panel the value
$\hat t=0.057$ gives a probability distribution that peaks at $\tau =0.06$.
When using the simulations that include systematic effects (bottom panel), the
value $\hat t=0.065$ gives a peak value of $\tau =0.06$. When a
data power spectrum gives this value of $\hat t=0.065$ we check on simulations
 that not including the systematic effects would produce a positive bias of about 0.008 in $\tau$.
When we include the systematic effects in the
simulations, we show our method is able to recover unbiased estimates of $\tau$ with an accuracy
small compare to the noise.

\subsection{Noise variance}
\label{sec:simbalnoise}
\begin{figure}[!ht]
\includegraphics[width=\columnwidth]{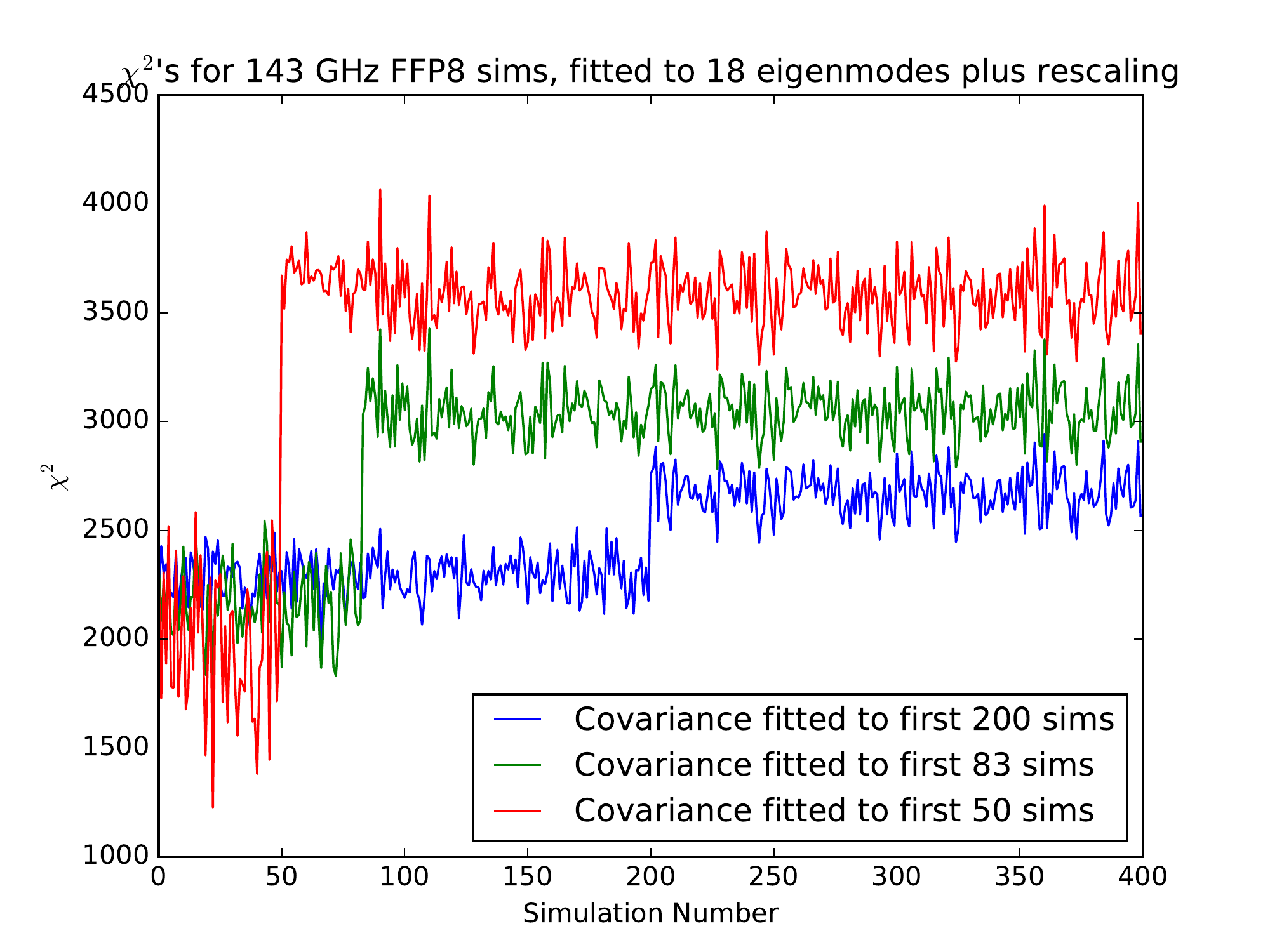} 
\caption{Pixel-pixel matrix $\chi^2$ values of the first 400 of the 143\,GHz
FFP8 simulations, evaluated against 18-arbitrary-eigenmode fits to either the
first 50, 83, or 200 simulations. The $\chi^2$ of the simulation data set used
to compute the pixel-pixel covariance matrix has been normalized here.
Due to the limited number of simulations, the real variance, estimated from the
simulations that were {\it not\/} used to compute the pixel-pixel matrix, is
significantly higher (i.e., the red line jumps at about 50, the green line at
about 83, and the blue line at about 200).}
\label{fig:plsim}
\end{figure}

Figure~\ref{fig:plsim} shows the $\chi^2$ values of low resolution maps from
FFP8 noise simulations \citep[see][]{planck2014-a14} evaluated with
covariance matrices built with an increasing fraction of the simulations. This illustrates the difficulty
induced by having only a limited number of simulations with which we estimate both a
low-resolution noise covariance matrix and then evaluate properties of
statistical distributions involving this covariance matrix. Each covariance
matrix is computed from a subset of the available simulations, rescaling a
fiducial covariance matrix and then adding extra modes to capture the additional
systematic effects. It can be seen that the simulations that are {\it not\/} used to build
the covariance matrix have a higher $\chi^2$ than those used to build the
matrix.  Hence using the same simulations twice gives a misleadingly low
indication of the scatter (note,  that increasing the number of
simulations used to fit the noise matrix decreases the $\chi^2$ discrepancy,
as one would expect). Thus, to take into account this feature in the
likelihood, {\tt SimBaL} uses two different simulation sets, one to compute the
pixel-pixel matrix and the other to measure the noise variance. 

\subsection{Dependence on the model}
\label{sec:simbalmodel}
\begin{figure}[!ht]
\includegraphics[width=\columnwidth]{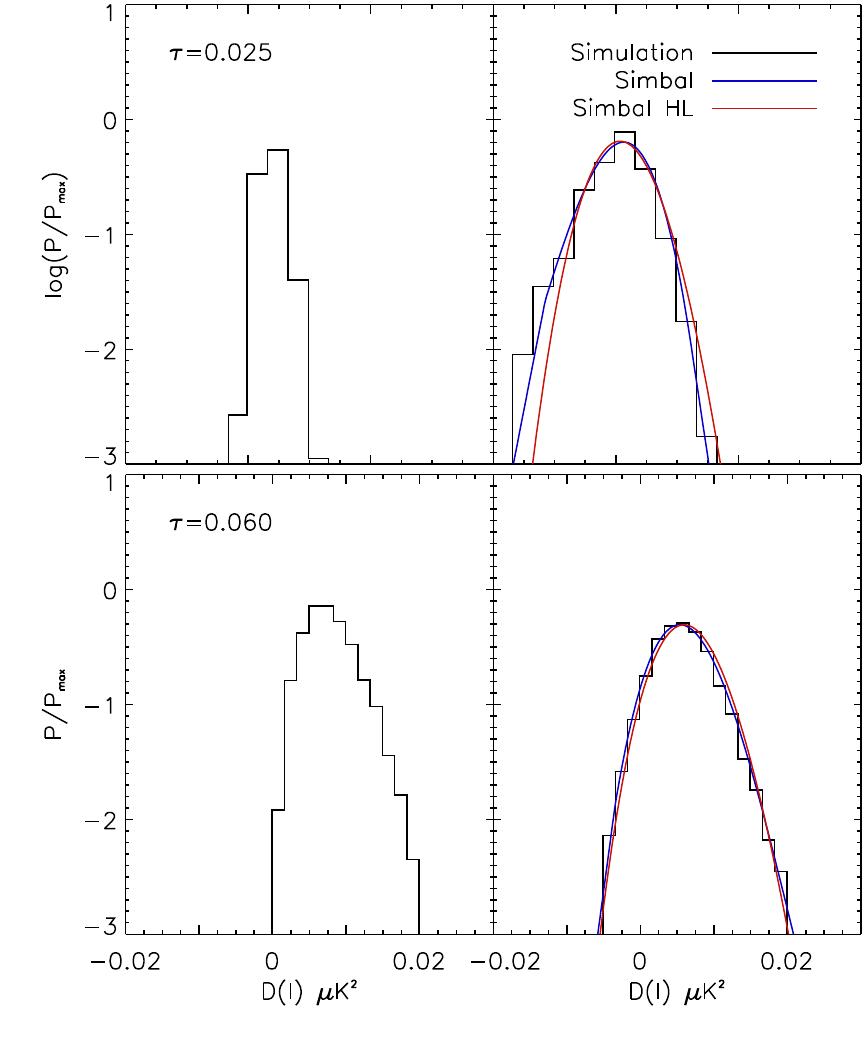}
\caption{Probability slice at $\ell=4$ for two different $\tau$ values,
$\tau=0.025$ in the top panels and $\tau=0.06$ on the bottom. The left panels
show the signal-only distribution, while the right ones show the signal plus
noise and systematic effects. For small $\tau$ values, the distribution is
dominated by noise and systematic effects and is poorly fitted by the
Hamimeche-Lewis-type model (red line), but instead is well captured by the
{\tt SimBaL} polynomial approximation (blue curve).}
\label{fig:sliceprob}
\end{figure}

\begin{figure}[!ht]
\includegraphics[width=\columnwidth]{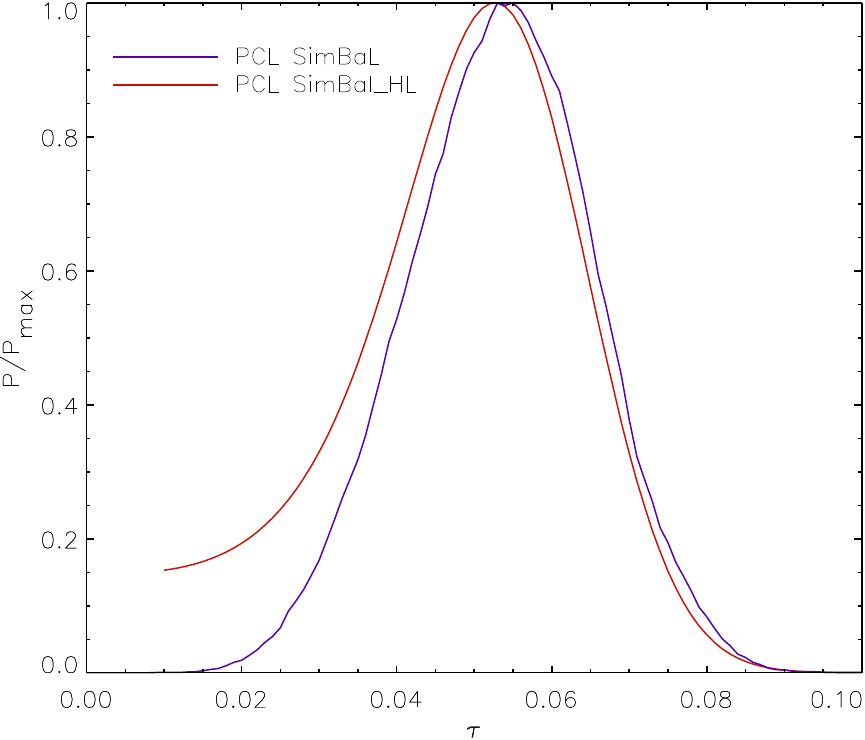} 
\caption{Posterior distributions of $\tau$ from {\tt SimBaL} (blue curve) and
{\tt SimBaL}\_HL (red curve), showing how they differ at low $\tau$. The red
line is an attempt to mimic in {\tt SimBaL} the modified ``HL'' approach and
demonstrates that the divergence is due to residual non-Gaussian behaviour of
the statistics at low multipoles (see text).}
\label{fig:plhmvssimbal}
\end{figure}

{\tt SimBaL} uses simulations of noise and systematic effects, to which added
100 CMB realizations, in order to model the statistical distribution of
$\ln\left( P(\widehat C_\ell^{\rm data}\mid\ell,\tau,\,\Omega) \right)$ for each $C_{\ell}$.  We fit this
distribution for each multipole using an asymmetrical polynomial function
detailed in Sect.~\ref{sec:simbaldescription}.  Figure~\ref{fig:sliceprob}
shows how the {\tt SimBaL} polynomial approximation is efficient in capturing
the shape of the probability distribution for low $\tau$ values when
systematic noise effects dominate, while a {\tt SimBaL}\_HL approximation
\citep[based on a Hamimeche \& Lewis model,][]{2008PhRvD..77j3013H} poorly
fits the probability distribution tails for low values of $\tau$. Figure~\ref{fig:plhmvssimbal} shows
the effect of these two models on the $\tau$ posterior computation.
{\tt SimBaL}\_HL modelling overestimates the probability of small $\tau$
values, where the distribution is dominated by noise and systematic effects.

\subsection{Dependence on masks}
\label{sec:simbalmasks}
\begin{figure}[!ht]
\includegraphics[width=\columnwidth]{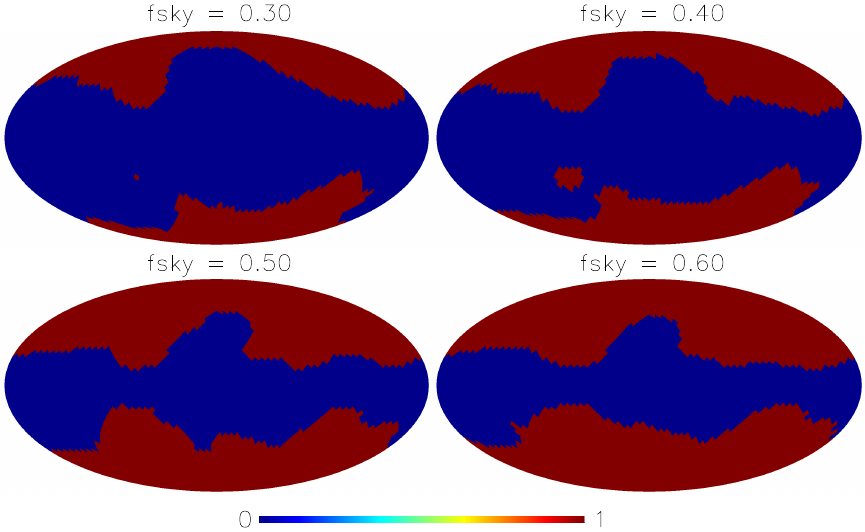} 
\caption{Masks for different values of $f_\mathrm{sky}$. The mask shown in
the bottom-left panel (50\,\% sky fraction) is the one used in the
{\tt Lollipop} and the {\tt SimBaL} analyses. The other three masks are used
to evaluate the dependence of the results on changing sky fractions from 30
to 60\,\%.}
\label{fig:tp:mask}
\end{figure}

We also explore the dependence on the masks used, by using several masks shown in
Fig.~\ref{fig:tp:mask}.
To this end, power spectra are computed using different masks. These are built
by thresholding the intensity of the Galactic polarized emission (the
50\,\% sky fraction mask is used for the $\tau$ determination in all
likelihood methods). 

\begin{figure}[!ht]
\includegraphics[width=\columnwidth]{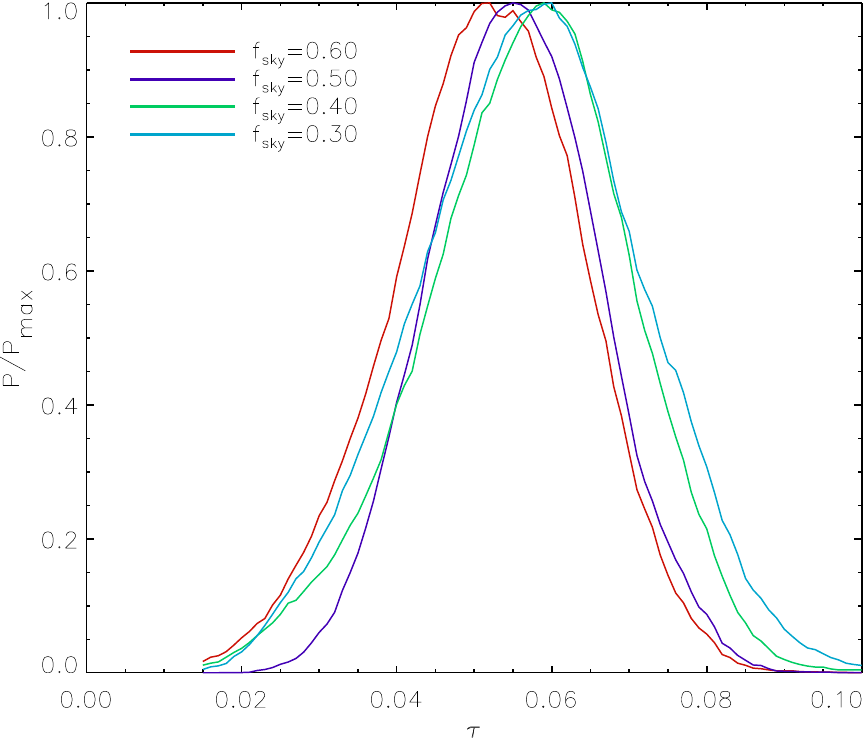} 
\caption{Posterior distributions from {\tt SimBaL} PCL $100\times143$
cross-power spectra obtained with different sky fractions, showing that
variations caused by masking the Galaxy have limited impact on the $\tau$
posterior likelihood for a range of mask sizes.}
\label{fig:pltestmask}
\end{figure}

We illustrate the effects of the sky masking on PCL $\tau$ posteriors for the
data in Fig.~\ref{fig:pltestmask}.  This shows that the effect of varying the
retained sky fraction is small. In the main likelihood analysis, the mask used
is the one with $f_{\rm sky}=0.50$, which is in fact the sky fraction that
yields the best accuracy  (smaller posterior width).
 Even for the more extreme sky fraction used there is a small bias (of around $0.4\,\sigma$)
towards lower peak values for the $\tau$ probability distribution.
The generally very good consistency between these curves around the used $f_{\rm sky}=0.50$
 shows that the component-separation process is not a
limitation in the determination of $\tau$ in this paper.

\subsection{Dependence on multipole range}
\label{sec:simbalmultipole}
\begin{figure}[!ht]
\includegraphics[width=\columnwidth]{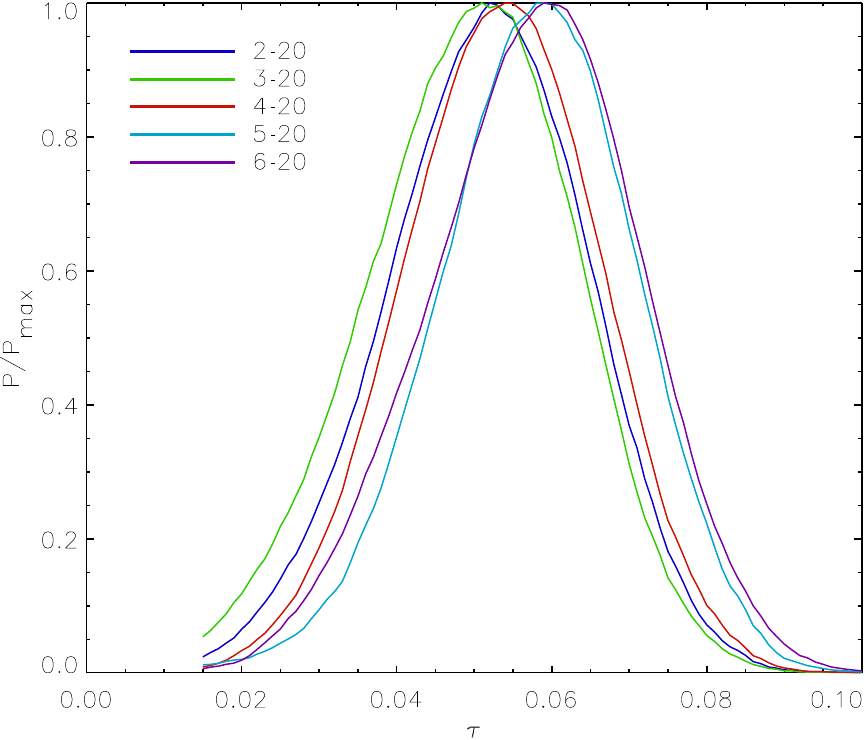} 
\caption{Posteriors on $\tau$ when using different ranges of multipole in the
$100\times143$ PCL cross-power spectrum. The $\ell$ ranges are plotted in
different colours, as indicated in the legend of the figure.}
\label{fig:pltestlmin}
\end{figure}

Figure~\ref{fig:pltestlmin} shows the effect of changing the range of
multipoles in the $100 \times 143$ PCL cross-power spectrum. As expected, using
a minimum $\ell$ of 5 or 6 starts to bias the result, because of the low
level of the reionization feature above these multipoles values. Although we know that the
ADC NL effects on the dipole are concentrated at $\ell=2$ and $3$ before removal,
 including these multipole range does not affect the result substantially. This is an important result as , it confirms that this systematic effect residuals in the cross power spectra are small as discussed in Sect.~\ref{sec:crossspectra} and accurately simulated in the HFPS and
accounted for when used in the {\tt SimBaL} likelihood.

\begin{figure}[!ht]
\includegraphics[width=\columnwidth]{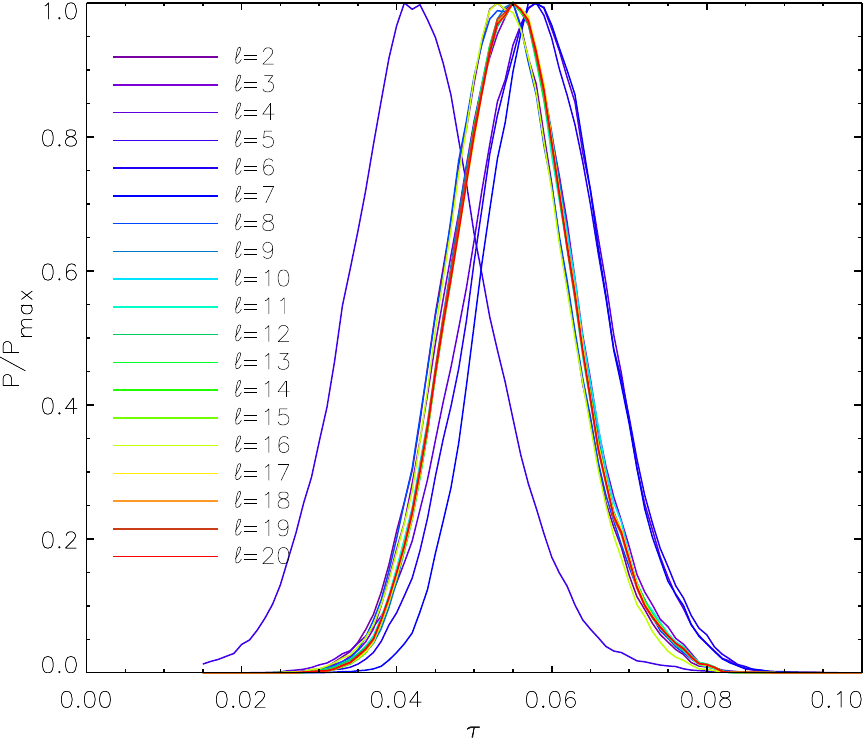} 
\caption{Posteriors in $\tau$ when removing one multipole at a time in the
$100\times143$ QML cross-power spectra.}
\label{fig:pltestell}
\end{figure}

\begin{figure}[!ht]
\includegraphics[width=\columnwidth]{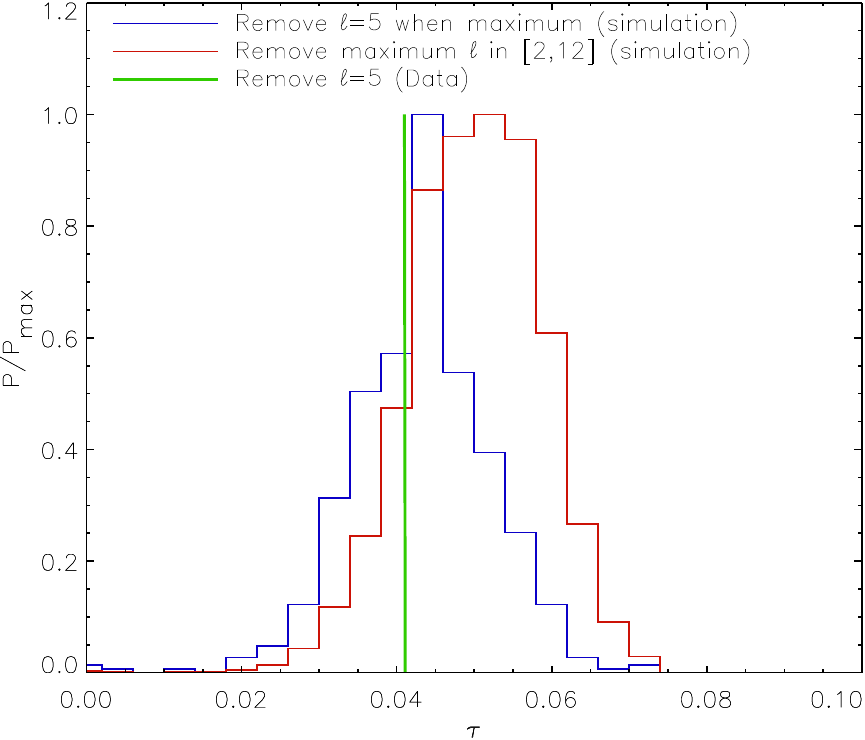} 
\caption{Histograms (computed from 8300 simulations with $\tau=0.055$) of the
peak values of the $\tau$ posterior distributions after removing one
multipole each time. For the red curve the maximum value of the $100\times143$
QML power spectrum in the range $\ell=2$--12 is removed, and for the blue curve
$\ell=5$ is removed when it happens to be the maximum. The peak value of the
$\tau$ posterior when removing $\ell=5$ from the data is shown in green and is
consistent with the blue curve, as expected.}
\label{fig:pll5}
\end{figure}

Figure~\ref{fig:pltestell} shows the effect of removing one multipole at a time
from the QML results.  Except for the case of removing $\ell=5$, which is the
multipole where the $EE$ cross-power spectrum is maximum and the probability to exceed near $2~\sigma$ (see Fig.~\ref{fig:100143}), we obtain very stable results.
Figure~\ref{fig:pll5} shows how removing one multipole  from the $\tau$ posterior computation
where the power spectrum is  maximum does bias the result low as expected;
however, the $\tau$ value obtained when removing $\ell=5$ is statistically
consistent with what is expected from simulations.

{\it As a consequence of this discussion, the likelihood analysis in this work
uses the multipole range 2--20.}
 
\subsection{{\tt SimLow} a low-$\ell$ likelihood based on {\tt SimBaL}}
\label{sec:simlow}
{\tt SimBaL} provides a $\tau$ posterior using the other parameters based on the best cosmology
from $TT$, $TE$, and $EE$ power spectra at higher multipoles. Thus,
{\tt SimBaL} is not usable as a low-$\ell$ likelihood for cosmological
parameters that are not the ones used to build {\tt SimBaL}.  Because of this,
we have developed {\tt SimLow} to estimate a set of
$P(C_{\ell}^{\rm fid}\,|\,\widehat C_{\ell}^{\rm data})$ based on the
{\tt SimBaL} paradigm, by computing
$P(\widehat C_{\ell}^{\rm sim}\,|\, C_{\ell}^{\rm fid})$ using simulations.
The QML evaluation of the power spectra minimizes correlations between different $\ell$s, and
{\tt SimLow} computes the likelihood independently between multipoles.

\begin{figure}[!ht]
\includegraphics[width=\columnwidth]{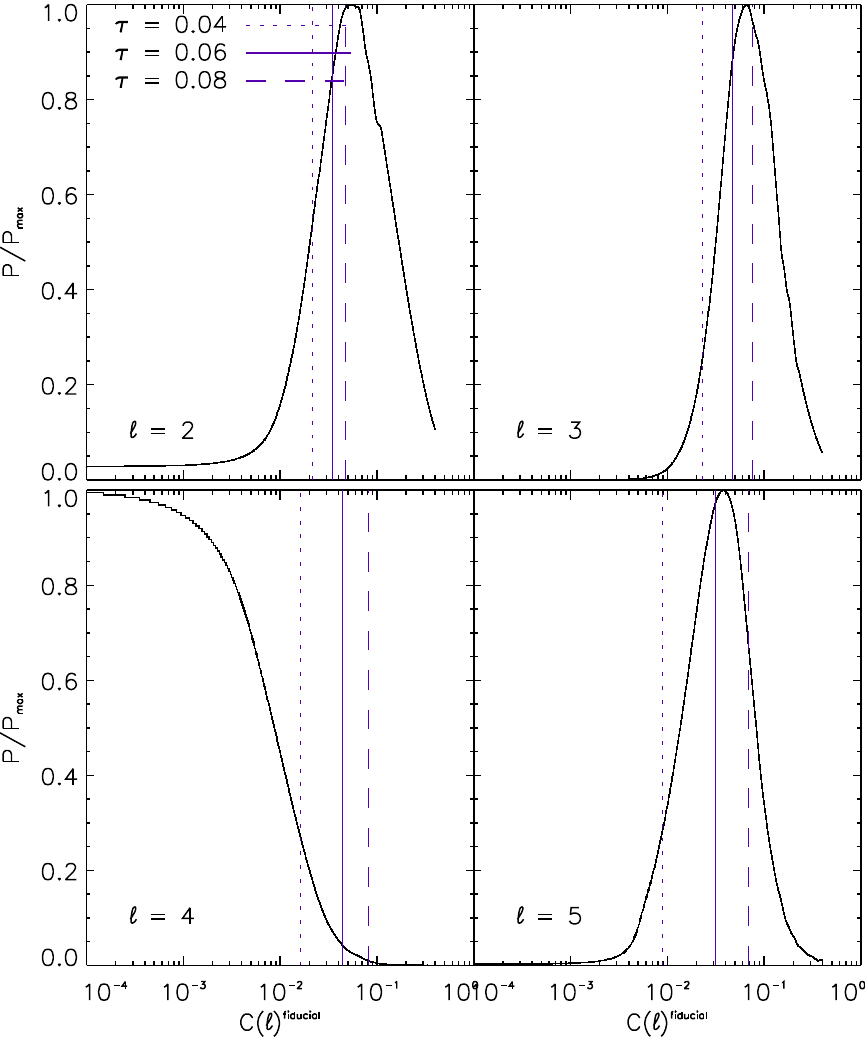} 
\caption{$P(C_{\ell}^{\rm fid}\mid\widehat C_{\ell}^{\rm data})$ computed for the data using
{\tt SimLow}.  Power spectrum values computed for different $\tau$ are shown in
blue.  For $\ell=3$ and $\ell=5$ we can see that very small values of $\tau$
are excluded, while for $\ell=4$ large values of $\tau$ have a low
probability.}
\label{fig:simlow}
\end{figure}

In order to measure $P(\widehat C_{\ell}^{\rm sim}\mid C_{\ell}^{\rm fid})$
for the $100\times143$ power spectrum, we compute 100 realizations of each
$C_{\ell}^{\rm fid}$, with 3000 steps of $10^{-4}\mu$K$^2$ added to HFPS1 and
HFPS3 simulations using the QML2 estimator for the power spectra.
Figure~\ref{fig:simlow} shows the results of running {\tt SimLow} for the
first few multipoles of the data.

\begin{figure}[!ht]
\includegraphics[width=\columnwidth]{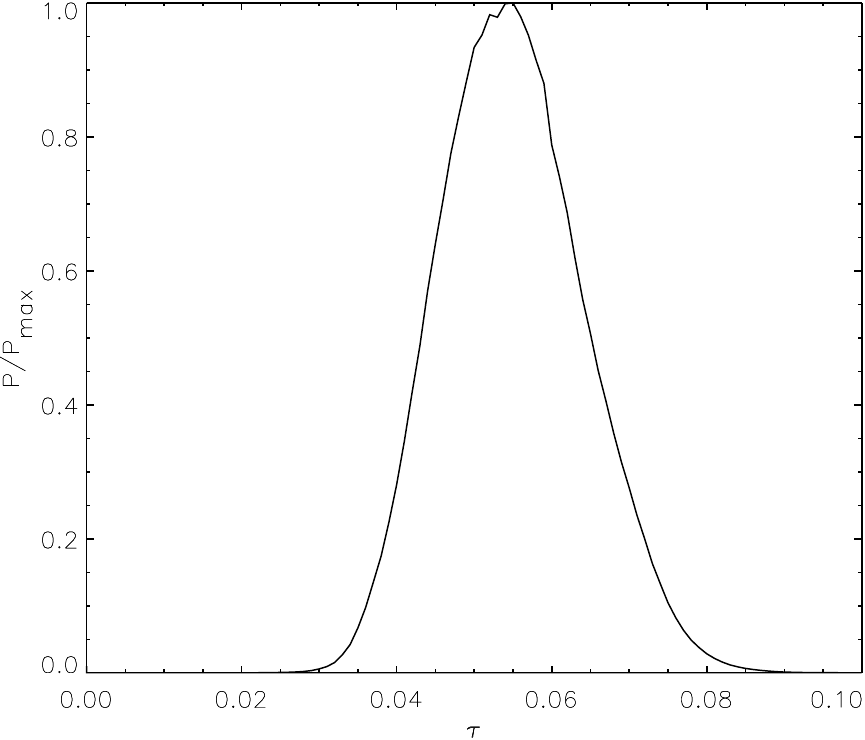} 
\caption{Posterior distribution for $\tau$ computed with the {\tt SimLow}
likelihood using the same cosmological parameters as for {\tt SimBaL}.
The posterior is consistent with the {\tt LowEH} result.}
\label{fig:simlowpost}
\end{figure}

{\tt SimLow} uses posteriors for multipoles in the range
$\ell=2$--20. Using {\tt SimLow} to compute the  $\tau$ posterior with  the same cosmological parameters as {\tt SimBaL} shown in  Fig.~\ref{fig:simlowpost} gives $\tau = 0.055\pm0.009$, which is fully consistent with the value from {\tt SimBaL} and referred to as {\tt lowEH}. 

\section{Glossary}
\label{dictionary}

This appendix gathers definitions of acronyms and other terms widely used in this paper, in addition to those more general terms listed in the glossary and acronym list of \citet{planck2014-ES}.
\paragraph{ADC NL:} Analogue-to-digital converter nonlinearities. The HFI bolometer electronics readout includes a 16-bit ADC that has a very loose tolerance on the differential nonlinearity (the maximum deviation from one least significant bit, LSB, between two consecutive levels, over the whole range), specified to be not worse than one LSB. The implications of this feature for HFI performance proved to be a major systematic effect on the flight data. A wide dynamic range at the ADC input was needed to both measure the CMB sky and foregrounds, and properly characterize and remove the tails of glitches from cosmic rays. Operating HFI electronics with the necessary low gains increased the effects of the ADC scale errors on CMB data \citep[see section~2 of][for further details]{planck2014-a09}.
\paragraph{ADU:} analogue-to-digital unit.
\paragraph{Complex transfer function:} an empirical function that captures residuals of the bolometer/electronics time response deconvolution (including VLTC) and residuals from far sidelobe effects.
\paragraph{Detset or ds:} ``detector set,'' i.e., a combination of sets of polarization-sensitive bolometer pairs with both orientations. Specifically, 100\,GHz ds1 combines 100-1a/b and 100-4a/b, 100\,GHz ds2 combines 100-2a/b and 100-3a/b, 143\,GHz ds1 combines 143-1a/b and 143-3a/b, and 143\,GHz ds2 combines 143-2a/b and 143-4a/b.
\paragraph{Distorted dipole:} the difference between the actual dipole signal (which is affected by the ADC NL, like {\it all\/} signals) and the sine wave that would have been measured without the nonlinearity.
\paragraph{F-TT, F-EE, and F-BB models:} CMB fiducial power spectra, based on best-fit \Planck\ cosmological parameters \citep{planck2014-a15}, with $\tau=0.066$ and $r=0.11$.
\paragraph{FSL:} far sidelobe effects, i.e., the pickup of the sky signal, dominated by the spillover of radiation around the edges of the secondary and the primary telescope mirrors after being reflected by the secondary mirrors and main baffle (with the usual convention following the light from the detectors outwards).
\paragraph{HFPS:} HFI focal plane simulations, built with the pre-2016 E2E software pipeline. They contain realizations of the noise described in Sect.~\ref{sec:annexnoise}, with the systematics dominated by the additional ADC NL model described in Sect.~\ref{sec:perfadc}. HFPS1 contains 83 realizations, while HFPS2 and HFPS3 each contain 100 realizations.
\paragraph{HPR:} {\tt HEALPix} ring, i.e., a partial map produced from projection onto the sky pixelization of a single pointing period.
\paragraph{PCL:} Pseudo-$C_{\ell}$ estimator, a specific method for deriving a cross-spectrum.
\paragraph{Pre-2016 E2E:} end-to-end simulation pipeline, based on the 2015 E2E simulation pipeline, where the ADC NL correction in the mapmaking process has been added to the TOI. The 4-K line and convolution effects are not simulated, because that could affect the higher multipoles.
\paragraph{QML:} quadratic maximum likelihood estimator method, specific method used to derive a cross-spectra. The pixel-pixel covariance matrices (QML1 and QML2) are built using three different QML estimators. For QML1 (or QML2), the pixel-pixel matrix has been computed directly with the HFPS1 (or HFPS2) data set. For QML3, in order to minimize the effect of overfitting, the off-diagonal terms are reduced to the first four eigenmodes of a principal component analysis from HPFS1.
\paragraph{SimBaL1, 2, 3:} three posterior distributions for $\tau$. They differ in the choice of the power spectrum estimator and the data set simulation used. SimBaL1 (or SimBaL2) uses QML1 (or QML2) as power spectrum estimator and HFPS2 (or QML1) as the simulation data set. SimBaL3 uses QML3 as power spectrum estimator and the full set of simulations (HFPS1+HFPS2+HFPS3) as the simulation data set.
\paragraph{SRoll:} a new polarized destriping algorithm, which removes residuals via template fitting, using HPR to compress the time-ordered information.
\paragraph{VLTC:} very long time constant, i.e., bolometer time response longer than 1\,s, which even at low amplitude may bias the calibration by distorting the dipole signal and causing leakage of the Solar dipole into the orbital dipole.

\end{document}